\documentclass[structabstract]{aa}
\usepackage{times}
\usepackage{paralist}
\usepackage{float}
\usepackage{array}
\usepackage{tabularx}
\usepackage{changepage}
\usepackage{color}
\usepackage[latin1]{inputenc}
\usepackage{graphicx}
\usepackage{natbib}
\usepackage{mathabx}
\usepackage{multirow}
\usepackage{txfonts}
\usepackage{bm}
\usepackage{xcolor}  

\def\degr{\hbox{$^\circ$}}
\def\arcmin{\hbox{$^\prime$}}
\def\arcsec{\hbox{$^{\prime\prime}$}}

\def\farcs{\hbox{$.\!\!^{\prime\prime}$}}

\begin{document}
   \title{CHANG-ES XXII:}

   \subtitle{Coherent Magnetic Fields in the Halos of Spiral Galaxies}

   \author{Marita Krause\inst{1}, Judith Irwin\inst{2}, Philip Schmidt\inst{1}, Yelena Stein\inst{3,4}, Arpad Miskolczi\inst{4}, Silvia Carolina Mora-Partiarroyo\inst{1},
   Theresa Wiegert\inst{2}, Rainer Beck\inst{1}, Jeroen M. Stil\inst{5}, George Heald\inst{6}, Jiang-Tao Li\inst{7}, Ancor Damas-Segovia\inst{8}, Carlos J. Vargas\inst{9},
   Richard J. Rand\inst{10}, Jennifer West\inst{11}, Rene A. M. Walterbos\inst{12}, Ralf-J\"urgen Dettmar\inst{4}, Jayanne English\inst{13}
              \and
              Alex Woodfinden\inst{14}
%
%
          }

   \institute{Max-Planck-Institut f\"ur Radioastronomie, Auf dem H\"ugel 69, 53121 Bonn, Germany\\
              \email{mkrause@mpifr-bonn.mpg.de}
          \and
              Dept. of Physics, Engeneering Physics, \& Astronomy, Queen's University,
              Kingston, Ontario, Canada, K7L 3N6\\
              \email{irwin@astro.queensu.ca}
          \and
              Observatoire astronomique de Strasbourg, Universit\'e de Strasbourg, CNRS, UMR 7550, 11 rue de l'Universit\'e,
              F-67000 Strasbourg France
              \email{yelena.stein@astro.unistra.fr}
          \and
               Ruhr Universit\"at Bochum, Fakult\"at f\"ur Physik und Astronomie, Astronomisches Institut, 44780 Bochum, Germany
          \and
               Institute for Space Imaging Science,, and Department of Physocs and Astronomy, University of Calgary, Canada
          \and
               CSIRO Astronomy and Space Science, 26 Dick Perry Avenue, Kensington, WA 6151, Australia
          \and
               Department of Astronomy, University of Michigan, 311 West Hall, 1085 S. University Ave, Ann Arbor, MI, 48109-1107, U.S.A.
          \and
               Instituto de Astrof\'isica de Andaluc\'ia, Glorieta de la Astronom\'ia sn, 18008 Granada, Spain
          \and
               Department of Astronomy and Steward Observatory, University of Arizona, Tucson, AZ, U.S.A.
          \and
               Department of Physics and Astronomy, University of New Mexico, 800 Yale Boulevard, NE, Albuquerque, NM, 87131, U.S.A.
          \and
               Dunlap Institute for Astronomy and Astrophysics, University of Toronto, Toronto, ON M5S 3H4, Canada
          \and
               Department of Astronomy, New Mexico State University, PO Box 30001, MSC 4500, Las Cruces, NM, 88003, U.S.A.
          \and
               Department of Physics and Astronomy, University of Manitoba, Winnipeg, Manitoba, R3T 2N2, Canada
          \and
               Waterloo Centre for Astrophysics, Department of Physics and Astronomy, University of Waterloo, 200 University Ave W,
               Waterloo, ON N2L 3G1, Canada
%
%
%
             }

   \date{Received ; accepted }


 \abstract
   {The magnetic field in spiral galaxies is known to have a large-scale spiral structure along the galactic disk and is observed as X-shaped
   in the halo of some galaxies. While the disk field can be well explained by dynamo action, the 3-dimensional structure of the halo field and
   its physical nature is still unclear.
   }
   {As first steps towards understanding the halo fields, we want to clarify whether the observed X-shaped field is a wide-spread pattern in the
   halos of spiral galaxies and whether these halo fields are just turbulent fields ordered by compression or shear (anisotropic turbulent fields), or
   have a large-scale regular structure.
   }
   {The analysis of the Faraday rotation in the halo is the tool to discern anisotropic turbulent fields from large-scale magnetic fields. This, however,
   has been challenging until recently because of the faint halo emission in linear polarization. Our sensitive VLA broadband observations C-band and L-band
   of 35 spiral galaxies seen edge-on (called CHANG-ES) allowed us to perform RM-synthesis in their halos and to analyze the results. We further accomplished
   a stacking of the observed polarization maps of 28 CHANG-ES galaxies at C-band.
  }
   {Though the stacked edge-on galaxies were of different Hubble types, star formation and interaction activities, the stacked image clearly reveals an
   X-shaped structure of the apparent magnetic field. We detected a large-scale (coherent) halo field in all 16 galaxies that have extended polarized intensity
   in their halos. We detected large-scale field reversals in all of their halos. In six galaxies they are along lines about vertical to the galactic midplane
   (vertical RMTL) with about 2~kpc separation. Only in NGC~3044 and possibly in NGC~3448 we observed vertical giant magnetic ropes (GMRs) similar to those detected
   recently in NGC~4631.
  }
   {The observed X-shaped structure of the halo field in spiral galaxies seems to be an underlying feature of spiral galaxies. It can be regarded as the 2-dimensional
   projection of the regular magnetic field which we found to have scales of typically 1~kpc or larger, observed over several kpc. The ordered magnetic field extends
   far out in the halo and beyond.
   We detected large-scale magnetic field reversals in the halo that may indicate GMRs being more or less tightly wound. With these discoveries, we hope to stimulate
   model simulations for the halo magnetic field that should also explain the determined asymmetry of the polarized intensity.
   }


               \titlerunning{Coherent magnetic fields in galactic halos}

               \authorrunning{M. Krause et al.}
               \maketitle

\section{Introduction}
\label{sec:introduction}

Magnetic fields have been extensively studied in the disks of spiral galaxies \citep{wielebinski+2010, krause2019, beck+2019}. They consist of a
{\em turbulent}
magnetic field component of scales ranging from a few to a few hundred pc and a {\em large-scale} magnetic field that is found to form a spiral pattern parallel
to the midplane of the galaxy. This large-scale, regular field can only be ordered and maintained by a large-scale dynamo action. The $\alpha - \Omega$ mean-field
dynamo theory predicts an axisymmetric spiral structure (ASS) along the galactic plane of the galaxy to be excited most easily \citep{ruzmaikin+1989}. A compression
or shearing of turbulent magnetic fields may lead to an additional magnetic field component (see e.g. \citealt{beck+2019}).

The magnetic fields in spiral galaxies can best be studied in the radio continuum and its linear polarization. All magnetic field components with scales larger than
the beam size of the radio observations and with non-random orientations contribute to the linear polarization of the signal. Hence, the observed linear polarization
usually traces the large-scale magnetic field and compressed or sheared small-scale  fields which together are called {\em ordered} magnetic fields.
The large-scale magnetic field is synonymously named by {\em regular} or {\em coherent} magnetic field.

The linear polarization just provides the orientation, not the direction, of the ordered magnetic field component that is perpendicular to the line of sight
(LoS). Their directions can only be determined by measuring the Rotation Measures (RM) which is proportional to the line-of-sight integral over the thermal electron
density and the strength of the magnetic field component parallel to the LoS: $\rm{RM} \propto \int \, n_e \,  B_\| \, dl $. Different magnetic field directions give
different signs of RM. If the field direction changes within the beam size and/or along the LoS the observed RM is reduced or even canceled out. By this the RM allows
to distinguish observationally between a regular and an anisotropic turbulent field: the detection of significant RM that vary smoothly over several beam sizes is a clear
indicator of a regular (coherent) magnetic field.

Observations of spiral galaxies seen edge-on reveal a plane-parallel magnetic field along the midplane which is the expected projection of the spiral magnetic field
in the disk. In the halo, the ordered magnetic field is found to be X-shaped, sometimes accompanied by strong vertical components above and/or below the central region
\citep{golla+1994,tuellmann+2000,krause+2006,krause2009,heesen+2009b,soida+2011,stein+2019a}. The magnetic field strength in large parts of the halo is comparable to,
or only slightly lower, than the magnetic field strength in the disk \citep{krause2019}. The physical 3-dimensional structure of the halo fields, however, is still
unclear.

The spiral disk field generated by the $\alpha - \Omega$ mean-field dynamo is accompanied by a poloidal magnetic field in the halo. However, its strength is by a factor of about 10 weaker than the disk field strength, hence it cannout explain the observed halo field alone. Also the action of a spherical turbulent dynamo in the halo \citep{sokoloff+1990} cannot explain the high magnetic field strength that is presently observed.

Dynamo action in the disk together with a galactic outflow \citep{brandenburg+1993, elstner+1995, moss+2010} indicated
a field configuration in the halo that looks similar to the observed X-shaped halo fields as mentioned above. Interestingly, the existence of galactic winds has been reported for many of the CHANG-ES galaxies
\citep{krause+2018, miskolczi+2019, stein+2019a, schmidt+2019, mora+2019a}, and extraplanar ionized gas emission can be seen in many H$\alpha$ images taken for the CHANG-ES
sample \citep{vargas+2019}. Alternatively, \cite{henriksen+2018} presented an analytical mean-field dynamo model that assumes self-similariy. Its solutions indicate large-scale helical magnetic spirals that emerge from the disk far into the halo.  The present theoretical understanding of halo magnetic fields and their problems was recently published by \cite{moss+2019}. Specific strategies for improving dynamo models so that they can
be compared with the observations were presented by \cite{beck+2019}.

Before the CHANG-ES survey we even had no reliable RM measurements of the halo to clearly conclude whether the halo contains regular, large-scale fields or just
anisotropic turbulent magnetic fields. Just recently \cite{mora+2019b} detected for the first time a large-scale, smooth RM pattern in the halo of an external galaxy,
NGC~4631, as evidence for a regular magnetic field there. They also dicovered large-scale magnetic field reversals in the northern halo, indicating giant magnetic ropes (GMR)
with alternating directions.

With CHANG-ES (Continuum HAlos in Nearby Galaxies -- an EVLA Survey) we performed an unprecedented deep radio continuum and polarization survey of a
sample of 35 nearby spiral galaxies seen edge-on and observed with the Karl G. Jansky Very Large Array in its commissioning phase \citep{irwin+2012a}.
This sample enables us, for the first
time, to observe halos in radio continuum and polarization in a statistically meaningful sample of nearby spiral galaxies of various Hubbles types, star
formation rates, and nuclear or interaction activities. The galaxies were observed in all Stokes parameters in C-and L-band at B (only L-band), C, and D array
configurations (see \citealt{irwin+2012a} for details). The total and polarized intensity maps at D-array are published in Paper IV \citep{wiegert+2015}
and are available in the CHANG-ES Data Release I \footnote {The CHANG-ES data releases are available at www.queensu.ca/changes}.
The observations at C-array are going to be published by Walterbos et al. (in preparation).

The CHANG-ES polarization results at D-array presented in the data release Paper IV by \cite{wiegert+2015} are made with uniform uv-weighting
(robust=0).
They show the apparent magnetic field orientation that is not corrected for Faraday
rotation. Here in this paper we present the {\it remade} $Q$ and $U$ images for all galaxies using a robust-weighting of $+2$  which considerably
improved the signal-to-noise ratio compared to a robust=0 weighting. We also display our results of Rotation Measure Synthesis (RM-synthesis) \citep{brentjens+2005}
of the broad band observations at C-band of the CHANG-ES galaxies.

We used two completely different approaches to analyze the magnetic fields in the halo of the CHANG-ES spiral galaxies.
The first approach draws on the power of the CHANG-ES survey in polarization by stacking the polarization-related
information from many galaxies in order to test whether there is a common characteristic in the polarization of spiral galaxies. This is presented in Sect.~\ref{sec:stacking}.
The second approach uses the power of RM-synthesis to deliver reliable rotation measures and Faraday-corrected, hence intrinsic magnetic field orientations,
within the halos in order to reveal possible large-scale magnetic fields there. Our RM-synthesis is described in Sect.~\ref{sec:rm}, the results are presented and
discussed in Sect.~\ref{sec:results}. The large-scale magnetic fields are discussed in Sect.~\ref{sec:large-scale} and the degree of polarization is examined
in Sect.~\ref{sec:uniformity}. A general discussion is presented in Sect.~\ref{sec:discussion}, followed by a summary and conclusions in Sect.~\ref{sec:summary}.

\section{Polarization Stacking}
\label{sec:stacking}

The magnetic field structures in the halo are, of course, seen best in edge-on galaxies. Since CHANG-ES galaxies are all edge-on to the
line-of-sight\footnote{Inclinations are greater than 75$^\circ$ as described in Sect.~\ref{sec:introduction}.}, this survey provides a unique opportunity to
see whether X-shaped fields are a common characteristic of galaxies.  If they are not, then an average, as described below, should show no
such structure in the result. We will discuss the strengths and limitations of our approach to polarization stacking in Sect.~\ref{sec:stackresults}.

The polarization angle of the observed electric vector, $\chi$,  and
the magnitude of the linearly polarized flux density, $PI$, are given, respectively, by
\begin{eqnarray}
\chi&=&\frac{1}{2} \,arctan\left(\frac{U}{Q}\right)\label{eqn:chi}\\
PI&=&\sqrt{Q^2\,+\,U^2}\label{eqn:P}.
\end{eqnarray}
$Q$ and $U$ are Stokes parameters, either of which could be positive or negative.  $Q$ and $U$ maps for all of our galaxies have been made,
both corrected and uncorrected for the primary beam (PB) as standard products of the CHANG-ES survey. The polarization angle, $\chi$, rotated by
90~$\degr$, gives the apparent magnetic field orientation in the sky plane, uncorrected for Faraday rotation.

Images that are corrected for the PB have
accurate flux densities but increasing rms noise values with increasing distance from the pointing center of the map.  Images that are uncorrected
for the PB have uniform noise across the map (except for possible irregularities due to imperfect cleaning) but lower flux densities due to the
down-weighting of the PB with distance from the pointing center.  For Eqn.~\ref{eqn:P}, we have assumed that circular polarization, Stokes $V$, is
zero, since $V\,\ne\,0$ only for AGNs in a few cases \citep{irwin+2018}. All steps were carried out using the Common Astronomy Software Application
(CASA) package\footnote{Version 5.1.1-5} \citep{mcmullin+2007}.

\subsection{Choice of Data Set and Preliminary Steps}

CHANG-ES data were observed at both L-band and C-band, but since the lower frequency L-band data is much more strongly affected by Faraday rotation, we
restrict ourselves to the C-band data for the stacking. The $Q$ and $U$ maps were made using data over the entire 2 GHz bandwidth as described in
\cite{irwin+2013}.
Since we are interested in broader scale emission so as to detect polarization that extends into the halo if possible, we choose the D-configuration data,
rather than the higher resolution C-configuration data. Note that for stacking, we use $Q$ and $U$ images that are {\it not} RM-corrected.  The RM-corrected
data that will be introduced in Sect.~\ref{sec:rm} are especially advantageous closer to the disk, but our goal in this section is to search for broad-scale
structure that may be common to all galaxies.  With these C-band D-configuration robust=+2 $Q$ and $U$ images, we have a uniform data set of 28  galaxies (see below).

Each $Q$ and $U$ map, both PB-corrected and PB-uncorrected, was first regridded to a common 2000 x 2000 pixel field with 2 arcsec-sized cells.
Thus each field was 33.3 arcmin across which is quite large so that any size  scaling does not lead to cut-offs at the edges of the field.  Note that the
full-width at half-maximum (FWHM) of the C-band PB is 7.0 arcmin at our central frequency of 6.0 GHz \footnote
{https://library.nrao.edu/public/memos/evla/EVLAM\_195.pdf}.

Maps of $\chi$  and $P$ (Eqns.~\ref{eqn:chi} and \ref{eqn:P}) were made for each galaxy for the purpose of initial inspection.  As a result of this
exercise, 7 galaxies were excluded from the stacking process.
These were:  NGC~2992 (core/jets, see \citealt{irwin+2017}), NGC~4244 (no emission), NGC~4438 (core plus radio lobe only), NGC~4594, NGC~4845 and
NGC 5084 (nuclear sources only), and UGC~10288 (dominated by a background double-lobed radio source, see \citealt{irwin+2013}). This left 28 galaxies
for stacking.
Each of these galaxies shows polarization in the disk although there is considerable structure in the polarization as Fig.~\ref{n660all} to
Fig.~\ref{n5907all} show.
Because of this, we do {\it not} exclude galaxies that have known radio structures, such as radio lobes, associated with AGNs (e.g. NGC~4388 and
NGC~3079).  Such peculiarities could perturb an average map, but as long as disk and halo PI exist, the galaxy is included under the assumption
that underlying disk- and halo-related polarization may still be present.

\subsection{Galaxy Alignment and Angular Scaling}
\label{sec:align_scale}

Table~\ref{scalerotate} provides galaxy distances, scaling and rotation data for the following discussion.
Each image ($Q$ and $U$ maps both PB-corrected and PB-uncorrected) had to be rotated and scaled in angular size in preparation for stacking.
However, since the
polarization images tend to be complex, it was not possible to scale/rotate them based on the appearance of the polarization.

For this reason, the orientation (position angle) of a galaxy was initially taken from the K$_s$ passband as given in the NASA Extragalactic Database
(NED)\footnote{The NASA/IPAC Extragalactic Database (NED) is operated by the Jet Propulsion Laboratory, California Institute of Technology,
under contract with the National Aeronautics and Space Administration.} but minor adjustments were made by eye (mostly zero but all $<$ 5 deg)
to fine-tune the disk alignment so that the total intensity radio images aligned with the x axis.

We next wish to scale all galaxies to have the same angular size.  Our argument for this scaling is based on the possibility that the underlying
polarization structure could be self-similar as described by \cite{henriksen+2018}. This means that galaxies of
different size could still show (for example) X-type fields but on a scale appropriate to the galaxy.  Without any such scaling, the 28 galaxies span a factor of 9 in
angular size and we would not expect to see any polarization structures that are common to them all in a
stacked image.  A factor of 4 in angular size is due to differences in the physical sizes of the galaxies and the remainder is due to different distances.

For the angular scaling, we examined the 22 $\mu$m WISE (Wide-field Infrared Survey Explorer) images \citep{wright+2010} with
resolution enhanced to a final value of 12.4 arcsec via the WERGA
process \citep{jarrett+2012,jarrett+2013}.
Measured diameters are given in \citet[][their Table 6]{wiegert+2015}.
We therefore scale the angular sizes of all galaxies (their $Q$ and $U$ images) to match that of the largest galaxy, namely NGC~4244 which was
11.53 arcmin in diameter\footnote{Note that NGC~4244 was not actually in the final sample but was a benchmark for scaling and rotation of the other
galaxies.\label{footnote:stacking}}. Which size to adopt for the angular scaling is arbitrary, but by choosing the largest galaxy, we can ensure that all other galaxies become larger and  no information is lost between pixels. In this process, we also scale the synthesized beam and ensure that the integrated flux density is conserved.
That is, the galaxies are scaled in angular size but retain their original integrated flux densities.

Once all images have the same angular size, we have actually adjusted for two effects: differences in the physical sizes of the galaxies, and
differences in their distances.  For example, suppose two galaxies are at the same distance but of different physical size.  In that case, if the smaller galaxy
is scaled up to the larger size but retains its original integrated flux density as we have done, then its surface brightness will be lower. We
take this as a reasonable approach to 'physical scaling'.  On the other hand, suppose that a distant galaxy is scaled up in angular size as if it were a nearby galaxy.
In that case, retaining the original integrated flux density as we have done is not correct and the integrated flux density should
actually be increased to account for the closer distance. Therefore this 'distance scaling' requires an adjustment to the integrated flux density
which we will make a correction for in Sect.~\ref{sec:weighting_scaling}.

The rotated scaled images were finally converted to units of Jy/pixel and rescaled on a common grid.
The PB for each galaxy was also scaled and rotated following the same procedure as above, and ensuring that the intensity value is 1.0 at the beam
center. Rotation is
necessary for the larger galaxies for which there were two pointings (NGC~891, NGC~3628, NGC~4565, NGC~4594, NGC~4631, and NGC~5907).  PB Rotation
is not strictly necessary for PBs of the single-pointing galaxies since the PB is symmetric but was carried out anyway for coding and header consistency.

  \begin{table}
      \caption[]{\label{scalerotate} Scaling and Rotation of the Galaxies}

     $$
         \begin{tabular}{lcccc}
            \hline
            \noalign{\smallskip}
            \multicolumn{1}{c}{Galaxy} & Key   & Distance $^\mathrm{a}$  & Angular  & Rotation $^\mathrm{c}$ \\
            &  for Fig.~2 &  & Scaling $^\mathrm{b}$   &   \\
            &   &  [Mpc]  & & [\degr] \\
            \noalign{\smallskip}
            \hline
            \noalign{\smallskip}
            NGC~660   & 28 & 12.3       & 3.82      &  49     \\
            NGC~891   & 10 & 9.1        & 1.21      &  66     \\
            NGC~2613  & 1  & 23.4       & 2.3       & -24     \\
            NGC~2683  & 11 & 6.27      & 2.24       & 49       \\
            NGC~2820  & 20 & 26.5      & 6.48           & 25       \\
            NGC~3003  & 2  & 25.4      & 3.07           & 12       \\
            NGC~3044  & 12 & 20.3      & 3.76           & -25       \\
            NGC~3079  & 21 & 20.6      & 2.66           & -79       \\
            NGC~3432  & 3 & 9.42      & 3.18          &  48     \\
            NGC~3448  & 13 & 24.5      & 6.59          &  25     \\
            NGC~3556  & 22 & 14.09      & 1.90         & 9      \\
            NGC~3628  & 4  & 8.5      & 1.16           & -14      \\
            NGC~3735  & 14 & 42      & 4.09            & -40      \\
            NGC~3877  & 23 & 17.7      & 3.22          & 56      \\
            NGC~4013  & 5  & 16      & 3.34            & 24       \\
            NGC~4096  & 15 & 10.32      & 2.94         & 73      \\
            NGC~4157  & 24 & 15.6     & 3.14           & 27      \\
            NGC~4192  & 6  & 13.55      & 1.86         & -62      \\
            NGC~4217  & 16 & 20.6      & 2.96          & 41      \\
            NGC~4302  & 25 & 19.41     & 3.02         & -88      \\
            NGC~4388  & 7  & 16.6     & 5.01          & 1      \\
            NGC~4565  & 17 & 11.9     & 1.11           & -44      \\
            NGC~4631  & 26 & 7.4     & 1.06           & 6      \\
            NGC~4666  & 8  & 27.5     & 2.86          & 50      \\
            NGC~5297  & 18 &40.4     & 5.24           & -59       \\
            NGC~5775  & 27 & 28.9     & 3.03          & -56      \\
            NGC~5792  & 9  & 31.7     & 4.49          & 4      \\
            NGC~5907  & 19 & 16.8     & 1.57          & -65     \\


         \noalign{\smallskip}
            \hline
         \end{tabular}
     $$

   \begin{list}{}{}
   \item[$^{\mathrm{a}}$] Galaxy distance from \citet{wiegert+2015}.
   \item[$^{\mathrm{b}}$] Factor by which galaxies are scaled in size, based on their 22 $\mu$m diameters as given in \citet{wiegert+2015}.
   \item[$^{\mathrm{c}}$] Required counter-clockwise (=positive) rotation for x-axis alignment.
   \end{list}
   \end{table}

\subsection{Galaxy Weighting}
\label{sec:weighting}

The PB-corrected images are required for stacking but, after scaling by angular size, each pixel corresponds to a different part of the PB
from galaxy to galaxy.
Consequently the rms noise at each pixel is different from galaxy to galaxy. Thus, we wish to weight each galaxy at a given pixel by the rms noise
at that location. The weighting at any position is
\begin{equation}
W_{i}\,=\,\frac{1}{{\sigma_i}^2}\label{eqn:weights}
\end{equation}
where $\sigma$ is the rms noise of the PB-corrected map at that position and $i$ refers to the galaxy. Also, the PB is cut off at 0.1 times the peak
of 1.0 (which is at the map center) and any point that might fall outside of the cutoff must be set to zero and not counted in a weighted sum.

In order to compute $\sigma_i$, we measure the value of $\sigma_i$ of the PB-uncorrected scaled and rotated $Q$ and $U$ maps which have uniform noise,
and divide the scaled and rotated PB by this noise value.  This generates maps of $\sigma_i$ that increase with distance from the map center for each
galaxy, $i$, as needed for Eqn.~\ref{eqn:weights}. We then form maps of $W_iU_i$ and $W_iQ_i$ for each galaxy, where $Q_i$ and $U_i$ have been
rotated and scaled in angular size as described above.  We also form a map of the sum of the weights, $\Sigma_i\,W_i$, for the purposes of normalization.

The weighted sum of the electric field vectors at any pixel is
\begin{equation}
\chi_W\,=\, \frac{1}{2}arctan\left[\frac{\Sigma_i \left(W_i U_i\right)}{\Sigma_i \left(W_i Q_i\right)}\right]\label{eqn:Xw}
\end{equation}
The weights, $W_i$ are the same for $Q_i$ and $U_i$ since the rms noise is approximately the same for the two maps at any position.  Hence a normalization
by the sum of the weights in the numerator and denominator of Eqn.~\ref{eqn:Xw} would cancel and is not required.
{Note that Eqn.~\ref{eqn:Xw} supercedes Eqn.~\ref{eqn:chi}.}

\subsection{Polarization Weighting and Scaling for Distance}
\label{sec:weighting_scaling}

\begin{figure*}
   \centering
   \includegraphics[width=2.0\columnwidth]{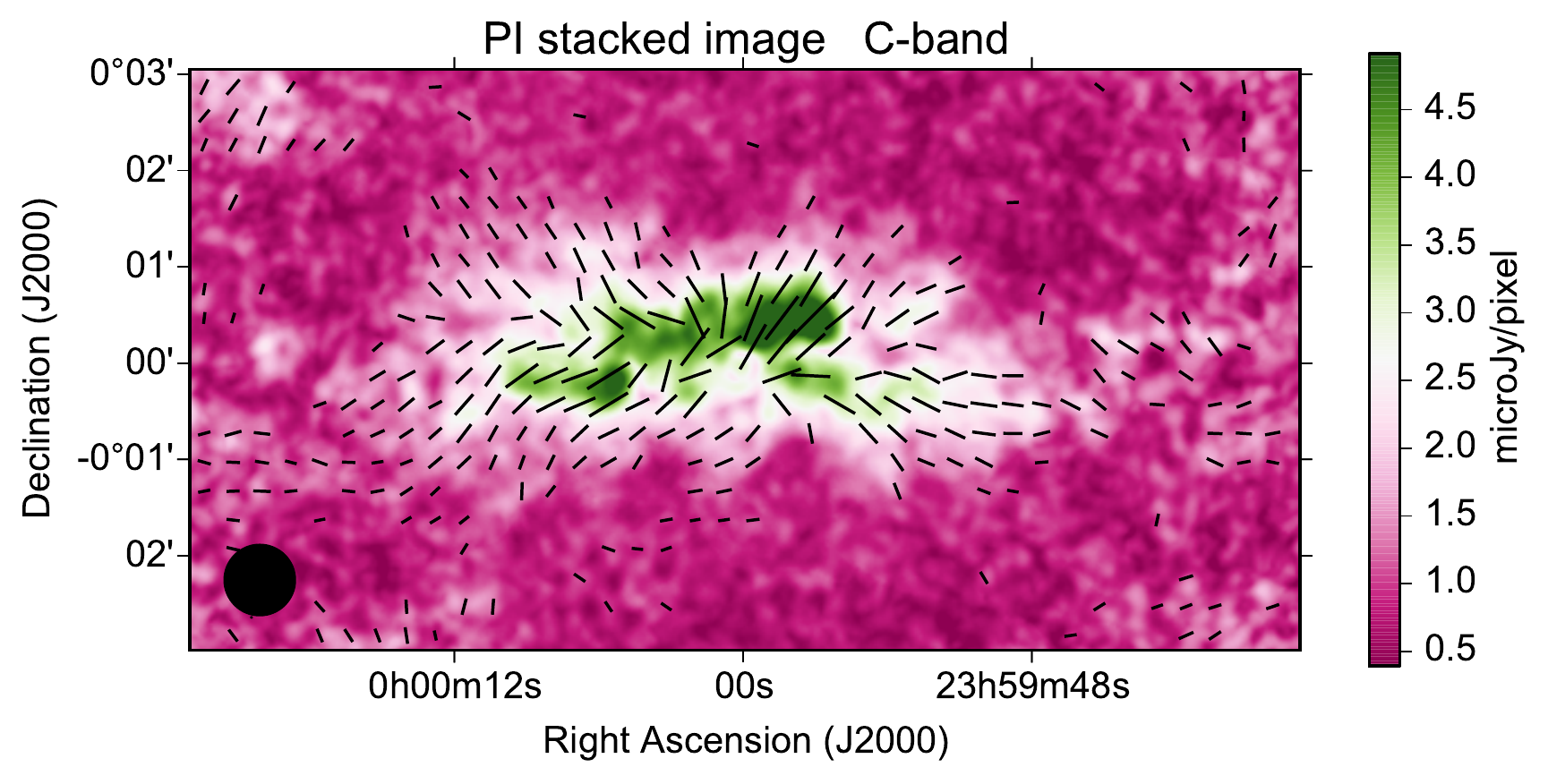}
   \caption{Weighted mean linear polarization (in colours), scaled to a common angular size (11.5 arcmin) and to a common distance (4.4 Mpc), which correspond to NGC~4244. The horizontal extent of the image shown is 11.5 arcmin, or 14.8 kpc at the common distance.
       The average scaled synthesized
       beam is shown as a circle at lower left.
       Vectors show the
      orientation of the weighted scaled apparent magnetic field. Note their `X-shaped' orientation.}
         \label{stackedimage}
\end{figure*}

We now have a series of galaxy $Q$ and $U$ images  that have been scaled in angular size and weighted by rms noise, but their integrated flux densities retain their original
values as explained in Sect.~\ref{sec:align_scale}.
 To correct for this, we now scale these flux densities by
distance. To accomplish this, we scaled each map to the
distance of the closest galaxy which is NGC~4244 at a distance of 4.4 Mpc (see Footnote~\ref{footnote:stacking}), that is, we apply a
scaling for galaxy, $i$, of  $(D_i/4.4)^2$ for $D$ in Mpc, as given in Table~\ref{scalerotate}. We finally form the weighted sum of this `polarization luminosity'
for any pixel according to
\begin{equation}
  {Q_W}\,=\,\frac{\Sigma_i\,\left[\left(\frac{D_i}{4.4}\right)^2W_i Q_i\right]}{\Sigma_i W_i},
  ~~~~~~~~
  {U_W}\,=\,\frac{\Sigma_i\,\left[\left(\frac{D_i}{4.4}\right)^2W_i U_i\right]}{\Sigma_i W_i}
  \label{eqn:QwUw}
  \end{equation}

\begin{equation}
\label{eqn:Pnew}
P_W\,=\,
\sqrt{{Q_W}^2\, \,+\,      {U_W}^2}
\end{equation}
{Note that Eqn.~\ref{eqn:Pnew} supercedes Eqn.~\ref{eqn:P}.} The final map showing $P_W$ in colour and $\chi_W$ as vectors (rotated by 90 degrees
to represent magnetic field orientations) is shown in Fig.~\ref{stackedimage}. All final quantities are in units of Jy/pixel.

\subsection{Checks for Dominant Galaxies}
\label{check}

\begin{figure*}
   \centering
   \includegraphics[width=1.5\columnwidth]{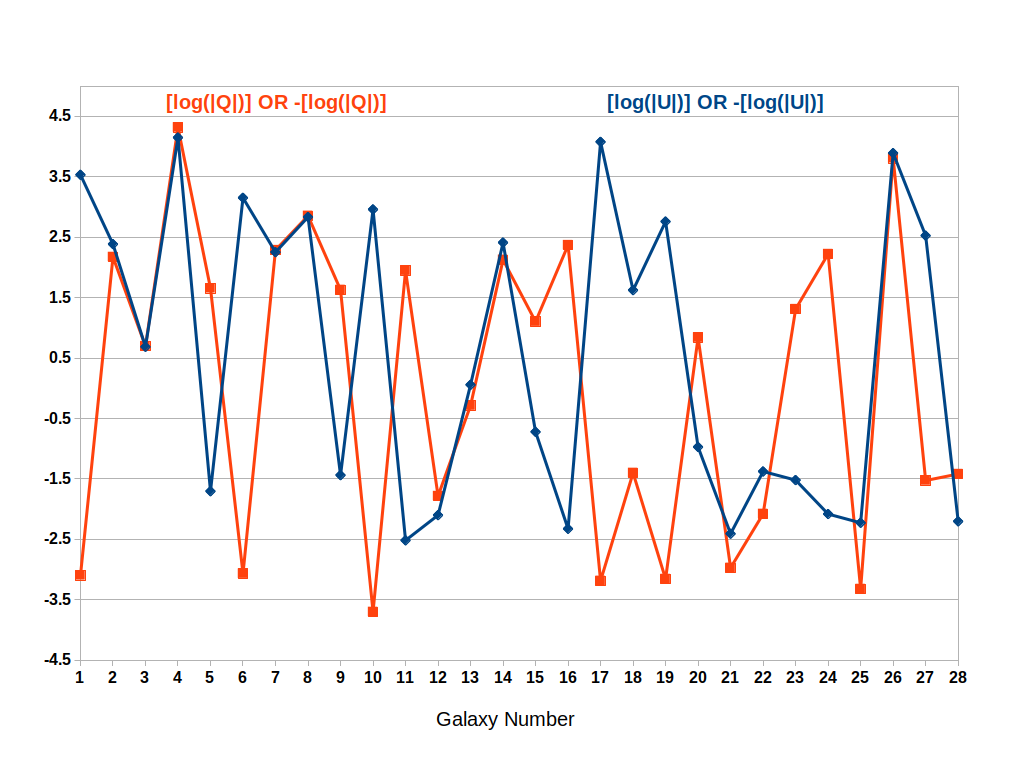}
   \caption{Scaled and weighted {\it Q} (red curve) and {\it U} (blue curve) for each galaxy individually.  See Sect.~\ref{check} for a description.  The numbers on the x-axis indicate the galaxies according to the Key given in Table~\ref{scalerotate}.}
\label{checkimage}
\end{figure*}

Before discussing the results of Fig.~\ref{stackedimage} we need an additional check to see whether the stacking has strongly favoured one or two galaxies in comparison to all others.  In other words, could a galaxy be so dominant that the stacking simply highlights the structure of that one galaxy?  To check this, we compute maps of scaled weighted {\it Q} and {\it U} for each galaxy individually
and then compute the average {\it Q} and {\it U} for each galaxy. For this calculation, there are no sums over galaxies in Eqn.~\ref{eqn:QwUw} so the calculation is determined from, $Q_{W_i}\,=\,(D_i/4.4)^2\,W_i\,Q_i$ and
$U_{w_i}\,=\,(D_i/4.4)^2\,W_i\,U_i$.  For the average of the region shown in Fig.~\ref{stackedimage}, we then plot the result in Fig.~\ref{checkimage}, presented as a log plot in order to see each galaxy more clearly. Since the plot is a diagnostic intended for comparison between galaxies only, we have first
multiplied $Q_{W_i}$ and $U_{w_i}$ by a constant for a simpler reading of the y-axis scale.  Finally, since $Q_{W_i}$ and $U_{w_i}$ can be positive or negative, we must take the log of the absolute value and then have restored the negative to the result.

Fig.~\ref{checkimage} shows that the individual galaxies vary in $Q$ and $U$ over several orders of magnitude.  Since the average of Fig.~\ref{stackedimage} is a linear average, it is possible that only the peaks (either positive or negative) strongly contribute to the average.  However, various galaxies dominate, depending on what parameter is considered.  For example, the most dominant galaxies in positive {\it Q} are NGC~3628, NGC~4631, while the most dominant galaxies in negative {\it Q} are NGC~891 and NGC~4302.  The most dominant galaxies in positive {\it U} are NGC~3628, NGC~4565, and NGC~4631, and those in negative {\it U} are NGC~2683, NGC~4217, and NGC~N3079.  If we repeat this exercise with a more restrictive field, then these dominant galaxies tend to change.  For example, NGC~4565 becomes the dominant galaxy in positive {\it Q}, depending on where the measurement is made.  Moreover, NGC~4565 does not show a particularly impressive halo or X-type field (Appendix~A).

We conclude that the stacked image shown in Fig.~\ref{stackedimage} to be discussed in the next section is, to the limitations of our data, a realistic description of a 'mean' magnetic field.

\subsection{Stacking results}
\label{sec:stackresults}

Fig.~\ref{stackedimage} displays the result of our polarization stacking at C-band.  The displayed horizontal size is approximately equal to the
scaled 22 $\mu$m size of our galaxies (11.53 arcmin).
Our first conclusion is that the region over which polarization occurs in the galaxies
is smaller than the 22 $\mu$m galaxy size.  This is likely a signal-to-noise (S/N) issue since linear polarization is weaker than total intensity.
The complexity of the emission is also evident.  There remain AGNs in the sample, some of which have radio lobes (e.g. NGC~3079, NGC~4388) and this
may contribute to the irregularity in the brightness.

The overall structure of the stacked apparent magnetic field vectors is indeed X-shaped,
with vectors more parallel to the galactic midplane along the disk (best visible in the western part of the disk), as expected from an inclined
plane-parallel spiral disk field. Note that the effects of Faraday rotation and depolarization (which are uncorrected in these C-band data) are
expected to be strongest along the disk and inner halo.
This explains why the stacked apparent magnetic field vectors look most regular in the outer parts of the halo and the polarized intensity
is slightly weaker along the thin galaxy disk direction (horizontal line at DEC = 0 deg).
In a sense, it is surprising that any structure is seen at all in the stacked magnetic field.  If we recall that $Q$ and $U$ can be positive
or negative, the sums in Eqn.~\ref{eqn:Xw} might have resulted in zero or, at least, a random field orientation.  The fact that X-type structure is visible in
a sum of 28 galaxies argues in favour of the scale invariance that was assumed in Sect.~\ref{sec:align_scale}.

Fig.~\ref{stackedimage} shows some scattered emission to the far right and left of the main emission region.  This is largely due to the fact that
some galaxies have
companions or strong background sources in the field that were not excised prior to stacking. This introduces a discussion as to alternate
methods of stacking the images.  One might have excised such emission prior to the stacking, for example.  We have retained any radio lobe structure near
AGNs (e.g. NGC~3079) and have also retained any background sources that might be seen through the disk (e.g. two background sources at the far end of the
disk of NGC~5907).  A future approach might be to subtract such known sources in advance.  Another approach is to do the angular scaling according to the
total intensity radio continuum emission extent, rather than according to the $22\mu$m emission size. Even the rotation could be honed so that the x-axis
alignment is oriented with all advancing sides together and all receding sides together.  A future effort could do a more complete search through parameter
space.  If the result improves, this could lead to more insight as to what is actually driving the X-types (or other) structures.  Therefore, we consider
our result to be a first step.

In spite of these caveats, it is remarkable that coherent
polarization structure remains when the $Q$ and $U$ maps are weighted and scaled as described above.  Although not perfect, the X-shaped structure of
the magnetic field can indeed be seen in the stacked image, arguing for self-similarity as described in \cite{henriksen+2018} with axially symmetric
fields as the dominant mode.

\section{RM-synthesis}
\subsection{Parameters and polarization imaging}
\label{sec:rm}

We performed RM-synthesis \citep{brentjens+2005} for each of our 35 CHANG-ES galaxies at C-band and L-band. The image cubes in Stokes Q and U were
prepared in the following way: at C-band we defined 16 spectral windows, corresponding to a frequency spacing of 128 MHz and $\lambda ^2$ spacings
between $0.7 \, \rm{and} \, 1.7 \, \rm{cm}^2$. To obtain a similar $\lambda ^2$ range at L-band, each
spectral window (among 29 in total\footnote{Out of initially 32 spectral windows, the first three were flagged entirely for all galaxies due to severe
RFI contamination. Hence the effective number of spectral windows is 29.})
was split into 4 sections of 13 channels each, assuming that the edge-channels 0 - 5 and 58 - 63 are always flagged. This corresponds to
a frequency spacing of 3.25 MHz within each spectral window and 6 MHz between spectral windows, or $\lambda ^2$ spacings between
$0.85 \, \rm{and} \, 2.7 \, \rm{cm}^2$.

As a first step, the CASA task $statw$ was run on each individual data set, to make sure all visibilities have the correct weighting when combining
different array configurations. For each of the frequency intervals specified above, the $clean$ task was run on Stokes Q and U of the polarization
calibrated data sets of the combined C- and D-array configurations at C-band and B-, C-, and D-array configurations at L-band. The pixel size was
chosen to be $1.5 \arcsec$ with an image size of 600 x 600 at C-band (720 x 720 for galaxies with two pointings)
and 2400 x 2400 at L-band. We used the multi-frequency synthesis (MFS) with nterms = 1 and used a maximum of 2000 clean components that were mostly
sufficient to reach our stopping threshold of $25 \, \mu \mathrm {Jy/beam}$ at C-band and $300 \,\mu \mathrm {Jy/beam}$ at L-band, usually
corresponding to 2 - 4 times the noise rms of the clean images. For all maps we used natural weighting (robust = 2) in order to be more sensitive to
extended faint emission, without applying any uv tapering. The maps were convolved to a common beam size of $12 \arcsec$, primary-beam corrected, and
concatenated into a single data cube file, each for Q and U.

These data cubes were used as input for the RM-synthesis code \citep{heald+2009} to produce the Faraday cubes in Q, U, and Faraday depth ($\phi$).
With our spectral windows we cover a range in Faraday depth between
$+8192 \,  \mathrm {rad/m}^2$ and $-8192 \,\mathrm {rad/m}^2$ with channel separations of $ 64 \, \mathrm {rad/m}^2$ at C-band and
between $ -2048 \,\mathrm {rad/m} ^ 2 \le \phi \le +2048 \,\mathrm {rad/m}^2$ with channel separations of $ 8 \, \mathrm {rad/m}^2$ at L-band.

The Faraday spectra were cleaned down to a $3 \,\sigma$-noise level of the uncleaned Q and U cubes using the algorithm by \citet{heald+2009}, extended
by \citet{adebahr+2017}. These cleaned Q and U Faraday cubes were used to estimate the polarized intensity PI pixel wise. PI is not corrected for Ricean bias.
We finally determined a PI map, consisting of the maximum value along the PI cube for each pixel in RA-DEC space, a map of the peak RM which for each
pixel corresponds to the $\phi$ value at which the maximum in the PI cube occurs. Further, a map of the observed polarization angle was computed at
each position from the Q and U values in the $\phi$ plane where the maximum in the PI cube occurs. The RM and polarization angle maps were computed in
regions where PI is higher than 5 times the noise rms in the clean Faraday Q and U image planes where the (Ricean) bias in the polarized intensity only
plays a minor role.

The observed polarization angles were Faraday rotation corrected with the corresponding RM values and the corresponding central wavelength in order
to receive a map of the intrinsic polarization angles PA. In addition, the RM maps have been corrected for their Galactic foreground contribution
RM$_\mathrm{fg}$ which was determined from the Galactic RM map of \citet{oppermann+2012} at the position of each galaxy (see Table~\ref{RM}). We also
calculated the error maps of the PI, RM, and PA maps.

\begin{table*}
      \caption[]{\label{RM} Foreground Rotation Measure $\mathrm{RM_{fg}}$, distances, linear resolutions that correspond to $12 \arcsec$, and range
      of observed RMs observed at C-band in the disk and/or halo of the galaxies}

     $$
         \begin{tabular}{lccccc}
            \hline
            \noalign{\smallskip}
            \multicolumn{1}{c}{Galaxy}    & $RM_\mathrm{fg}$  & distance & lin. res.@12\arcsec & RM$_{disk/halo}$ range & remarks \\
                           &  [$\rm rad/m^2$]  & [Mpc] & [pc] &  [$\rm rad/m^2$] & \\
            \noalign{\smallskip}
            \hline
            \noalign{\smallskip}
            NGC~660   &  $-4 \pm 10$ & 12.3 & 720 &  -204 to 330 & polar ring galaxy \\
            NGC~891   & $-80 \pm 18$ &  9.1 & 530 & -209 to 466 \\
            NGC~2613  & $209 \pm 27$ & 23.4 & 1360 & -328 to 242 \\
            NGC~2683  &  $21 \pm 7$ &  6.3 & 360 &  & PI too weak \\
            NGC~2820  & $-20 \pm 9$ & 26.5 & 1540 & -173 to 156 \\
            NGC~2992  & $-20 \pm 16$ & 34.0 & 1980 &  & central source only\\
            NGC~3003  &  $23 \pm 7$ & 25.4 & 1480 &  & PI too weak \\
            NGC~3044  & $-13 \pm 8$ & 20.3 & 1180 & -238 to 372 & \\
            NGC~3079  &  $10 \pm 6$ & 20.6 & 1200 & -331 to 239 & strong central source \\
            NGC~3432  &  $13 \pm 5$ &  9.4 & 550  &  & PI too weak \\
            NGC~3448  &  $16 \pm 5$ & 24.5 & 1430 & -232 to 217 \\
            NGC~3556  &  $11 \pm 6$ & 14.9 & 820 & -226 to 362 \\
            NGC~3628  &  $ 9 \pm 4$ &  8.5 & 490 & -339 to 319 & strong central source \\
            NGC~3735  & $-14 \pm 7$ & 42.0 & 2440 & -216 to 249 &  \\
            NGC~3877  &  $19 \pm 5$ & 17.7 & 1030 &  & PI too weak \\
            NGC~4013  &  $ 3 \pm 3$ & 16.0 & 930 & -248 to 163 \\
            NGC~4096  &  $14 \pm 4$ & 10.3 & 600 &  & PI too weak\\
            NGC~4157  &  $20 \pm 4$ & 15.6 & 910 & -422 to 255 \\
            NGC~4192  &  $-8 \pm 4$ & 13.6 & 790 & -184 to 263 \\
            NGC~4217  &  $ 7 \pm 5$ & 20.6 & 1200 & -179 to 189 &  \\
            NGC~4244  &  $-8 \pm 3$ & 4.4 & 260 &  & no polarization at all \\
            NGC~4302  &  $-6 \pm 4$ & 19.4 & 1130 & -287 to 214 \\
            NGC~4388  &  $-2 \pm 4$ & 16.6 & 970 & -164 to 295 & nuclear outflow $\,^\mathrm{a}$ \\
            NGC~4438  &  $-3 \pm 4$ & 10.4 & 600 &  & core plus radio lobe only\\
            NGC~4565  &  $ 6 \pm 3$ & 11.9 & 690 & -204 to 342 & \\
            NGC~4594  &  $-2 \pm 6$ & 12.7 & 740 &  & central source only\\
            NGC~4631  &  $ 0 \pm 3$ &  7.4 & 430 & -551 to 823 & \\
            NGC~4666  &  $-7 \pm 6$ & 27.5 & 1600 & -190 to 358 & \\
            NGC~4845  & $-15 \pm 4$ & 17.9 & 990 & & central source only $\,^\mathrm{b}$ \\
            NGC~5084  &  $-8 \pm 10$ & 23.4 & 1360 & -49 to 322 & central source only\\
            NGC~5297  &  $ 4 \pm 4$ & 40.4 & 2350 &  & PI too weak\\
            NGC~5775  &  $17 \pm 6$ & 28.9 & 1680 & -271 to 335 & \\
            NGC~5792  &  $ 1 \pm 7$ & 32.7 & 1840 &  & PI too weak \\
            NGC~5907  &  $ 2 \pm 7$ & 16.8 & 980 & -152 to 126\\
            UGC~10288 &  $ 3 \pm 11$ & 34.1 & 1980 &   & background radio source only $\,^\mathrm{c}$ \\
   \noalign{\smallskip}
            \hline
         \end{tabular}
               $$
\begin{list}{}{}
\item[$^{\mathrm{a}}$] See CHANG-ES VII paper \citep{damas+2016}
\item[$^{\mathrm{b}}$] See CHANG-ES V paper \citep{irwin+2015}
\item[$^{\mathrm{c}}$] See CHANG-ES III paper \citep{irwin+2013}
\end{list}

\end{table*}

\subsection{Results and Faraday depolarization}
\label{sec:results}

\begin{figure*}
\centering
\includegraphics[width=9 cm]{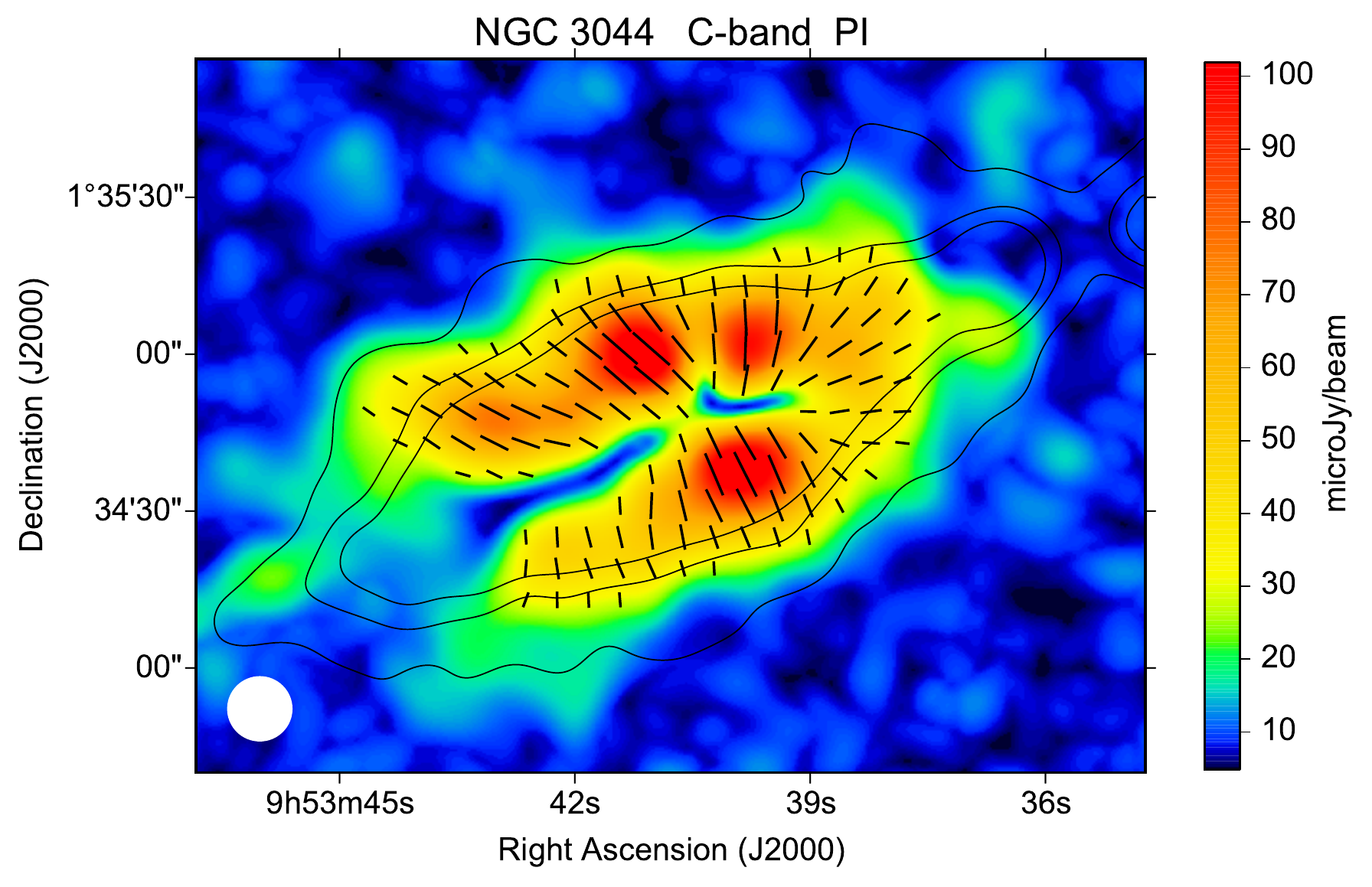}
\includegraphics[width=9 cm]{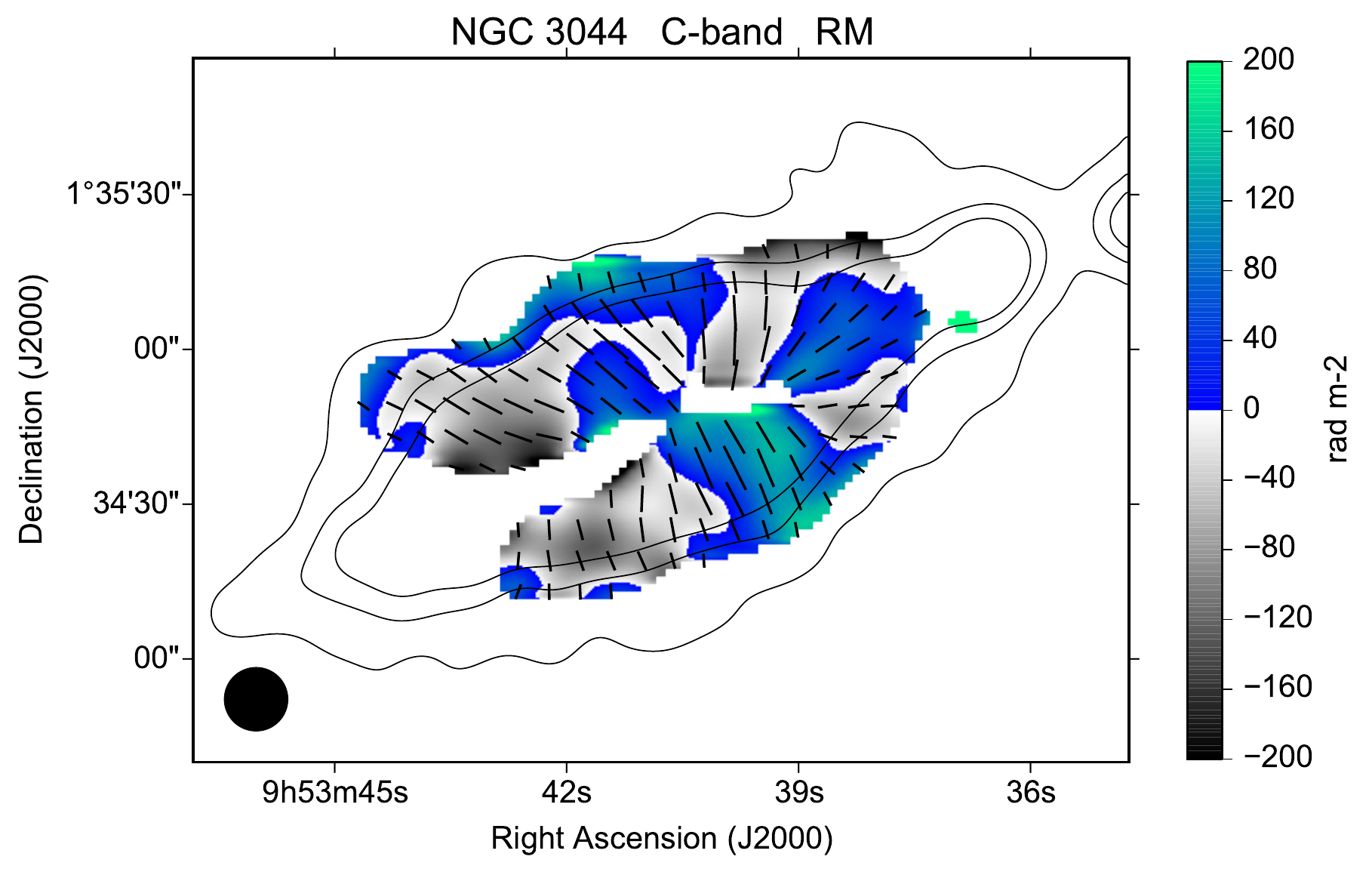}
\includegraphics[width=9 cm]{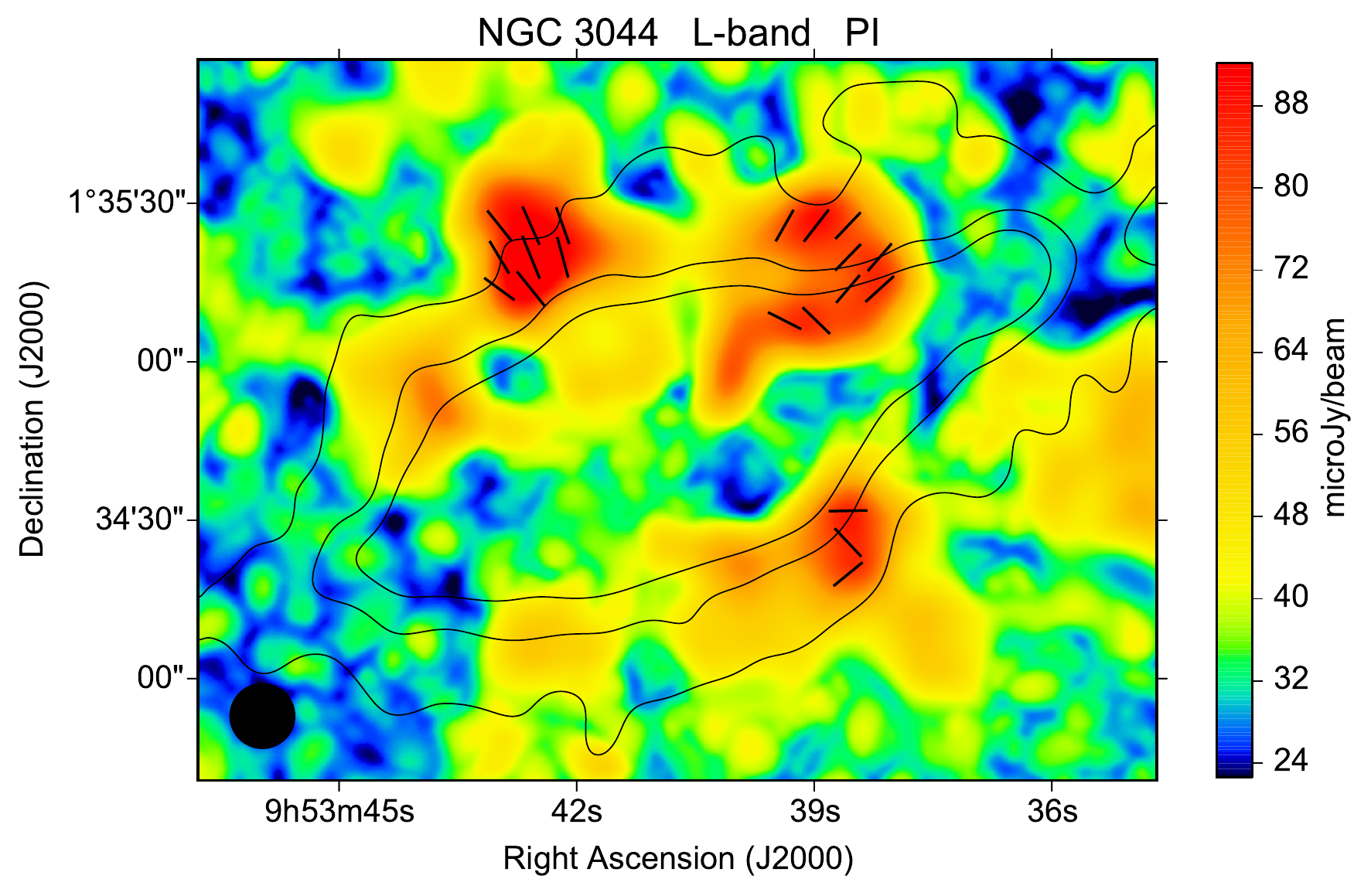}
\includegraphics[width=9 cm]{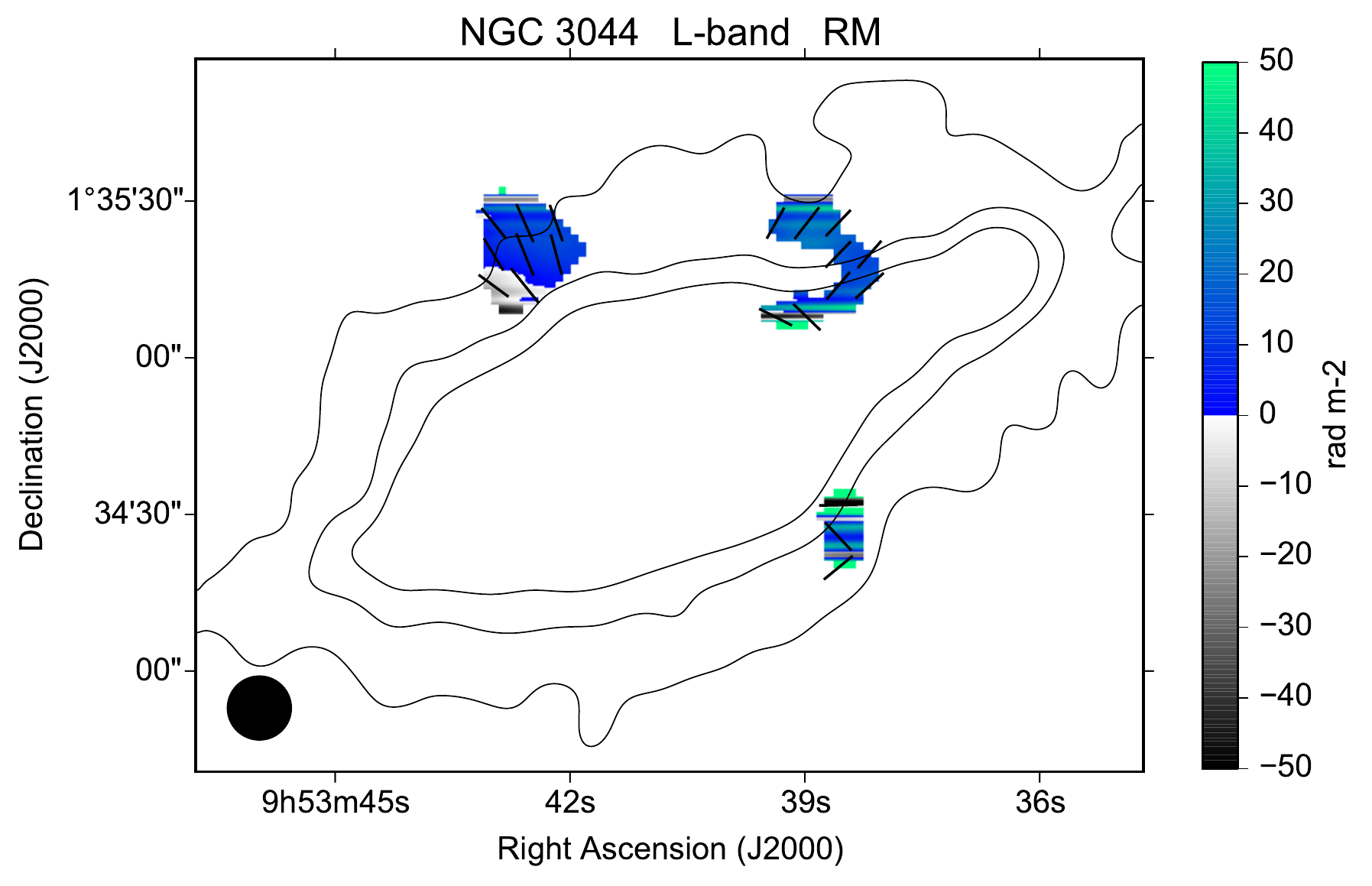}
\caption{Polarized intensity (\textbf{left}) and Rotation Measure (\textbf{right}) of the spiral galaxy NGC~3044 at C-band (5 - 7 GHz)
(\textbf{upper panels}) and L-band (\textbf{lower panels}) with a resolution of $12 \arcsec$~HPBW presented within the same field on the sky.
The vectors are corrected for the observed Faraday rotation at the corresponding frequency band which give the intrinsic magnetic field
orientation at C-band. The contours indicate the total intensity at 50, 150, and 250 $\mu$Jy/beam at C-band and 125, 375, and 625 $\mu$Jy/beam
at L-band, corresponding to $5 \, \sigma$, $15 \, \sigma$, and $25 \, \sigma$ r.m.s. each.
}
\label{n3044CLband}
\end{figure*}

We discuss NGC~3044 as an example galaxy first, before presenting the others.
The resultant maps of the linear polarization (PI) and RM of the NGC~3044 are shown in Fig.~\ref{n3044CLband} for C-band in the upper
panels and for L-band in the lower panels. At C-band the linear polarization is observable within most parts along the disk and the halo. It is concentrated
within the $25 \, \sigma$ contours of total intensity. There is only a thin area along the central and eastern disk of the galaxy without linear
polarization. At L-band, the distribution of the polarized intensity looks completely different: The peaks in PI are located far away from the
disk in the outer halo of NGC~3044, while the projected galactic disk and inner halo are completely depolarized.

The observed RM values in NGC~3044 are very different at both frequency bands: at C-band more than 99\% of the values are in the range between
$+200\,\mathrm {rad/m}^2$ and $-200 \,\mathrm {rad/m}^2$ (max$= 372\,\mathrm {rad/m}^2$, min$= -238\,\mathrm {rad/m}^2$) while at L-band they are in a
the small range between $+24 \,\mathrm {rad/m}^2$ and $-14 \,\mathrm {rad/m}^2$. These RM ranges should be similar in a source without Faraday
depolarization. Also the magnetic field vectors corrected for the observed RM at the
corresponding frequency band also look somehow different at L-band when compared to those at C-band. In general, the vectors at C-band show a very
smooth, regular structure, while those in two of the three blobs at L-band look rather irregular.

All these findings strongly indicate that the L-band data are strongly affected by Faraday depolarization and even completely depolarized over large areas in
NGC~3044. Faraday depolarization can
be caused by thermal electrons together with a regular magnetic field within the emitting source (differential Faraday rotation) and/or by random
magnetic fields within the source (internal Faraday dispersion) or between the source and the observer (external Faraday dispersion)
\citep{burn1966, sokoloff+1998}. From the observations of a smooth, large-scale pattern in RM at C-band together with the regular pattern of the
magnetic field vectors, we conclude that the magnetic field in NGC~3044 is regular (see Sect.~\ref{sec:large-scale}). Hence, we expect
differential Faraday depolarization in NGC~3044.

RM-synthesis can be used to identify the effects of differential Faraday depolarization, however,
it is unable to recover emission that is completely depolarized by this process. A first complete depolarization by differential
Faraday rotation at L-band occurs already around $ RM  \approx \pm 40 \, \mathrm {rad/m}^2$ for the mid frequency of the observed total band
while at C-band this occurs at a much larger
value around $ RM  \approx \pm 630 \, \mathrm {rad/m}^2$ at the mid frequency \citep[][their Eq. 4]{arshakian+2011}.
While the observed RM values in NGC~3044 at C-band are below the range of complete differential Faraday depolarization, they
are far above the value at L-band. This indicates that the observed polarized intensity at L-band is depolarized over wide ranges along the line of sight.
This means that NGC~3044 is Faraday thick at L-band along its disk and huge parts of its halo while the galaxy is Faraday thin at C-band in the entire
halo. There is just a thin depolarization lane along parts of the galactic disk of NGC~3044 where the thermal electrons and also the total magnetic field strength
are expected to be highest.

NGC~3044 is just an example for the strong depolarization at L-band. Generally, we find that the observed RM values at L-band are in the range
$+50 \,\mathrm {rad/m}^2$ and $-50 \,\mathrm {rad/m}^2$, hence limited by strong Faraday depolarization which means that they are Faraday thick over wide areas
of their halos if they host large-scale magnetic fields. Only the observed RM values at C-band are below the range of complete differential
depolarization. They can be regarded as a reliable tracer of the parallel magnetic field component along the
whole line-of-sight of the galaxy, especially outside the midplane where the thermal electron density is expected to be largest. Hence, we only
used the C-band data for our further analysis.

After RM-synthesis, the vectors are corrected for the observed RM and the Galactic foreground rotation measure $\mathrm{RM_{fg}}$ (see Sect.~\ref{sec:rm}) and
hence are the intrinsic magnetic field vectors.

When analyzing the C-band D-array maps we detected polarized intensity PI in each of the 35 CHANG-ES galaxies except in NGC~4244. This galaxy is extremely faint even
in total intensity (TP) with values of $\lesssim 12\, \mu$Jy/beam as extended emission at 12$\arcsec$~HPBW. Hence we cannot expect to observe any PI from NGC~4244
with the sensitivity of our survey. In addition, seven other galaxies show extended PI that is $\lesssim 40 \, \mu$Jy/beam for which our RM-synthesis with the
parameters given in Sect.~\ref{sec:rm} does not give any polarization vectors or RM values. These galaxies are: NGC~2683, NGC~3003, NGC~3432, NGC~3877, NGC~4096,
NGC~5297, and NGC~5792. These 7~galaxies are, however, included in the stacked PI image of Fig.~\ref{stackedimage}. In 5~other galaxies (NGC~2992, NGC~4438, NGC~4594,
NGC~4845, and NGC~5084) the polarized intensity is dominated by the strong emission of the central source and no polarization can (clearly) be ascribed to the disk/halo
emission. The same is valid for UGC~10288 where the polarized intensity originates from a background double-lobed radio source \citep{irwin+2013}. These latter
6~galaxies were also excluded from the stacked PI image (Fig.~\ref{stackedimage}) and will not further be discussed here. Instead, they will be published in a separate
CHANG-ES paper.

Altogether, we are left with a sample of 21 spiral galaxies with significant linear polarization for which the CHANG-ES observations
allowed for the first time to determine reliable RM-values in galactic halos.

We present the polarization results at C-band for these 21 galaxies in Fig.~\ref{n660all} to Fig.~\ref{n5907all} \footnote {The corresponding fits-maps are available for
download at www.queensu.ca/changes}.
Each galaxy is presented on a separate page with six panels. They contain: total intensity (TP, Stokes I) with contours at 5, 15, 25 $\sigma$
r.m.s. and intrinsic polarization vectors (upper left, \emph{panel~1}), polarized intensity (PI) with intrinsic polarization vectors and the contours of TP as given in
panel~1 (upper right, \emph{panel~2}), percentage polarization (referred to as Perc in the Figures) with TP contours of panel~1 (mid left, \emph{panel~3}), rotation measure
(RM) with TP contours of panel~1 (mid right, \emph{panel~4}), errors in RM with TP contours of panel~1 (lower left, \emph{panel~5}), rotation
measure with intrinsic polarization vectors and TP contours of panel~1 (lower right, \emph{panel~6}). The total intensity maps show the observations at C-band D-array only,
while the polarization maps are made from C-band, D- and C-array combined as described in Sect.~\ref{sec:rm}. The maps in the 6 panels for each galaxy are equal in size given
by the extent of the total emission (5~$\sigma$ r.m.s.)
and are centered on the nucleus of the galaxy. All maps have an angular resolution of 12$\arcsec$ and are corrected for the primary beam. The latter leads to an increase
of the r.m.s. noise towards the map edges which is visible in some of the images, especially those that are observed with two pointings (given in
Sect.~\ref{sec:align_scale}). Corresponding high values at the edges of the PA, RM- and RM error maps were excluded.
In the percentage map, values $\le 0 \%$ and $\ge 70 \%$ are excluded as they are unphysical. The TP contours of panel 1 are presented in all
panels of each galaxy to serve as orientation. The length of the intrinsic magnetic field vectors is proportional to PI at this position.

Similar to NGC~3044, the majority of the intrinsic RM-values of our sample galaxies are in the range between $+200\,\mathrm{rad/m}^2$ and $-200 \,\mathrm{rad/m}^2$. The
RM ranges found within the disk and halo (i.e. excluding other areas) are explicitly given in Table~\ref{RM} for all galaxies. NGC~4631 is the galaxy with the
highest RM values, comparable to the values found by \cite{mora+2013}. These high values are mainly found along the disk and disk/halo interface of NGC~4631 and are related
to the high thermal electron density there \citep{mora+2019a, mora+2019b}.

Though the polarized intensity is in general quite weak in some parts of the galaxies presented in Fig.~\ref{n660all} to Fig.~\ref{n5907all}, it leads to remarkable percentage
polarization within the outer contours of total intensity in all galaxies. This will be discussed in Sect.~\ref{sec:uniformity}. The intrinsic magnetic
field vectors and RM values are only given in those areas where PI is larger than about $40 \, \mu \mathrm {Jy/beam}$, due to the cut-off levels used
for RM-synthesis as explained in Sect.~\ref{sec:rm}.

\section{Large-scale magnetic fields}
\label{sec:large-scale}

The majority of the 21 galaxies presented in Fig.~\ref{n660all} to Fig.~\ref{n5907all} show intrinsic magnetic field vectors within a significant part of their
halos and/or projected disks as summarized in Table~\ref{pattern}. This is not the case only for four of these galaxies, namely NGC~2613, NGC~3628, NGC~4302, and NGC~5907.
The reason is for NGC~2613
(Fig.~\ref{n2613all}), NGC~4302 (Fig.~\ref{n4302all}), and NGC~5907 (Fig.~\ref{n5907all}) that they are too faint in linear polarization within most parts of their
halos/disks, while NGC~3628 (Fig.~\ref{n3628all}) has, in addition, a strong central source. The intrinsic magnetic field orientations exhibit a regular and smooth
pattern in 14 of the remaining 17 galaxies. Only NGC~891 (Fig.~\ref{n891_12all}), NGC~3556 (Fig.~\ref{n3556all}), and NGC~4631 (Fig.~\ref{n4631_12all})
show a more patchy structure in their intrinsic magnetic field vectors.

\begin{table}
      \caption[]{\label{pattern} Magnetic field angles and RM pattern analysis}

     $$
         \begin{tabular}{lcc}
            \hline
            \noalign{\smallskip}
            \multicolumn{1}{c}{regular}    &  more patchy  & only visible in small areas  \\
            \noalign{\smallskip}
            \hline
            \noalign{\smallskip}

            NGC~660   &  NGC~891$\,^\mathrm{a,b}$   &  NGC~2613 \\
            NGC~891   &  NGC~3556  &  NGC~3628 \\
            NGC~2820  &  NGC~4631$\,^\mathrm{c}$  &  NGC~4302 \\
            NGC~3044  &     &         NGC~5907 \\
            NGC~3079  &     &                  \\
            NGC~3448  &     &                  \\
            NGC~3735  &     &                  \\
            NGC~4013  &     &                  \\
            NGC~4157  &     &                  \\
            NGC~4192  &     &                  \\
            NGC~4217  &     &                  \\
            NGC~4388  &     &     \\
            NGC~4565  &     &                  \\
            NGC~4631$\,^\mathrm{d}$  &     &                  \\
            NGC~4666  &     &                  \\
            NGC~5775  &     &                  \\

   \noalign{\smallskip}
            \hline
         \end{tabular}
               $$
\begin{list}{}{}
\item[$^{\mathrm{a}}$] at $20 \arcsec$~HPBW (corresponding to 880~pc)
\item[$^{\mathrm{b}}$] at $12 \arcsec$~HPBW (corresponding to 530~pc)
\item[$^{\mathrm{c}}$] at $20 \farcs 5 $~HPBW (corresponding to 730~pc)
\item[$^{\mathrm{d}}$] at $12 \arcsec$~HPBW (corresponding to 430~pc)
\end{list}

\end{table}

A smooth pattern of the intrinsic magnetic field vectors may indicate a large-scale magnetic field but can also originate from anisotropic random magnetic fields,
e.g., compressed fields with reversing directions. However, as anisotropic random magnetic fields have different directions along the line of sight (LoS), the differential
RMs along the LoS have different signs and at least partly cancel each other. Only the detection of RMs of reasonable strengths along the LoS indicates a regular magnetic
field component. If, in addition, the different RM observed along different LoS presented in a map have a smooth distribution of regions with positive and negative RM values
on scales significantly larger than the beam size, a regular (coherent) magnetic field is indicated. In this case, the intrinsic polarization vectors are also expected to
be ordered on scales larger than the synthesized beam size.

Remarkably, in all these areas of regular magnetic field vectors in the galaxies mentioned above, we also find significant RM values that vary smoothly and show the same sign
over huge areas when compared to the telescope's beam of $12 \arcsec$~HPBW. This already indicates a regular magnetic field component parallel to the LoS.
The RM values in the Figures are presented in two different color wedges, one for positive RM values (blue-green), and one for negative RM values (gray-scale). This makes it
easier to distinguish between the different magnetic field directions. The errors in RM are between a few and $40\,\mathrm {rad/m}^2$, and much smaller than most of the
observed RMs. Hence, the observed large-scale changes of sign in RM are real and can only be interpreted as a change of the direction of the parallel component of the regular
magnetic field in the corresponding regions of the galaxy. Altogether, we conclude that we detected coherent magnetic fields in the halos of our sample spiral galaxies.

In three galaxies the magnetic field vectors and RM values could only be detected along the projected disk. These are NGC~5907 and NGC~4013 (see also
\cite{stein+2019b}), and also probably NGC~4565 (see \citealt{schmidt+2019}) where we detect magnetic field vectors and RM values only up to about 1.0~kpc distance
from the midplane. For the other galaxies, the distances from the midplane up to which the large-scale magnetic field could be detected, are given in Table~\ref{structure}.
As they are measured in the sky plane, they are lower limits. However, with the inclination $i  \ge  75\degr$ of our CHANG-ES galaxies, their deprojected values may only be
larger by 10\% at most. We stress that these values are limited by the sensitivity of our polarization observations (and the cut levels we used for a reliable RM-synthesis)
and hence are smaller than the physical extent of the large-scale fields in the halo.

\begin{table*}
      \caption[]{\label{structure} Large-scale magnetic fields in the halo}

     $$
         \begin{tabular}{lcccccc}
            \hline
            \noalign{\smallskip}
            \multicolumn{1}{c}{source} &  vertical & RMTL$\,^\mathrm{b}$ &  magnetic field   & $\Delta$ RMTL$\,^\mathrm{c}$ & Hubble type$\,^\mathrm{d}$  & SFR$\,^\mathrm{e}$ \\
                                      &   detection$\,^\mathrm{a}$ &     &  orientation & \\
                                      &   [kpc] & & & [kpc]  \\
            \noalign{\smallskip}
            \hline
            \noalign{\smallskip}

            NGC~660   &  1.7   &      &              &       &  SBa  &  3.3 \\
            NGC~891   &  2.5   &      &              &       &  Sb   &  1.9 \\
            NGC~2820  &  2.9   &  yes & horizontal   &  2.0  &  SBc  &  1.4 \\
            NGC~3044  &  3.3   &  yes & vertical     &  2.2  &  SBc  &  1.8 \\
            NGC~3079  &  4.9   &      &              &       &  SBcd &  5.1 \\
            NGC~3448  &  2.7   & possible & possibly vertical & &  S0-a &  1.8 \\
            NGC~3556  &  2.5   &      &              &       &  SBc  &  3.6 \\
            NGC~3735  &  4.6   & possible & horizontal &     &  Sc   &  6.2 \\
            NGC~4157  &  2.5   &  yes & horizontal   &  2.2  &  SABb &  1.8 \\
            NGC~4192  &  1.3   &  yes & horizontal   &  1.3  &  SABb &  0.8 \\
            NGC~4217  &  3.1   &  yes & horizontal   &  2.4  &  Sb   &  1.9 \\
            NGC~4388  &  2.4   &      &              &       &  Sb   &  2.4 \\
            NGC~4631  &  2.8   &  yes & vertical &   1.9$\,^\mathrm{f}$ & SBcd & 2.6 \\
            NGC~4666  &  5.3   &      &              &       &  SABc  & 10.5 \\
            NGC~5775  &  5.4   &      &              &       &  Sbc   & 7.6  \\

            \noalign{\smallskip}
            \hline
         \end{tabular}
               $$

\begin{list}{}{}
\item[$^{\mathrm{a}}$] in the sky plane
\item[$^{\mathrm{b}}$] Rotation measure transition line
\item[$^{\mathrm{c}}$] Projected distance between RMTLs
\item[$^{\mathrm{d}}$] From HYPERLEDA, as used in \cite{irwin+2012a}
\item[$^{\mathrm{e}}$] this is SFR$_{revised}$ in \cite{vargas+2019}
\item[$^{\mathrm{f}}$] In the northern half (as detected in \citealt{mora+2019b})
\end{list}

\end{table*}

\begin{figure*}
\centering
\includegraphics[width=9.0 cm]{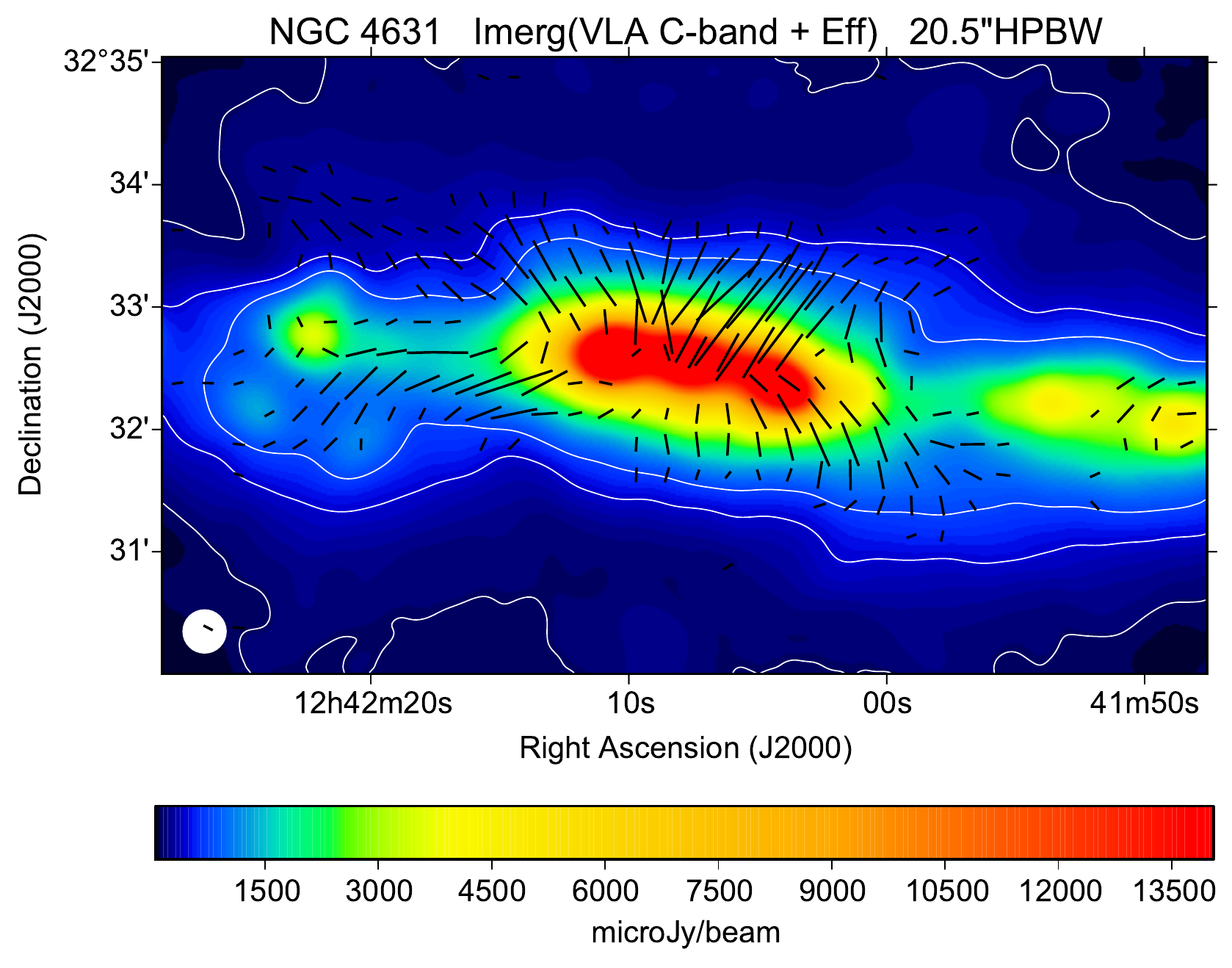}
\includegraphics[width=9.0 cm]{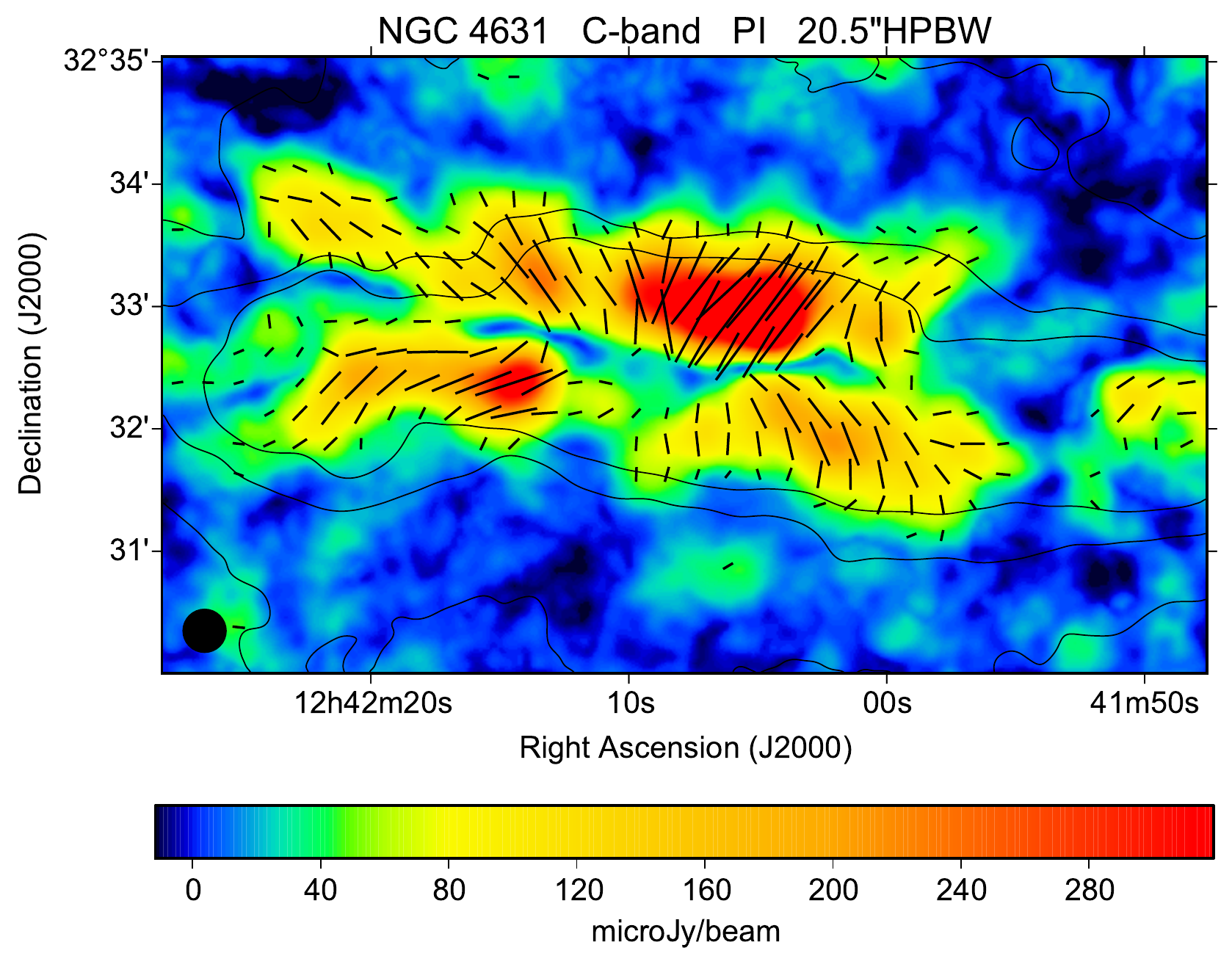}
\includegraphics[width=9.0 cm]{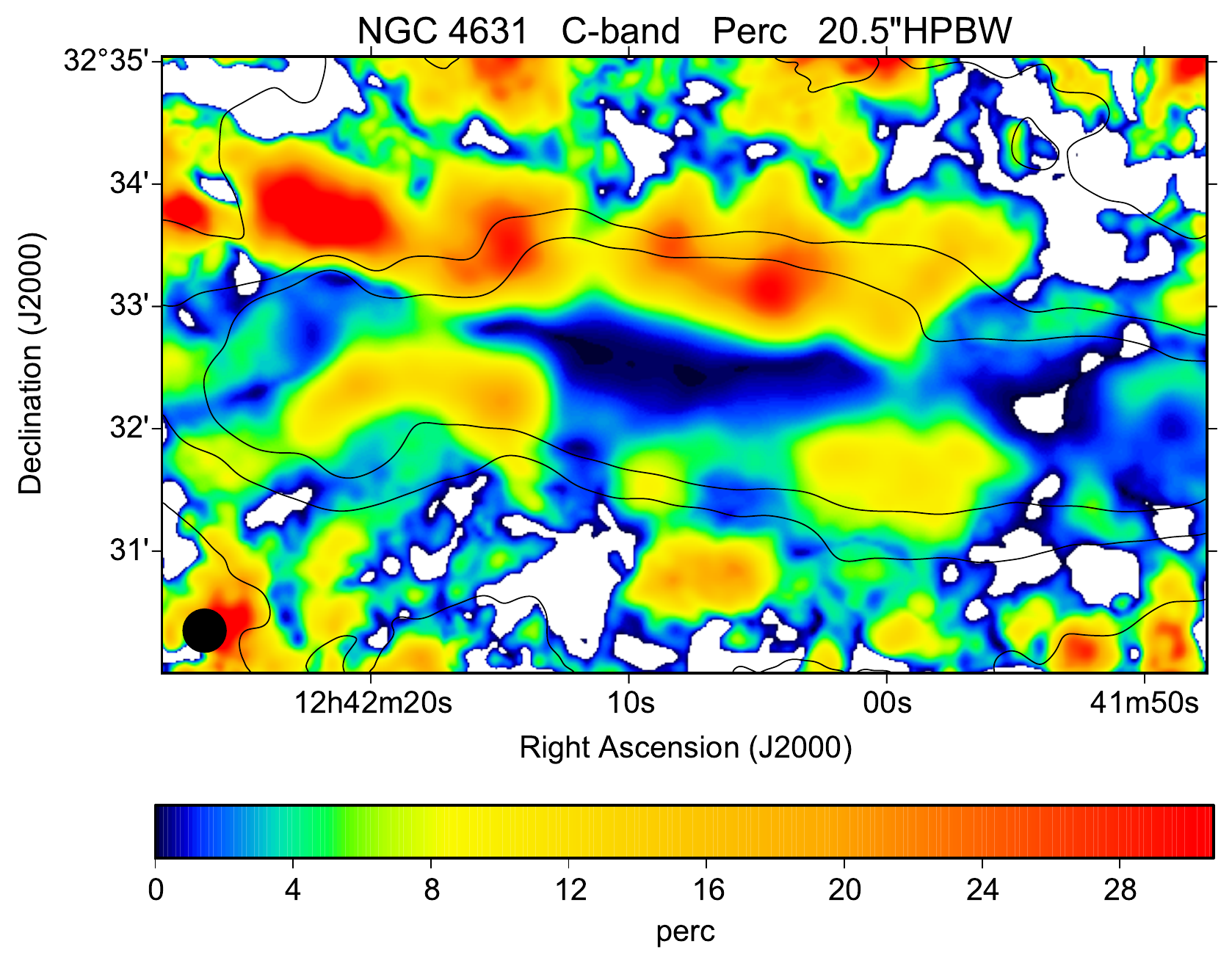}
\includegraphics[width=9.2 cm]{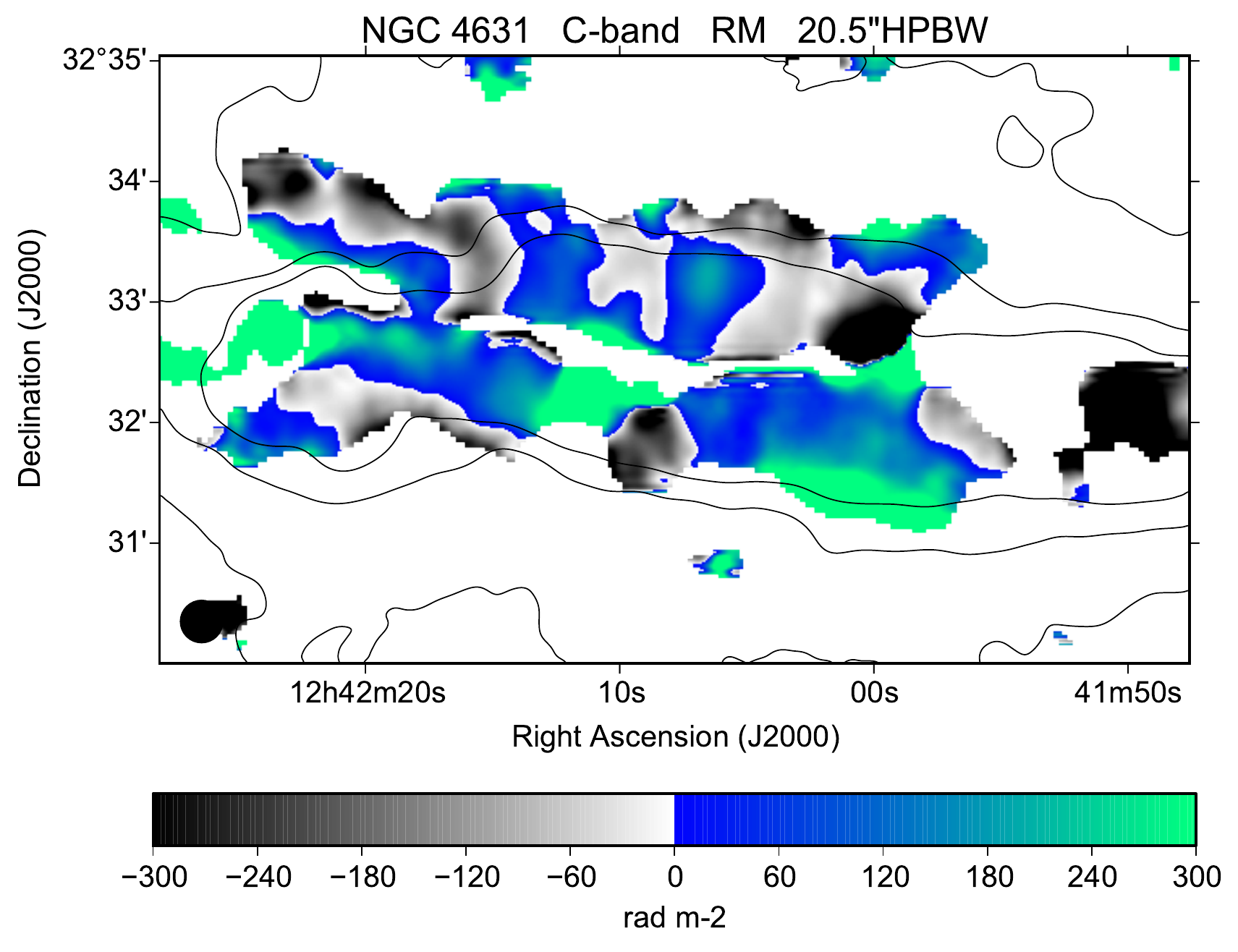}
\includegraphics[width=9.0 cm]{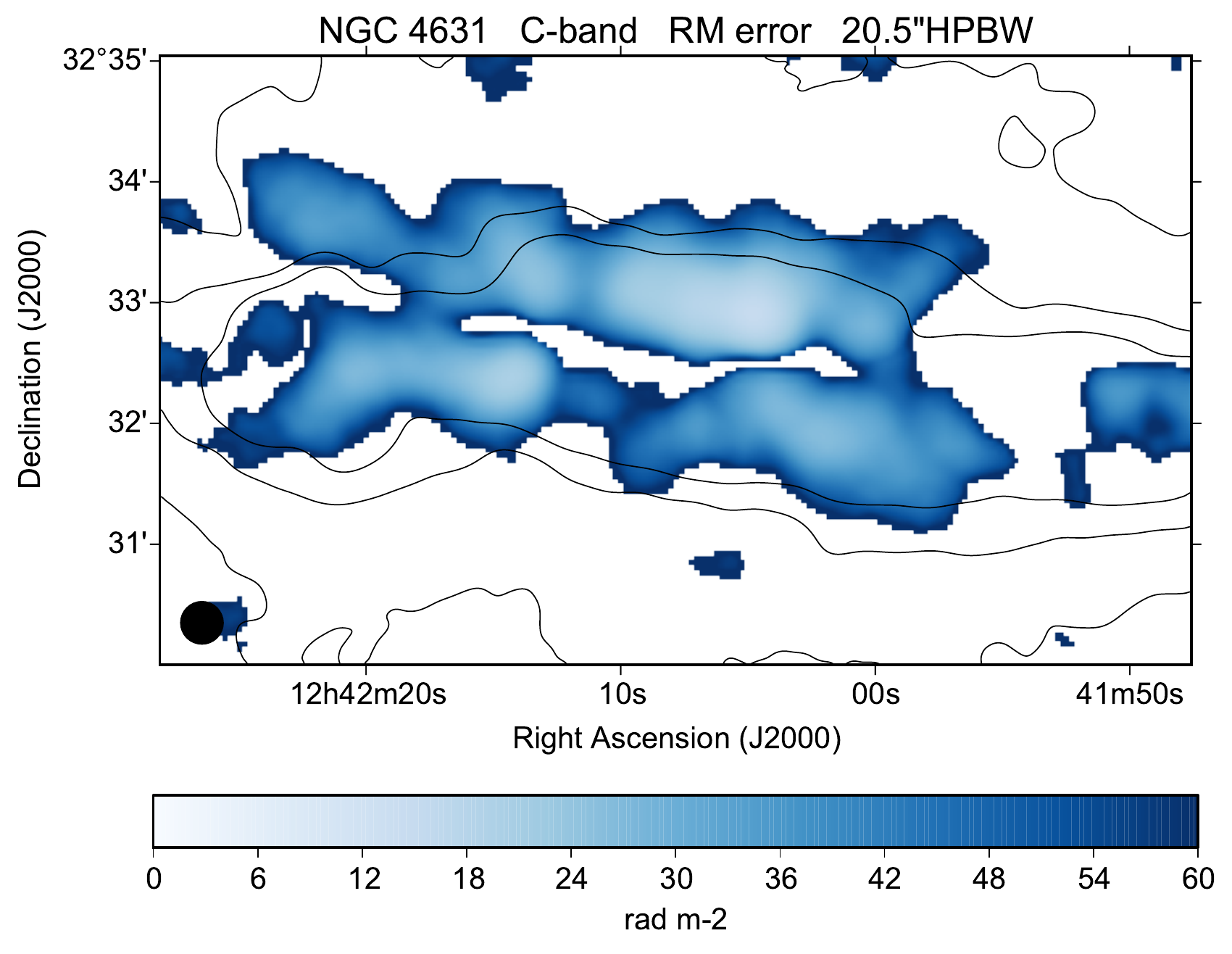}
\includegraphics[width=9.1 cm]{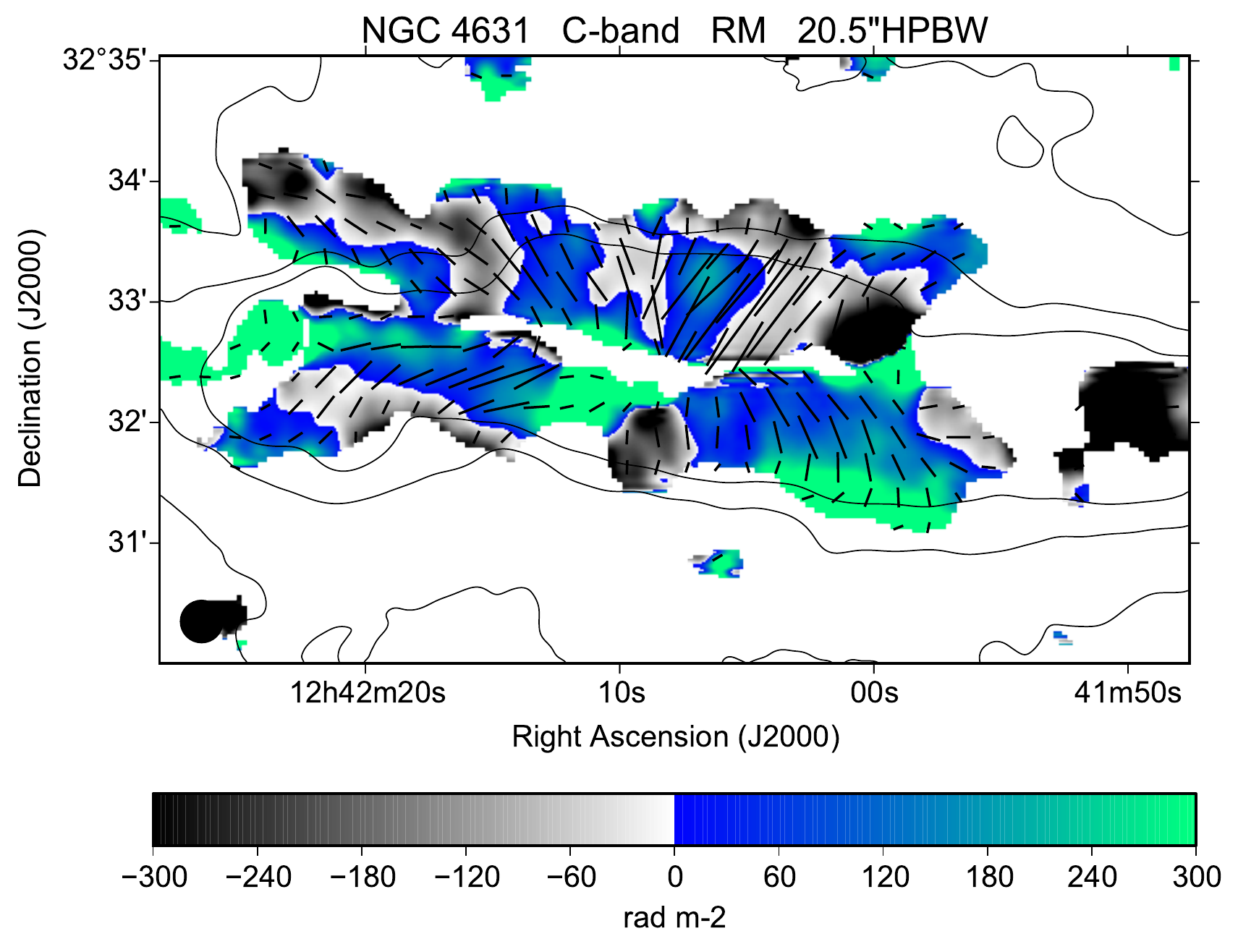}
\caption{Polarization results for NGC~4631 at C-band and 20~\farcs 5 HPBW. The contour levels (TP) are 150, 450, and 750 $\mu$Jy/beam.
}
\label{n4631_20all}
\end{figure*}

It remains the question why NGC~4631 does not show such a smooth, regular pattern in the magnetic vectors and RM values in Fig.~\ref{n4631_12all} though this galaxy has been
reported recently to show large-scale magnetic field reversals \citep{mora+2019b}. These authors presented the same polarization observations with an angular resolution
of $20 \farcs 5$, while the maps in Fig.~\ref{n4631_12all} are made with $12\arcsec$~HPBW. At the distance of NGC~4631, $12\arcsec$~HPBW correspond to 430~pc while
$20 \farcs 5$ correspond to 740~pc. For comparison, all maps of NGC~4631 with a resolution of $20 \farcs 5$ are presented again in our notations in Fig.~\ref{n4631_20all}.
And indeed, at this linear resolution of about 800~pc also NGC~4631 shows a smooth and regular pattern in the magnetic field vectors as well as in the RMs. This example
indicates that the larger-scale coherent magnetic field component in NGC~4631 has typical scales larger than about 800~pc, and is accompanied by a smaller-scale magnetic field
component. As soon as the resolution is better than the scale of the coherent magnetic field, we start to resolve the smaller-scale magnetic field and hence reduce the
uniformity of the vectors and the RM pattern.

Though the galaxies are presented with the same angular resolution ($12\arcsec$~HPBW), the detected linear polarization corresponds to different linear scales
for each single galaxy due to their different distances. These distances are listed in Table~\ref{RM}, leading to linear resolutions between 260~pc (for NGC~4244) and
2.44~kpc (for NGC~3735) for the angular resolution of $12 \arcsec$ ~HPBW.

\begin{figure*}
\centering
\includegraphics[width=7.5 cm]{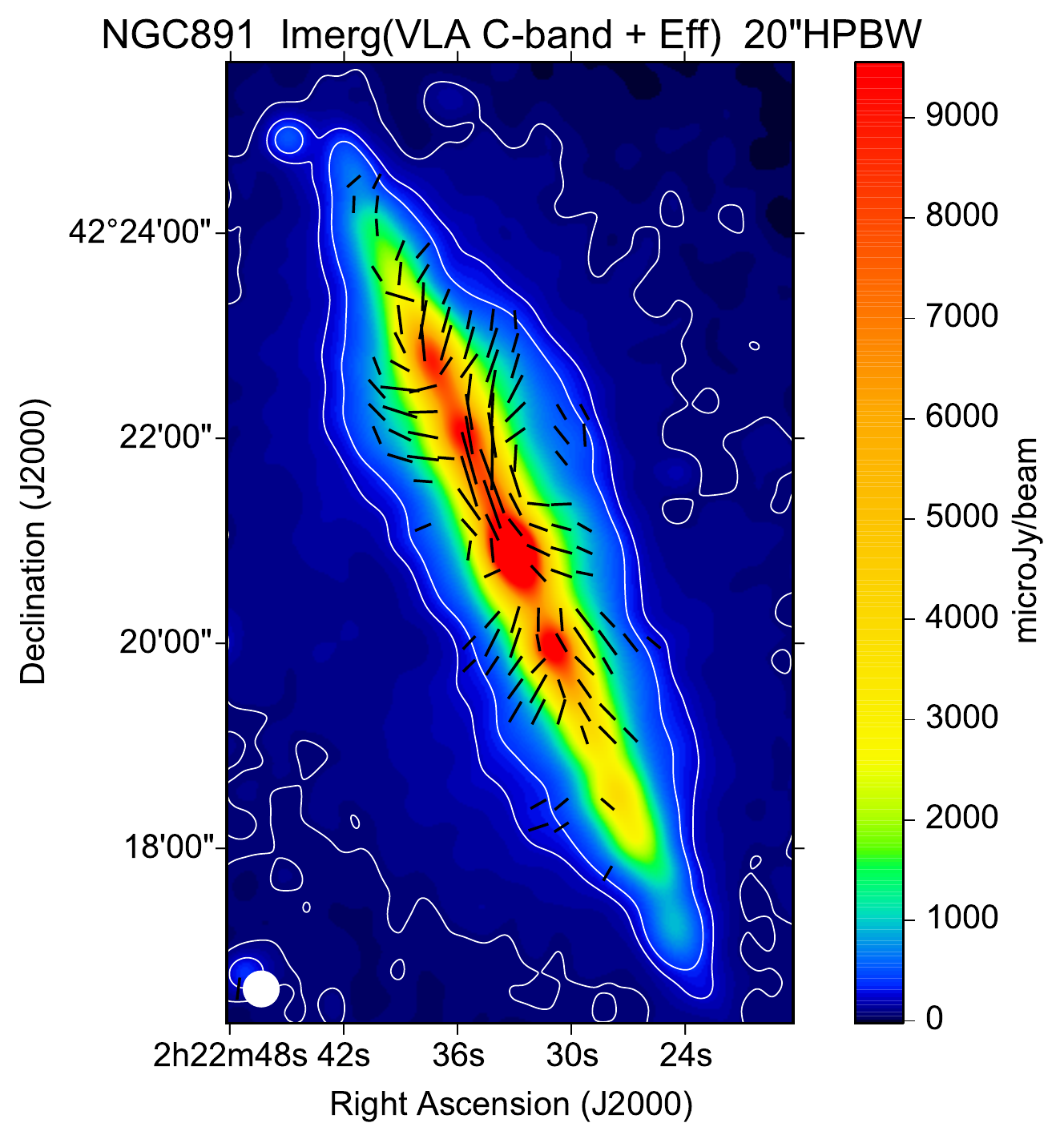}
\includegraphics[width=7.4 cm]{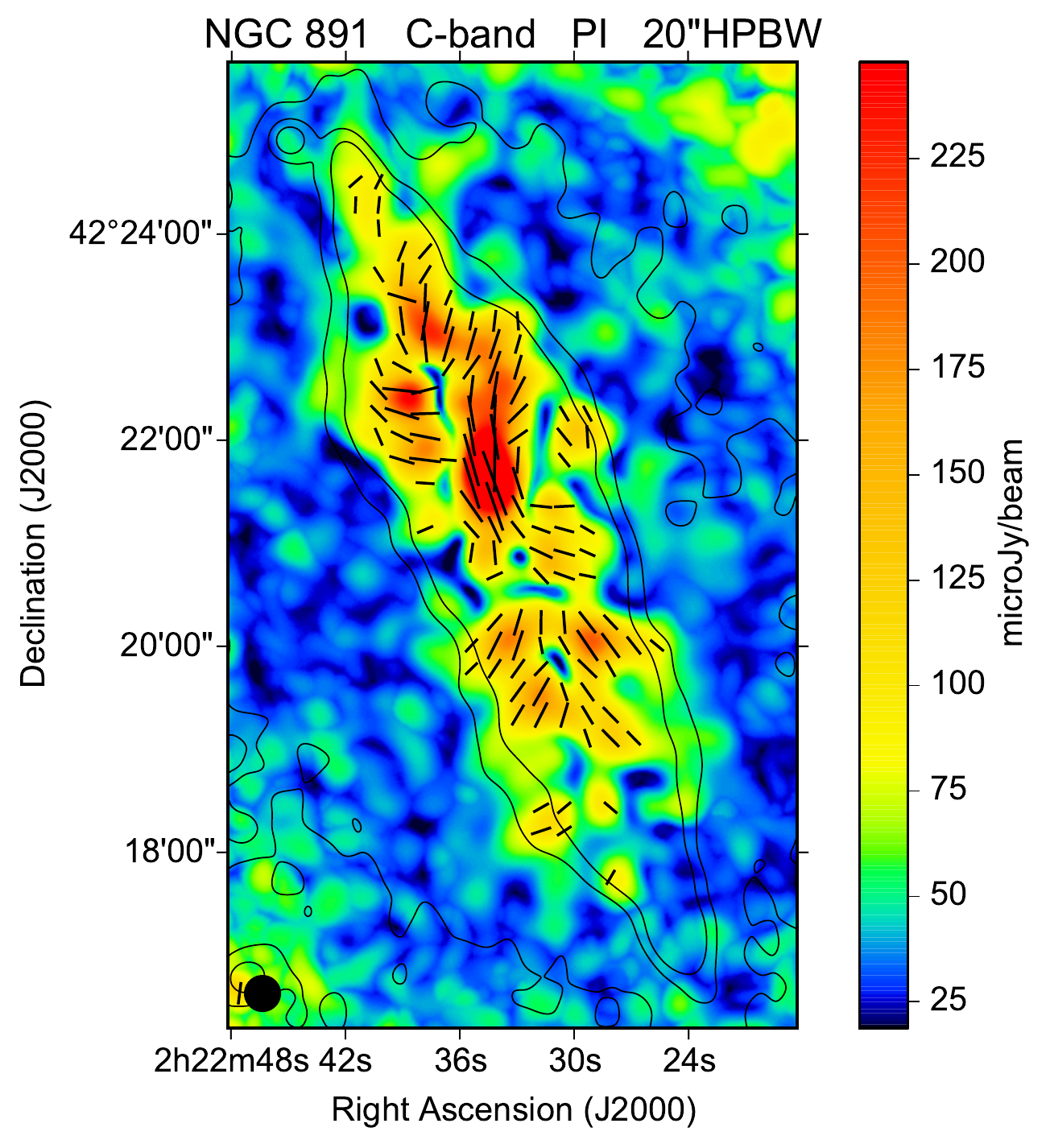}
\includegraphics[width=7.5 cm]{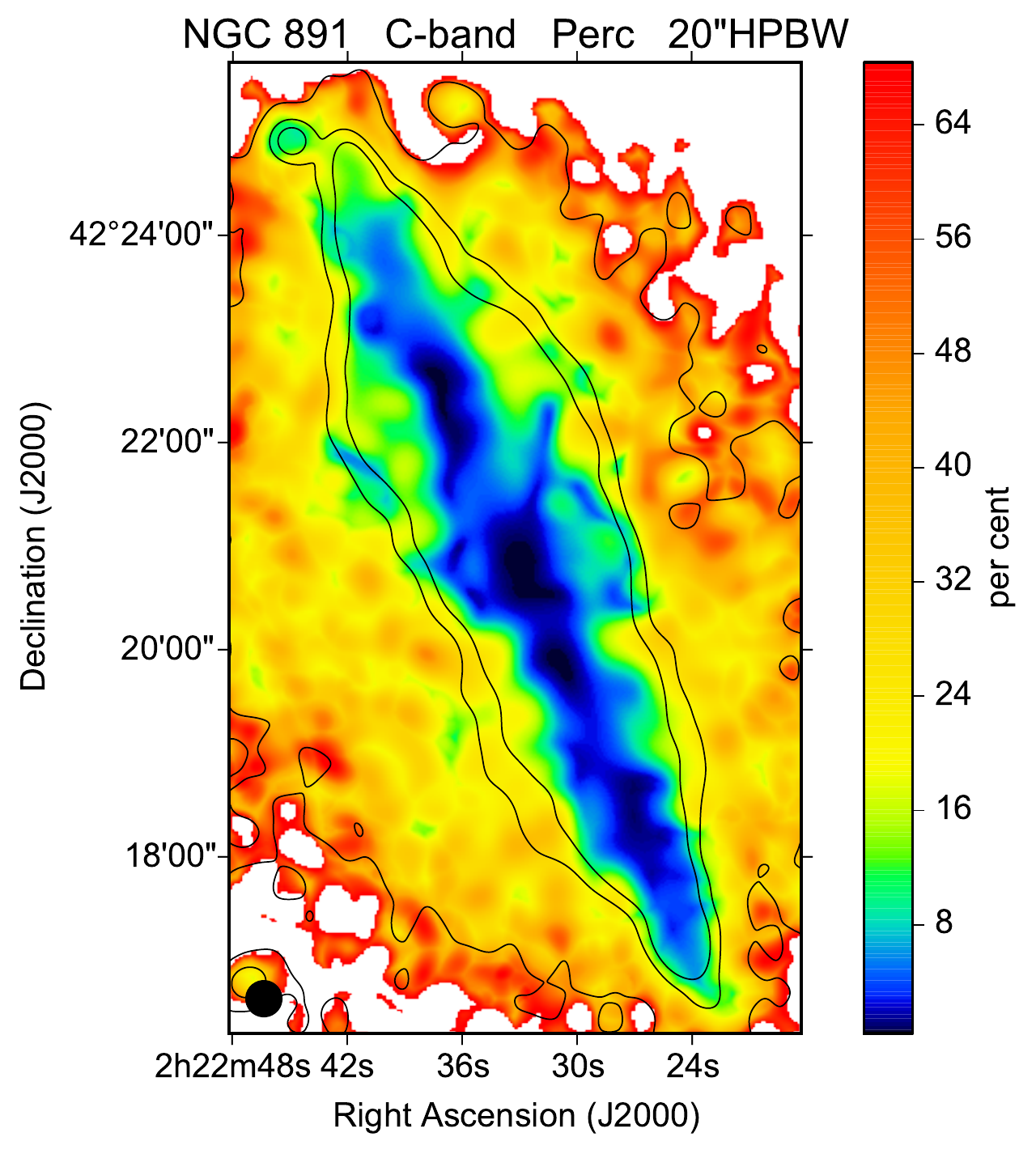}
\includegraphics[width=7.7 cm]{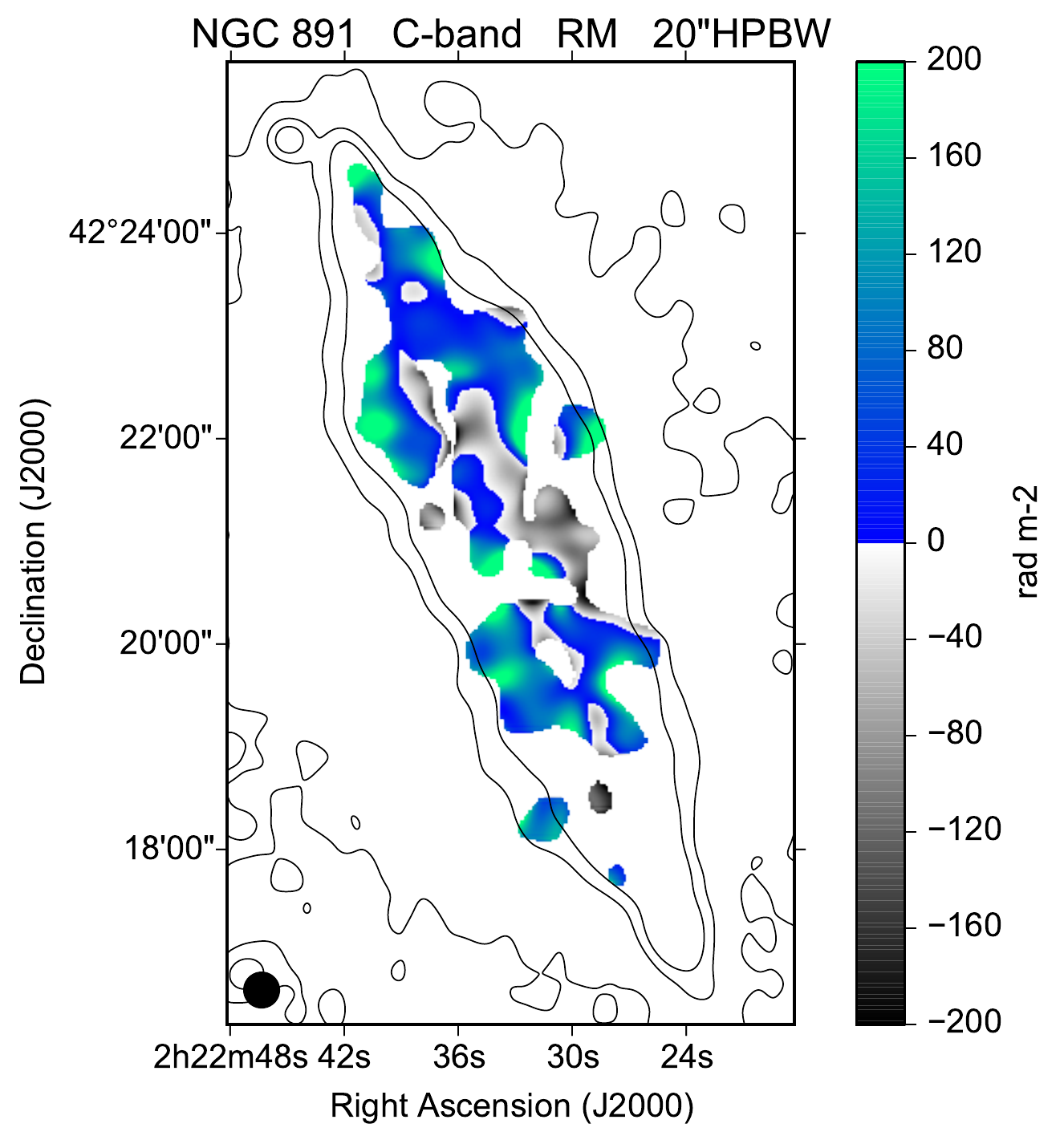}
\includegraphics[width=7.5 cm]{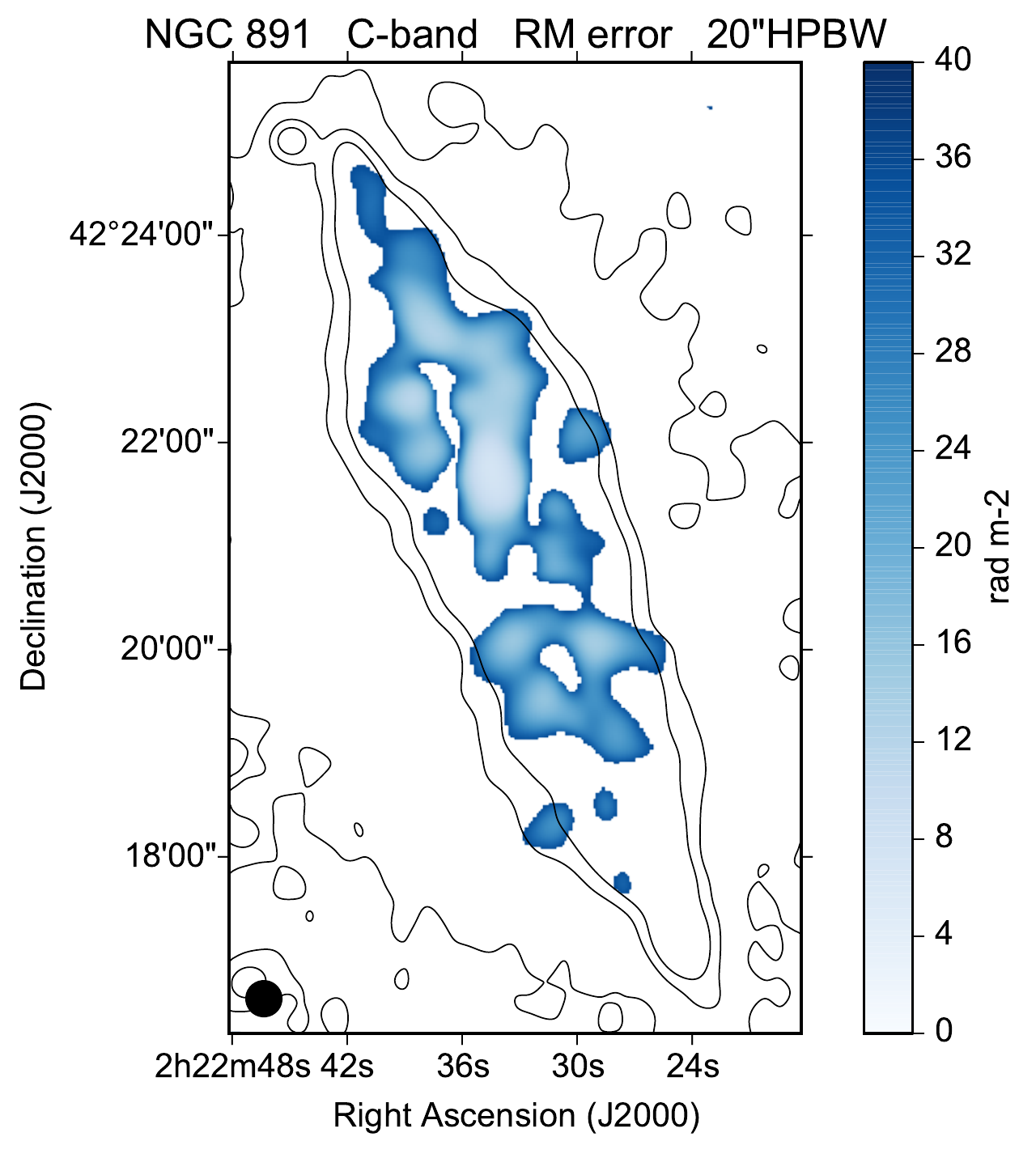}
\includegraphics[width=7.7 cm]{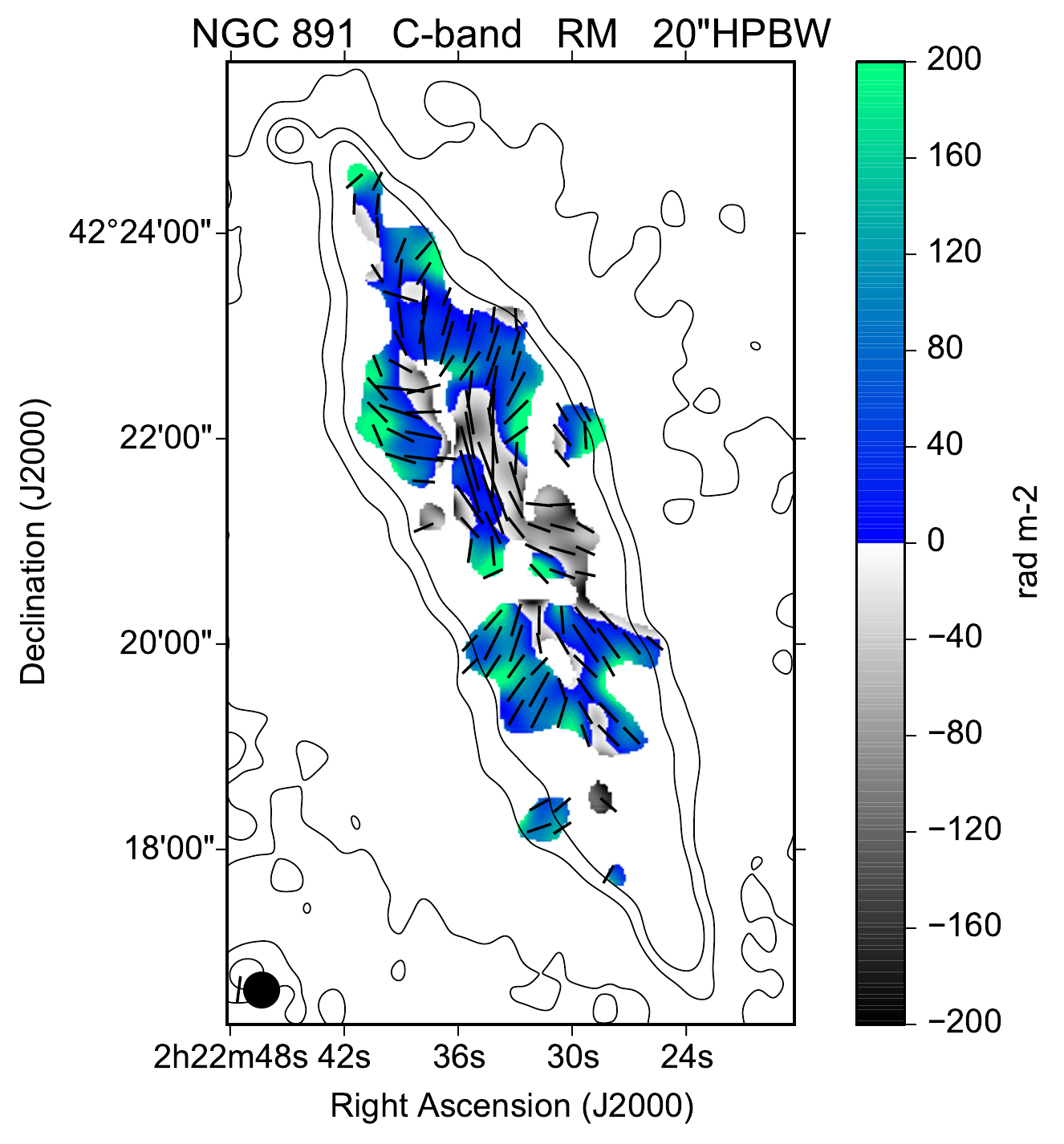}
\caption{Polarization results for NGC~891 at C-band and $20 \arcsec$ HPBW. The contour levels (TP) are 75, 225, and 375 $\mu$Jy/beam.
}
\label{n891_20all}
\end{figure*}

Also NGC~891 shows a patchy structure at scales larger than the HPBW of $12 \arcsec$ (corresponding to 530~pc) in their magnetic field vectors and their RM values, although it was referred
to as a kind of prototype of an X-field structure by observations at the 100-m telescope at Effelsberg at much lower resolution of $84 \arcsec$ ~HPBW \citep{krause2009}.
Again, we smoothed our Q- and U-cubes to $20 \arcsec$~HPBW (corresponding to 880~pc) before RM-synthesis. The corresponding results are presented in Fig.~\ref{n891_20all}.
The magnetic field angles as well as the RM-values show again a more regular pattern that is not only due to a higher signal-to-noise in the smoothed maps. This is best
visible in the RM-maps (panel~4 and panel~5) in the central part of NGC~891.

Nevertheless, the global structure in NGC~891 is still somewhat patchy even at 880~pc linear
resolution. There is also NGC~3556 observed near the detection limit that shows a somewhat patchy pattern. This galaxy has a linear resolution of 820~pc, hence comparable to NGC~891 at
$20 \arcsec$~HPBW. Also NGC~3556 may look more regular at a lower angular resolution which in this case may also be due to a higher signal-to-noise ratio.

\begin{figure*}
\centering
\includegraphics[width=8.8 cm]{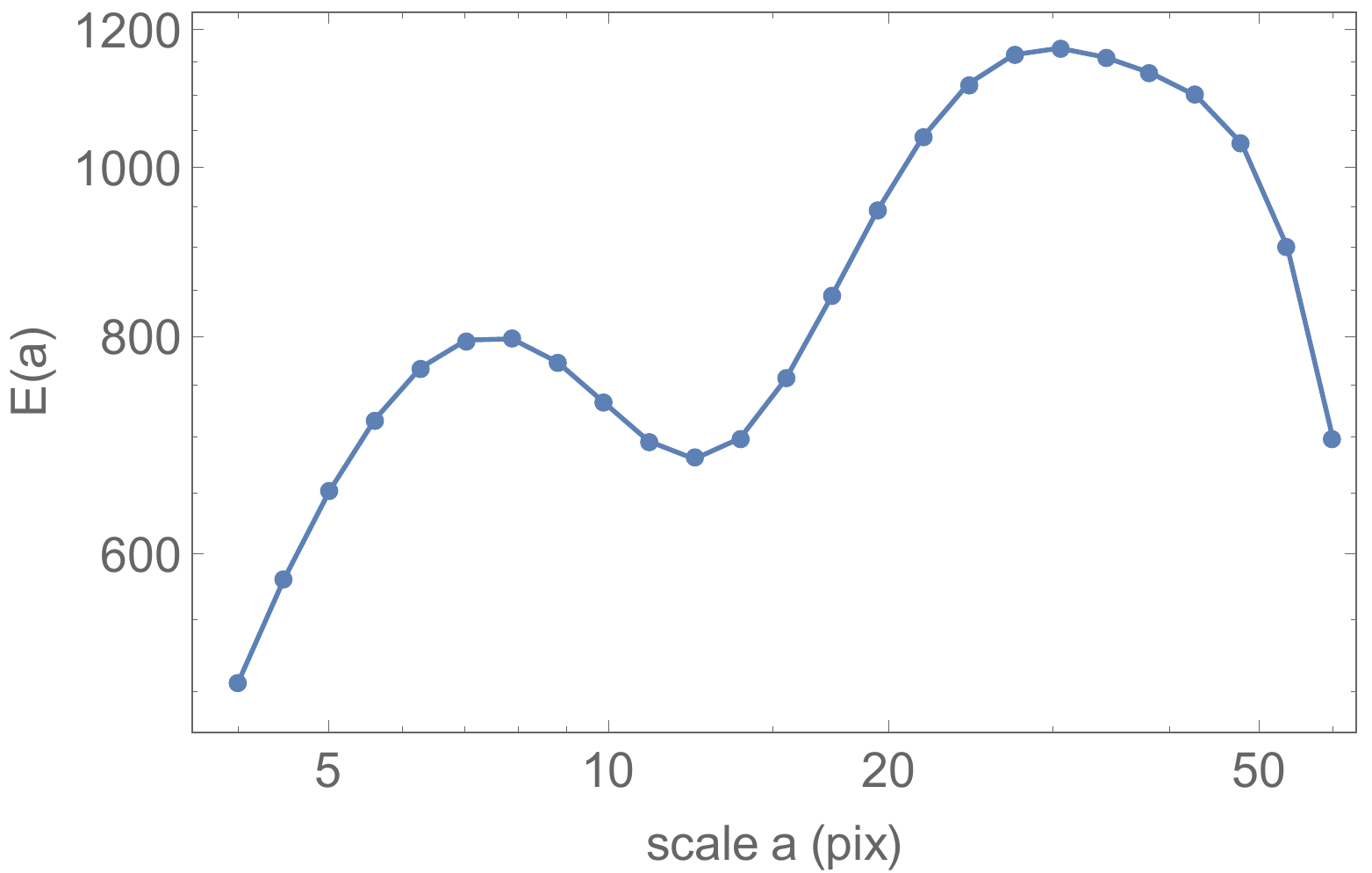}
\includegraphics[width=8.8 cm]{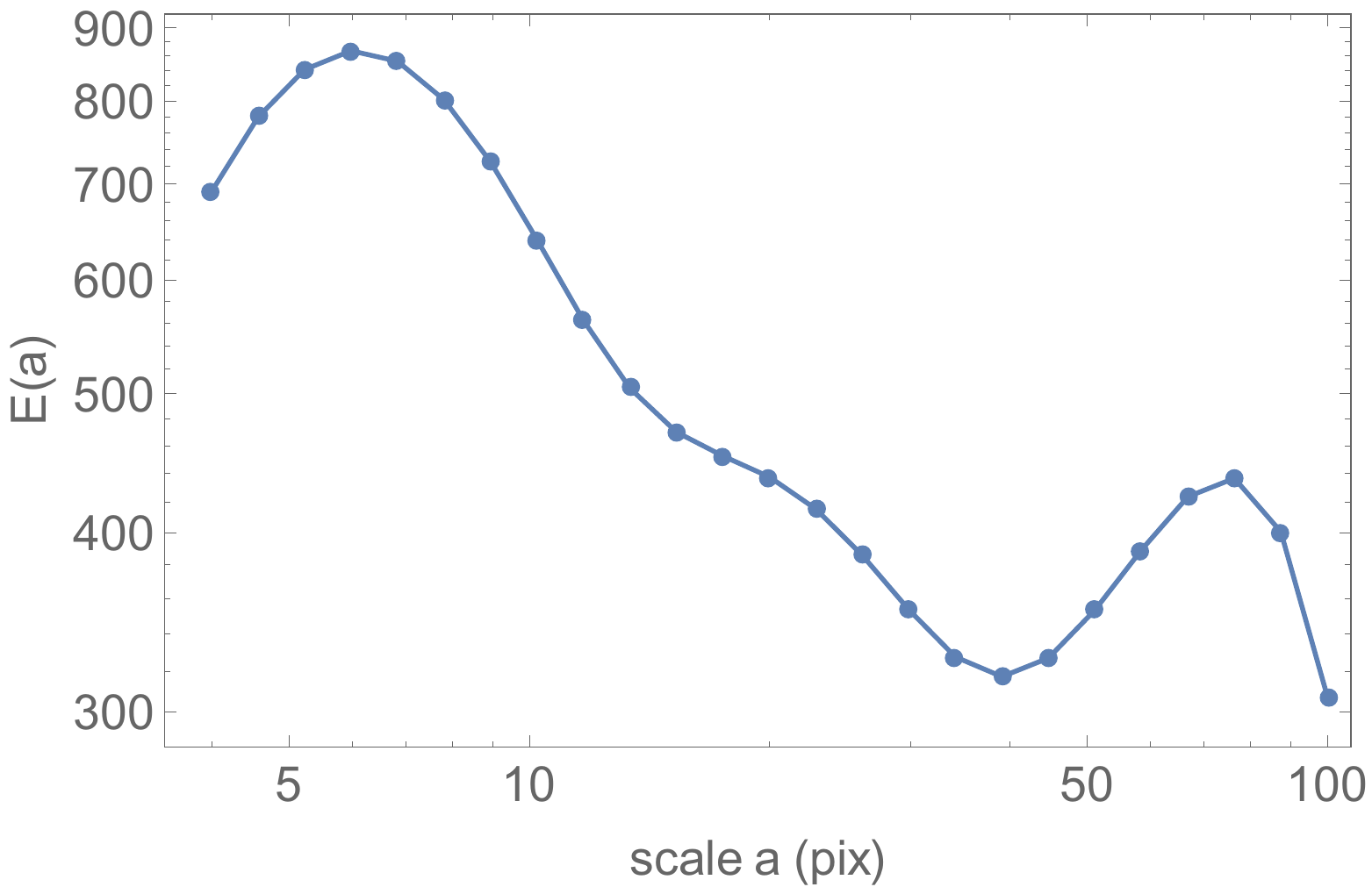}
\caption{Wavelet power spectra of the maps of polarized intensity ($PI$) at C-band for NGC~891 (\textbf{left panel}) and for NGC~4631 (\textbf{right panel}) at $12 \arcsec$~HPBW.
Scale $a$ is the scale of the wavelet function in pixelsize ($1 \farcs 5$) which is half of the size of the structure in the map (see text for details). $E(a)$ is the energy density of thewavelet transform on scale $a$.
}
\label{wavelet}
\end{figure*}

To quantify the relative importance of angular scales, we applied wavelet transformations to the maps in polarized intensity of NGC~891 and NGC~4631 at $12 \arcsec$~HPBW,following \citet{frick+2016}. We used `Texan Hat' functions that provide a higher resolution in spatial frequency space than `Mexican Hat' functions.
The wavelet power spectra are shown in Fig.~\ref{wavelet}.
The smallest scale of $a=4$ corresponds to the HPBW of $12 \arcsec$. The power spectra show two prominent peaks. The first ones are located at small scales of $22\, \arcsec$
($\approx$~970~pc) and $18\,\arcsec$ ($\approx$~670~pc) in NGC~891 and NGC~4631, respectively. These scales represent structures of the magnetic field smaller than the
large-scale coherent field, but clearly larger than the telescope resolution.
The second peaks in the power spectra are located at scales of about $1.5\, \arcmin$ ($\approx$~4~kpc) and $3.8\, \arcmin$ ($\approx$~8~kpc) in NGC~891 and NGC~4631, respectively, are due to the large-scale polarized emission from the disk.

It is striking that we detect a smooth pattern in the magnetic field angles and
the RMs only in galaxies with a linear resolution corresponding to scales larger than about 700~pc. Hence we conclude that the large-scale (coherent) magnetic fields in the halos have typical scales of about 1~kpc or larger.

With our present observations we can trace the magnetic field vectors and RMs up to about 10~kpc (mean value $5 \pm 2$~kpc) distance from the
midplane of the disk (see Table~\ref{structure}), hence far into the halo. The intrinsic magnetic field vectors give the orientation of the large-scale (coherent) magnetic
field component within the sky plane (perpendicular magnetic field component), while the RMs give the magnetic field component along the line of sight (parallel magnetic
field component) of the large-scale magnetic field. Unfortunately, they are averaged along the long line of sight through the disk and halo in edge-on galaxies. Though
RM-synthesis partly corrects for differential depolarization, it can only be done 'on average' along the line of sight. This makes it difficult to identify the
3-dimensional structure of the large-scale magnetic field from the observations.

Our observations do not indicate a simple, large-scale magnetic field pattern in the halo like a dipole or quadrupole structure as expected by dynamo models.
They also do not exhibit a systematic single sign change of RM across the minor axis as would be expected for large-scale toroidal halo fields. The field patterns
seem to be more complicated. Only the RMs give the magnetic field direction of the parallel magnetic field component, while the
'vectors' just give the orientation of the perpendicular magnetic field components. At the transition line between positive and negative RMs (i.e. where the parallel
magnetic field component changes its direction) we observe that the perpendicular magnetic field components occur at any angles in our maps.

These RM-transition lines (RMTL) themselves show an interesting vertical pattern in NGC~4631 as detected by \cite{mora+2019b}: they are vertical to the galactic plane,
indicating several periodic large-scale field reversals in the northern halo of NGC~4631 (see Fig.~\ref{n4631_20all}). The intrinsic magnetic field vectors have only
small angles with respect to the RM-transition lines there. As presented and discussed in \cite{mora+2019b} the observations indicate giant magnetic ropes (GMR) rising perpendicular from the northern galactic disk into the halo (see their Fig.~7). In our sample galaxies we find another five (possibly seven) galaxies with vertical RM transition lines (see Table~\ref{structure}). In  two of them (NGC~3044 and NGC~3448) the magnetic field vectors are also oriented nearly in parallel with these lines. We conclude that NGC~3044
is another galaxy with GMR extending far into its halo. NGC~3448 can be considered as a candidate for GMR which needs to be confirmed by more sensitive polarization observations. In the remaining  four (possibly five) galaxies with vertical RM-transition lines, we find magnetic field vectors that are orientated roughly perpendicular to the RM-transition lines (best visible in NGC~2820 in Fig.~\ref{n2820all}).

We estimated the projected distances between the vertical RM-transition lines ($\Delta$~RMTL) parallel to the major axis. The values are
about 2~kpc (see Table~\ref{structure}).

\section{Degree of polarization and the ordered magnetic field}
\label{sec:uniformity}

We determined the degree of polarization (P$ = \rm{PI / TP} \cdot 100$) for our sample galaxies. This quantity may be affected by missing large-scale flux density as
expected in TP for galaxies that are larger than $4\arcmin$ in extent at C-band. As PI is usually structured on smaller scales, it is expected to be
less or barely affected by missing spacings.

For three of the large galaxies, NGC~891, NGC~4565 and NGC~4631, we used the VLA TP maps that were merged with Effelsberg 100-m single dish observations at 6~cm, as described
in \cite{schmidt+2019} for NGC~891 and NGC~4565, and \cite{mora+2019a} for NGC~4631. Most of the other galaxies are smaller than $4\arcmin$ in TP at C-band and are tested
to show no indication of obvious missing spacing problems \citep{krause+2018}. Only four other galaxies in our sample are larger than 4\arcmin~at C-band: NGC~3556,
NGC~3628, NGC~4192, and NGC~5907. We do not have single-dish observations for them. Hence, their degree of polarization may be affected by missing spacings in TP and can
be regarded as upper limits.

Values for which P is $< 0 \%$ and $\gtrsim 70 \%$ are unphysical and appear in map areas with low signal-to-noise ratios. Hence, they are excluded.
The resultant maps of the degree of polarization are presented in panels~3 of Fig.~\ref{n660all} to Fig.~\ref{n5907all} for each galaxy.

\begin{figure*}
\centering
\includegraphics[width=7.7 cm]{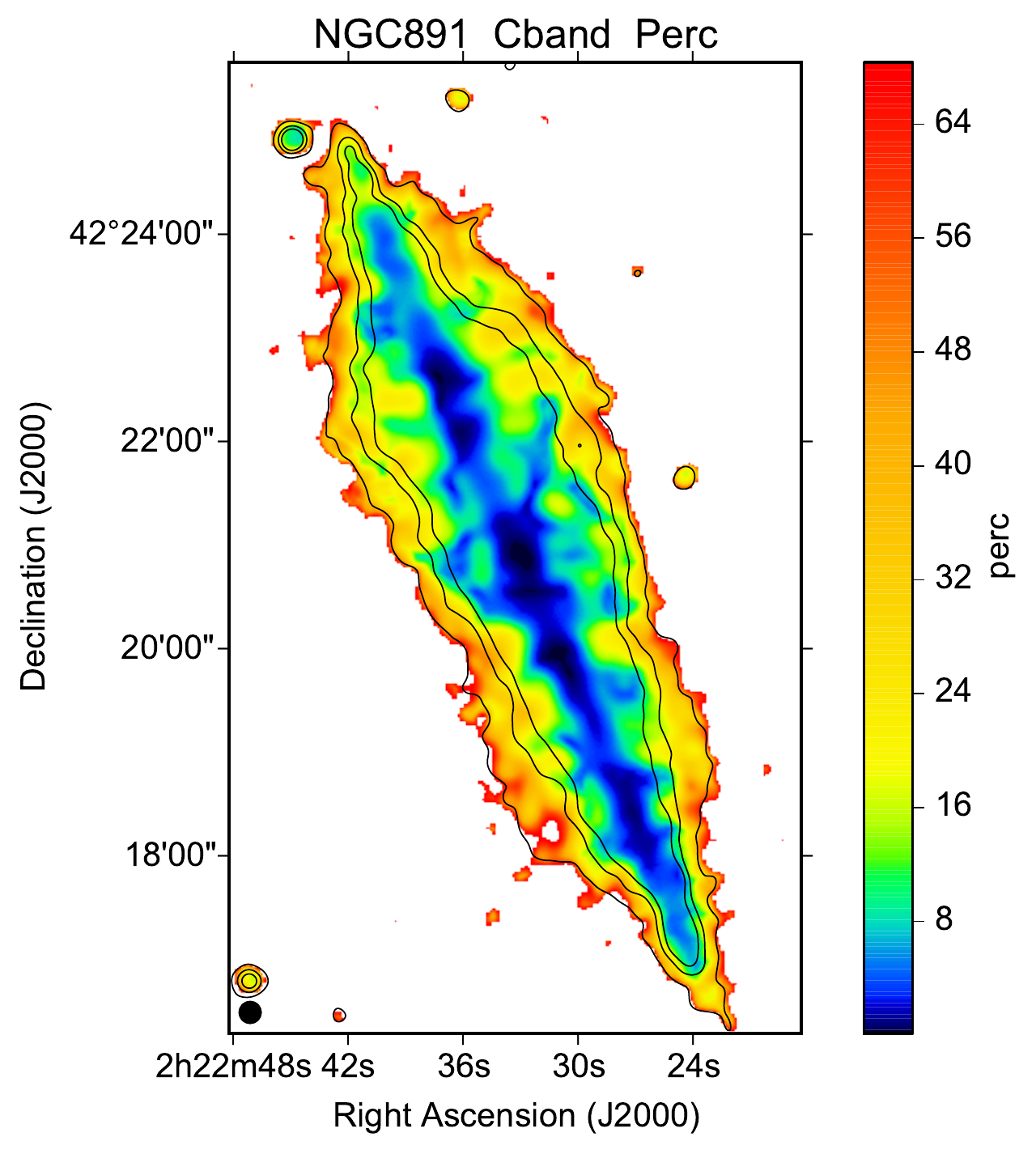}
\includegraphics[width=7.6 cm]{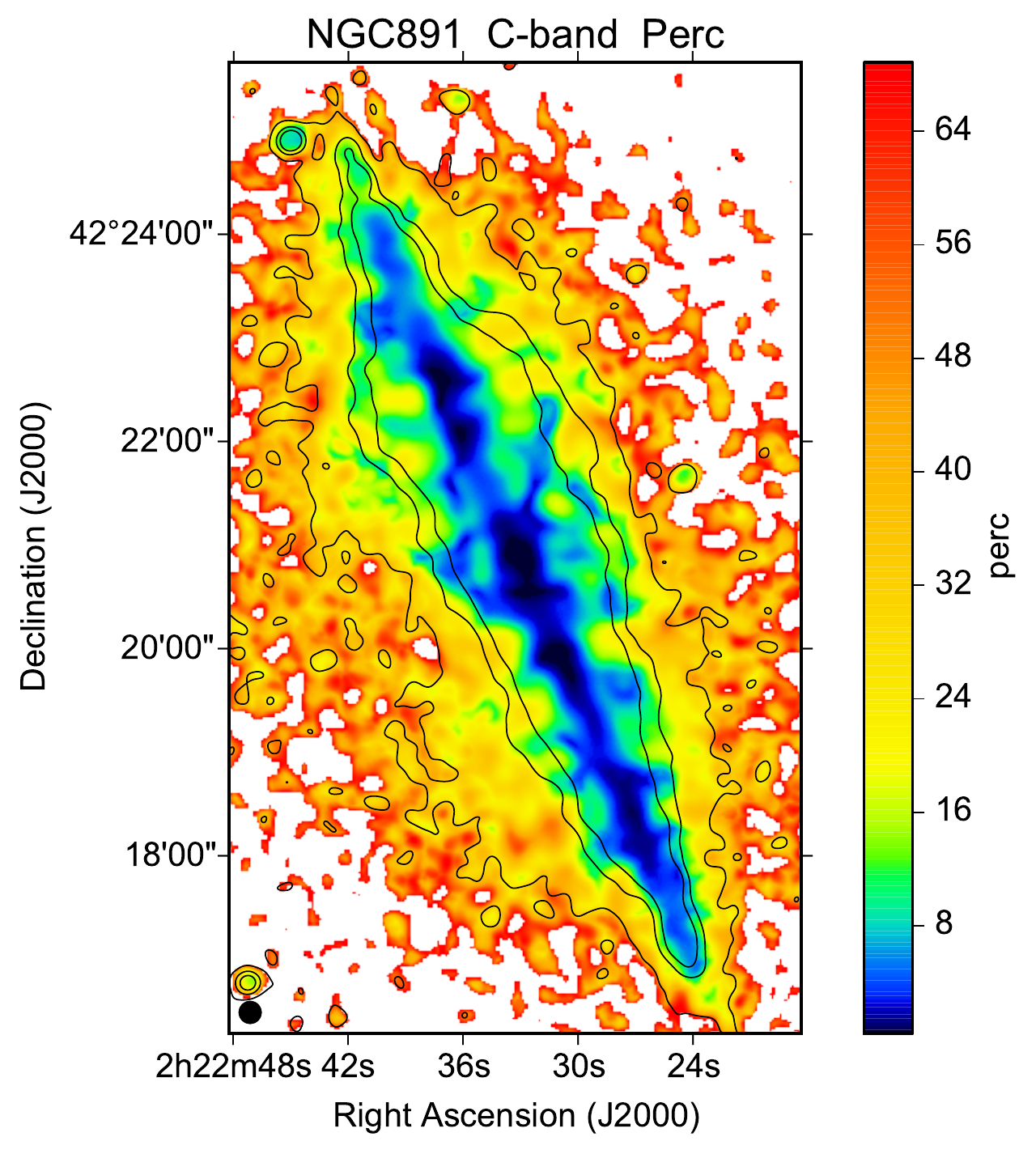}
\includegraphics[width=7.7 cm]{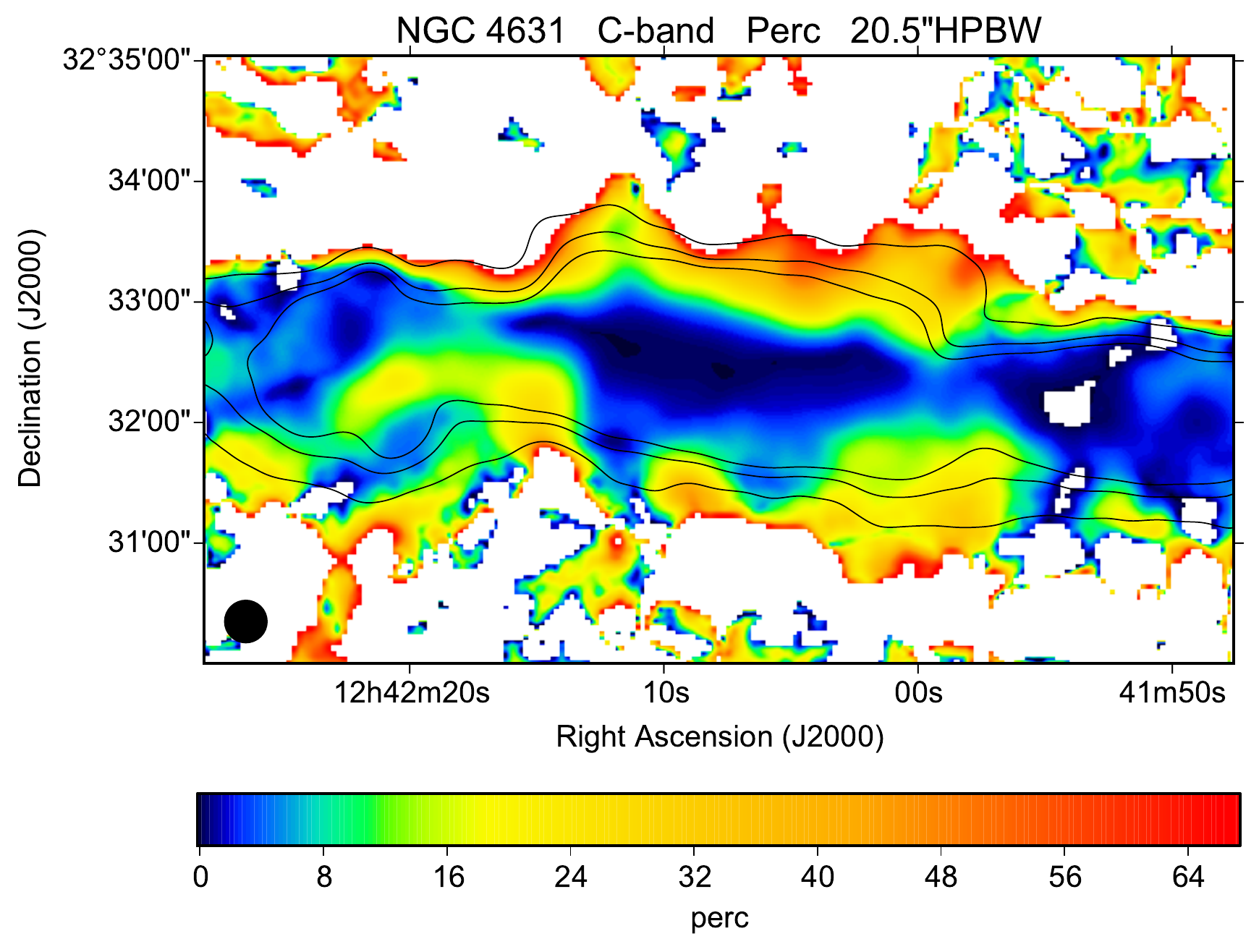}
\includegraphics[width=7.4 cm]{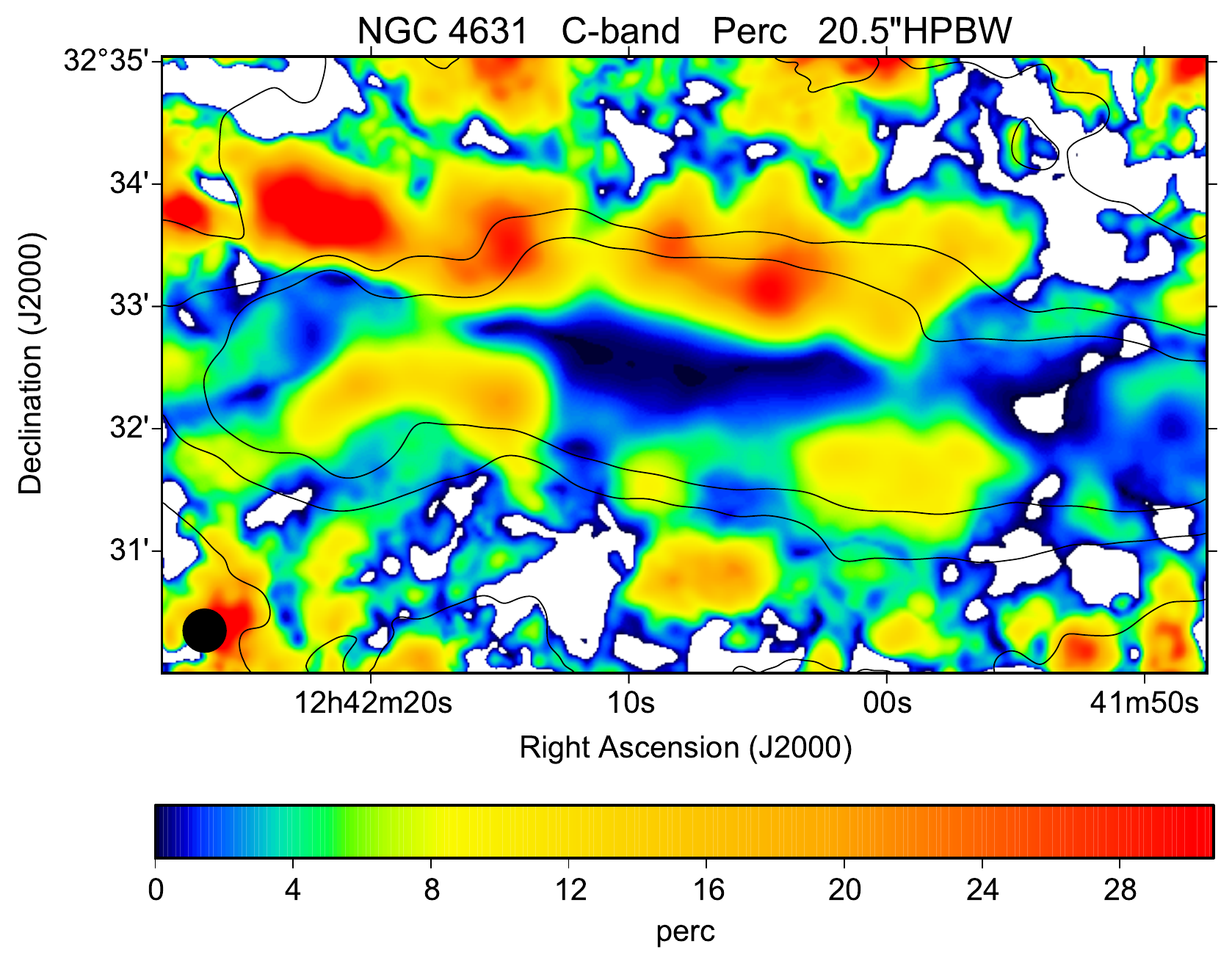}
\caption{Degree of polarization at C-band for NGC~891 at $12 \arcsec$~HPBW (\textbf{upper panels}) and for NGC~4631 at $20 \farcs 5$~HPBW (\textbf{lower panels}) with
VLA-data only (\textbf{left panels}) and VLA and 100-m Effelsberg data merged in TP (\textbf{right panels}). The contours indicate the total intensity at
$5 \, \sigma$, $15 \, \sigma$, and $25 \, \sigma$ r.m.s. of the corresponding TP.
}
\label{perc}
\end{figure*}

We observe that the degree of polarization increases monotonically from the galactic midplane towards the outer boundaries of the halo for all galaxies,
except for NGC~4631 at $20 \farcs 5$~HPBW. We estimated the mean values for P in the area where TP is between $ 5 \, \sigma $ and $ 15 \, \sigma$ r.m.s for all galaxies.
They are very similar for these galaxies with values in the range of about $ 10\% < \rm{P} < 40 \% $ there, with the highest values near the outer edges of the galaxies.
This large-scale increase of the degree of polarization cannot be explained just by a vertical decrease of the thermal fraction of the disk as the decrease is also observed
far in the upper halo. This can only be explained by a vertical increase of the degree of alignment of the magnetic field. This means that the scale height of the ordered
magnetic field is larger than that of the total magnetic field and that the ordered magnetic fields extend far out in the halo and beyond.

This observational result is fully consistent with what follows from the assumption of energy equipartition between the magnetic field and the nonthermal electrons
as usually applied for the magnetic field strength estimate \citep{beck+2005}). The assumption of equipartition implies $N_{CR} \propto B^2$ with $N_{CR}$ being the total
number density of cosmic ray particles. This leads to a scale length of the ordered magnetic field $l_{B} = 2 \, l_{CR}$, hence $l_{B} \simeq 4 \, h_{syn}$ (see
\citealt{krause2019}). 

The only different distribution in P is observed in NGC~4631 at $20 \farcs 5$~HPBW with the merged TP map. There, the degree of polarization increases with distance
from the galactic midplane up to about $35 \%$ at a vertical distance of 2~kpc but decreases again at larger distances from the midplane. This is different to the
other galaxies. Some of those with a smaller angular extent (hence without large-scale missing flux densities) like NGC~3735 or NGC~5775 show a P that can be traced in
vertical direction up to 8~kpc or 10~kpc and P is still monotonically increasing. Also the merged version of NGC~891 shows a vertical monotonically increasing degree of
polarization up to distances of 5~kpc.

It is unclear yet what causes the different appearance of NGC~4631, and whether it is real. NGC~4631 has been observed with two pointings along its major axis. The polarized
intensity in panel~2  Fig.~\ref{n4631_20all} decreases upwards of about 2~kpc ($1\arcmin$) above the midplane, but then seems to increase again at the upper boundary of
the map. This increase is probably artificial and due to the increasing noise due to the primary beam correction that had been applied to the maps. On the other hand, this
is in the area where the TP is significant only in the merged map (compare panel~1 in Fig.~\ref{n4631_20all} with that in Fig.~\ref{n4631_12all}) and may hence be less
reliable. We conclude that the large vertical extent of the halo in NGC~4631 requires
observations with more pointings in vertical direction to give reliable values for the degree of polarization. Only then we can decide whether NGC~4631 also needs a
correction for large-scale missing polarized flux density. This would imply that the large-scale polarization structure and hence the ordered magnetic field in NGC~4631
would be more extended than about 8~kpc.


\section{Discussion}
\label{sec:discussion}

The most striking result from the polarization stacking is that the stacked (apparent) magnetic field vectors reveal an underlying `X-shaped' structure as described in
Sect.~\ref{sec:stacking}. The 28 galaxies that were used for stacking are of very different Hubble types, with different star formation activities in their disks, and
at locations from
being rather isolated to various interacting phases. The X-shaped structure has been seen in various individual galaxies in the past but the results of this polarization
stacking appear to show that this structure is an underlying feature of many and likely most galaxies. This result also indicates scale invariance of the polarization
structure. Even the distribution of the stacked polarized intensity seems to form an X-shaped structure similar to the observation of NGC~4631 alone as visible in
Fig.~\ref{n4631_20all}.

We like to stress that the vectors in Fig.~\ref{stackedimage} are apparent magnetic field vectors that are not corrected for Faraday rotation. The largest effect of Faraday
rotation is expected in the midplane, decreasing from the inner halo outwards to negligible values. This is also reflected in Fig.~\ref{stackedimage} in the way that the
vectors in or near the midplane look less regular than further out. It may explain why we do not observe plane-parallel vectors along the central midplane
of the stacked image as is expected for a plane-parallel magnetic field structure as observed in most of the disks of spiral galaxies seen face-on.
Though, a plane-parallel field is indicated along the outer western midplane of the stacked image.

The magnetic field vectors in the maps of the individual galaxies (Fig.~\ref{n660all} to Fig.~\ref{n5907all}) are corrected for Faraday rotation, hence give the intrinsic
magnetic field orientation averaged along the whole line of sight through the galaxy. They are not less regular along the midplane and in most cases reveal a plane-parallel
disk-field in projection. In some galaxies there are thin lines without polarized intensity along the galactic disk as in  NGC~4631 (Fig.~\ref{n4631_20all}, panel~2) and
in NGC~5775 (Fig.~\ref{n5775all}, panel~2). In these cases we see, different to our general assumption in Sect.~\ref{sec:results}, that even at C-band some galaxies are
Faraday thick along some parts of their disk planes due to high thermal electron densities there. This has been extensively discussed for NGC~4631 in \cite{mora+2013}.

The decrease of the polarized intensity at the outer areas of the galaxies, however, is related to the decrease in synchrotron emission, as it is accompanied by the decrease
in total intensity. As the scale length of the ordered magnetic field is much larger than that of synchrotron emission (as discussed in
Sect.~\ref{sec:uniformity}) it means that the decrease of (polarized) radio intensity is mainly due to the decrease in number density of the relativistic electrons in the halo
or their energy losses by radiation. With more sensitive receivers or longer integration times we could probably observe galactic halos to much further extents.

Considerable theoretical effort has been put into models of dynamo action to explain regular halo fields \citep[e.g.][and references therein]
{sokoloff+1990,brandenburg+1993,moss+2010,henriksen+2018} that look X-shaped when observed. A current summary can be found in \cite{moss+2019} and Beck et al. 2019 (GALAXIES).
A galactic outflow is included in many of these models. This is no unrealistic constraint as the existence of galactic winds has been reported for many of the CHANG-ES galaxies
\citep{krause+2018, miskolczi+2019, stein+2019a, schmidt+2019, mora+2019a}, and extraplanar ionized gas emission can be seen in many H$\alpha$ images taken for the CHANG-ES
sample \citep{vargas+2019}.

NGC~4631 was the first galaxy in which a large-scale magnetic field in the halo was detected \citep{mora+2019b}. It even shows periodic large-scale field reversals in its
northern halo (see Fig.~\ref{n4631_20all}) which can be modeled by accretion models of the scale-invariant mean-field dynamo theory \citep{woodfinden+2019}. Vertical
RM-transition lines (RMTL) have been detected in at least six galaxies of our sample (as described in Sect.~\ref{sec:large-scale}). One of them, NGC~3044, exhibits
Giant Magnetic Ropes (GMR) as detected in NGC~4631. Similar to the latter, NGC~3044 shows indications for a strong disturbance, like e.g. by a past merger
\citep{zschaechner+2015}. Also NGC~3448 shows indications of GMRs.

The large-scale magnetic fields in the other 4 galaxies with vertical RMTL are mainly horizontally orientated. These could possibly be explained by 'GMR' that
spiral with rather tightly wound lines, either upwards from or downwards to the galactic disk through the halo along cylinders with a diameter given by $\Delta$~RMTL. We
call them {\em horizontal GMRs}. The vertical GMRs observed in NGC~3044 and NGC~4631 could eventually be understood as part of very loosely wound helices. Model simulations
are necessary to test these ideas.

We noticed an asymmetry in the distribution of the polarized intensity in the disk and halo: PI is usually stronger in one half of the galaxy along the major axis than
in the other. In order to quantify the asymmetry, PI was integrated on both sides of the major axis in boxes extending out to the weakest emission in vertical direction (Table~4)
and in radial direction. The bright polarized emission from the lobes of NGC~3079 and NGC~4388 is not related to the large-scale magnetic fields in the disk or halo
and hence was subtracted before the integration. Polarized background sources, not related to the galaxies, were found, one in each of the fields of NGC~3735,
NGC~5775, and NGC~5907, and were subtracted, too.

The asymmetry is measured by the parameter $q=(A-R)/(A+R)$ where $A$ and $R$ are the integrated polarized flux densities on the approaching and
the receding sides, respectively. $q=0$ means no asymmetry, $q=\pm 1$ maximum asymmetry (one side missing completely). The accuracy of the $q$ measurements was tested
by slightly varying the integration boxes and was found to be about 20\%.The $q$ values are given in Table~\ref{rotation}.
$q$ is positive in 13 out of 18 galaxies, with values ranging between 0.047 and 0.275, while four galaxies show negative values. In one galaxy (NGC~4631) no significant
asymmetry can be measured. The probability that the preference of positive $q$ values is by chance is estimated as $(17!\, / \,(13! 4!))\, / \, 2^{17} \cdot 100=1.8\,\%.$

We compared this PI asymmetry with the overall rotation of the galactic disk by determining the approaching side of the major axis from observed velocities
in HI or CO. These values were found in the literature for all galaxies except for NGC~3735. For this galaxy we reduced HI data observed with the VLA and determined a velocity map of NGC~3735 as described and shown in Fig.~\ref{n3735HI} in Appendix~B . It shows that the south eastern side is the approaching one.
The results are summarized in Table~\ref{rotation}. Most of the galaxies presented in this paper show the strongest PI in C-band on the
side of the major axis that is approaching with respect to the galaxy's global rotation. This was first recognized in NGC~4666 by \citet{stein+2019a}.

\begin{table*}
      \caption[]{\label{rotation} Global rotation and ordered magnetic field in the halo}

     $$
         \begin{tabular}{lcccc}
            \hline
            \noalign{\smallskip}
            \multicolumn{1}{c}{Galaxy} &  Side with      & Approaching  &  Asymmetry $q$  & Reference \\
                                       &  strongest PI   & side of \\
                                       &  in disk+halo   & major axis \\
            \hline
            \noalign{\smallskip}

            NGC~660   &  NE   &    NE   &  0.246  &   \cite{driel+1995}      \\
            NGC~891   &  NE   &    NE   &  0.147  &   \cite{garcia+1995}     \\
            NGC~2613  &  NW   &    NW   &  0.101  &   \cite{chaves+2001}     \\
            NGC~2820  &  SW   &    SW   &  0.100  &   \cite{kantharia+2005}  \\
            NGC~3044  &  NW   &    NW   &  0.047  &   \cite{lee+1997}        \\
            NGC~3079  &  NW   &    NW   &  0.275  &   \cite{irwin+1991}      \\
            NGC~3448  &  SW   &    NE   &  -0.131 &   \cite{noreau+1986}     \\
            NGC~3556  &  SW   &    NE   &  -0.042 &   \cite{king+1997}       \\
            NGC~3735  &  SE   &    SE   &  0.198  &    this paper            \\
            NGC~4157  &  SW   &    SW   &  0.110  &   \cite{verheijen+2001}  \\
            NGC~4192  &  NW   &    NW   &  0.087  &   \cite{cayatte+1990}    \\
            NGC~4217  &  NE   &    NE   &  0.106  &   \cite{verheijen+2001}  \\
            NGC~4388  &  E    &    W    &  -0.192 &   \cite{oosterloo+2005}  \\
            NGC~4565  &  SE   &    SE   &  0.055  &   \cite{neininger+1996}  \\
            NGC~4631  &  similar & W    &  0.011  &   \cite{wielebinski+1999}\\
            NGC~4666  &  NE   &    NE   &  0.130  &   \cite{walter+2004}     \\
            NGC~5775  &  SE   &    NW   &  -0.124 &   \cite{irwin1994}       \\
            NGC~5907  &  SE   &    SE   &  0.084  &   \cite{dumke+1997}      \\

            \noalign{\smallskip}
            \hline
         \end{tabular}
               $$


\end{table*}

The asymmetry in C-band is of similar sign but weaker compared to that observed in L-band in mildly inclined galaxies of the SINGS survey \citep{braun+2010}.
These authors explained the asymmetry by the superposition of a large-scale axisymmetric spiral field in the disk and a large-scale quadrupolar field in the halo.
The asymmetry arises if only the near side of the galaxy is visible in polarized emission due to strong Faraday depolarization.
The strength of the detected PI does not directly depend on the total ordered field strength (B$_t$) but on the strength of its perpendicular magnetic field component.
Hence, PI depends on our viewing angle onto the magnetic field and its observed asymmetry may give important information about the large-scale
structure of the regular field. The model by \citet{braun+2010} may also apply to edge-on galaxies
observed in C-band with smaller but still significant Faraday depolarization. Refined model calculations for edge-on galaxies are needed.

An alternative model was proposed by Stein et al. (submitted to A\&A) where the trailing spiral arms may explain the asymmetry in the disk and its correlation
with the galactic rotation if the polarized emission from the near side of the galaxy dominates. However, a large fraction of the polarized emission
in our sample galaxies emerges from the halo that does not host spiral arms. We hope that our comprehensive polarization study of edge-on galaxies triggers more
theoretical efforts.

As noted Sect.~\ref{sec:introduction}, the galaxies of the CHANG-ES sample are of various Hubble types and star formation rates (SFR). The determination of reliable SFRs in edge-on galaxies is more difficult than in face-on galaxies and has recently been reexamined by \cite{vargas+2019}. We present these values for the galaxies with an observed large-scale magnetic field in the halo in Table~\ref{structure}. There is no indication found that the existence of a large-scale magnetic is preferentially observed for galaxies with certain Hubble types or special SFR values.

Many of the CHANG-ES galaxies have been or are currently interacting with other galaxies (like the polar ring galaxy NGC~660) or are members of the Virgo cluster
(like NGC~4192, NGC~4388, and NGC~4438) and we still detected large-scale halo magnetic fields in many of them. This indicates that a large-scale magnetic field -
if once generated - cannot be destroyed easily and persists. Hence, it should be considered whether large-scale halo fields can be regarded as a link to intergalactic
magnetic fields.

\section{Summary and conclusions}
\label{sec:summary}

In this paper we used two different approaches (stacking and RM-synthesis) to analyze the magnetic fields in the halo of spiral galaxies. We present a stacked image
(C-band D-array configuration) of the linear polarization and apparent magnetic field vectors of all CHANG-ES galaxies that show polarization in their disks. These
are 28 galaxies of very different Hubble types, star formation and interaction activities. The result is shown in Fig.~\ref{stackedimage}. The striking result is that
it clearly reveals an underlying X-shaped structure of the apparent magnetic field which seems to be an underlying feature of many and likely most spiral galaxies.

Secondly, we performed RM-synthesis on combined array configurations at C-band and at L-band of all 35
CHANG-ES galaxies and argue that only the C-band results can be regarded as a reliable tracer of the
intrinsic magnetic field in the disk and halo. Polarized intensity was detected in all but one of the 35 galaxies. The outlier is NGC~4244 which is too faint, even in total
power.

In seven galaxies of the sample, the polarized emission is detected but their signal-to-noise is too low for a reliable RM-synthesis (as described in
Sect.~\ref{sec:results}). In six other galaxies, the polarized emission is clearly dominated by the central source or a background radio galaxy (UGC~10288) without clear
polarized disk/halo emission. In total, we are left with 21 spiral galaxies that show extended polarized intensity within their disks and/or halos and for which the
CHANG-ES observations allowed for the first time to determine reliable RM-values in galactic halos.

The resulting maps are presented in Fig.~\ref{n660all} to
Fig.~\ref{n5907all}.
Our main results are:

\begin{itemize}
 \item We detected a regular, large-scale magnetic field in the halo of 16 galaxies.

 \item The scale of the regular magnetic field in the halo is typically 1~kpc or larger, while it extends over several kpc.

 \item While all these galaxies show large-scale magnetic field reversals with respect to the line of sight, we observed that these magnetic field
 reversals occur where the field in the plane of the sky extends approximately perpendicular to the galactic midplane (vertical RMTL, see Table~\ref{structure})
 in six galaxies.

 \item In four of these six galaxies, the vertical RMTL have magnetic field vectors perpendicular to the lines. Only NGC~3044 and possibly NGC~3448 have the magnetic
 field vectors roughly parallel to these lines, similar to NGC~4631. Hence we detected at least one more galaxy with giant magnetic ropes (GMR) as defined in
 \citealt{mora+2019b}. The magnetic field vectors in the other four galaxies are mainly horizontally orientated.

 \item We observed an asymmetry in the distribution of the polarized intensity (PI) on both halves of the galaxy with respect to the minor axis. We found that the
 strongest PI is on the side of the major axis that is approaching with respect to the galaxy's global rotation. PI depends on our viewing angle onto the magnetic
 field and its observed asymmetry may give important informations about the large-scale structure of the regular field.

 \item The degree of polarization increases monotonically from the galactic midplane towards the outer boundary of the halo. This implies that the scale height
 of the ordered magnetic field is larger than that of the total magnetic field and that the ordered (and probably also the regular) magnetic fields extend far out
 in the halo and beyond.
 \\
\end{itemize}

Altogether, we showed that large-scale (coherent) magnetic fields are common in the halos of spiral galaxies. Our observations do not indicate a simple,
large-scale magnetic field pattern in the halo like a dipole or quadrupole structure as expected by $\alpha - \Omega$ mean-field dynamo models for the disk.
They also do not exhibit a systematic single sign change of RM across the minor axis as would be expected for large-scale toroidal halo fields. The field patterns
seem to be more complicated.

With more sensitive observations at C-band, we expect to detect these magnetic fields also in the halos of those galaxies whose
signal-to-noise ratio are currently too low to detect extended RM and intrinsic magnetic field vectors.
From our experience, we conclude that C-band is the best suited frequency band in terms of sensitivity and depolarization for a
polarization study of galactic halos in nearby spiral galaxies. The best angular resolution for future observations is the
one that corresponds to about 1~kpc in the galaxies, respectively.

We also observed large-scale magnetic field reversals with respect to the line of sight
indicating giant magnetic ropes (GMR) that may be more or less tightly wound and extend far out in the halo. We anticipate that this discovery will strengthen
the impact of large-scale dynamo theories for spiral galaxies. The regular halo fields may also be regarded as a link to intergalactic magnetic fields
(see e.g. \citealt{henriksen+2016}) and could help to understand their origin which is still a mystery.

\begin{acknowledgements}

We thank Peter M\"uller for several fast adjustments of the NOD3 software to the requirements for our plots and Simon Bauer for his help with the data processing during his internship at the MPIfR in Bonn. We acknowledge the unknown referee for valuable comments.
The Dunlap Institute is funded through an endowment established by the David Dunlap family and the University of Toronto.

\end{acknowledgements}

\bibliographystyle{aa}
\footnotesize
\bibliography{BIBLIO.Ma.bib}

\begin{appendix}

\section{Polarization maps with RM-synthesis}
\label{maps}

 \begin{figure*}[t]
\centering
\includegraphics[width=9.0 cm]{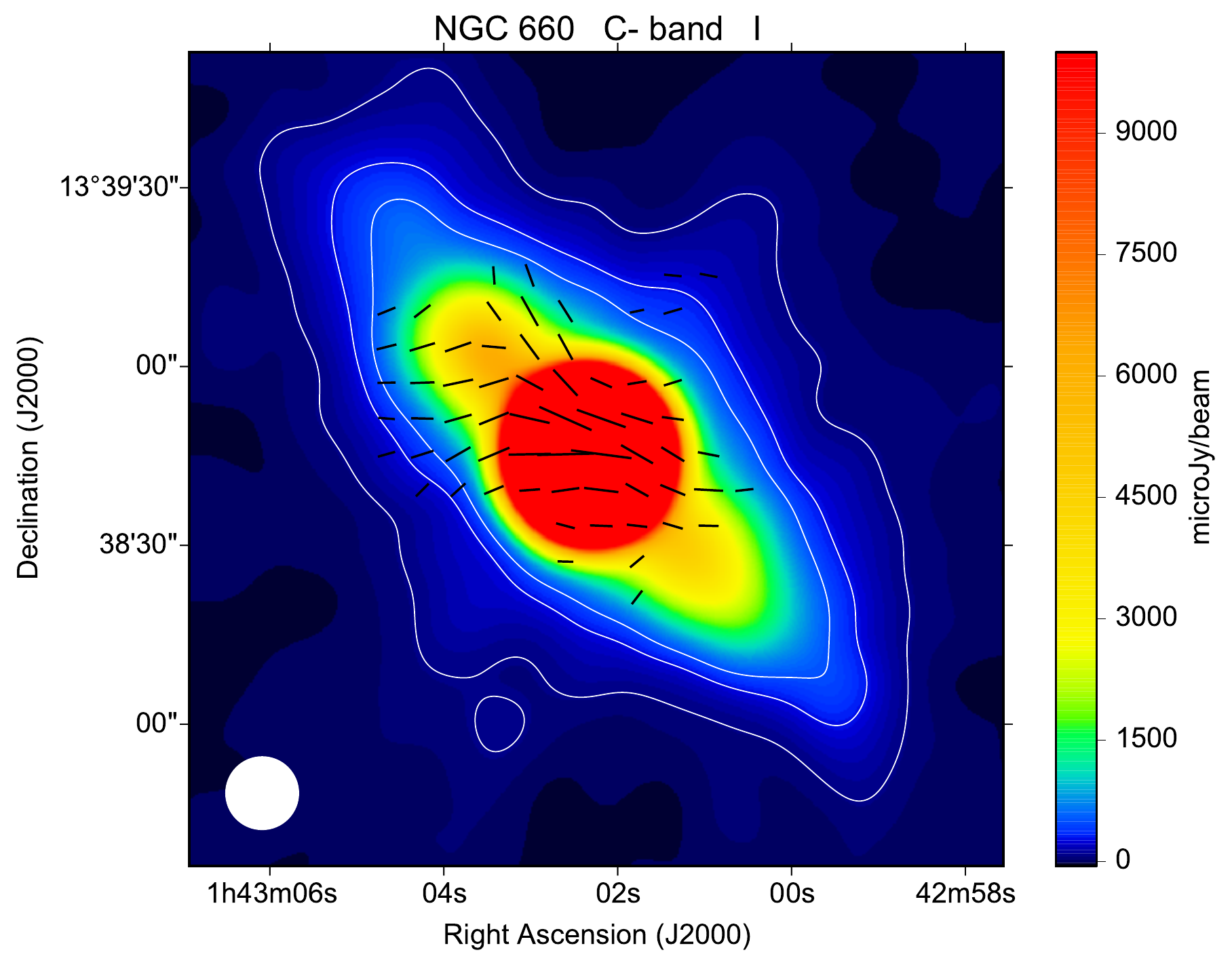}
\includegraphics[width=9.0 cm]{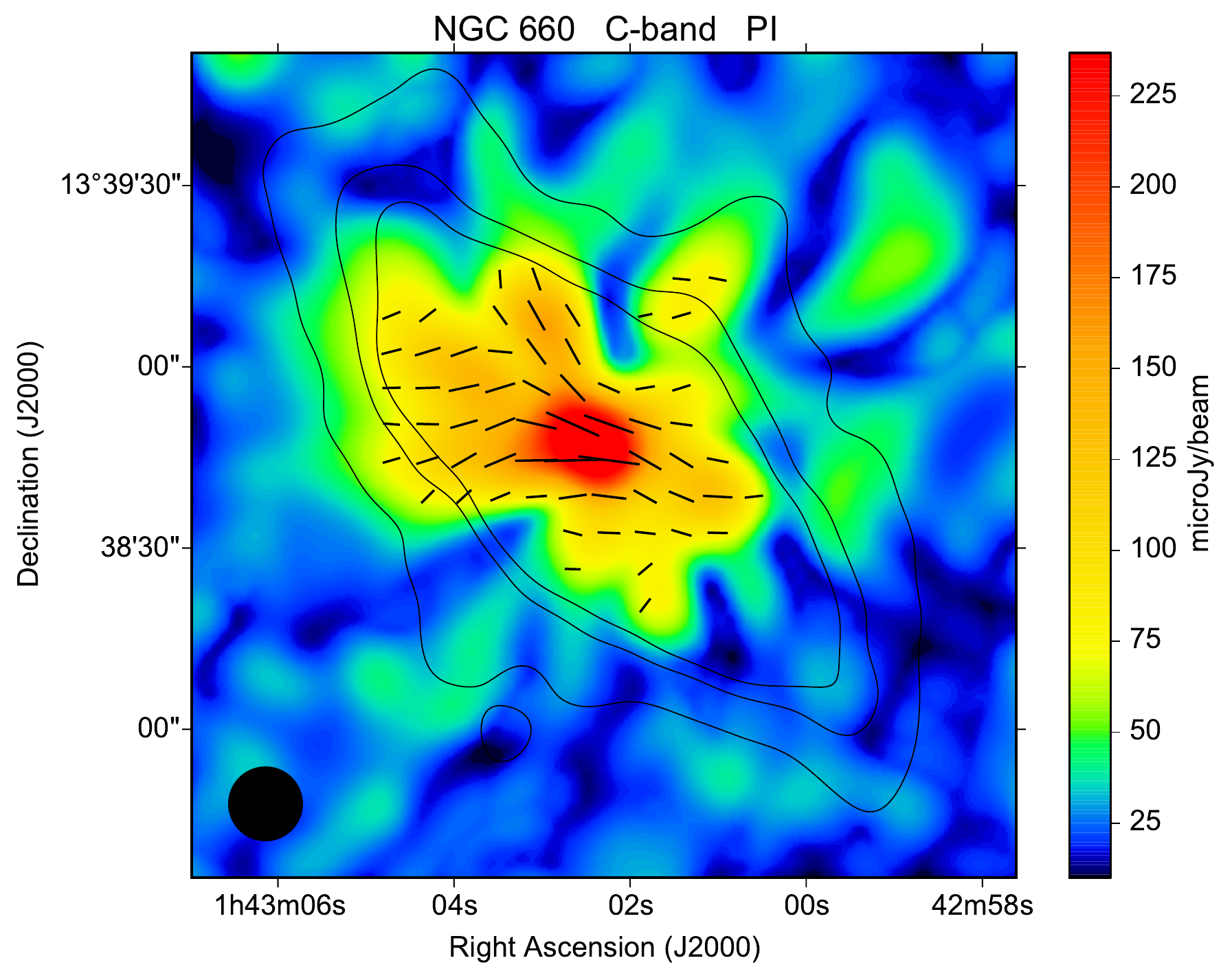}
\includegraphics[width=9.0 cm]{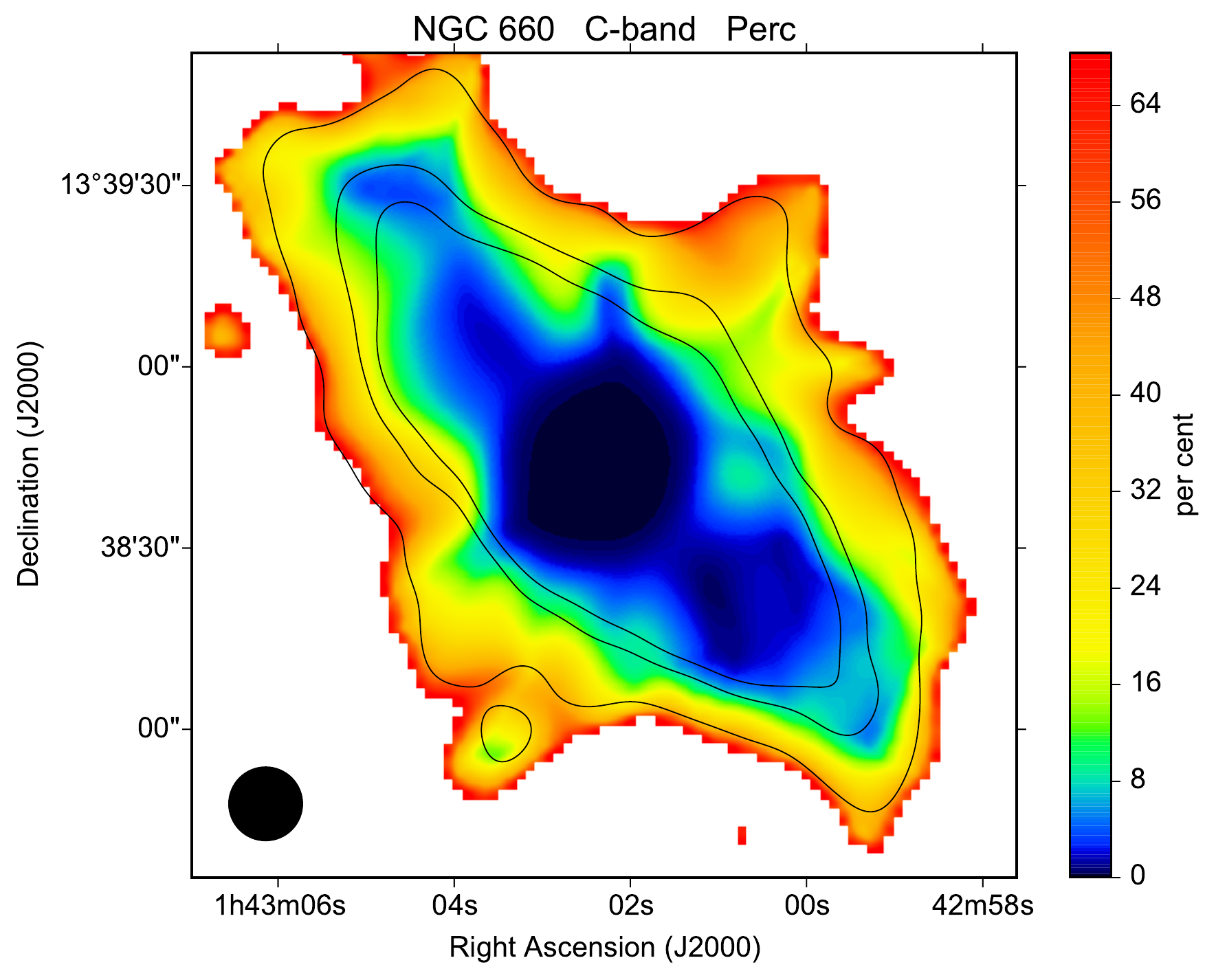}
\includegraphics[width=9.2 cm]{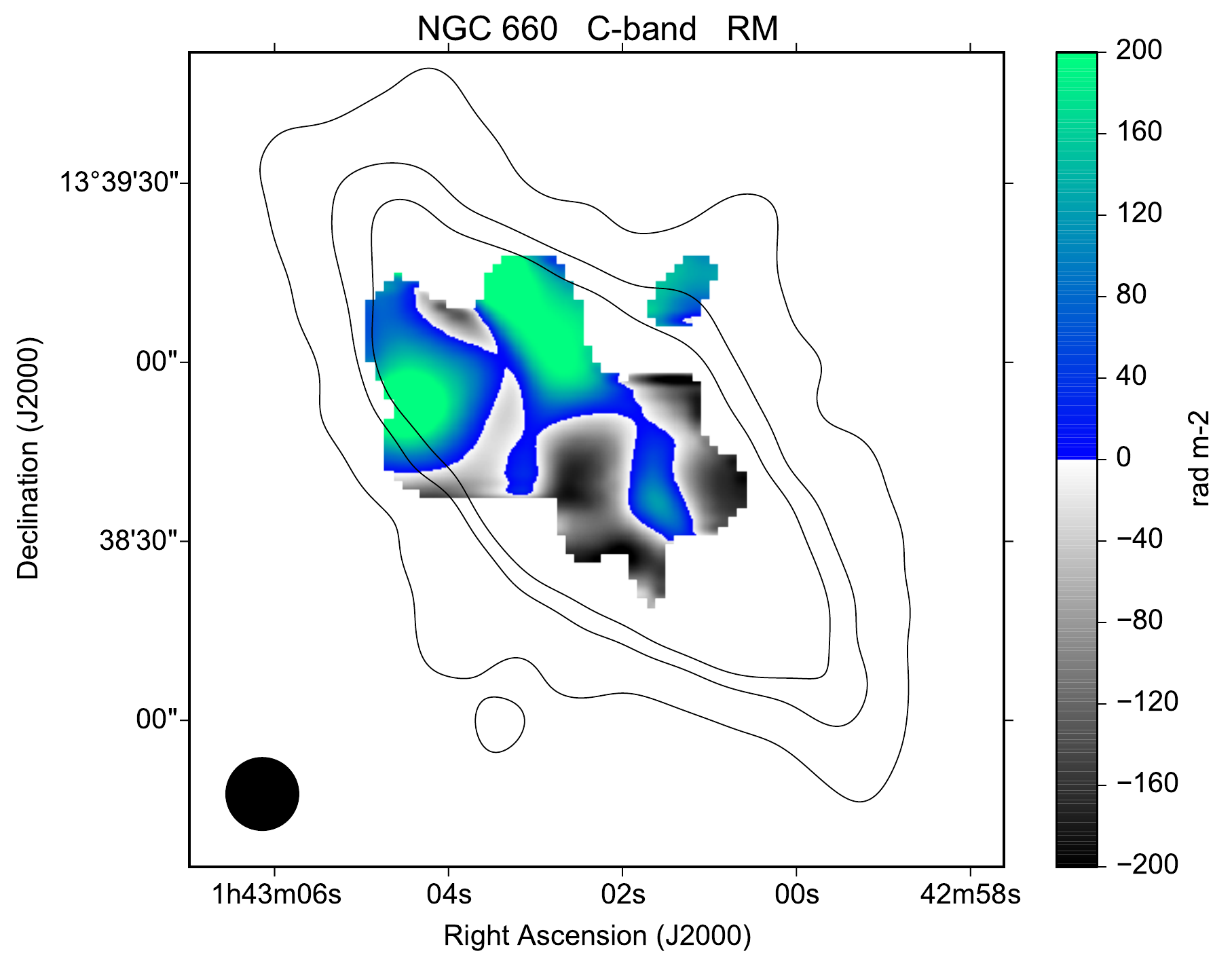}
\includegraphics[width=9.0 cm]{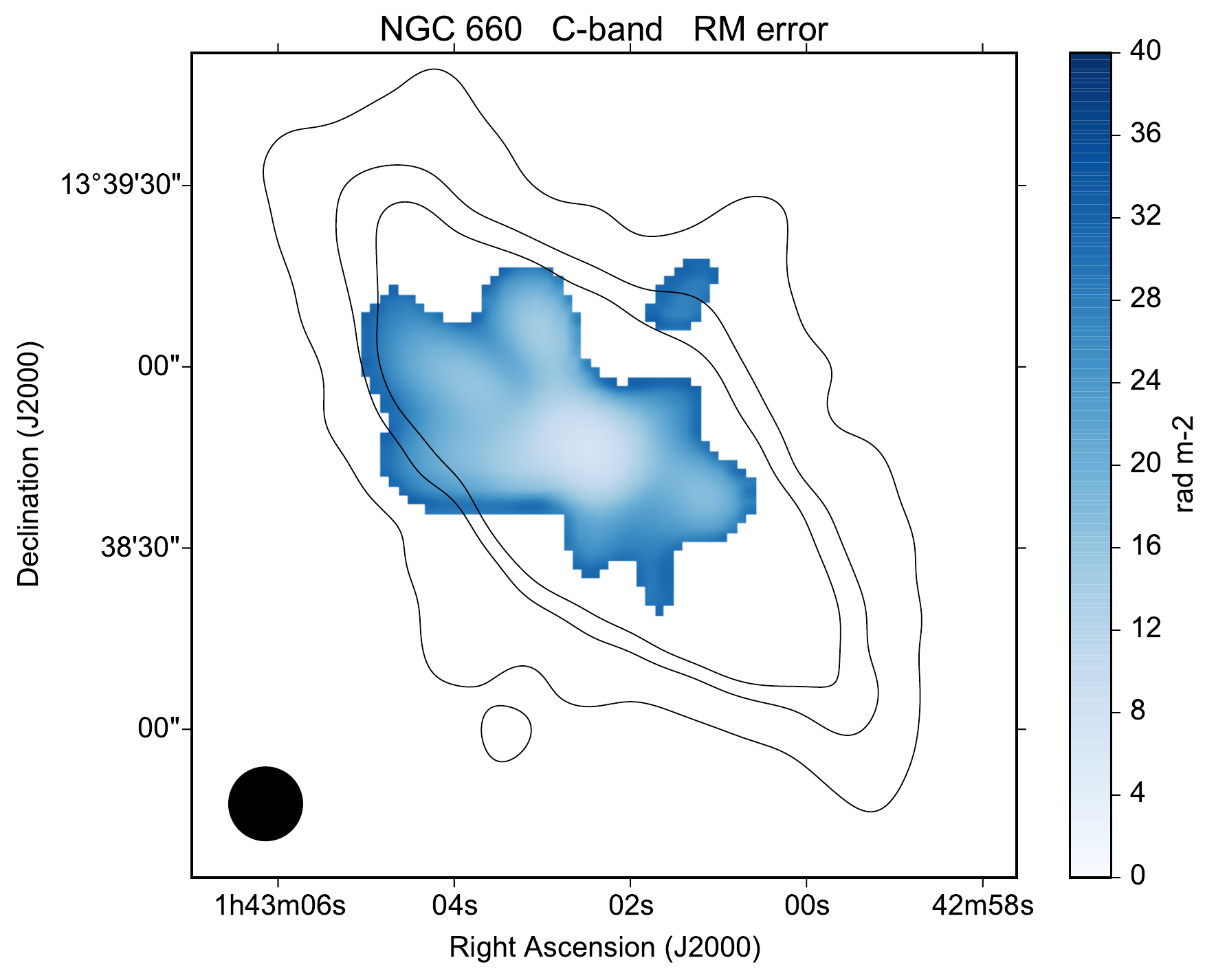}
\includegraphics[width=9.2 cm]{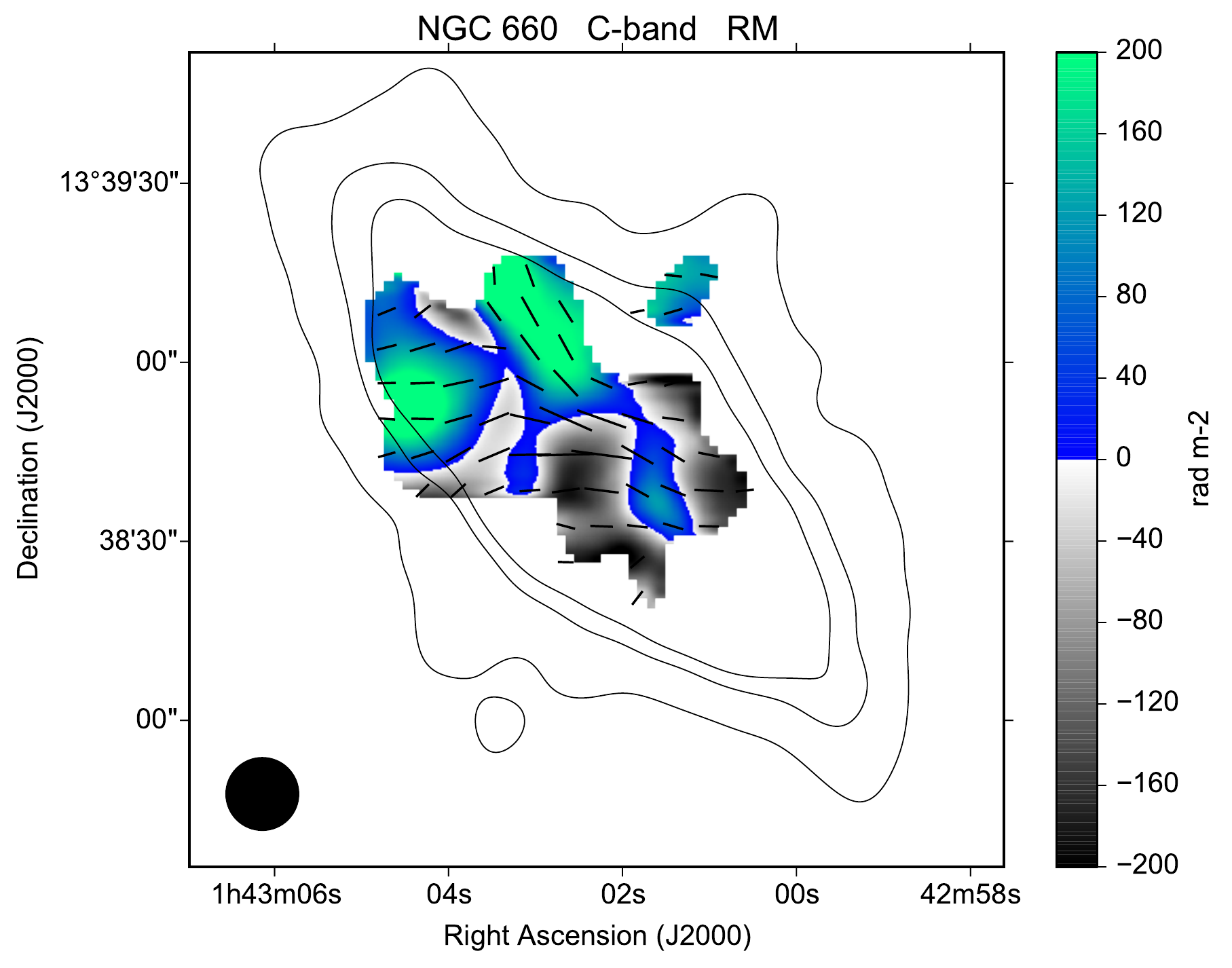}
\caption{Polarization results for NGC~660 at C-band and $12 \arcsec$ HPBW, corresponding to $720\,\rm{pc}$. The contour levels (TP) are 70, 210, and 350 $\mu$Jy/beam. The image of the TP map is cut at 10000~$\mu$Jy/beam in order to present the disk emission well.
The six panels show total intensity (TP, Stokes I) with contours at 5, 15, 25 $\sigma$
r.m.s. and intrinsic polarization vectors (upper left, \emph{panel~1}), polarized intensity (PI) with intrinsic polarization vectors and the contours of TP as given in
panel~1 (upper right, \emph{panel~2}), percentage polarization (referred to as Perc) with TP contours of panel~1 (mid left, \emph{panel~3}), rotation measure
(RM) with TP contours of panel~1 (mid right, \emph{panel~4}), errors in RM with TP contours of panel~1 (lower left, \emph{panel~5}), rotation measure with intrinsic
polarization vectors and TP contours of panel~1 (lower right, \emph{panel~6}).
}
\label{n660all}
\end{figure*}

\begin{figure*}[p]
\centering
\includegraphics[width=7.3 cm]{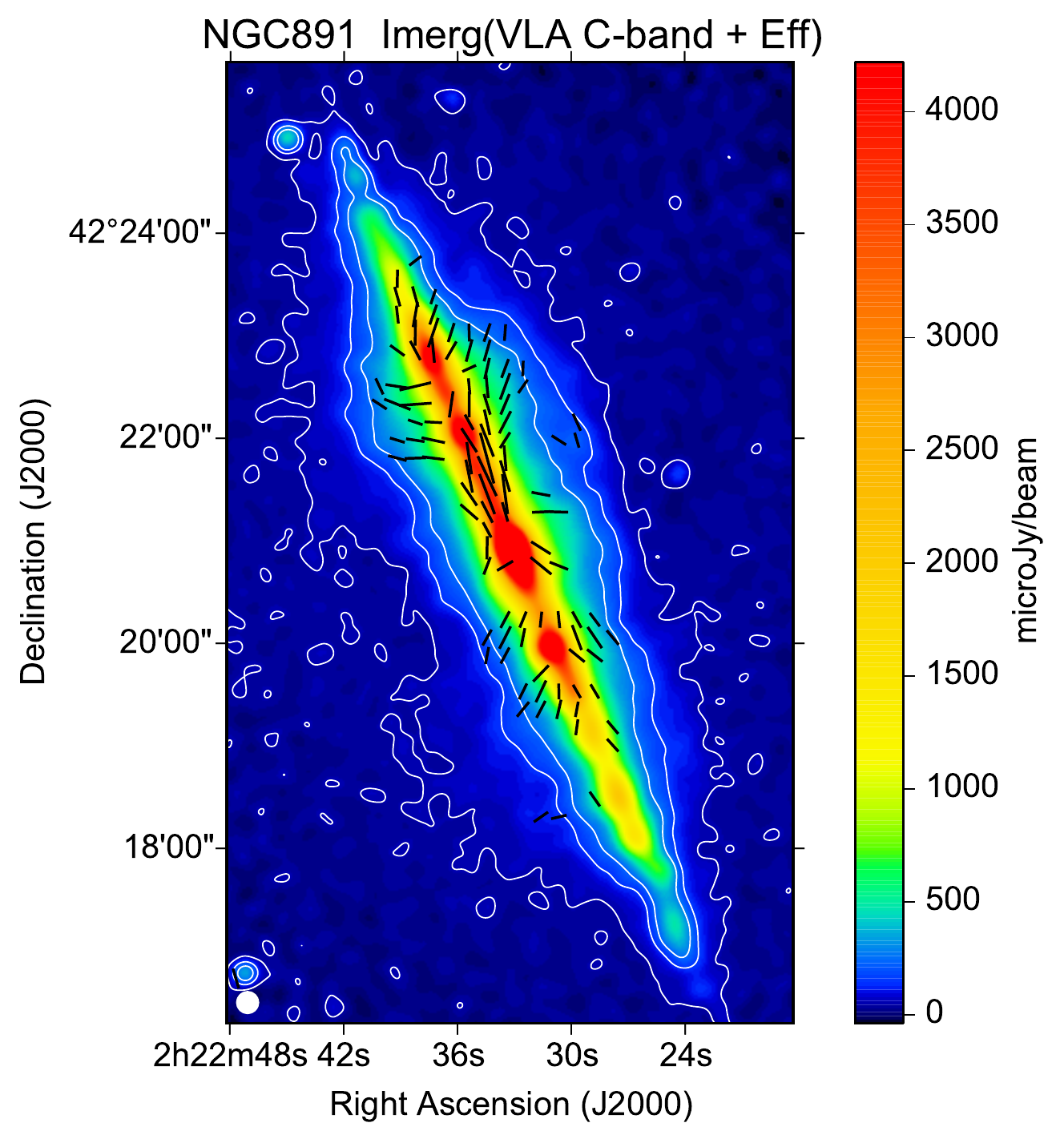}
\includegraphics[width=7.2 cm]{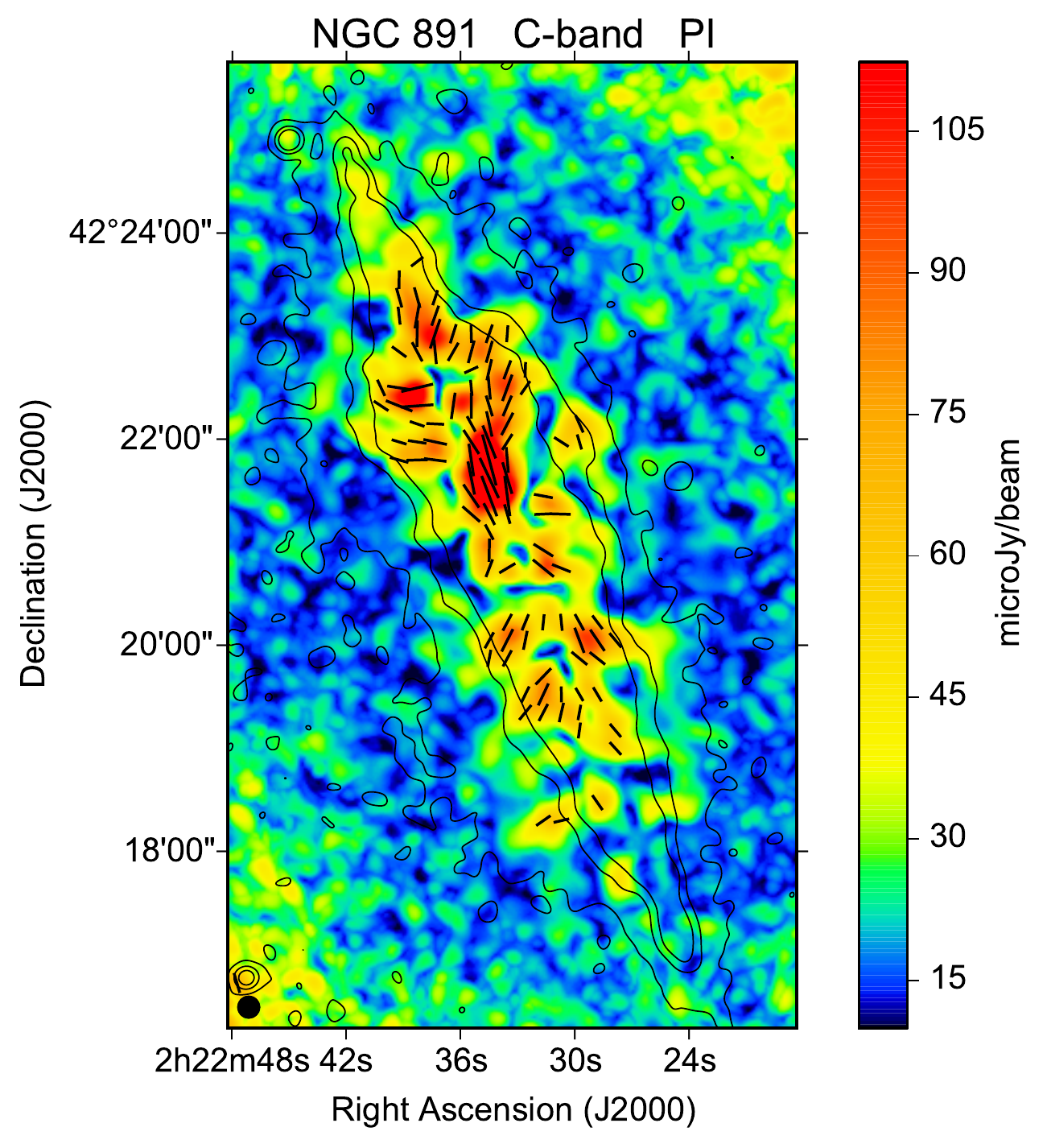}
\includegraphics[width=7.3 cm]{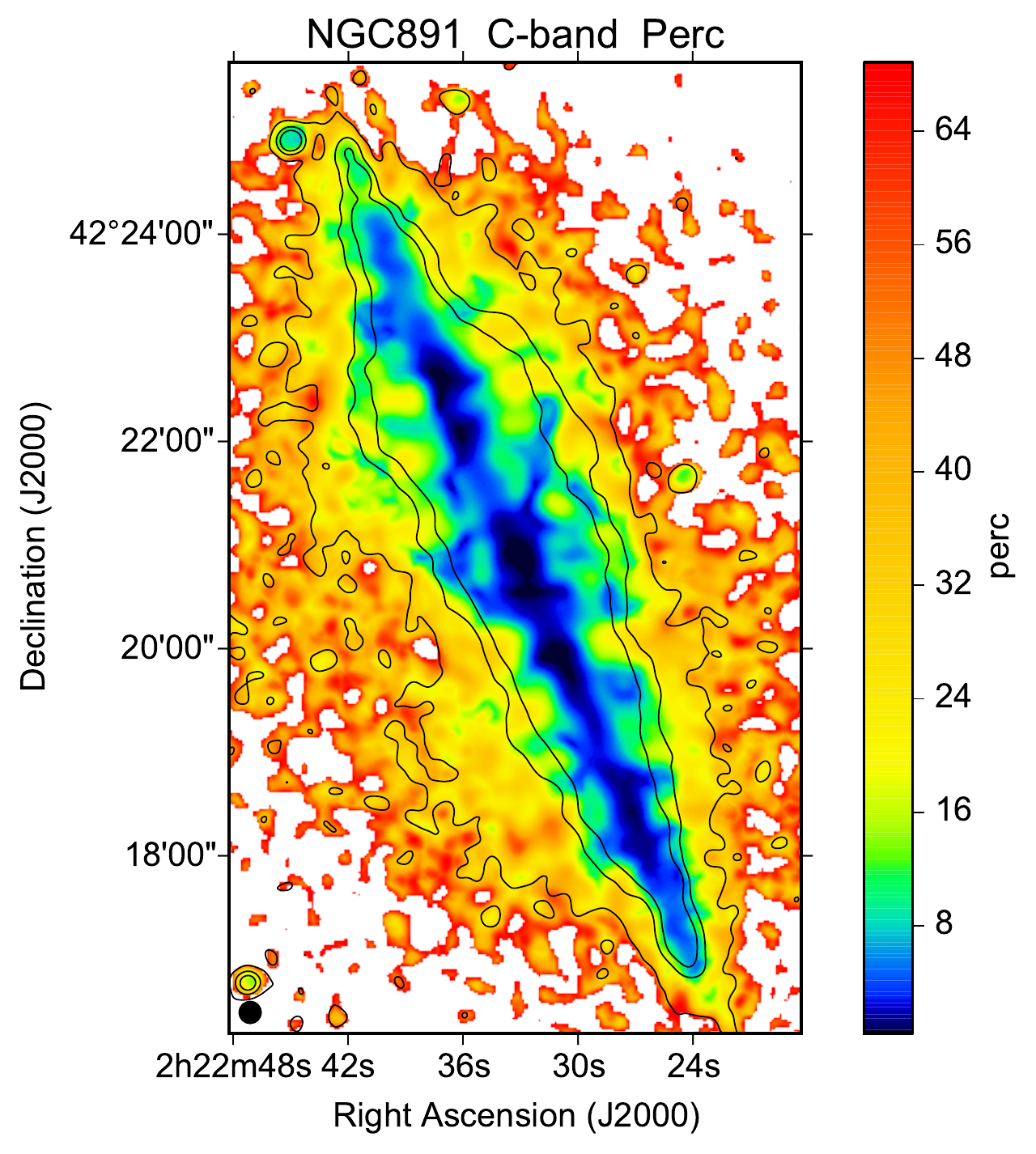}
\includegraphics[width=7.5 cm]{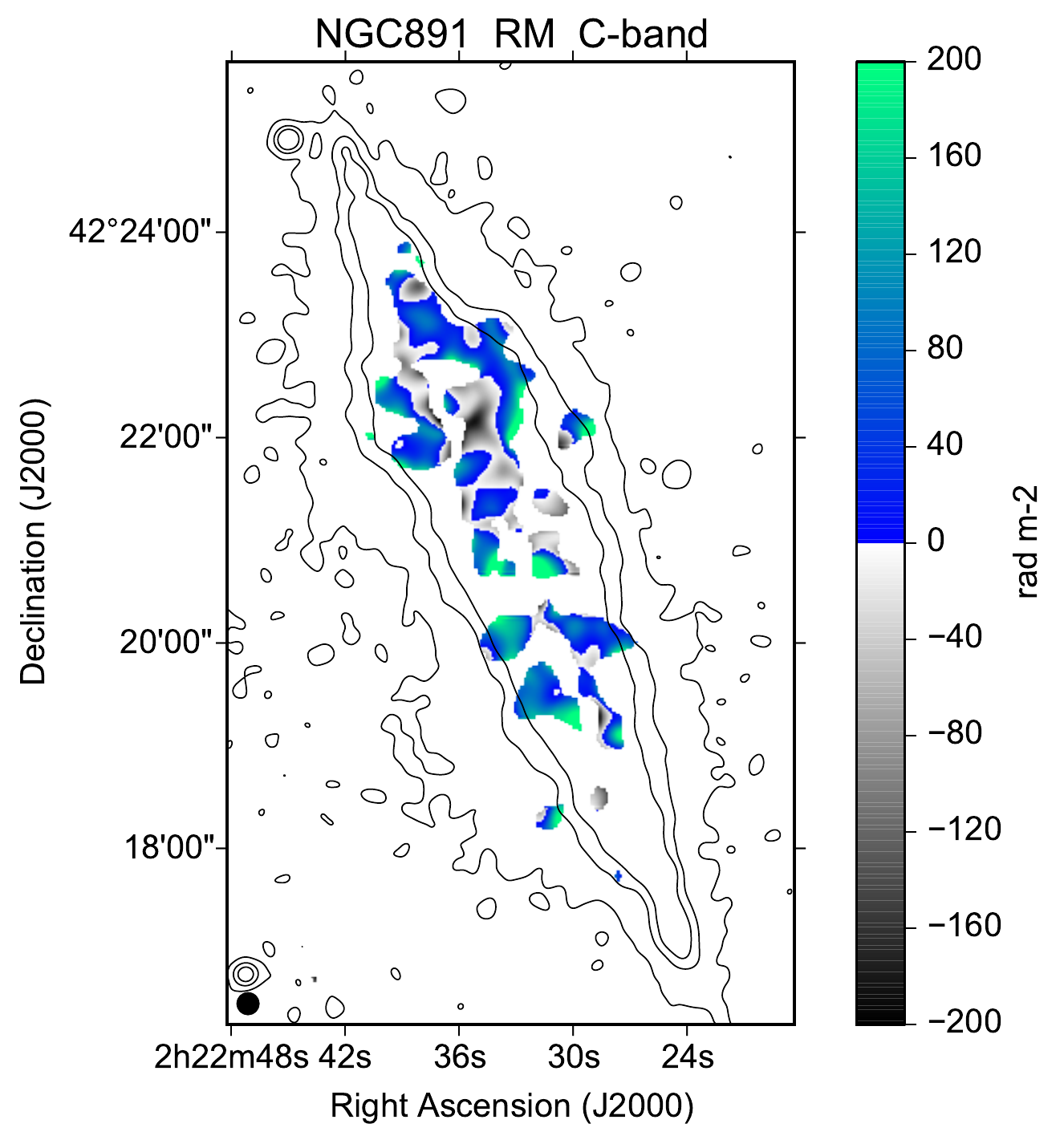}
\includegraphics[width=7.3 cm]{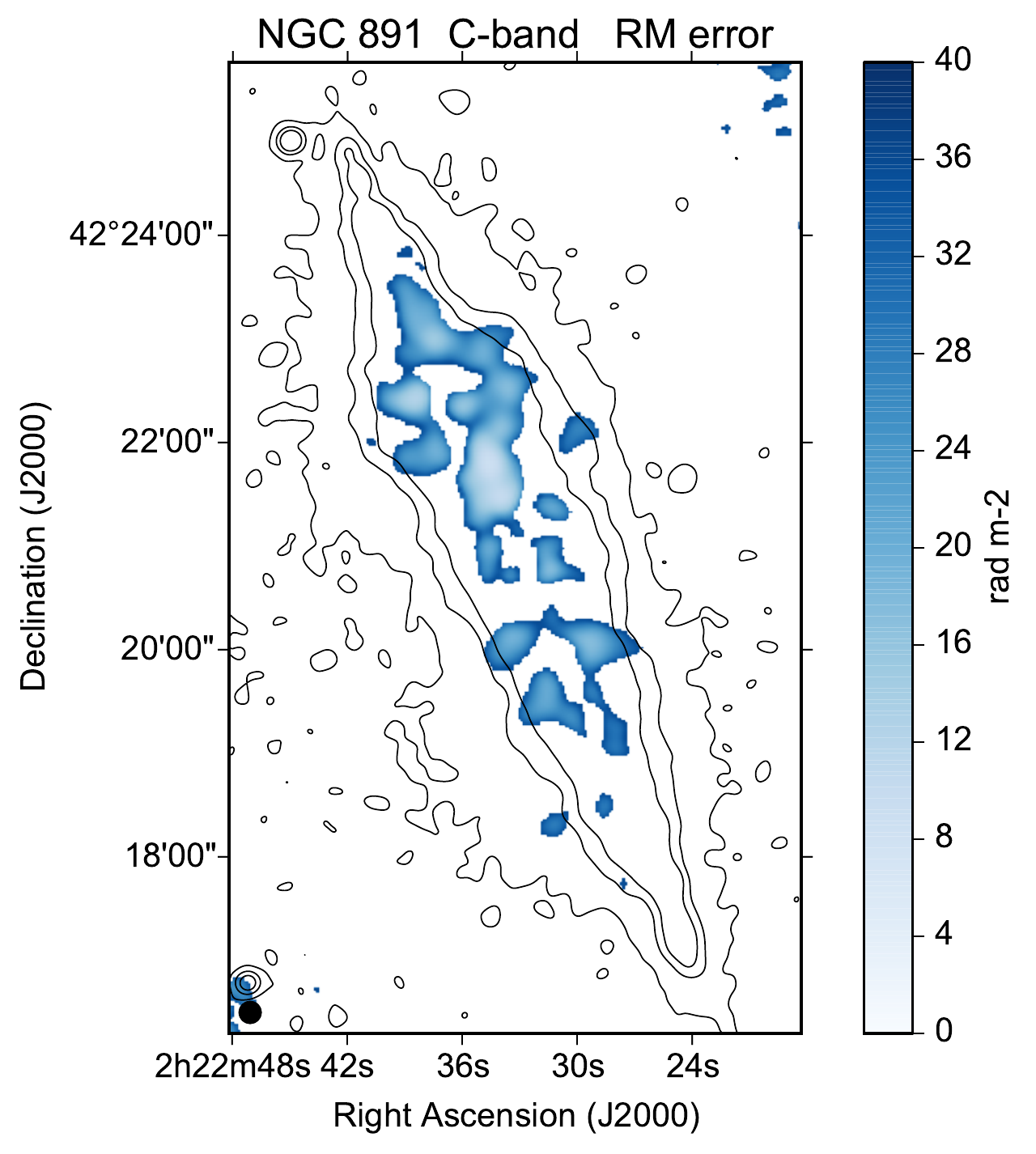}
\includegraphics[width=7.5 cm]{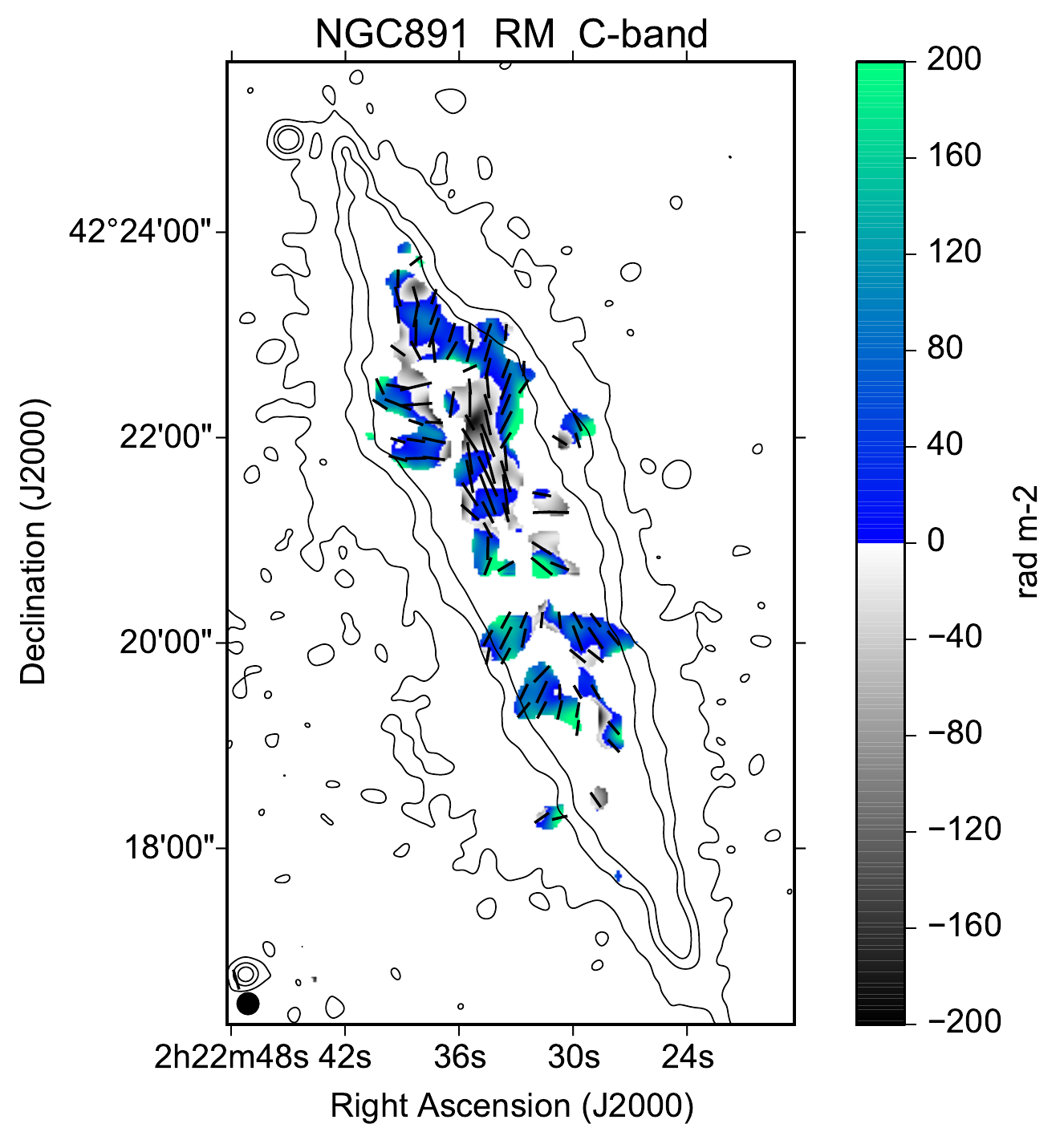}
\caption{Polarization results for NGC~891 at C-band and $12 \arcsec$ HPBW, corresponding to $530\,\rm{pc}$. The contour levels (TP) are 50, 150, and 250 $\mu$Jy/beam.
}
\label{n891_12all}
\end{figure*}

\begin{figure*}[p]
\centering
\includegraphics[width=9.0 cm]{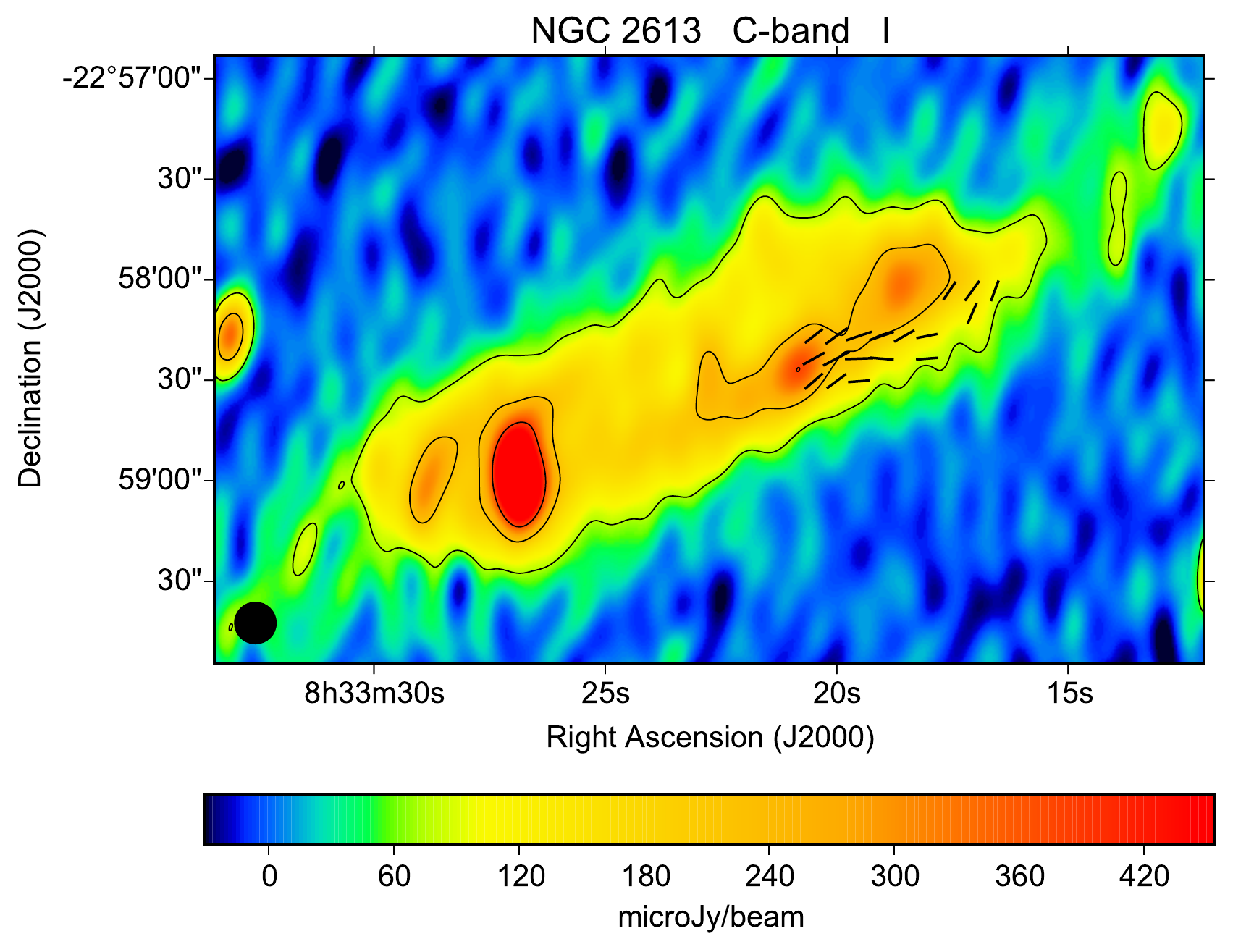}
\includegraphics[width=9.1 cm]{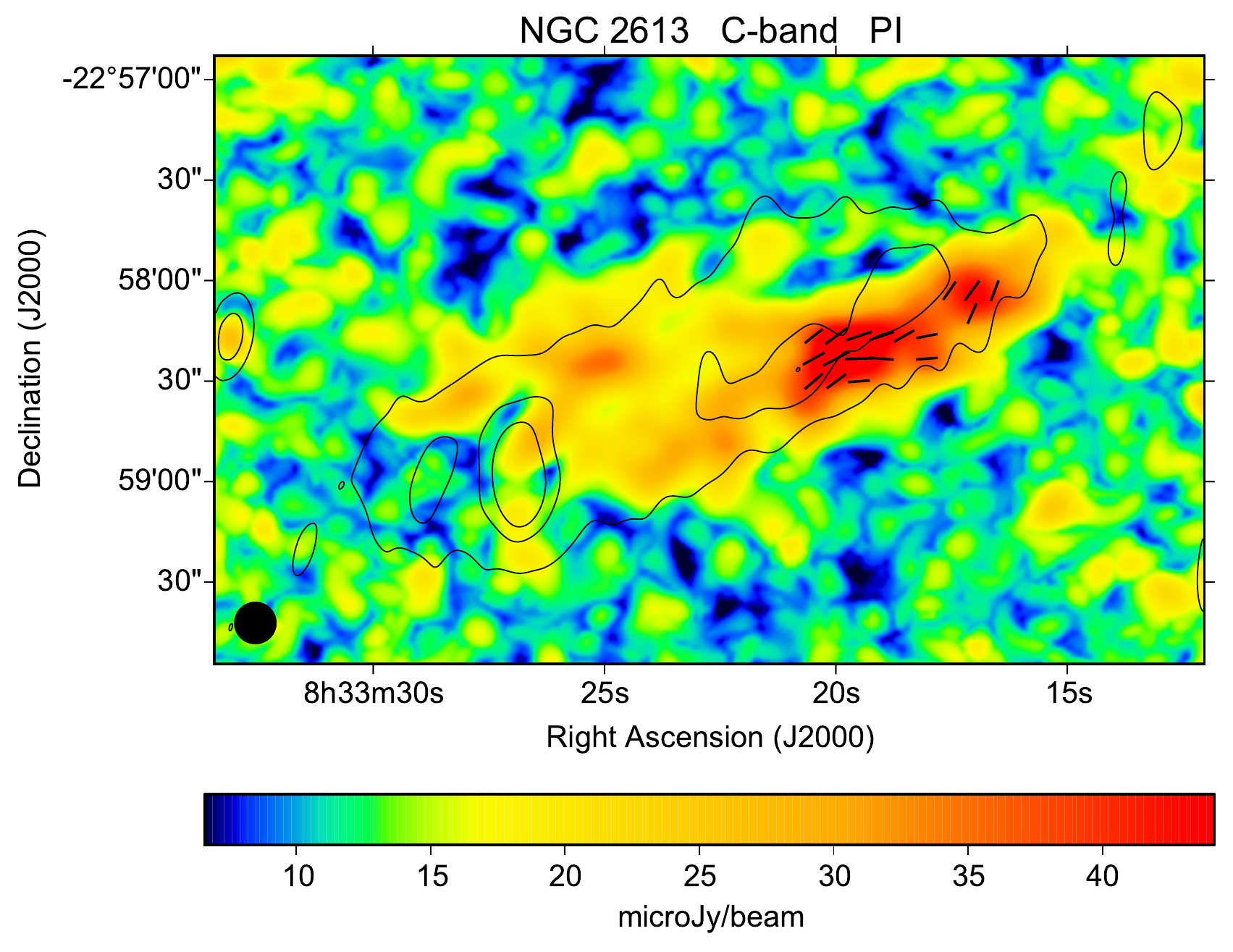}
\includegraphics[width=9.0 cm]{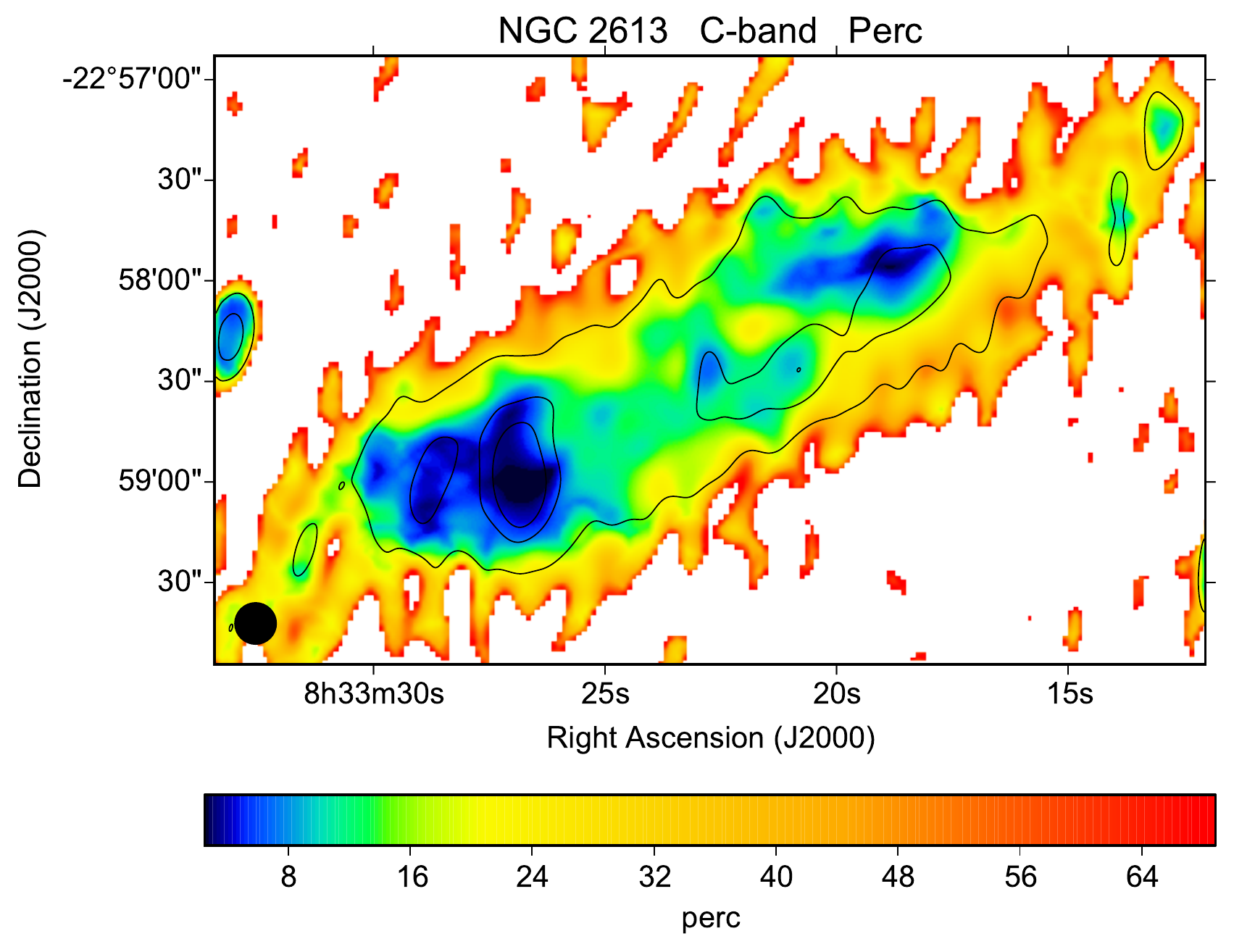}
\includegraphics[width=9.2 cm]{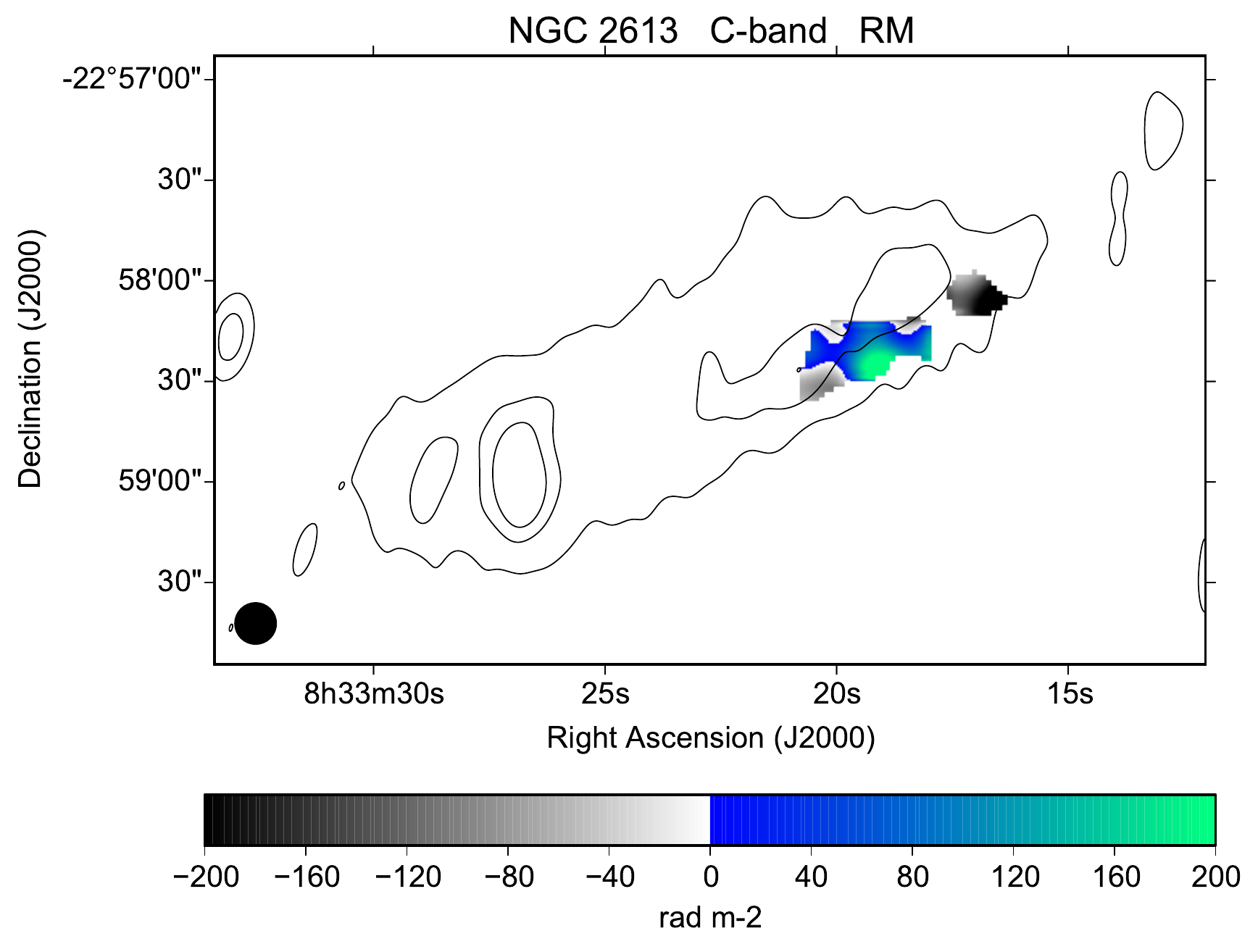}
\includegraphics[width=9.0 cm]{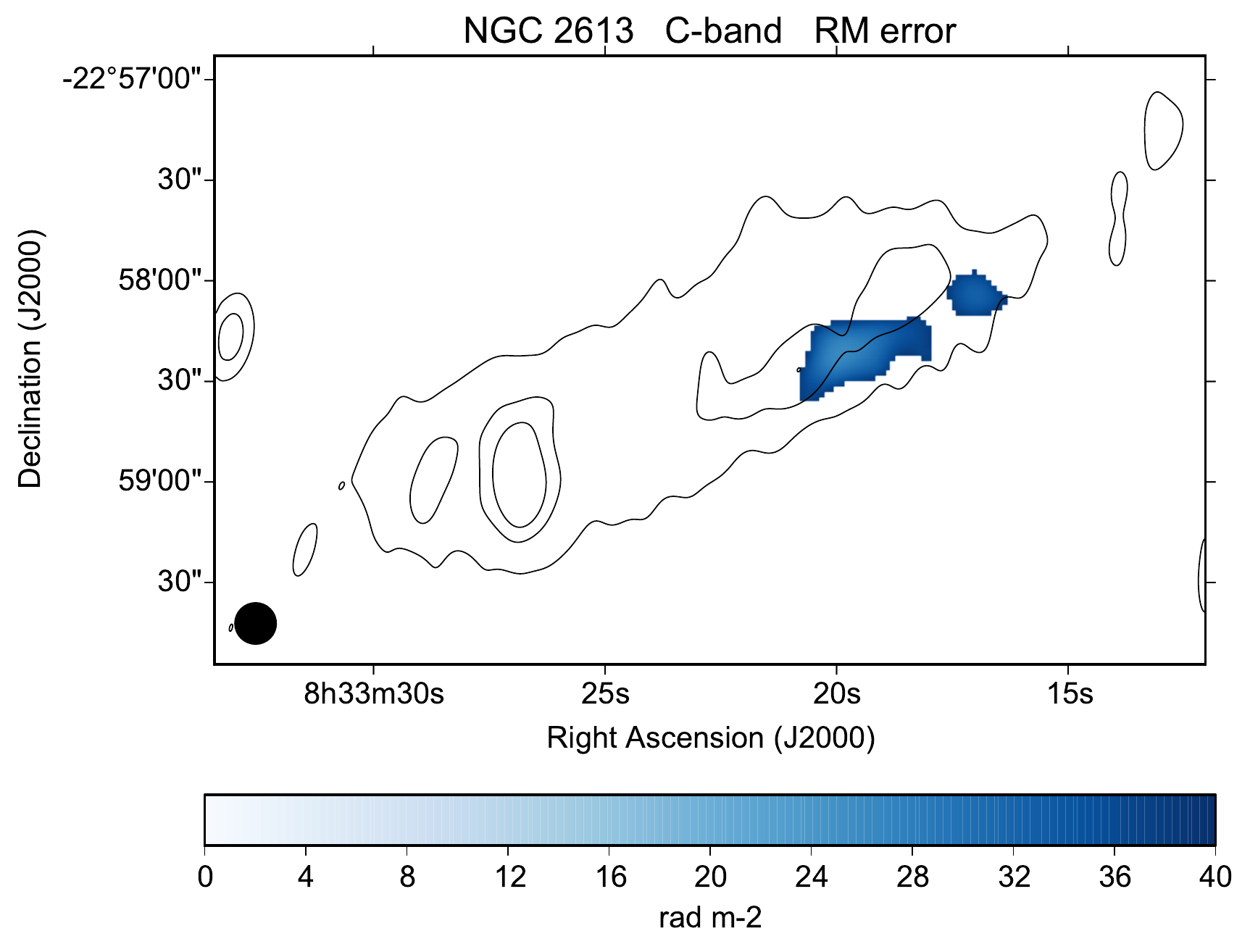}
\includegraphics[width=9.2 cm]{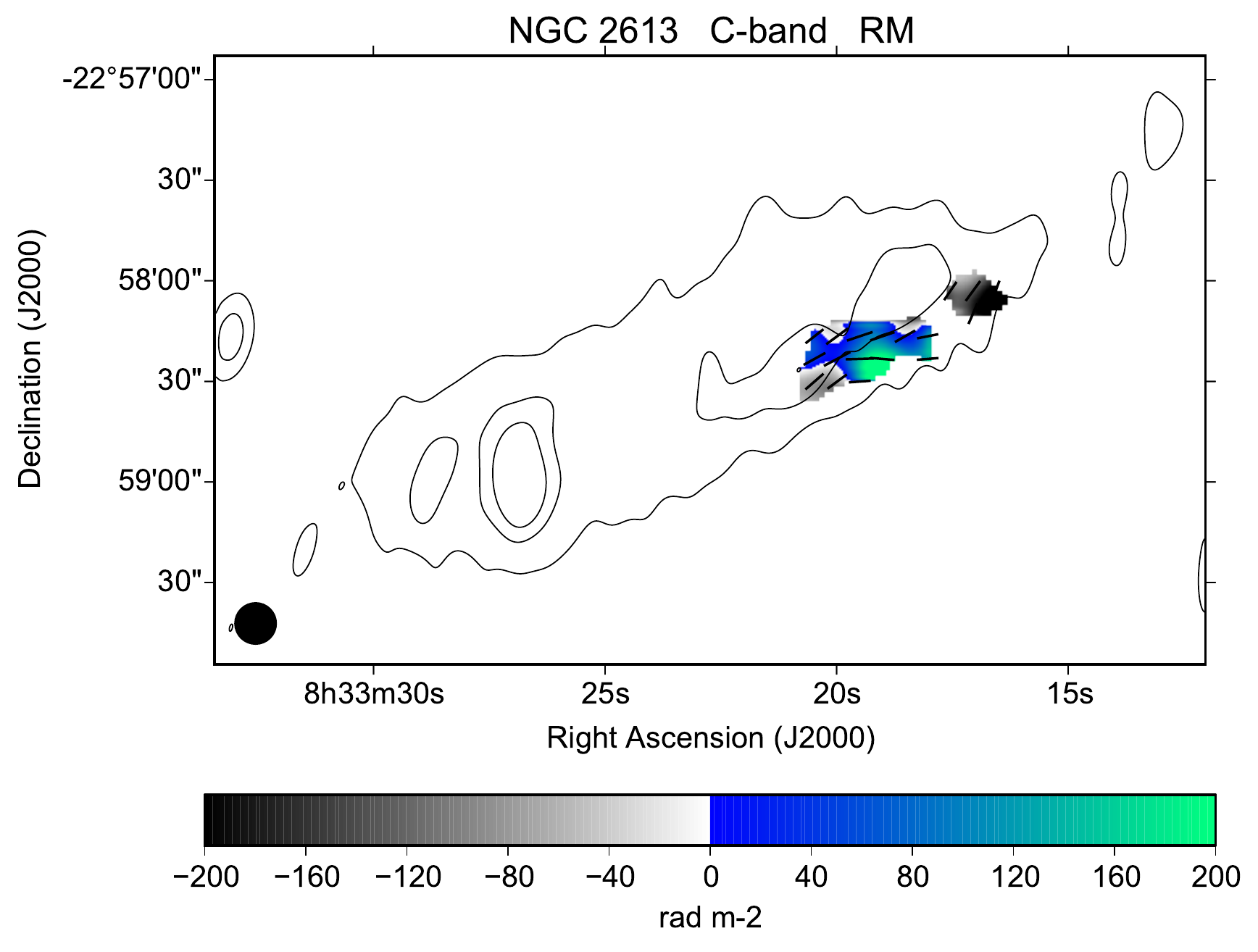}
\caption{Polarization results for NGC~2613 at C-band and $12 \arcsec$ HPBW, corresponding to $1360\,\rm{pc}$. The contour levels (TP) are 75, 225, and 375 $\mu$Jy/beam.
}
\label{n2613all}
\end{figure*}

\begin{figure*}[p]
\centering
\includegraphics[width=9.0 cm]{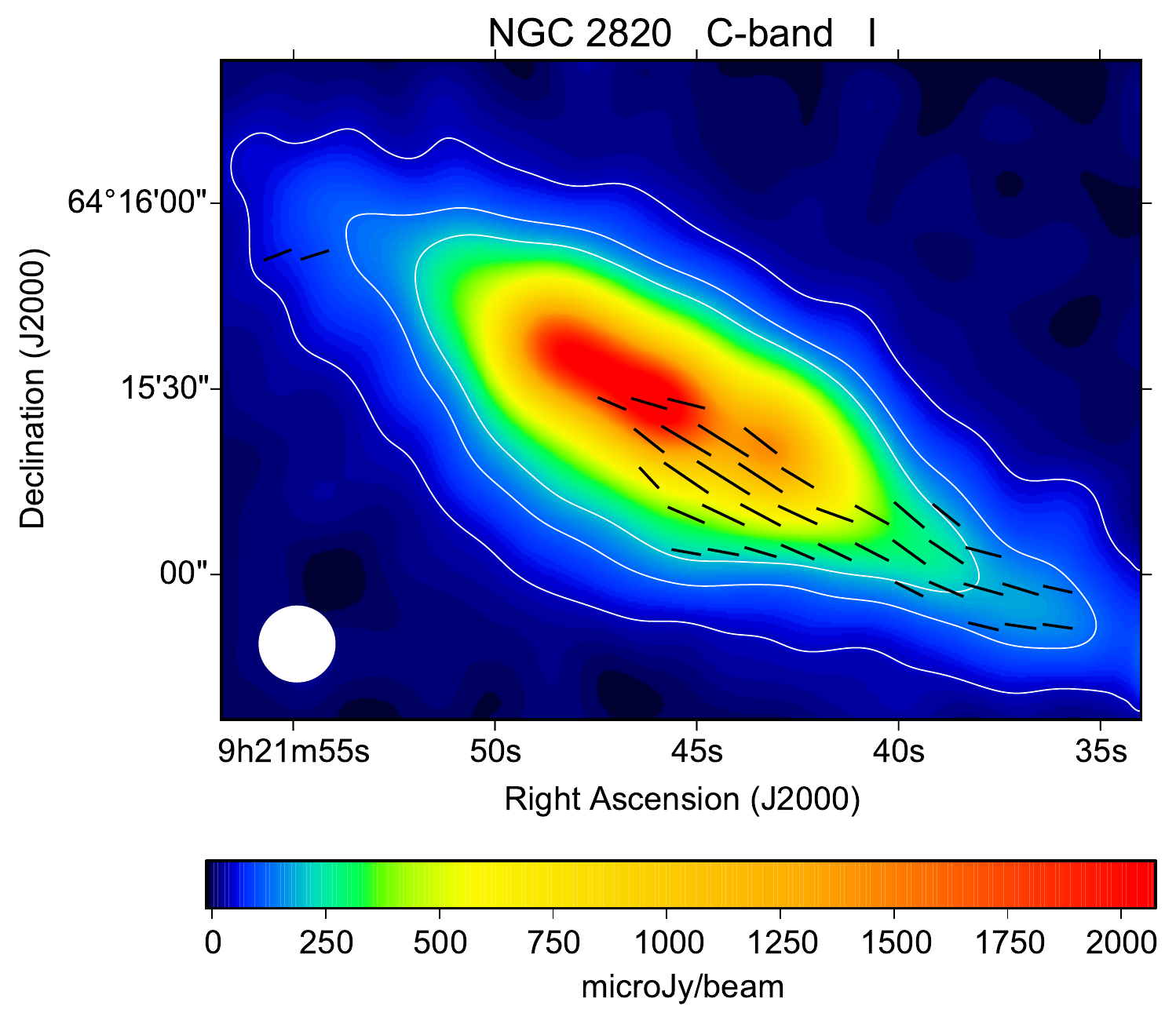}
\includegraphics[width=9.0 cm]{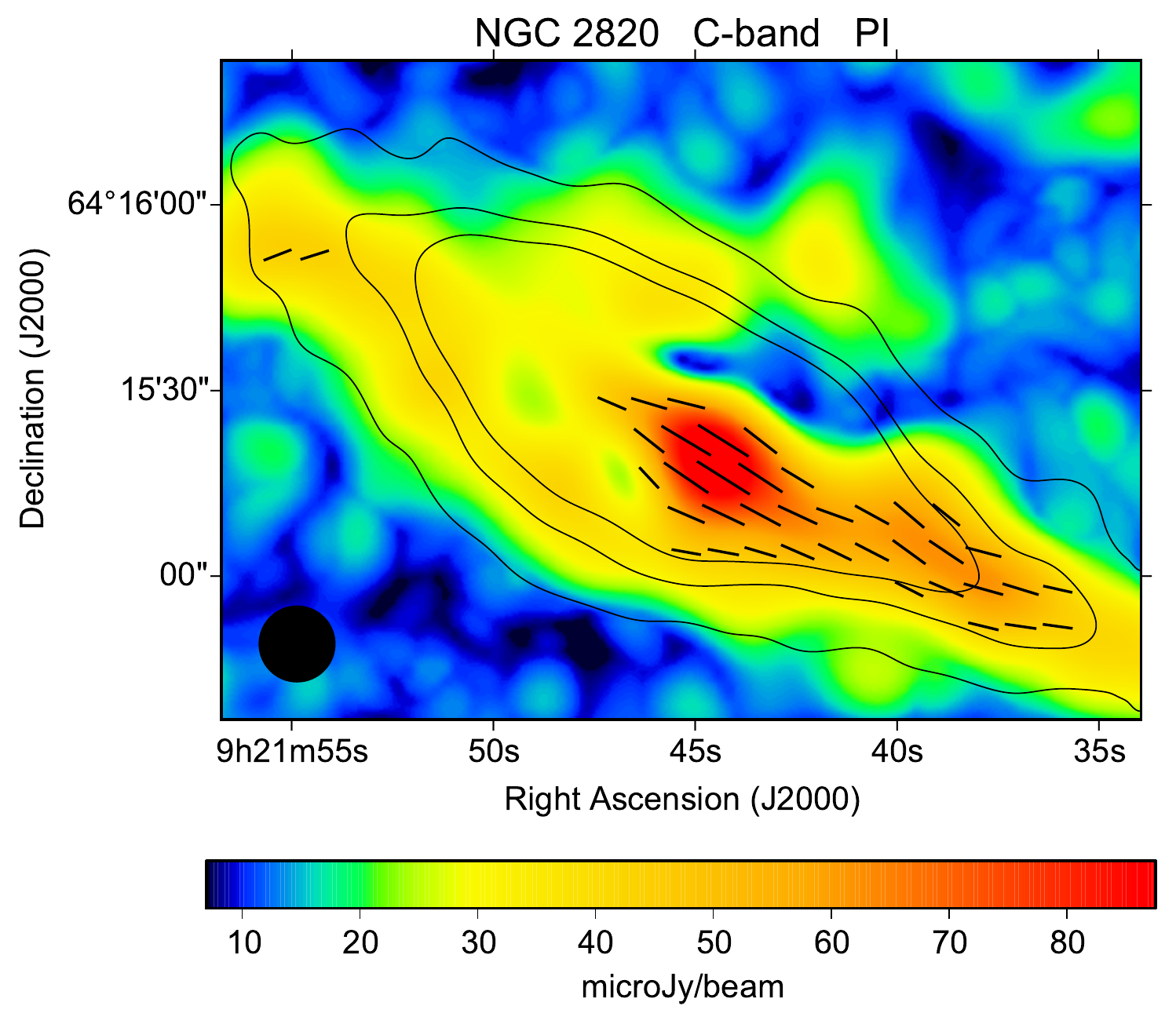}
\includegraphics[width=9.0 cm]{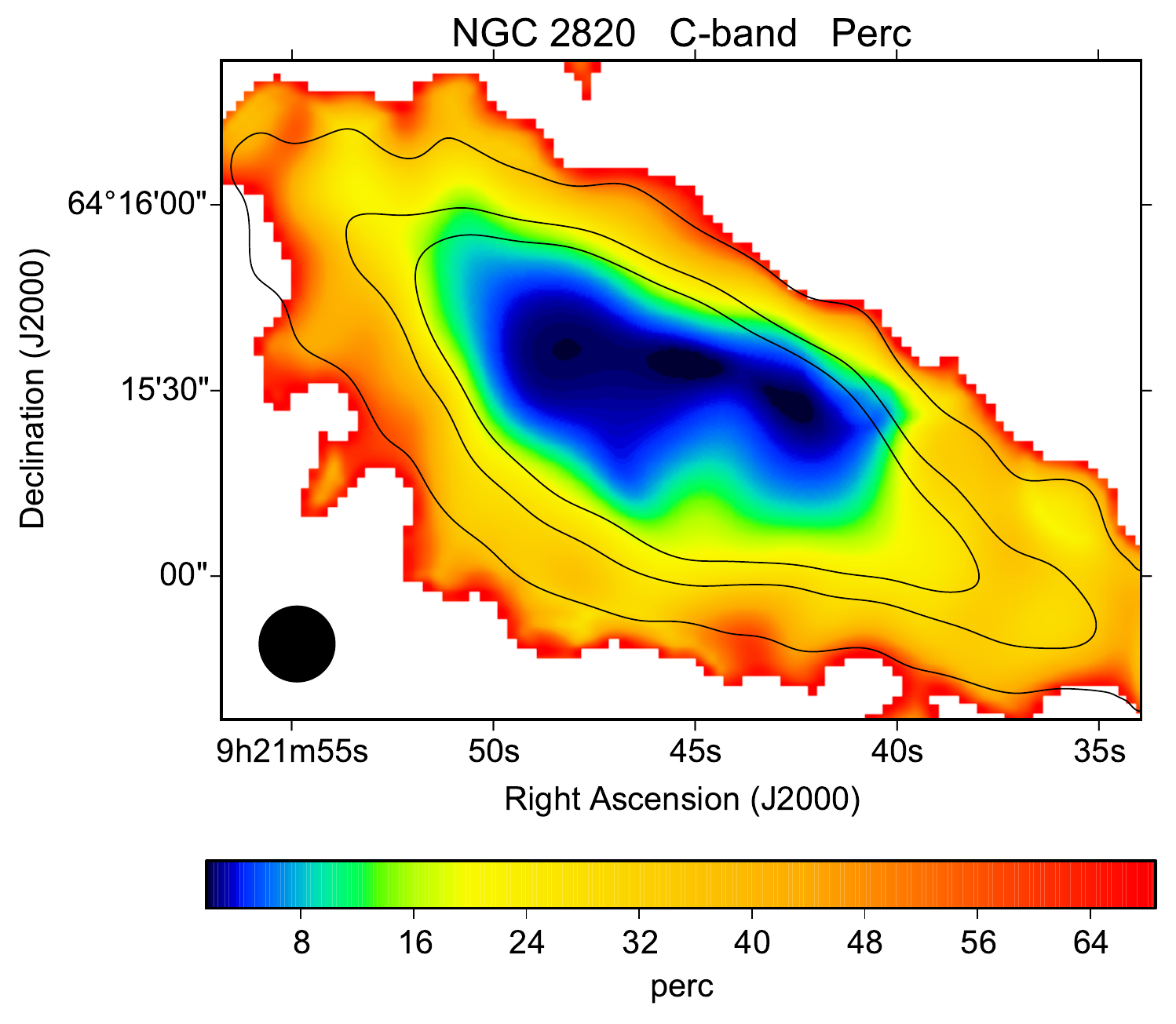}
\includegraphics[width=9.2 cm]{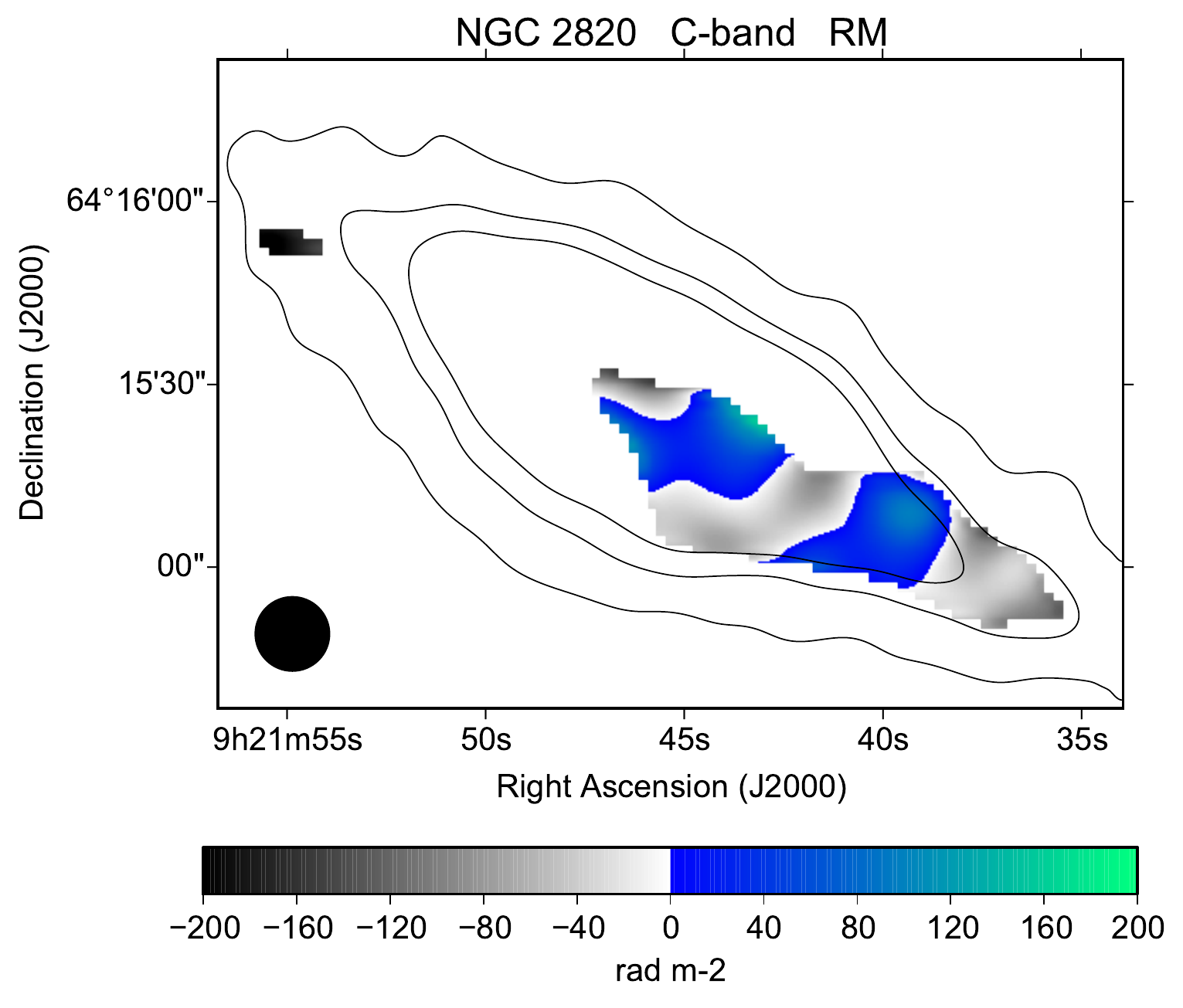}
\includegraphics[width=9.0 cm]{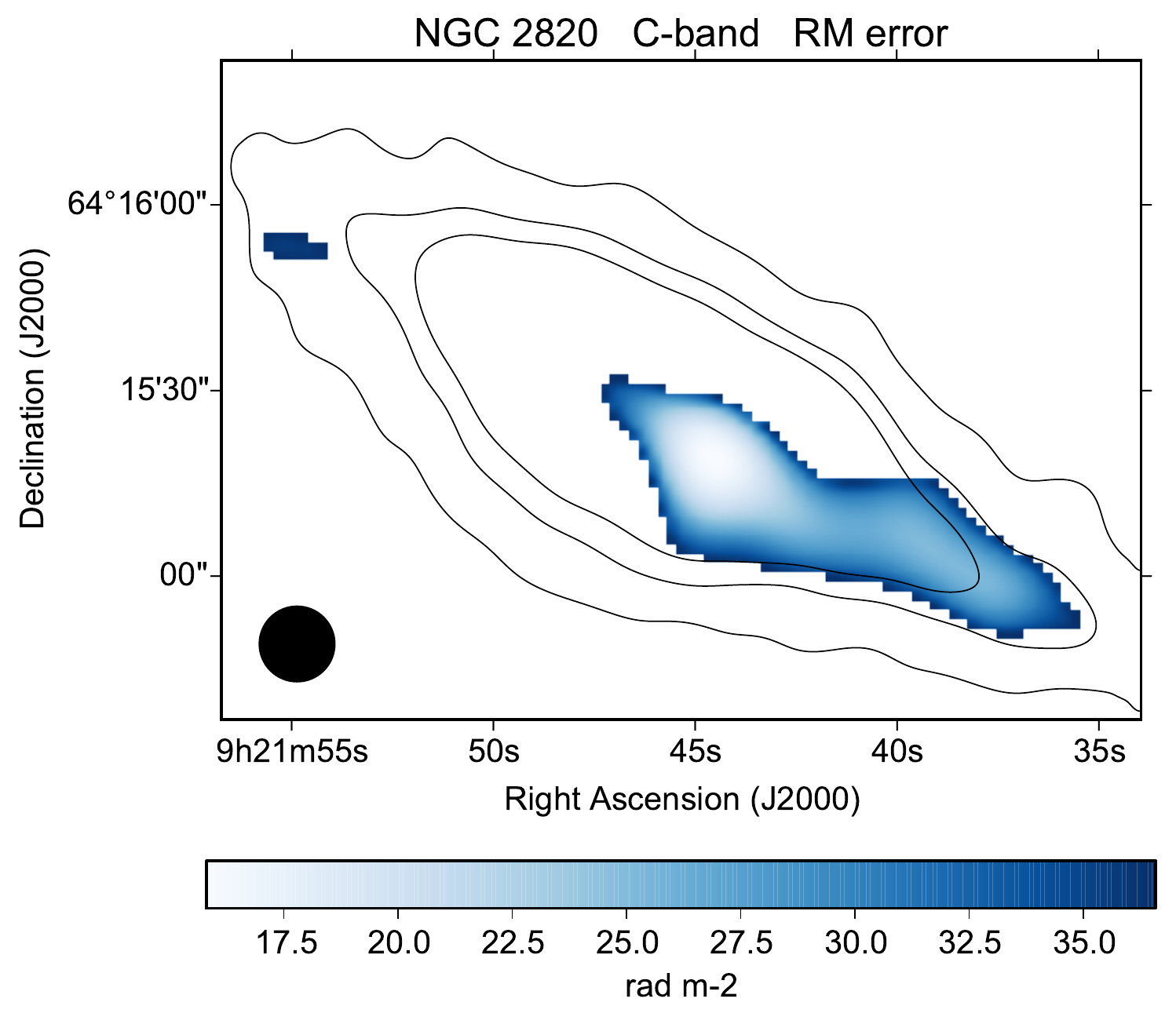}
\includegraphics[width=9.2 cm]{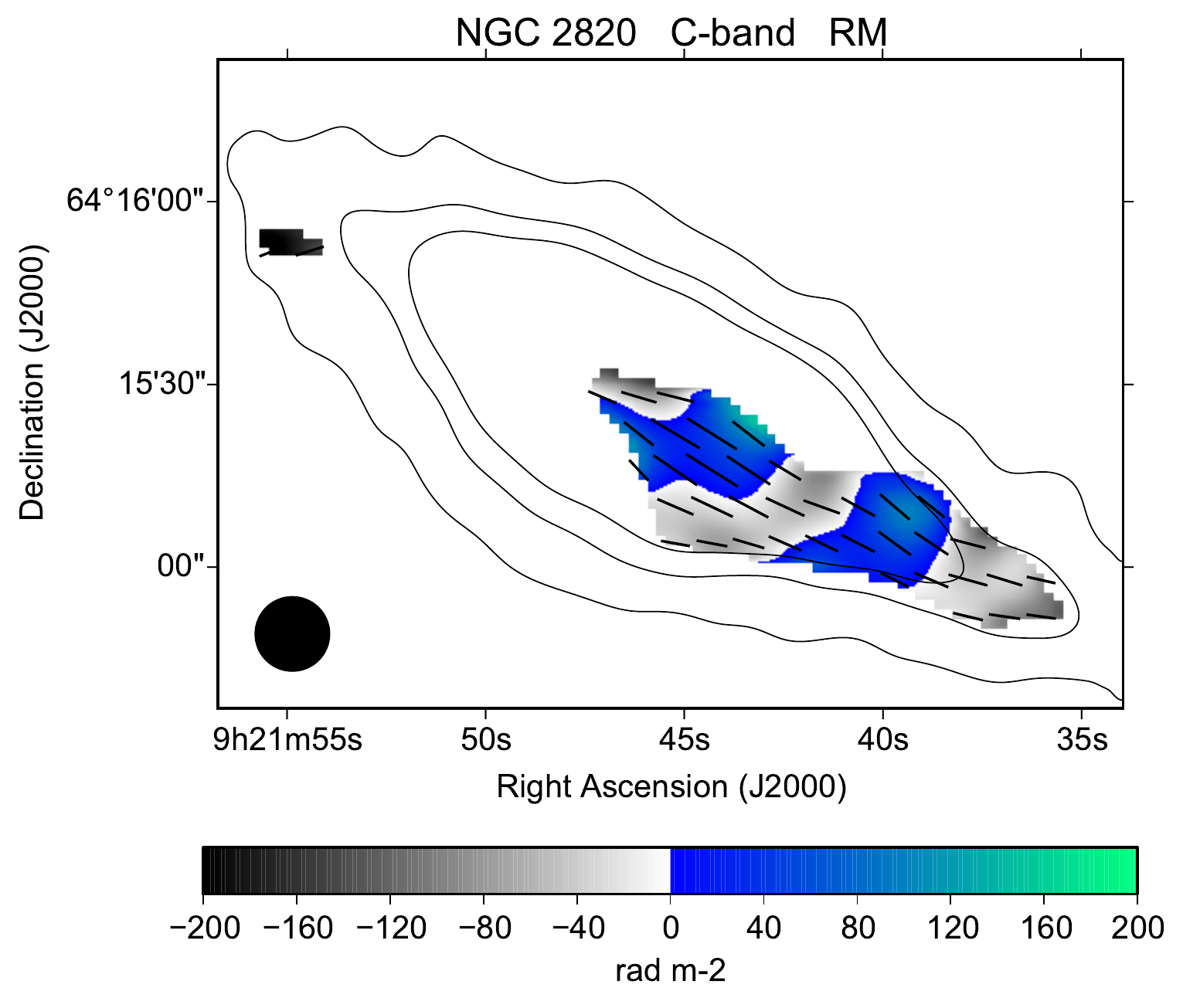}
\caption{Polarization results for NGC~2820 at C-band and $12 \arcsec$ HPBW, corresponding to $360\,\rm{pc}$. The contour levels (TP) are 40, 120, and 200 $\mu$Jy/beam.
}
\label{n2820all}
\end{figure*}

\begin{figure*}[p]
\centering
\includegraphics[width=8.5 cm]{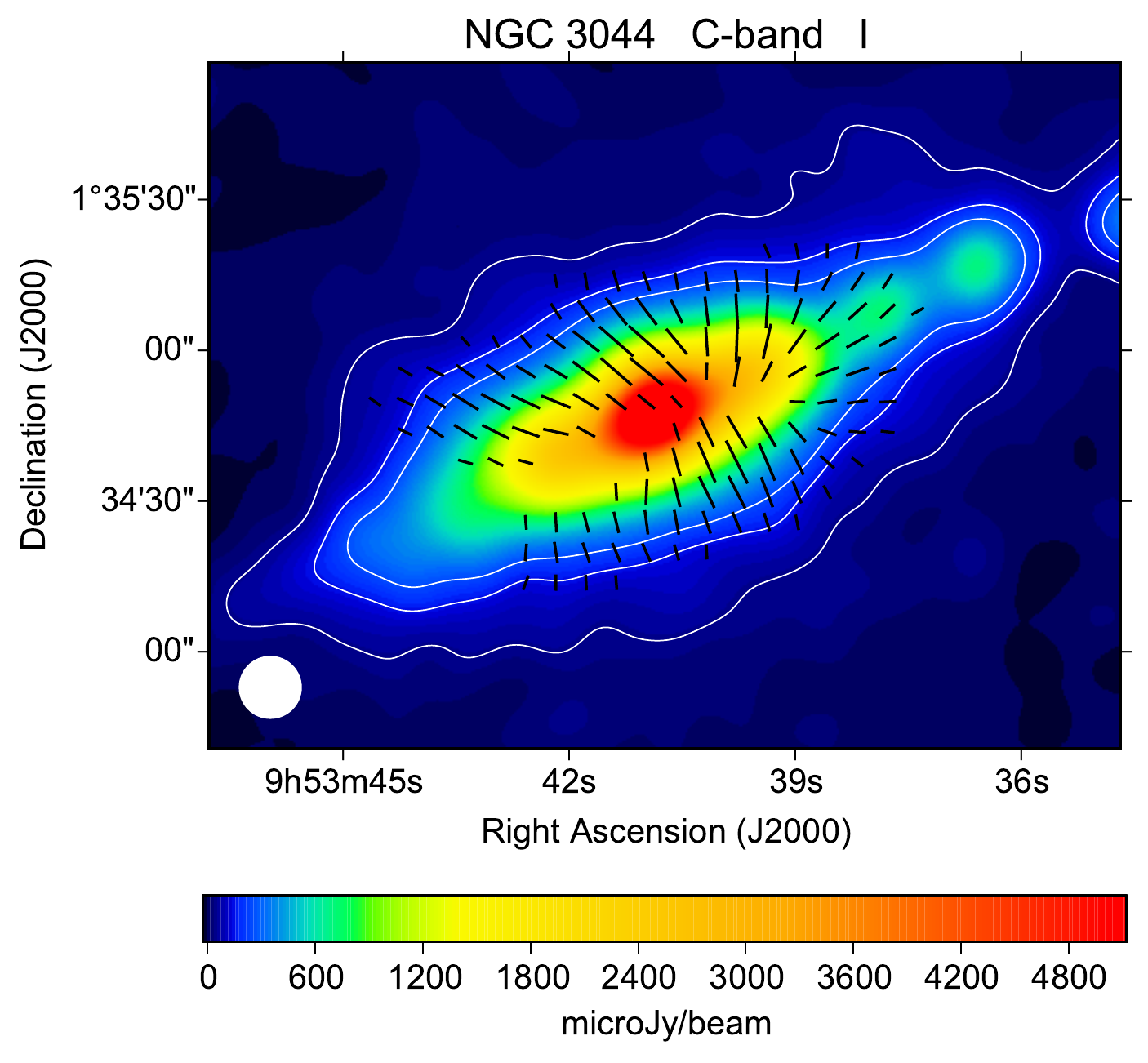}
\includegraphics[width=8.5 cm]{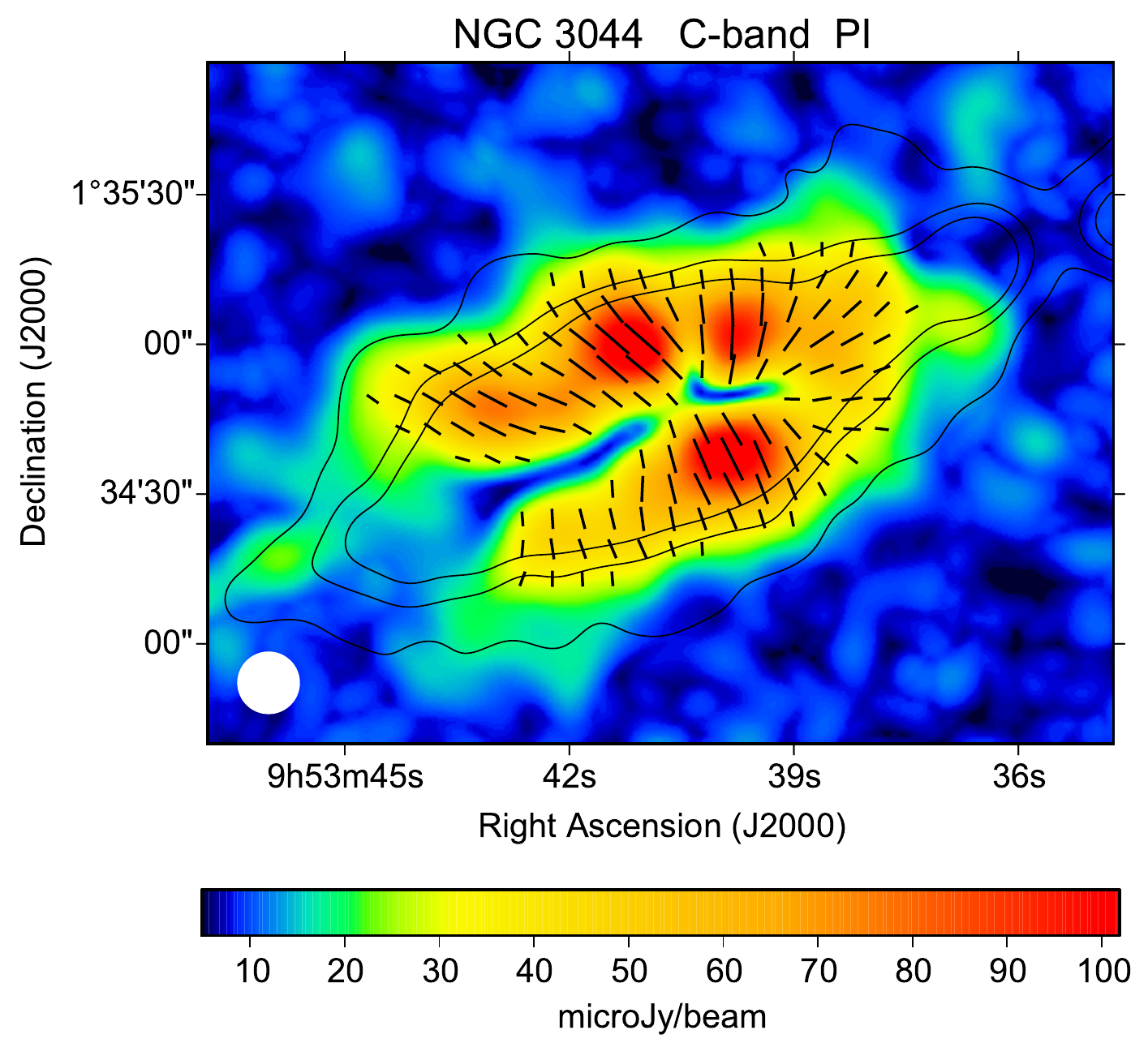}
\includegraphics[width=8.5 cm]{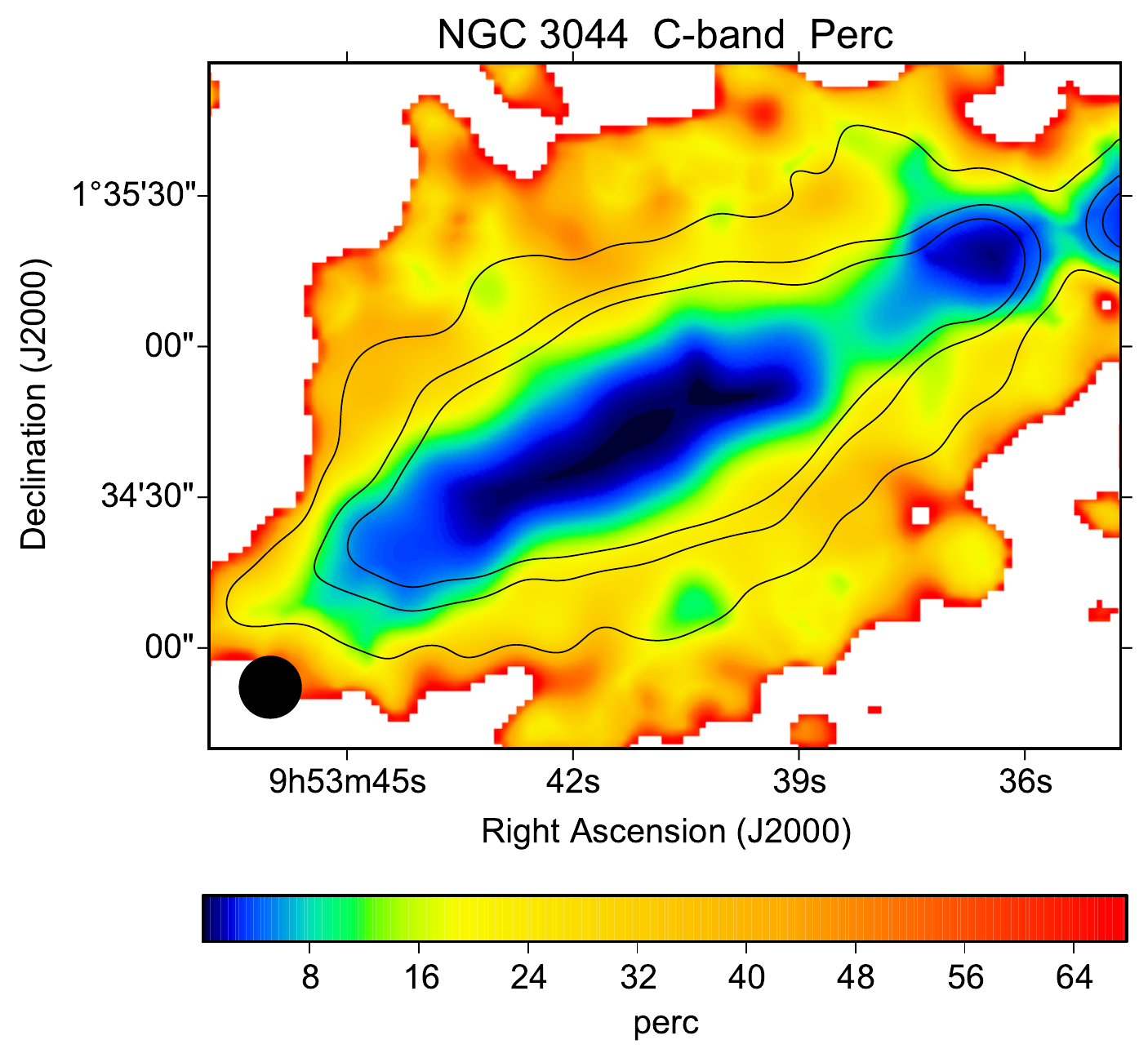}
\includegraphics[width=8.6 cm]{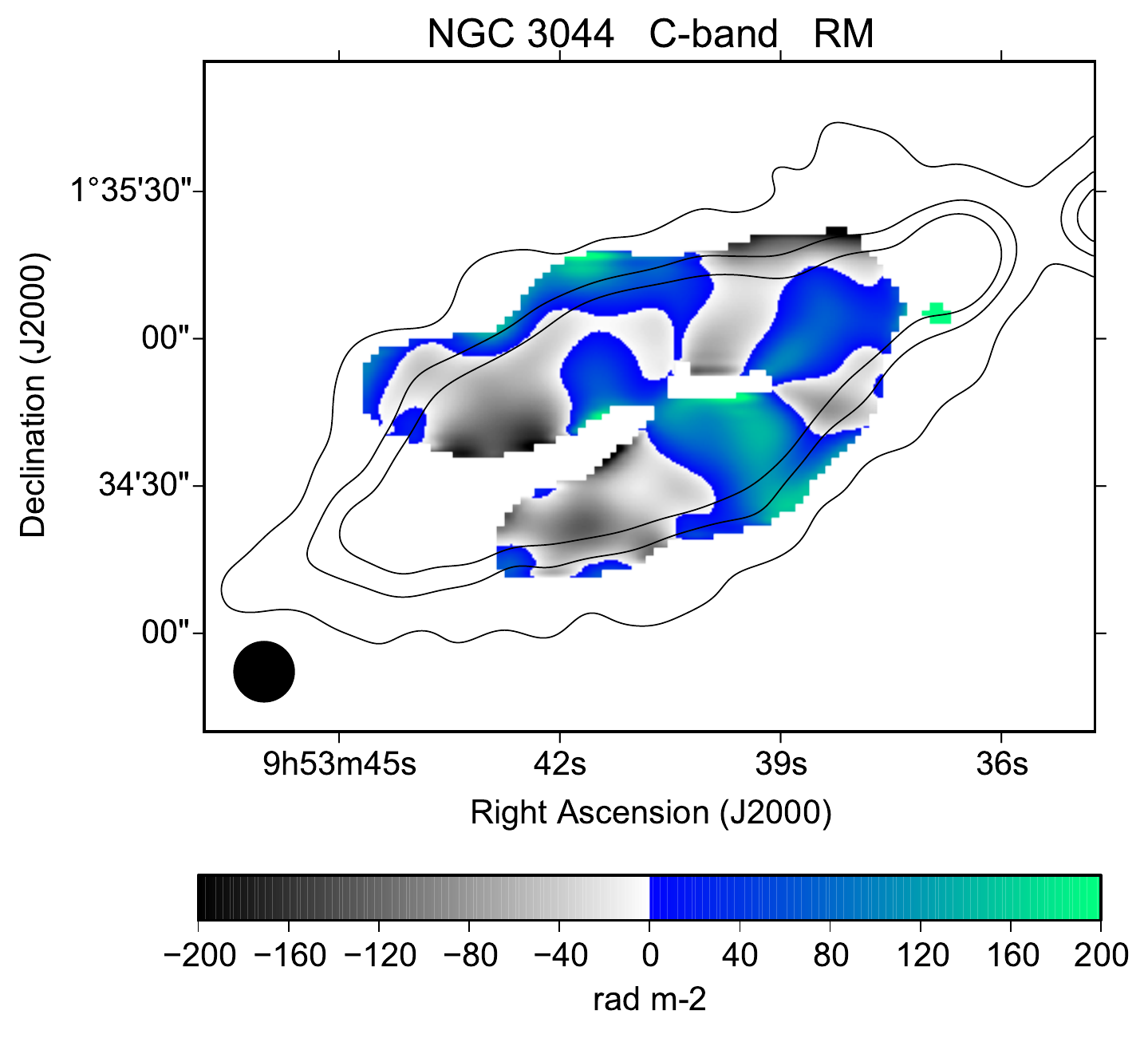}
\includegraphics[width=8.5 cm]{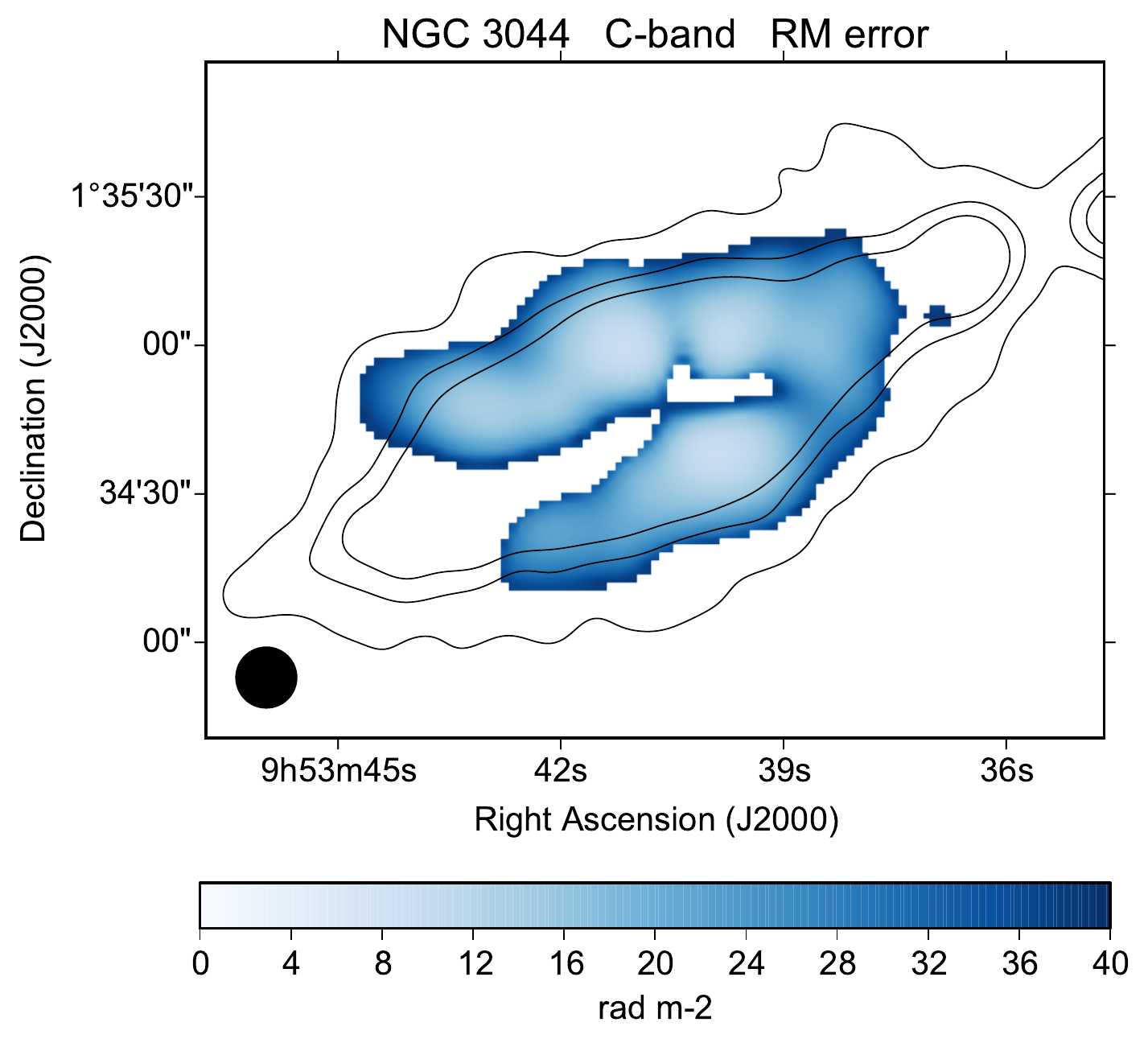}
\includegraphics[width=8.6 cm]{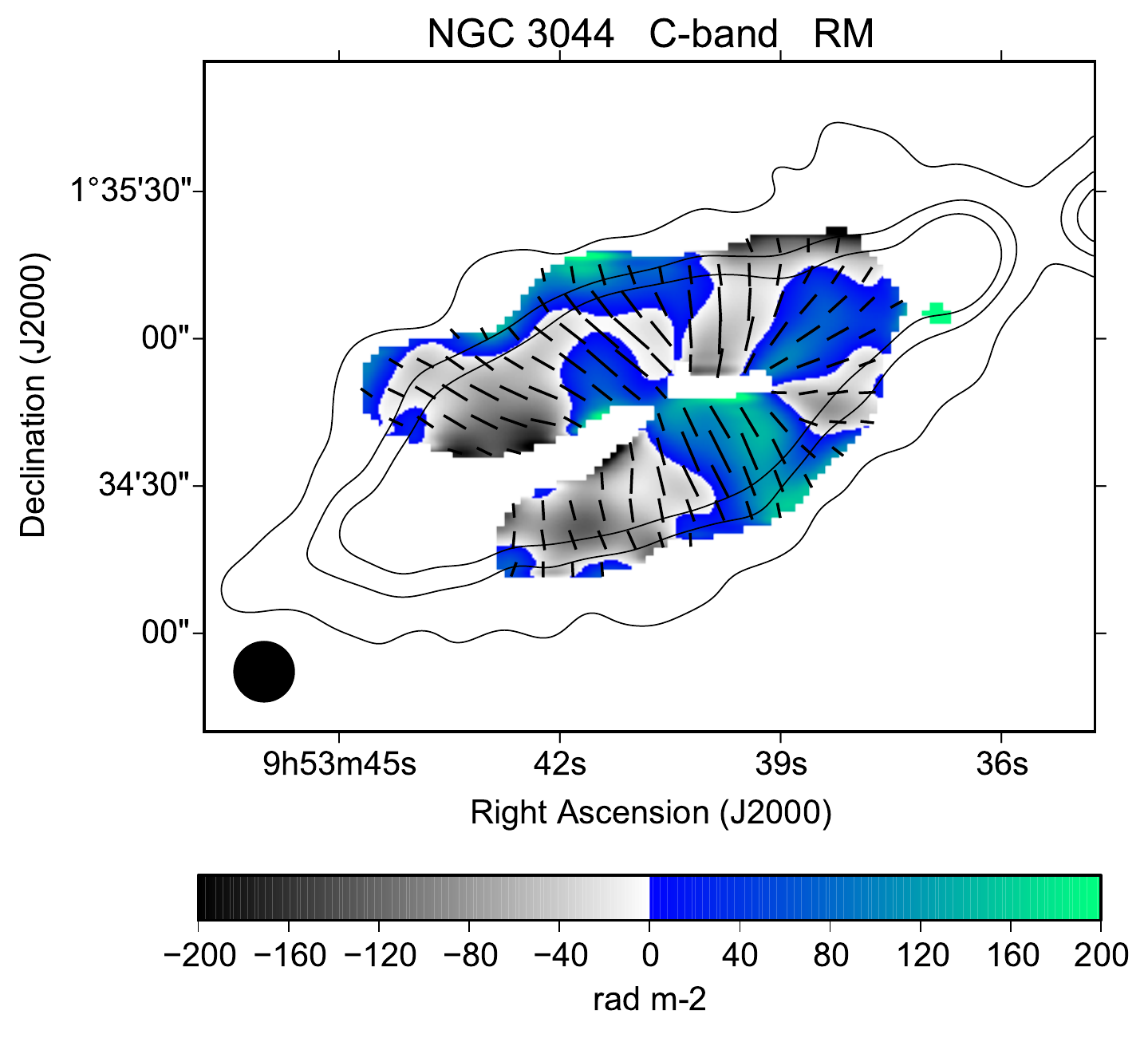}
\caption{Polarization results for NGC~3044 at C-band and $12 \arcsec$ HPBW, corresponding to $1180\,\rm{pc}$. The contour levels (TP) are 50, 150, and 250 $\mu$Jy/beam.
}
\label{n3044all}
\end{figure*}

\begin{figure*}[p]
\centering
\includegraphics[width=8.0 cm]{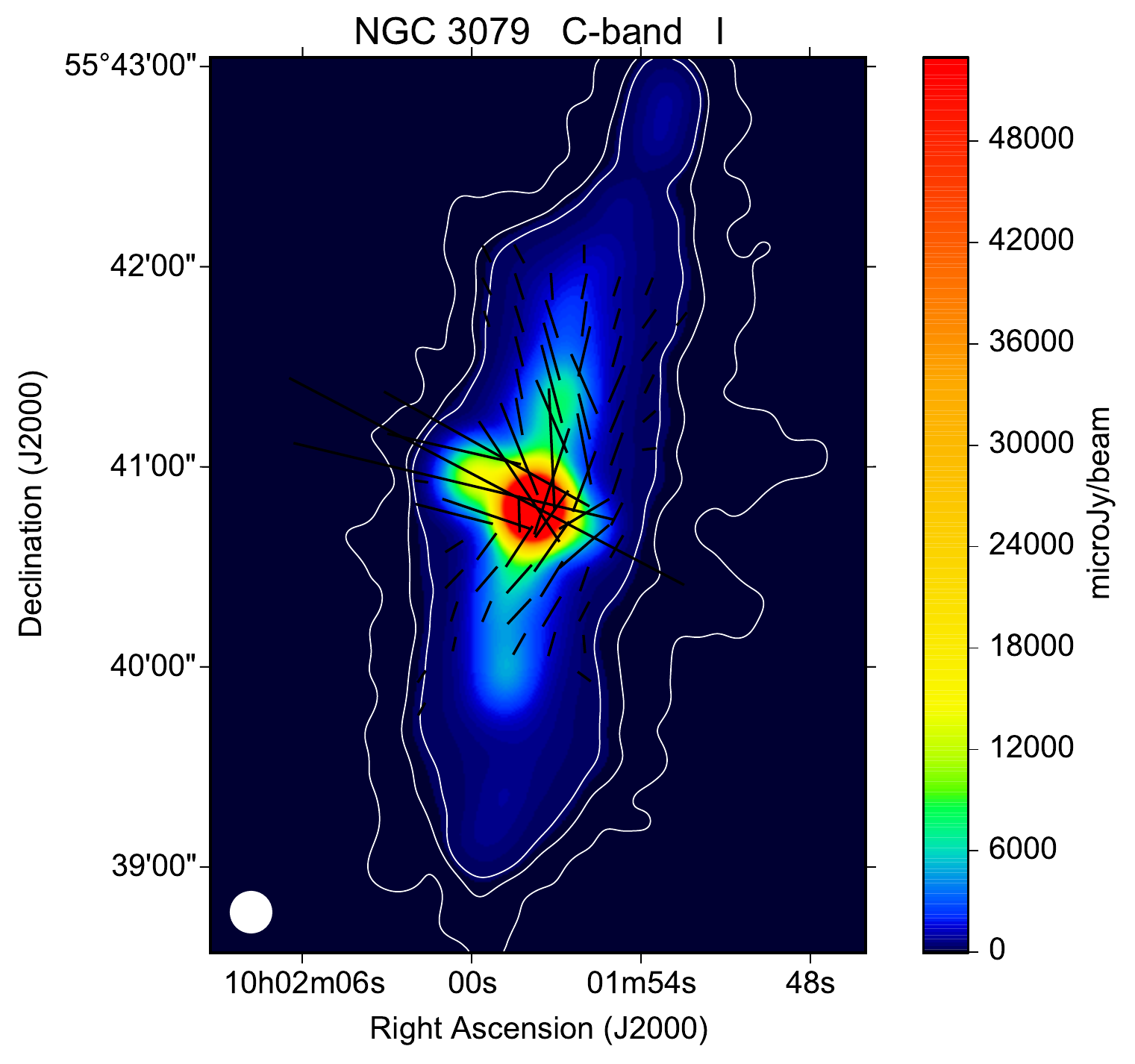}
\includegraphics[width=7.7 cm]{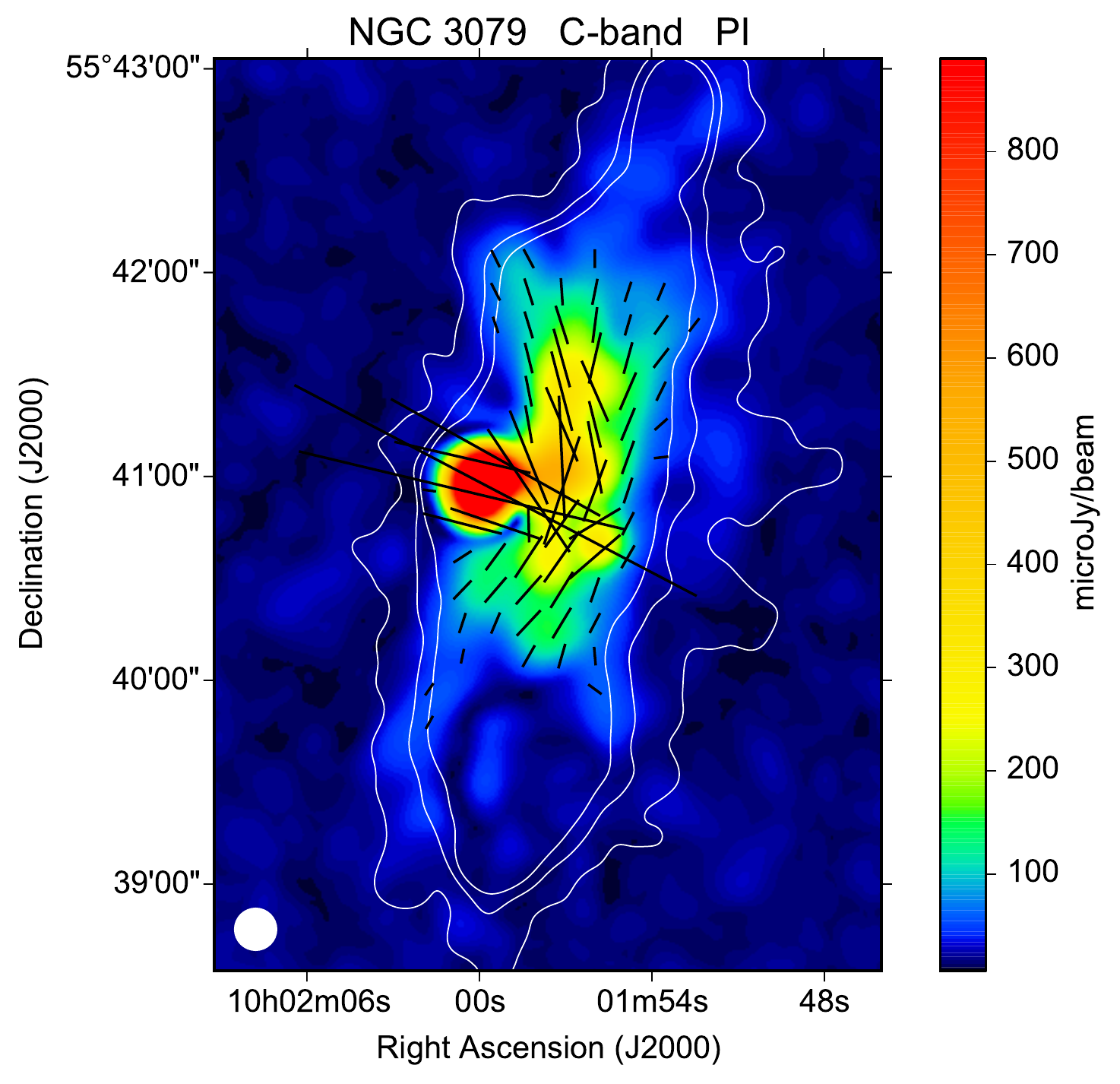}
\includegraphics[width=8.0 cm]{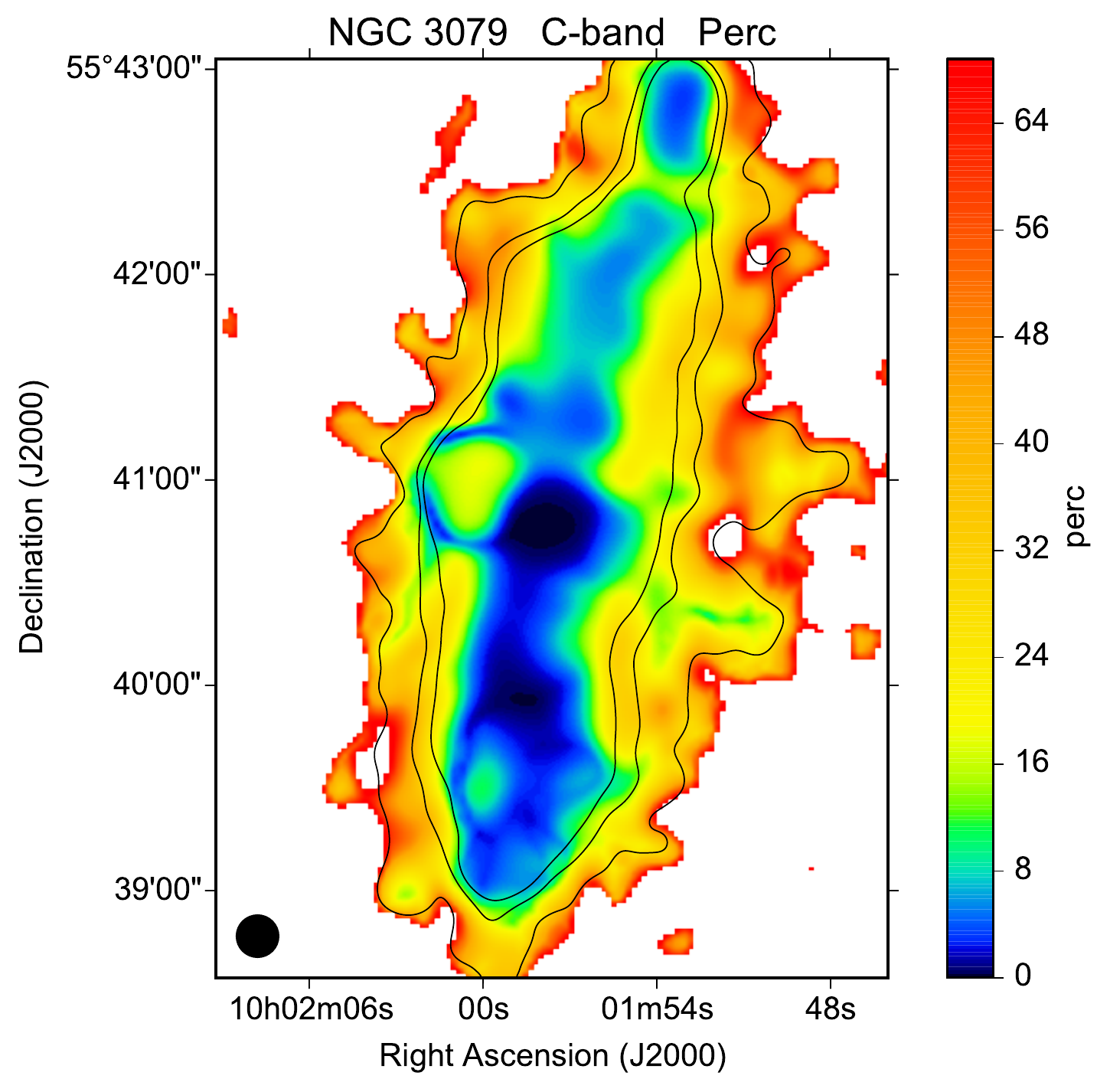}
\includegraphics[width=8.2 cm]{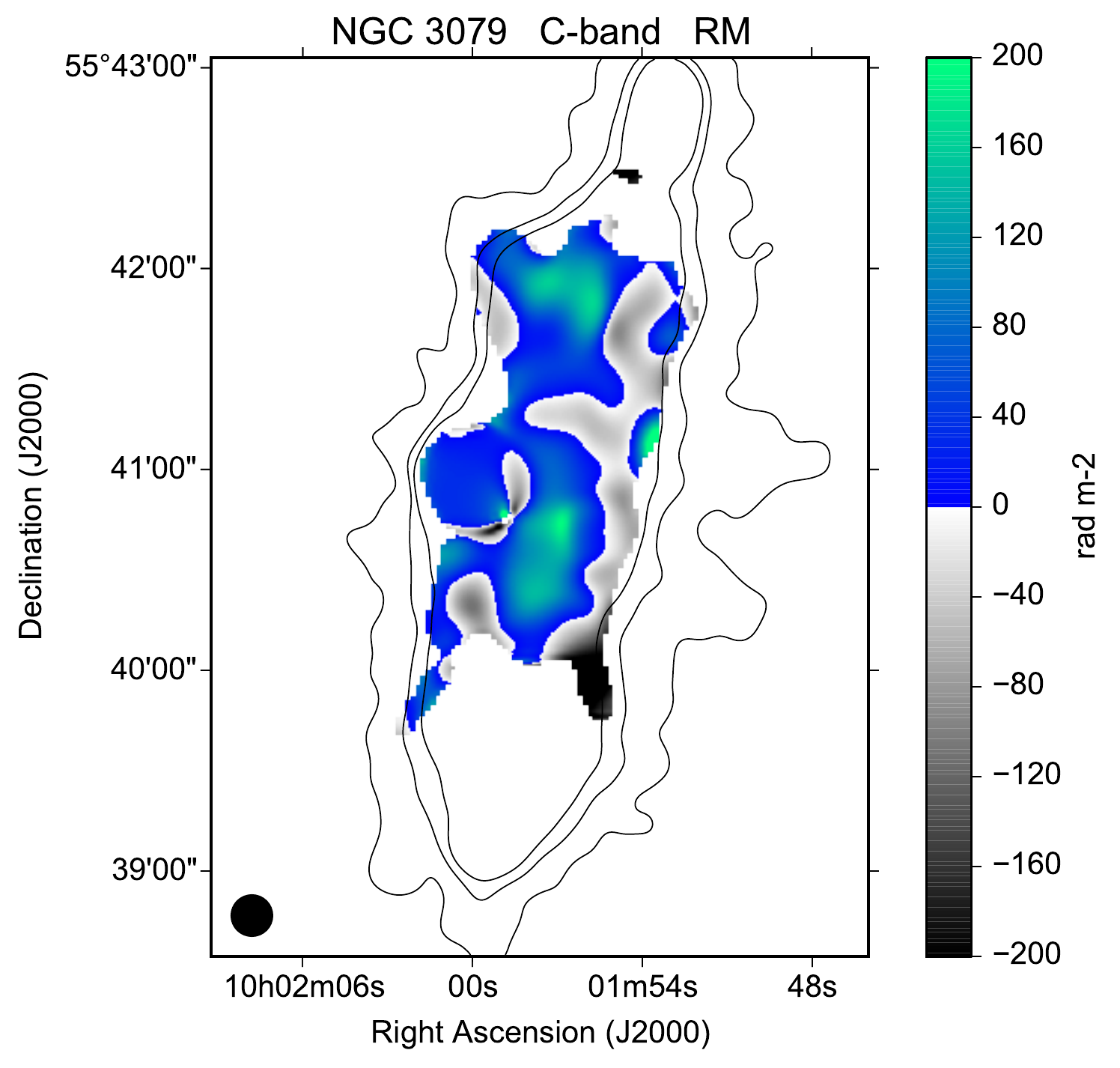}
\includegraphics[width=7.9 cm]{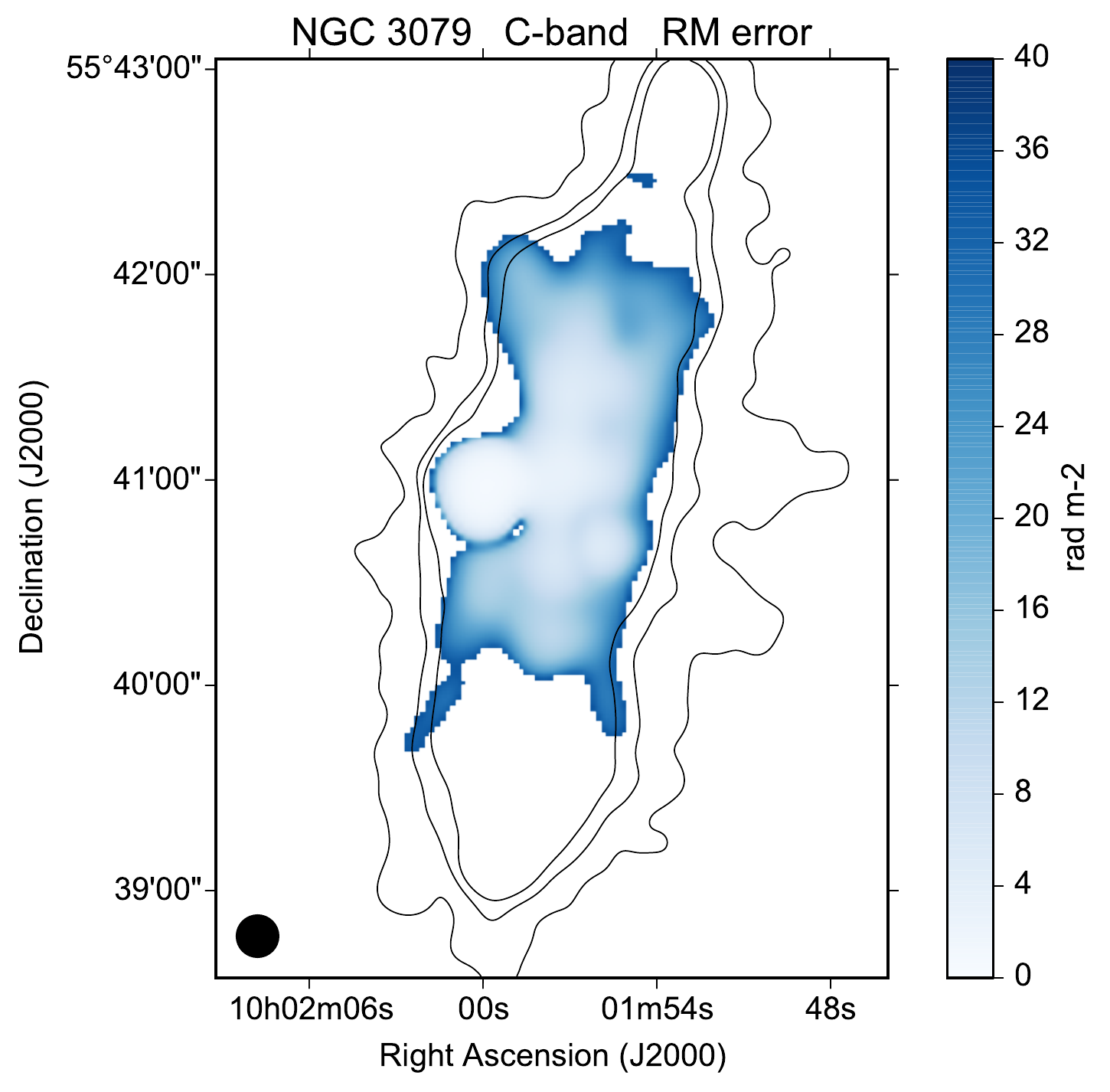}
\includegraphics[width=8.1 cm]{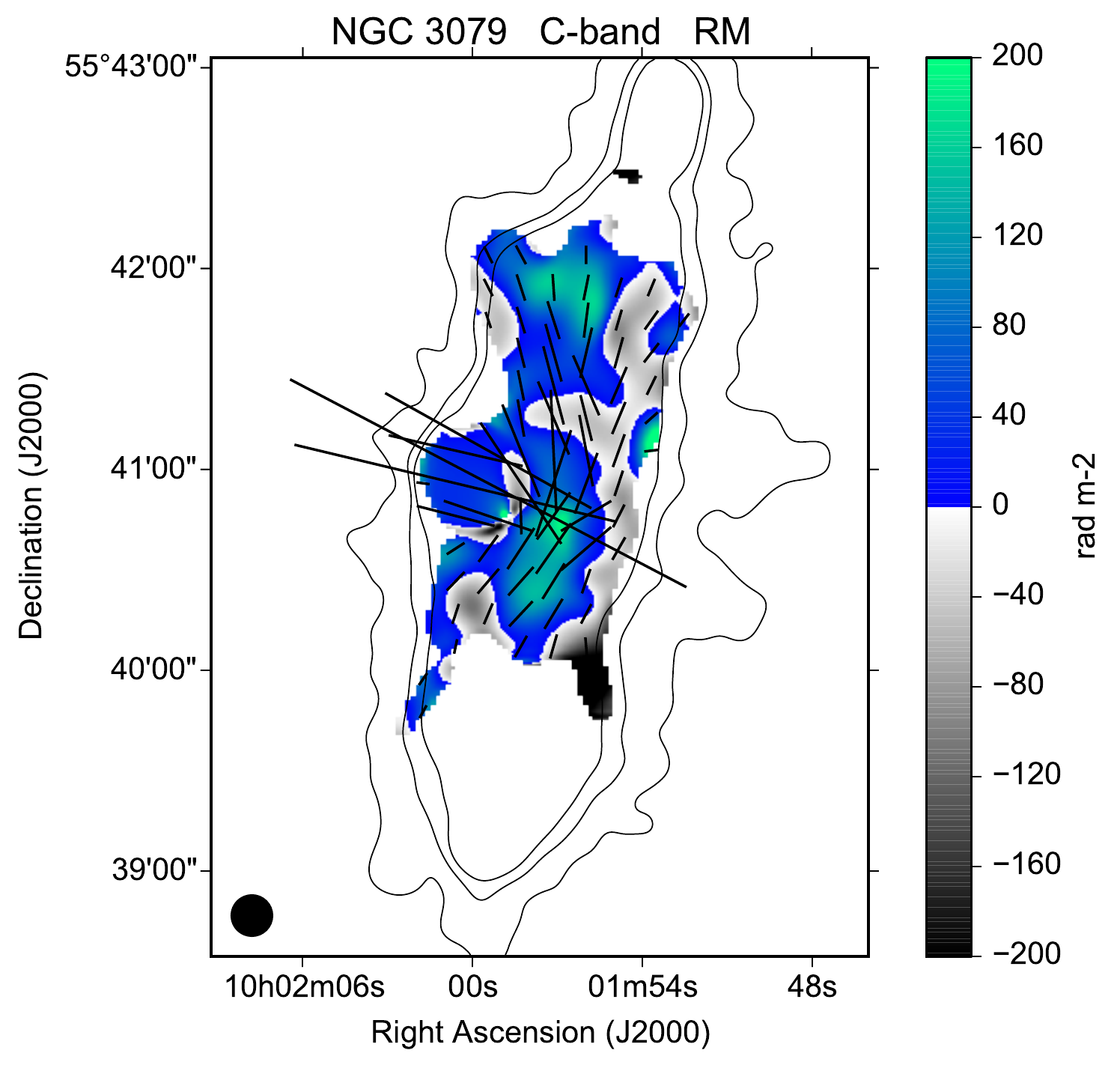}
\caption{Polarization results for NGC~3079 at C-band and $12 \arcsec$ HPBW, corresponding to $1200\,\rm{pc}$. The contour levels (TP) are 40, 120, and 200 $\mu$Jy/beam. The image of the TP map is cut
at 10000~$\mu$Jy/beam in order to present the disk emission well.
}
\label{n3079all}
\end{figure*}

\begin{figure*}[p]
\centering
\includegraphics[width=9.0 cm]{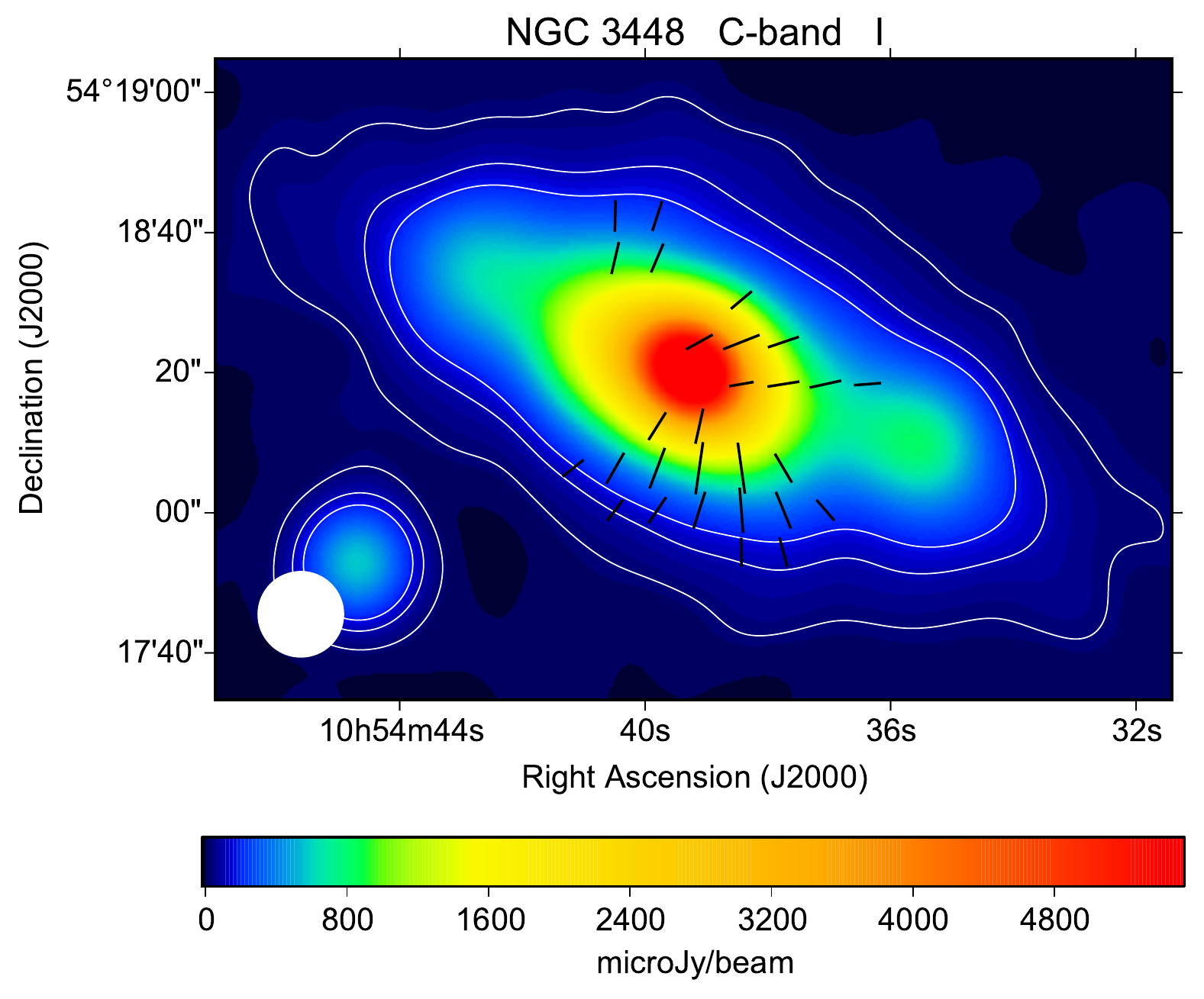}
\includegraphics[width=9.0 cm]{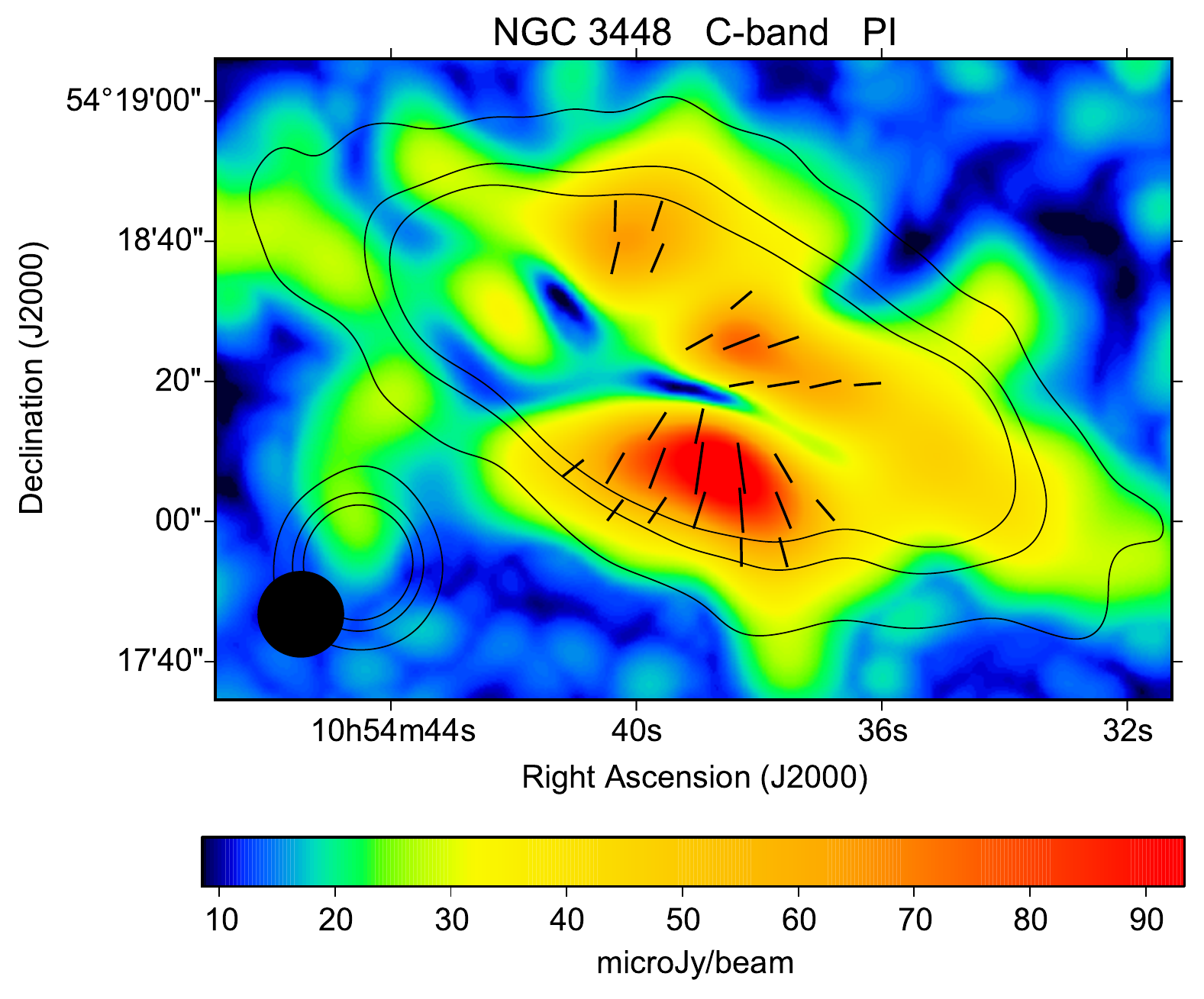}
\includegraphics[width=9.0 cm]{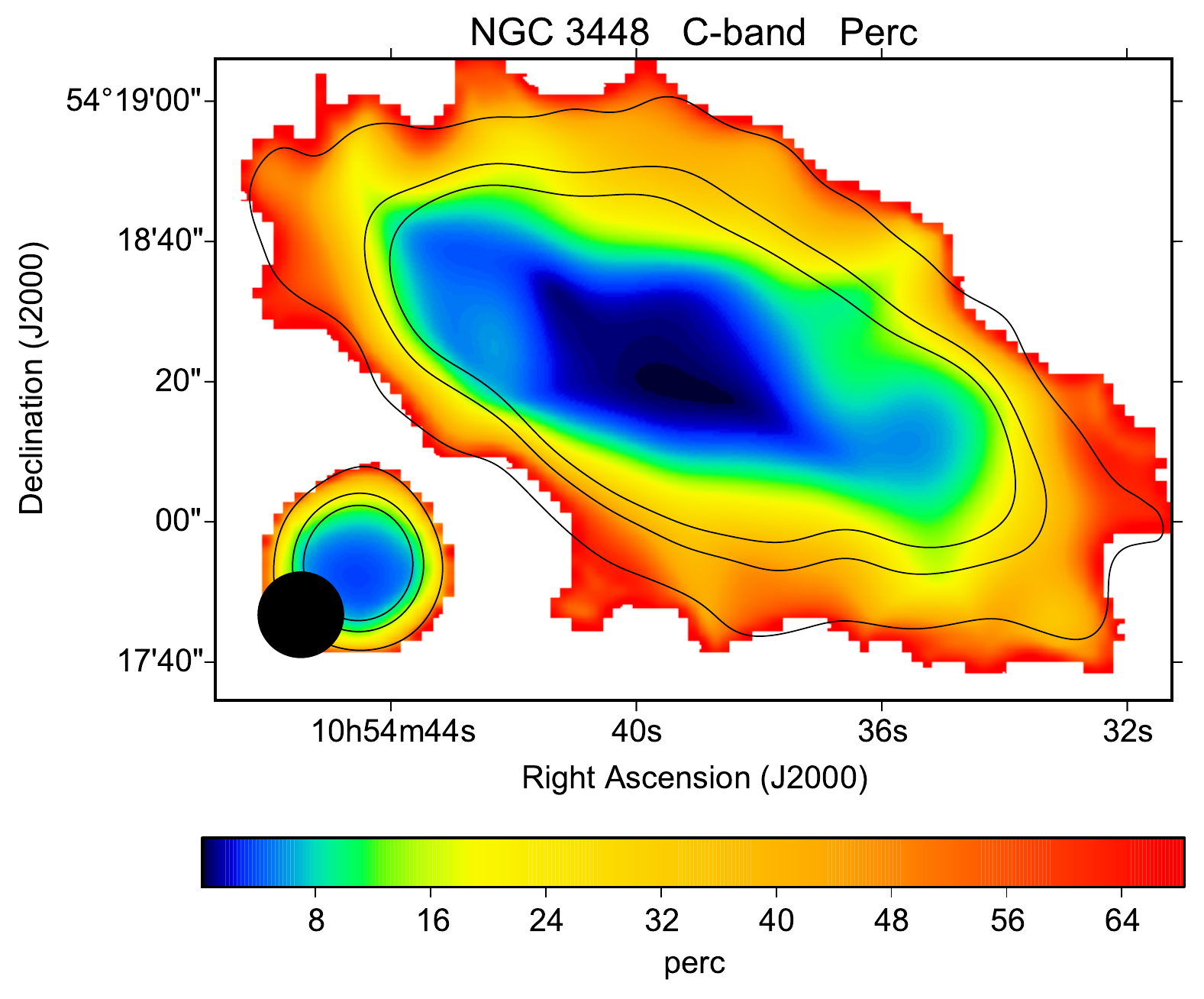}
\includegraphics[width=9.2 cm]{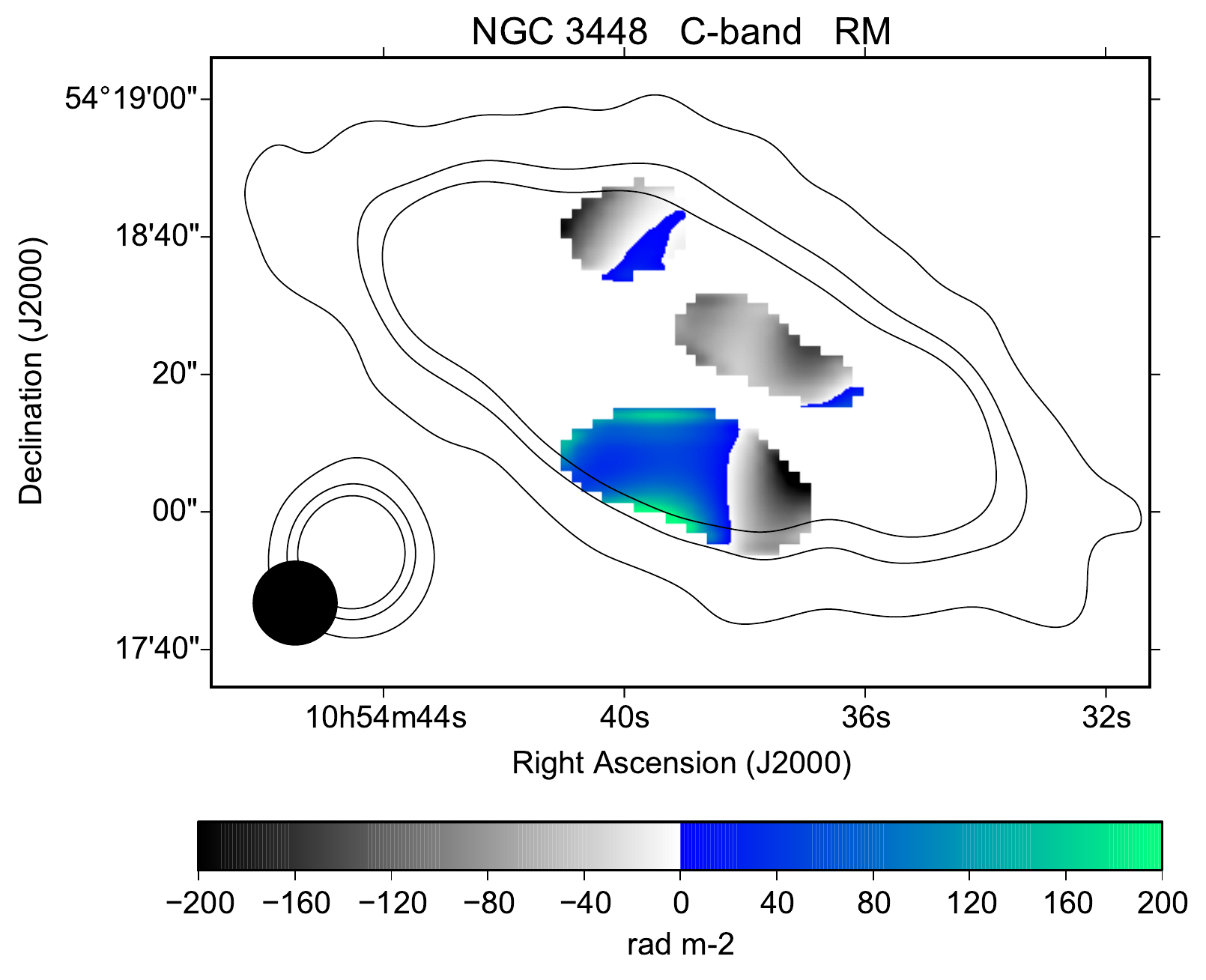}
\includegraphics[width=9.0 cm]{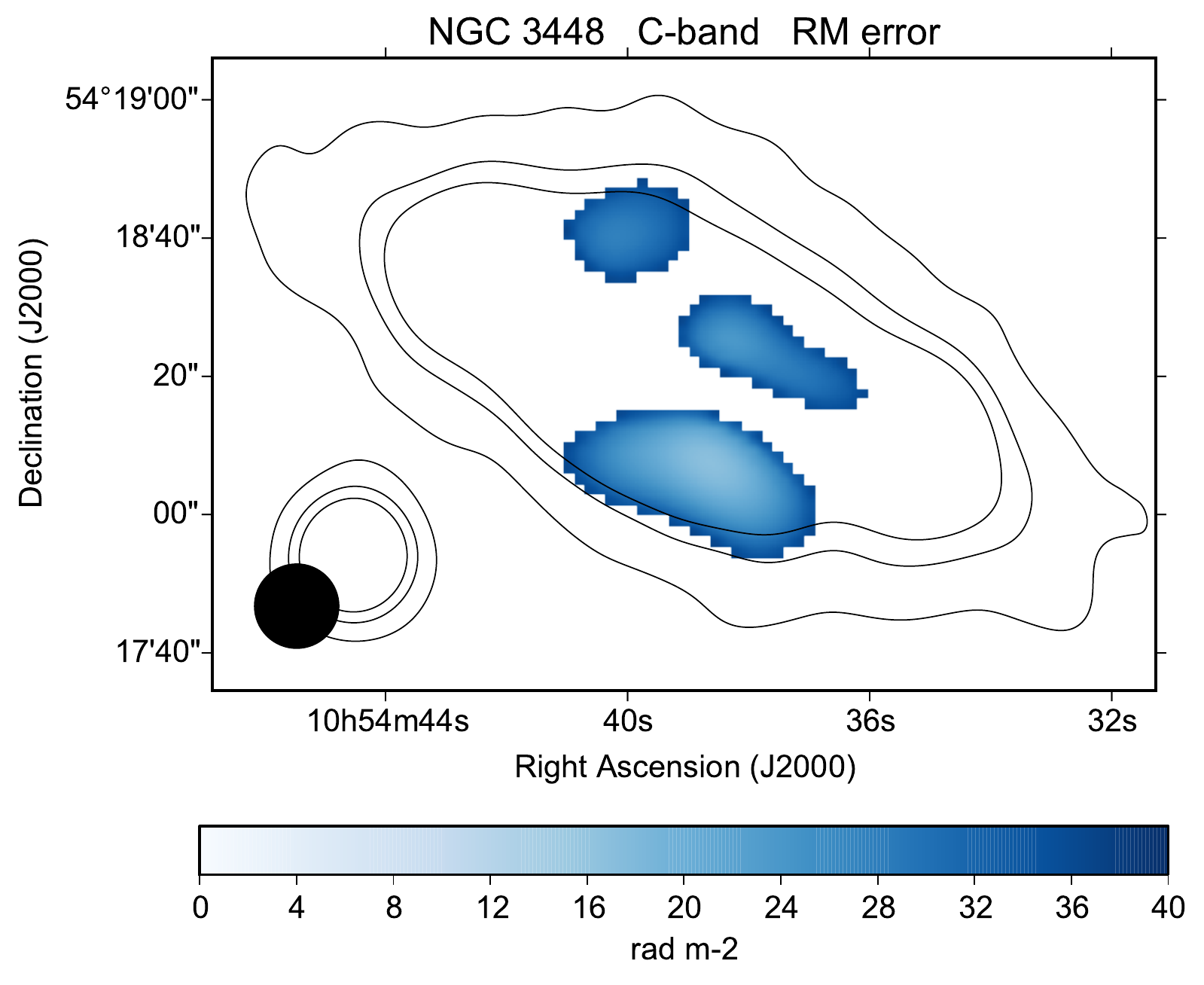}
\includegraphics[width=9.1 cm]{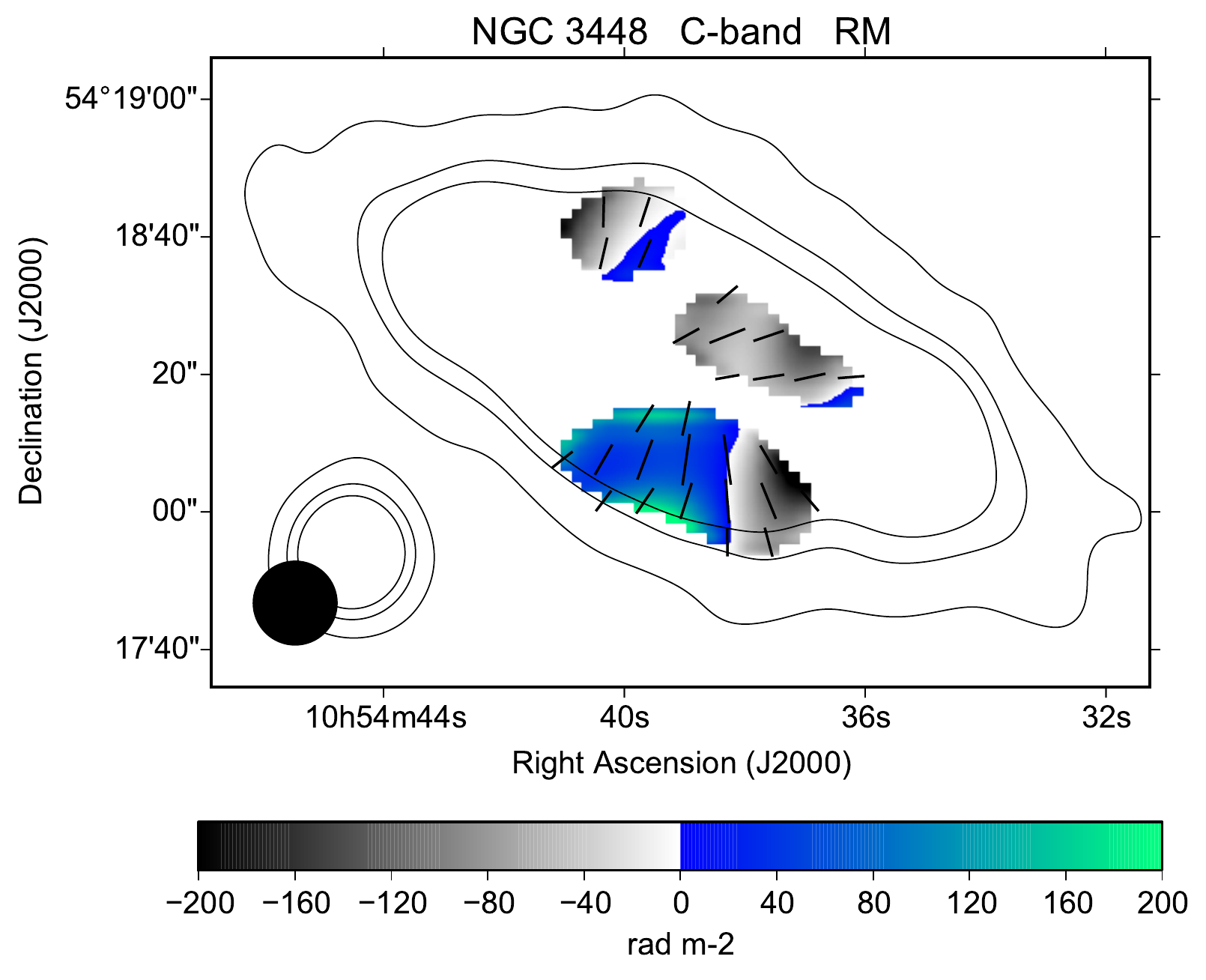}
\caption{Polarization results for NGC~3448 at C-band and $12 \arcsec$ HPBW, corresponding to $1430\,\rm{pc}$. The contour levels (TP) are 35, 105, and 175 $\mu$Jy/beam.
}
\label{n3448all}
\end{figure*}

\begin{figure*}[p]
\centering
\includegraphics[width=9.0 cm]{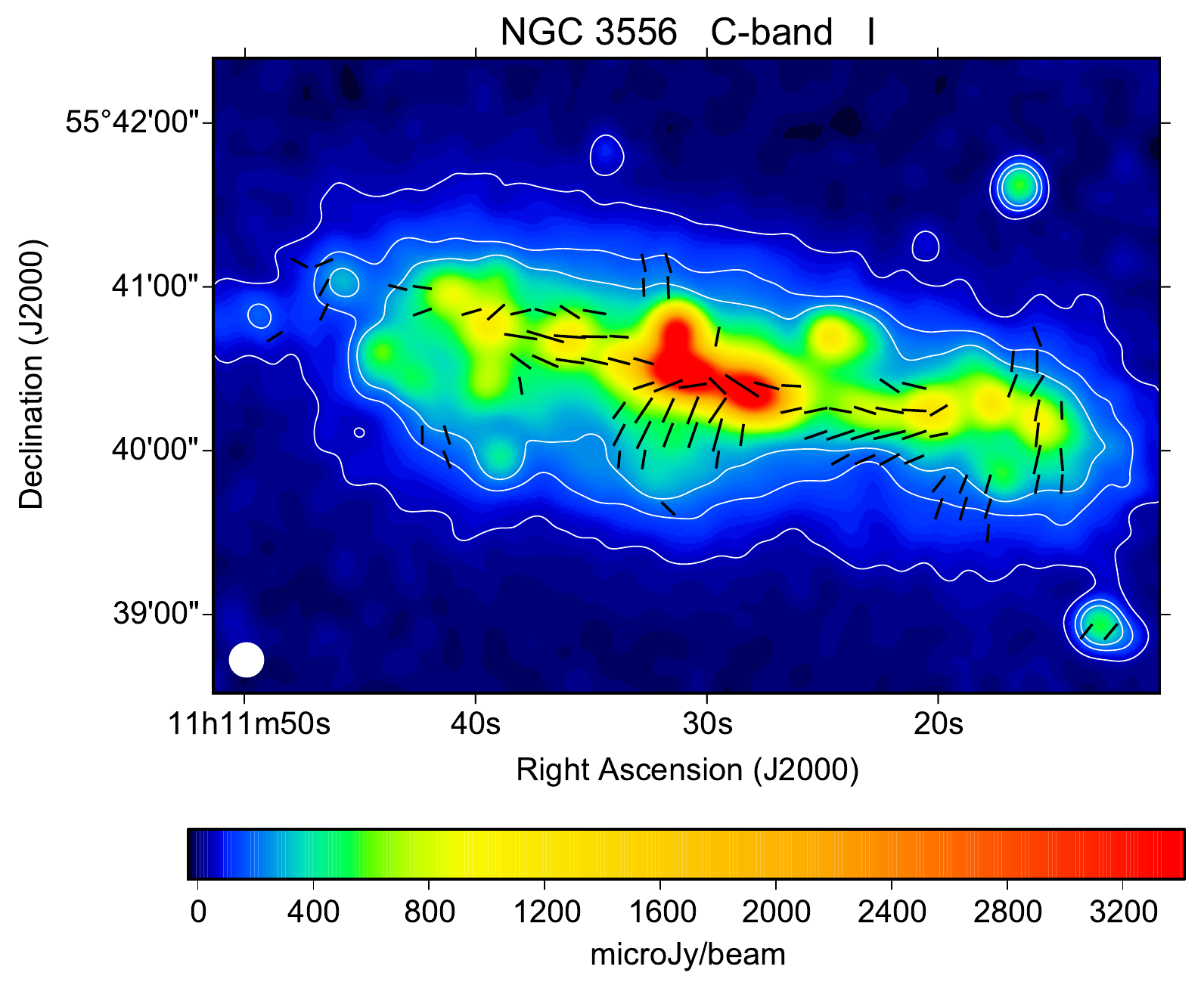}
\includegraphics[width=9.1 cm]{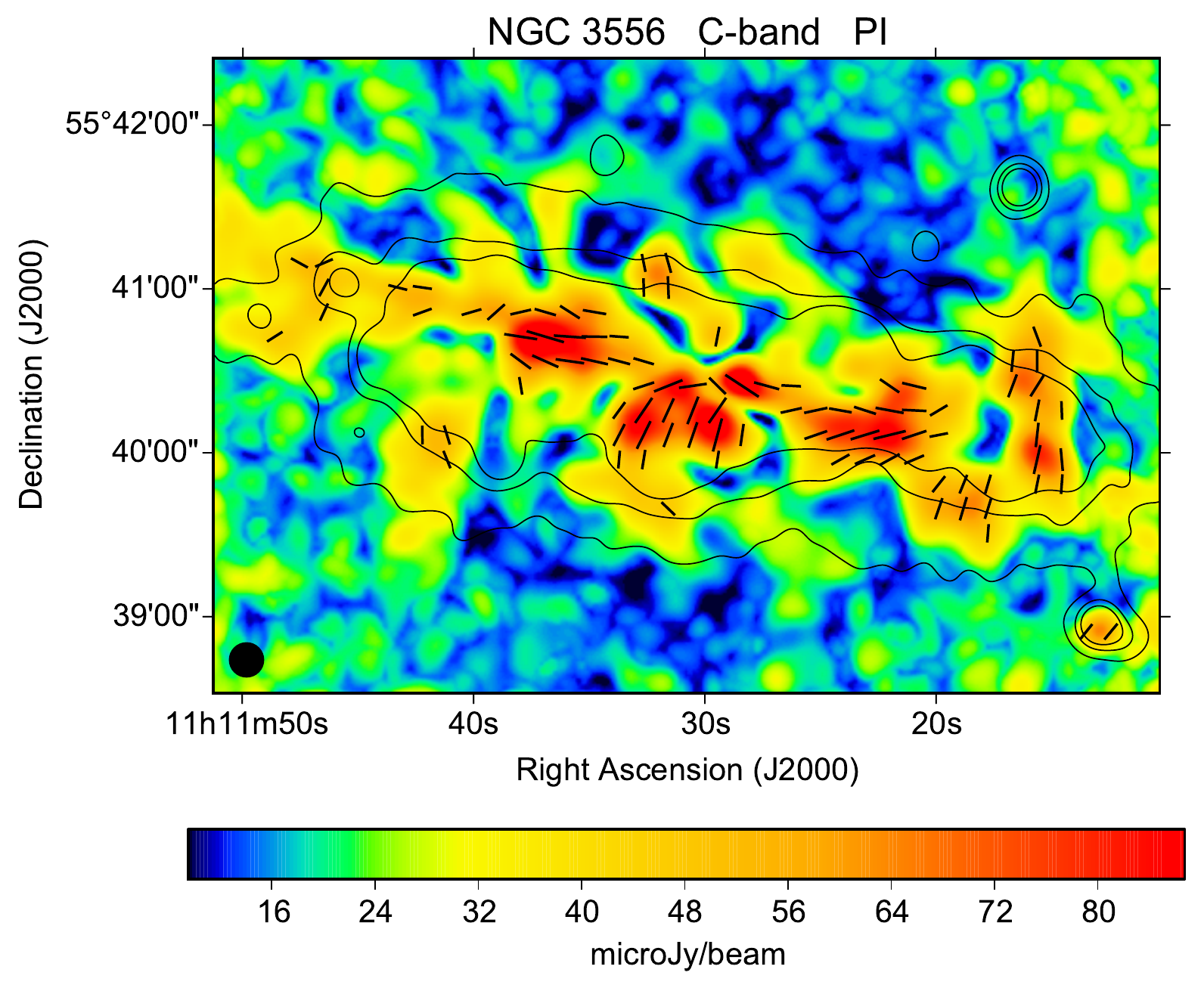}
\includegraphics[width=9.0 cm]{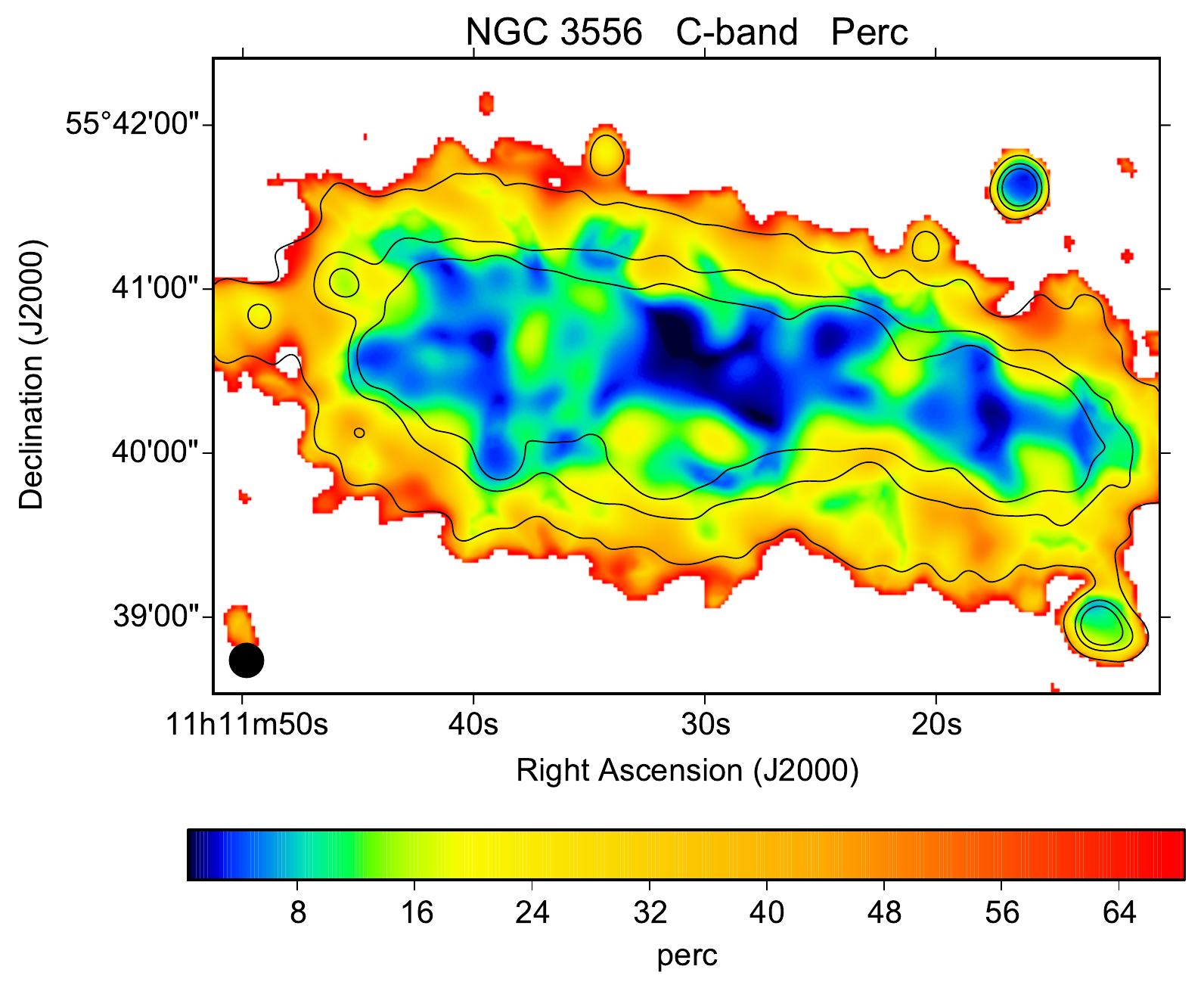}
\includegraphics[width=9.2 cm]{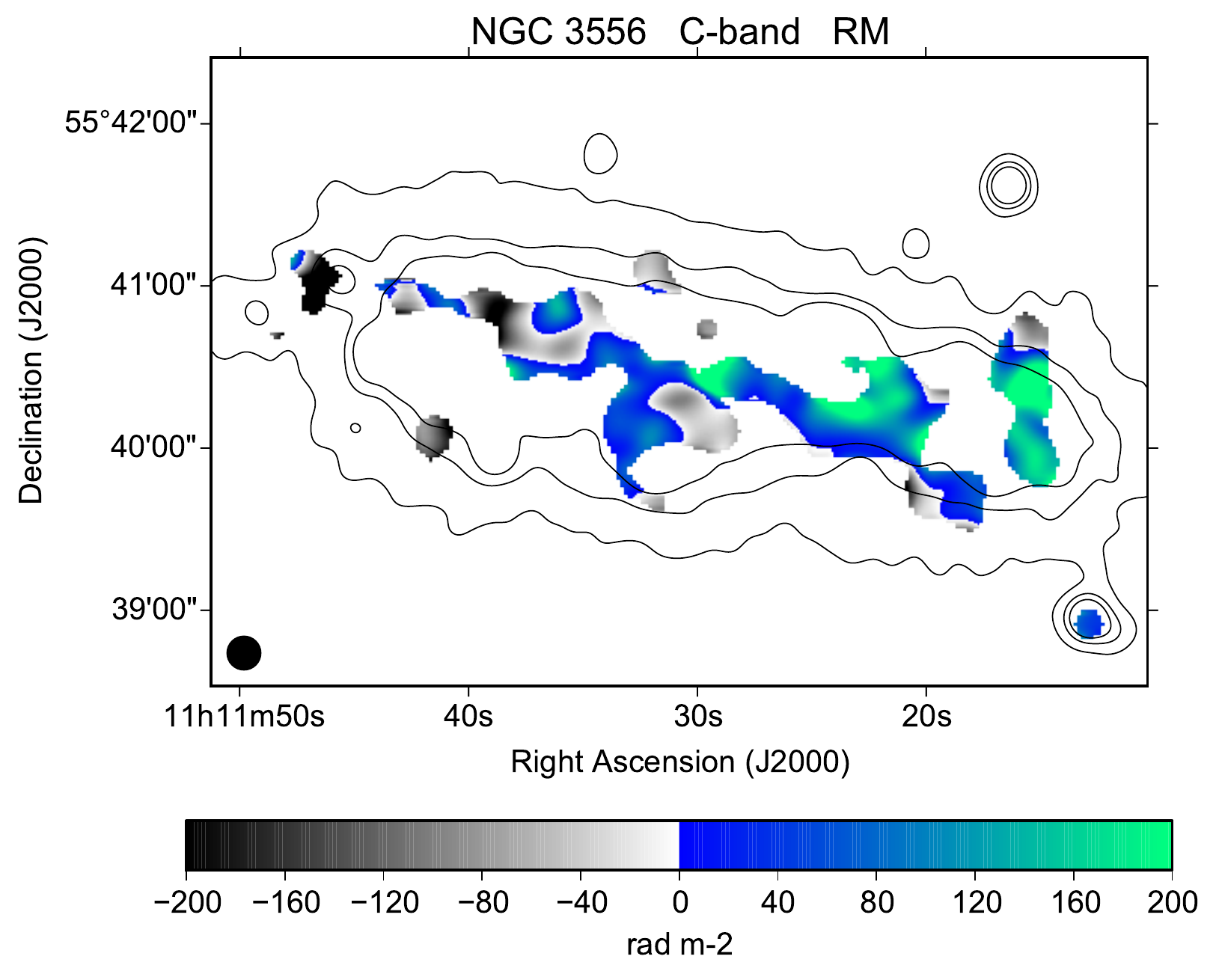}
\includegraphics[width=9.0 cm]{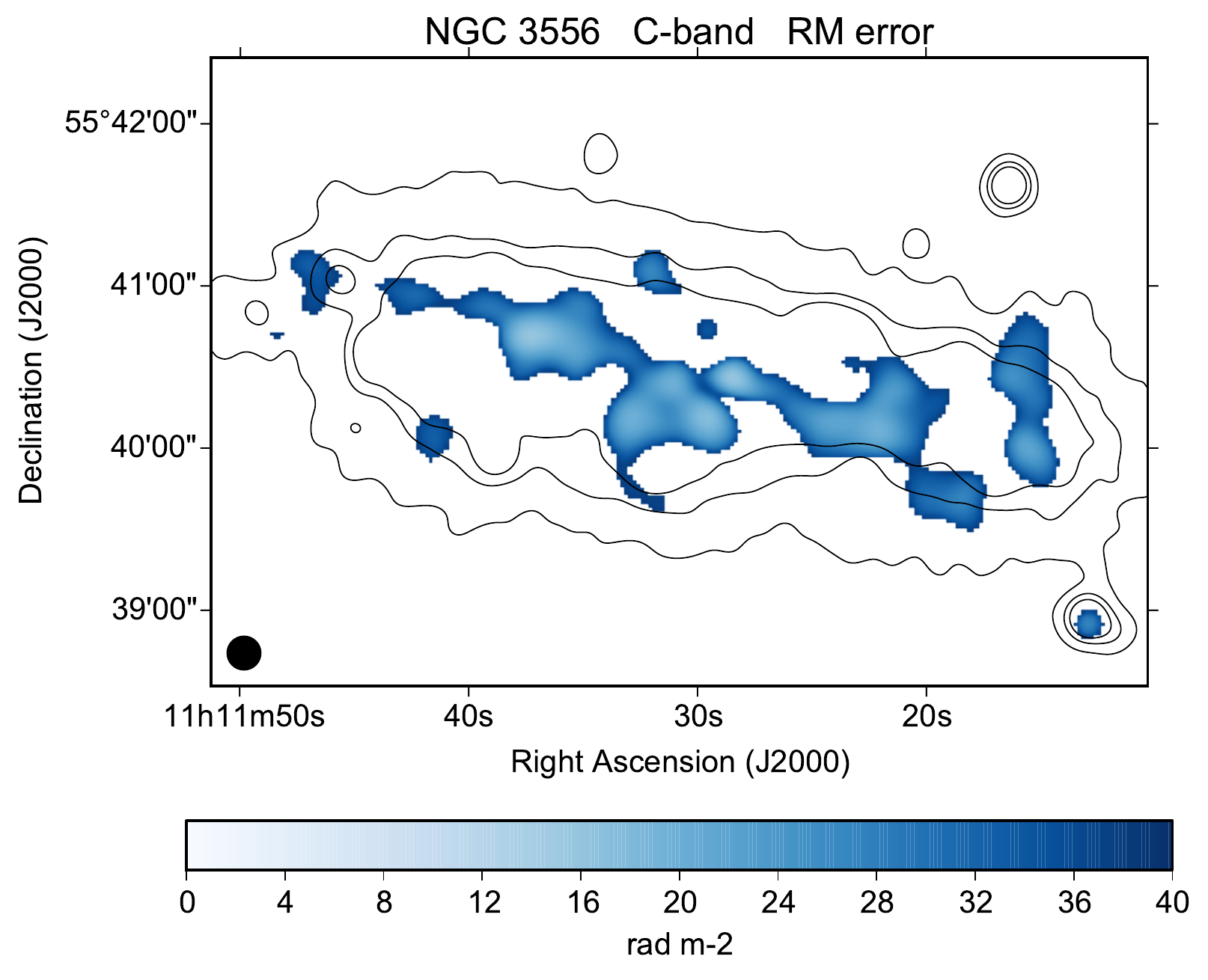}
\includegraphics[width=9.1 cm]{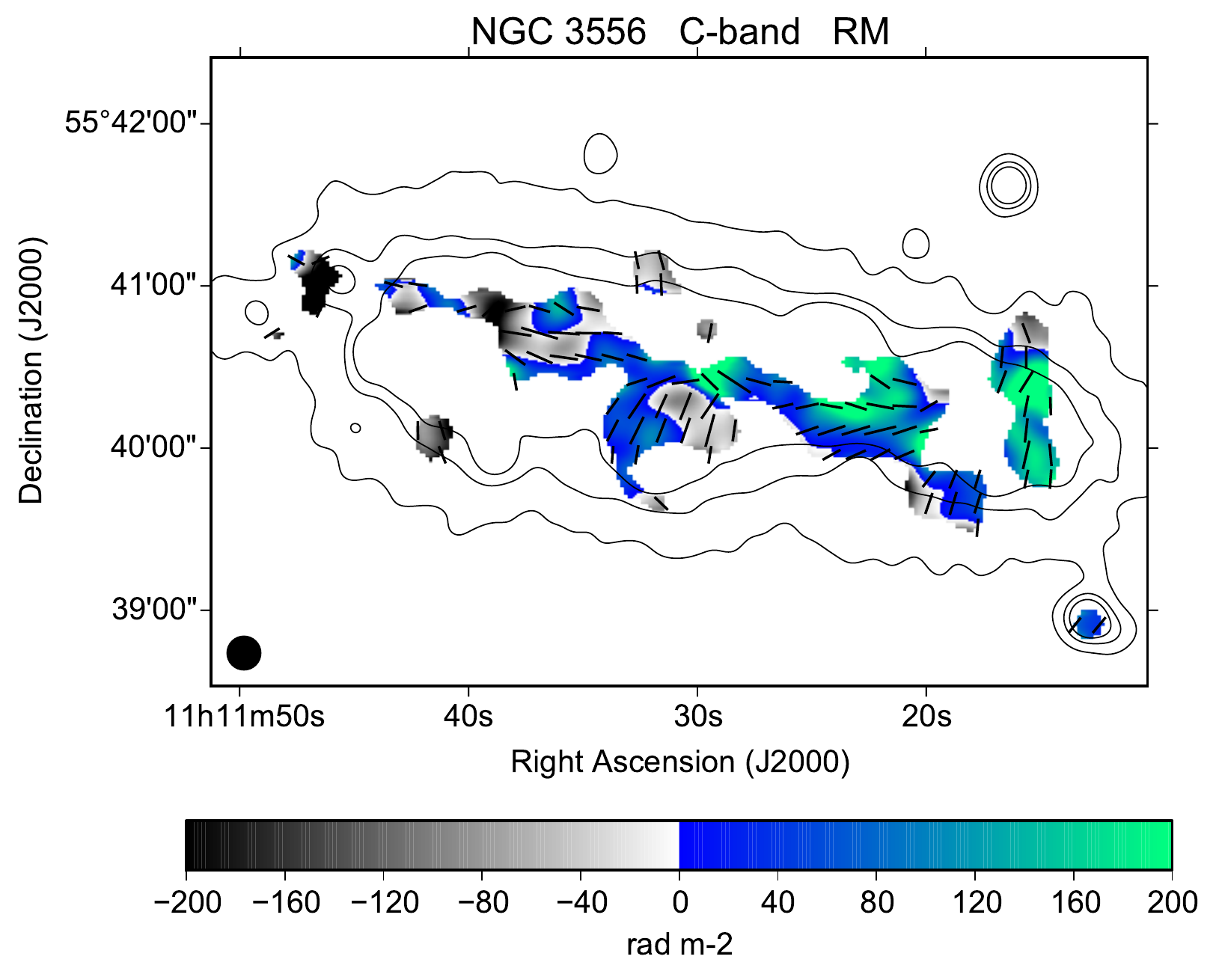}
\caption{Polarization results for NGC~3556 at C-band and $12 \arcsec$ HPBW, corresponding to $820\,\rm{pc}$. The contour levels (TP) are 50, 150, and 250 $\mu$Jy/beam.
}
\label{n3556all}
\end{figure*}

\begin{figure*}[p]
\centering
\includegraphics[width=9.0 cm]{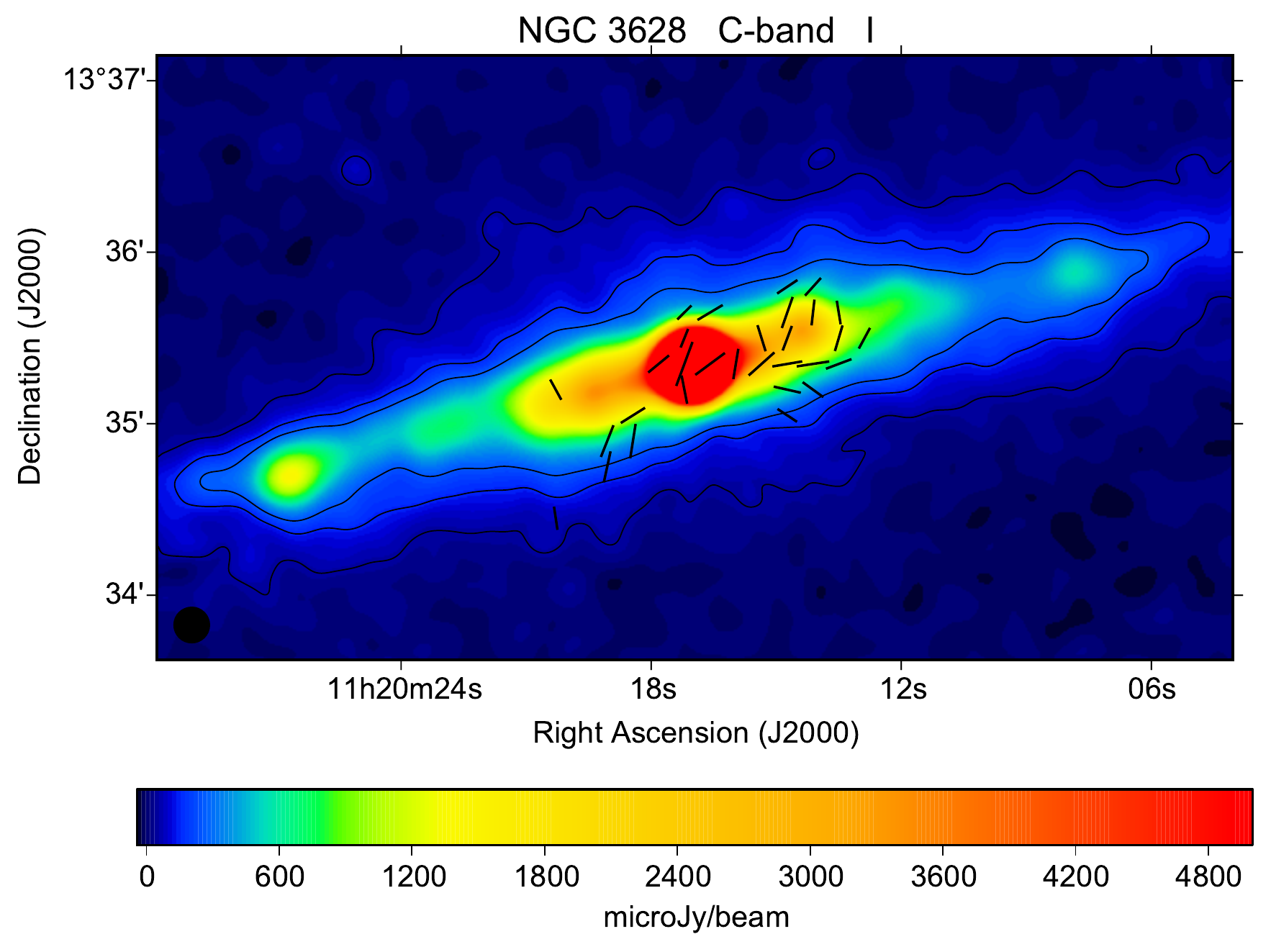}
\includegraphics[width=9.1 cm]{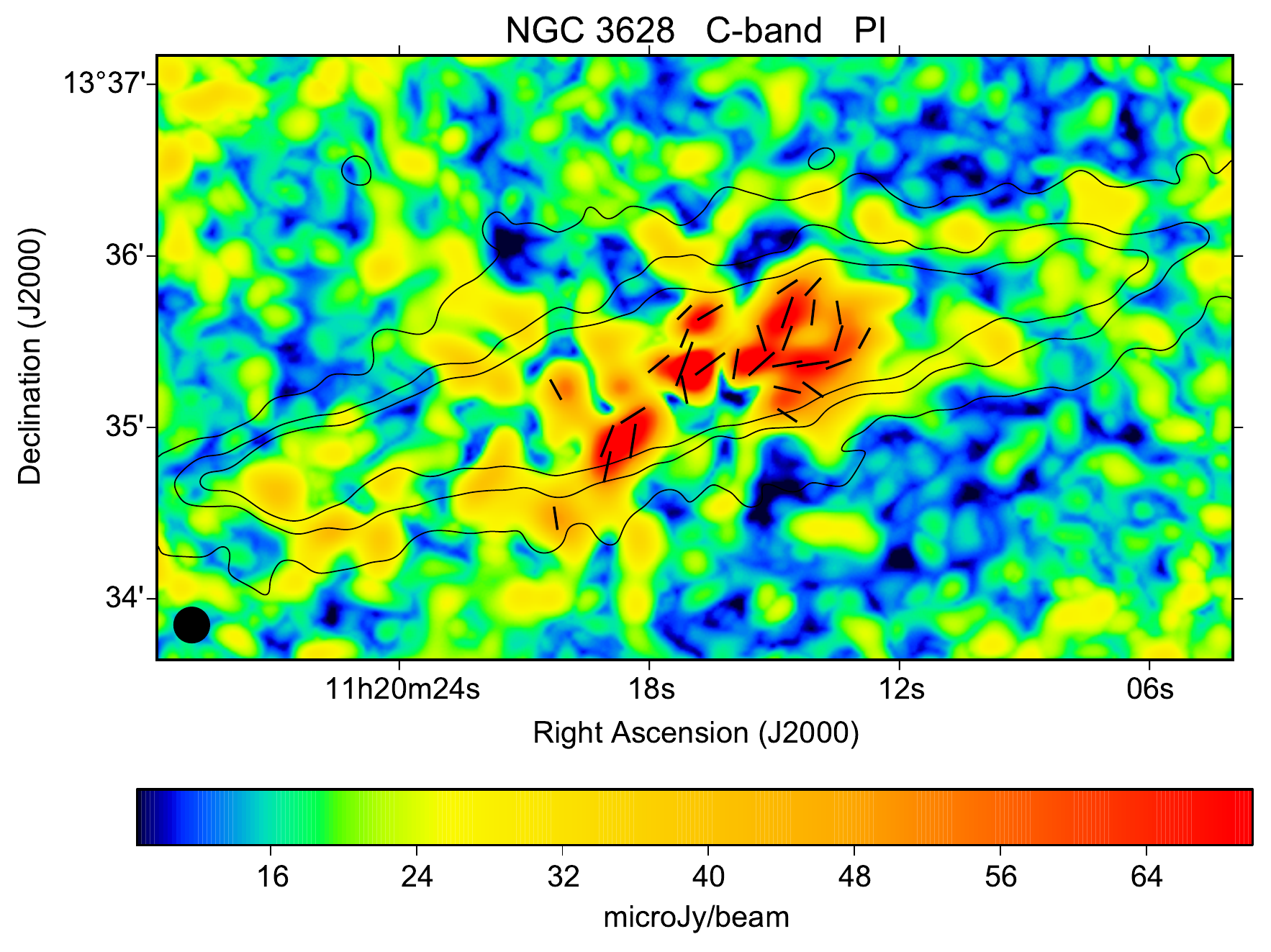}
\includegraphics[width=9.0 cm]{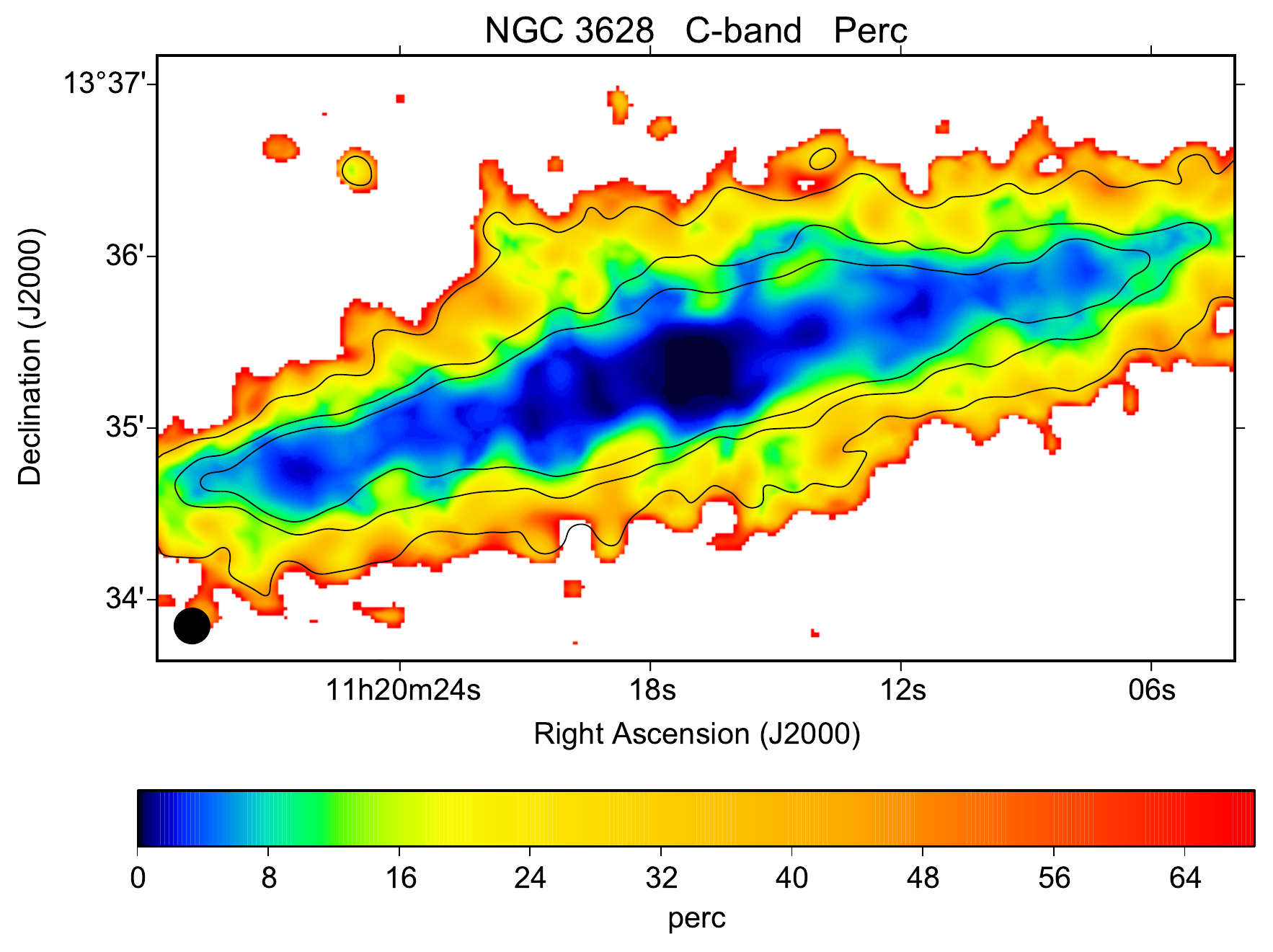}
\includegraphics[width=9.2 cm]{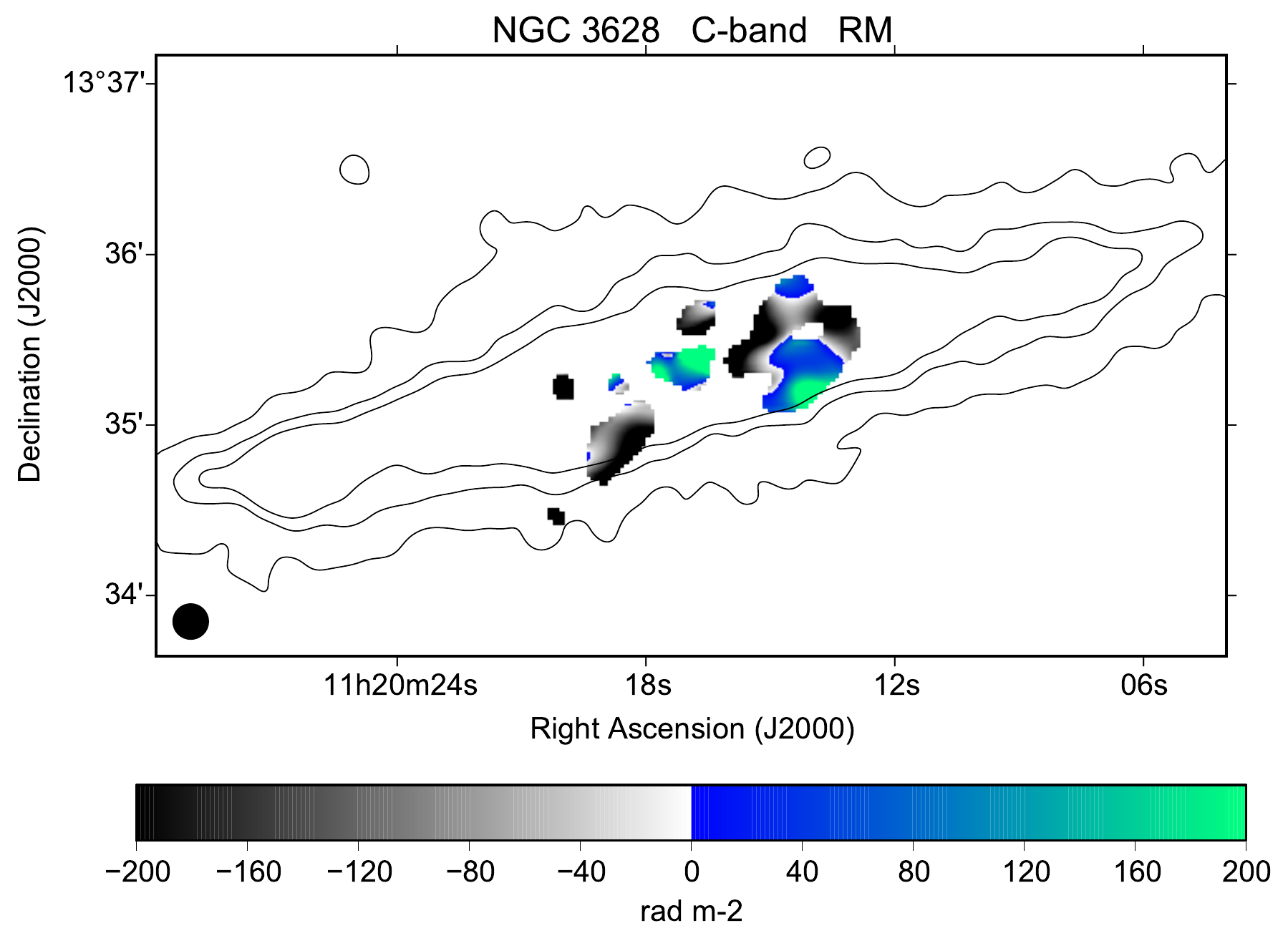}
\includegraphics[width=9.0 cm]{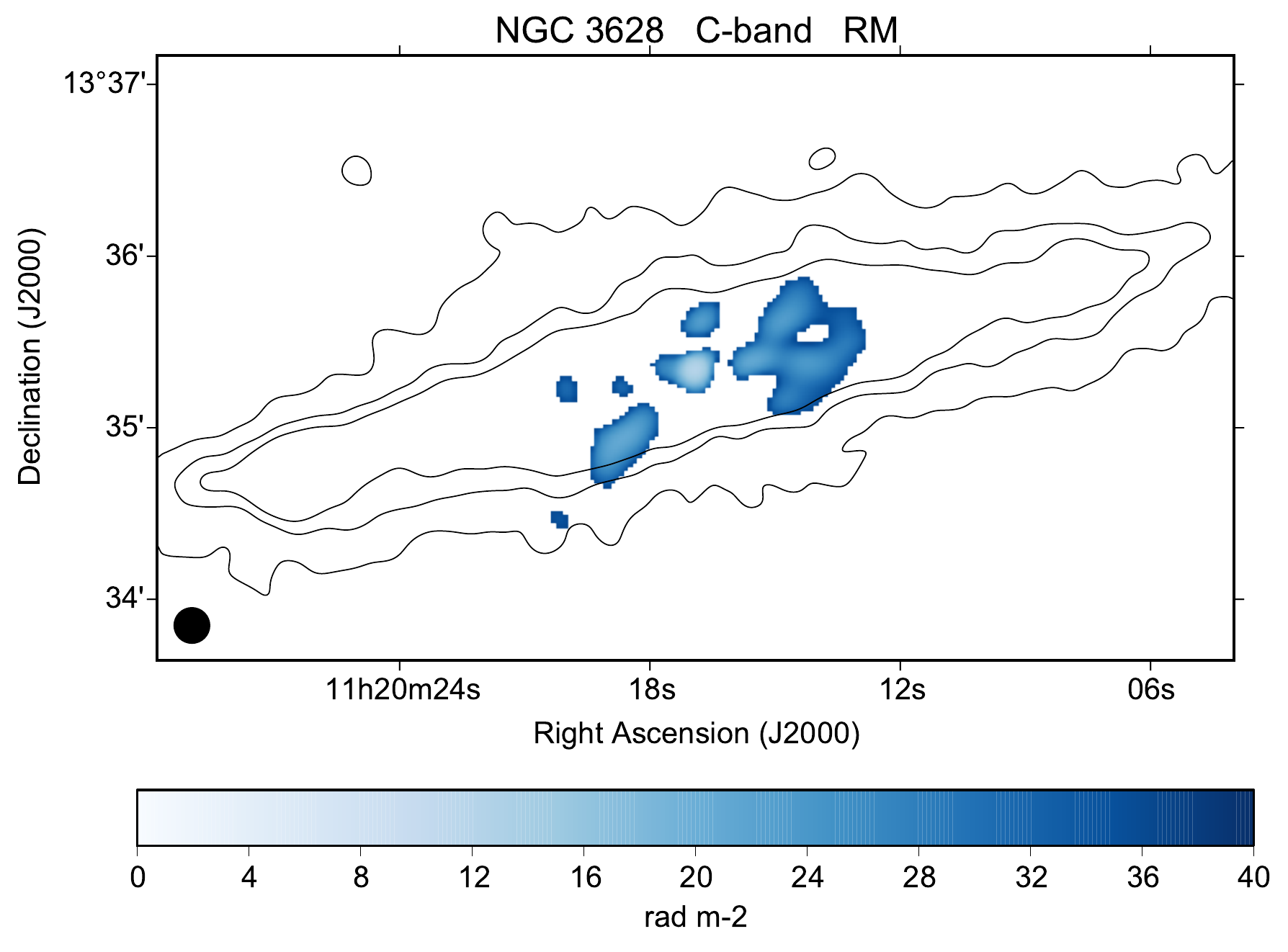}
\includegraphics[width=9.1 cm]{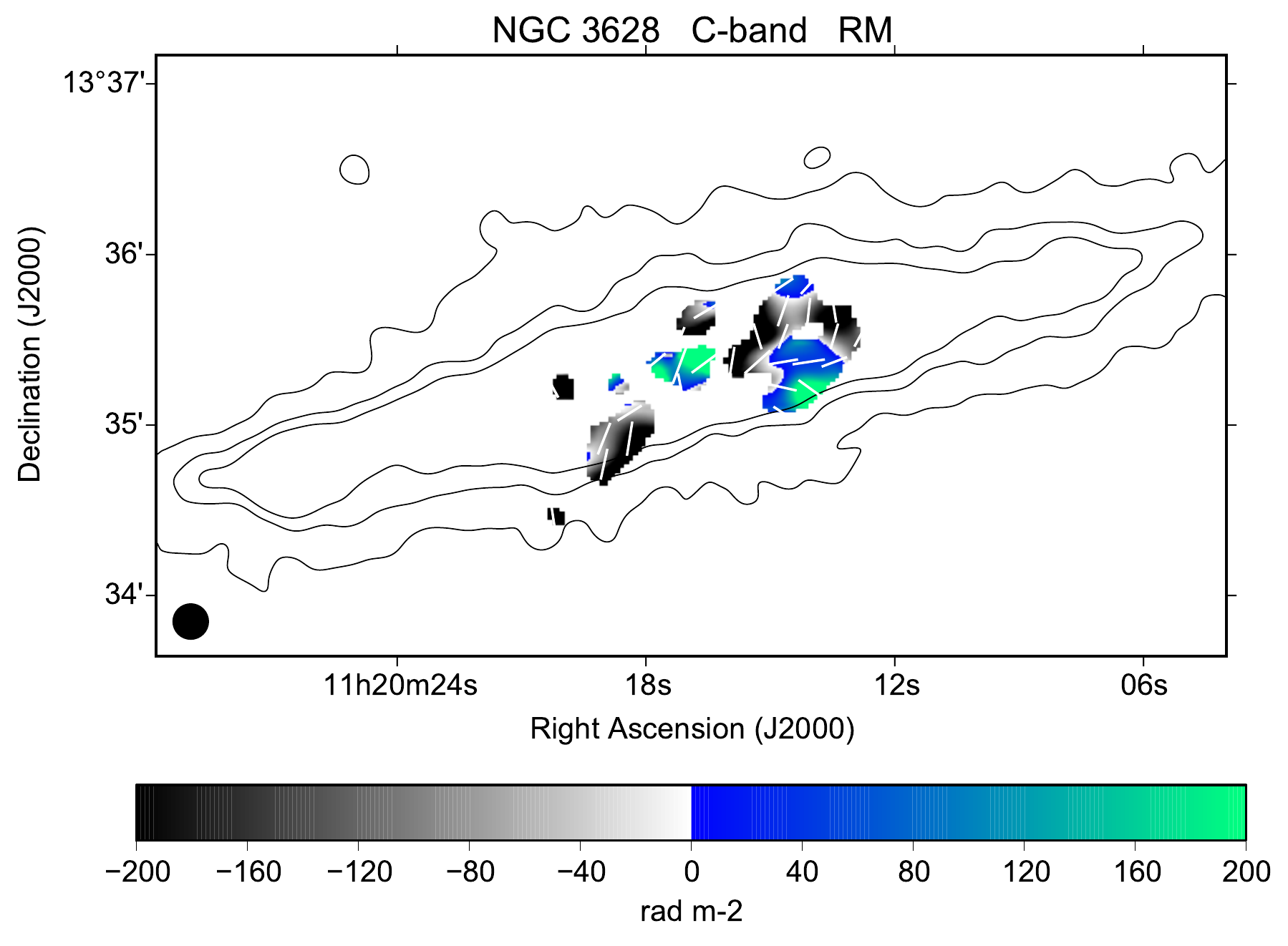}
\caption{Polarization results for NGC~3628 at C-band and $12 \arcsec$ HPBW, corresponding to $490\,\rm{pc}$. The contour levels (TP) are 50, 150, and 250 $\mu$Jy/beam. The image of the TP map is cut
at 5000~$\mu$Jy/beam in order to present the disk emission well.
}
\label{n3628all}
\end{figure*}

\begin{figure*}[p]
\centering
\includegraphics[width=9.0 cm]{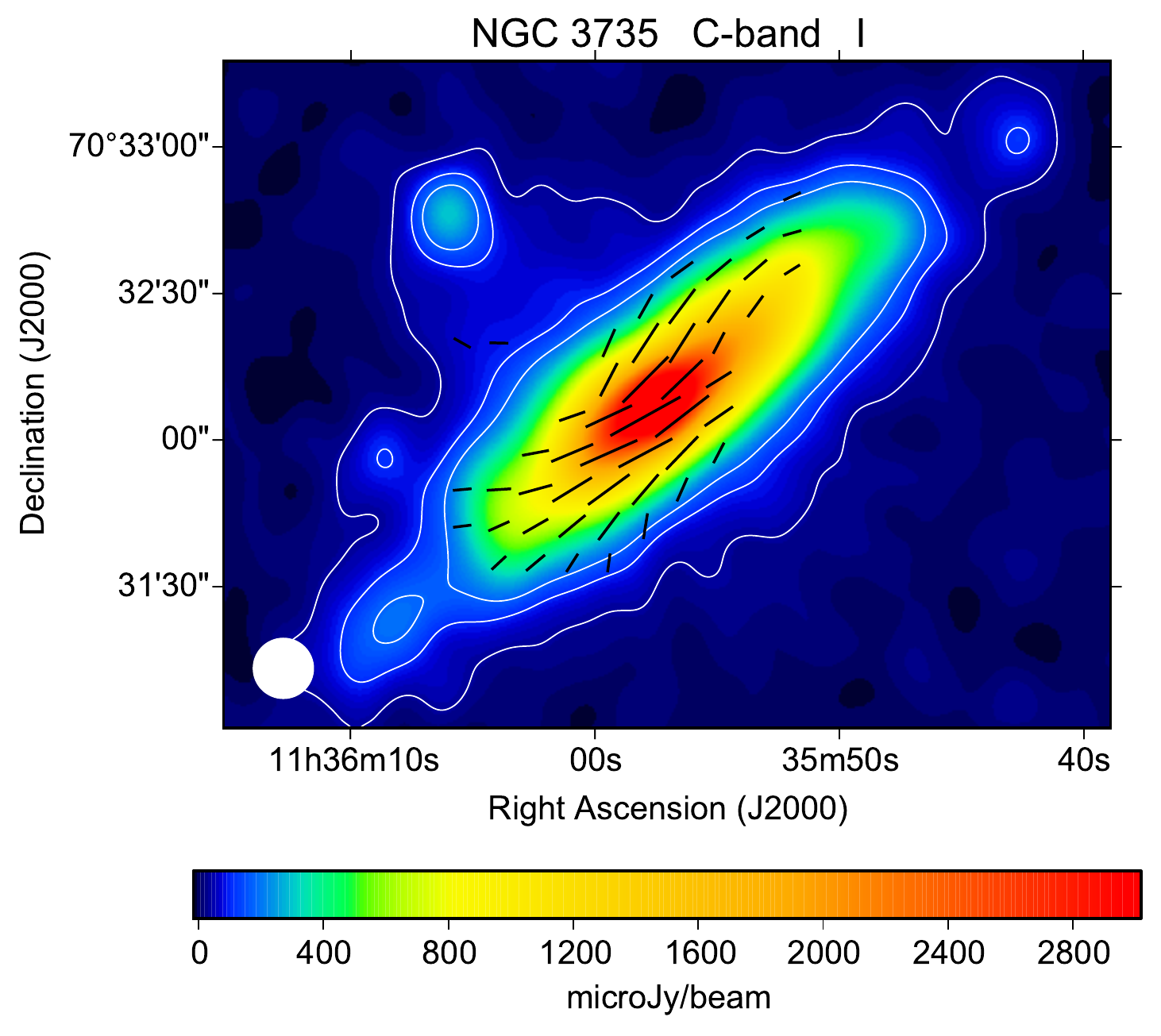}
\includegraphics[width=9.1 cm]{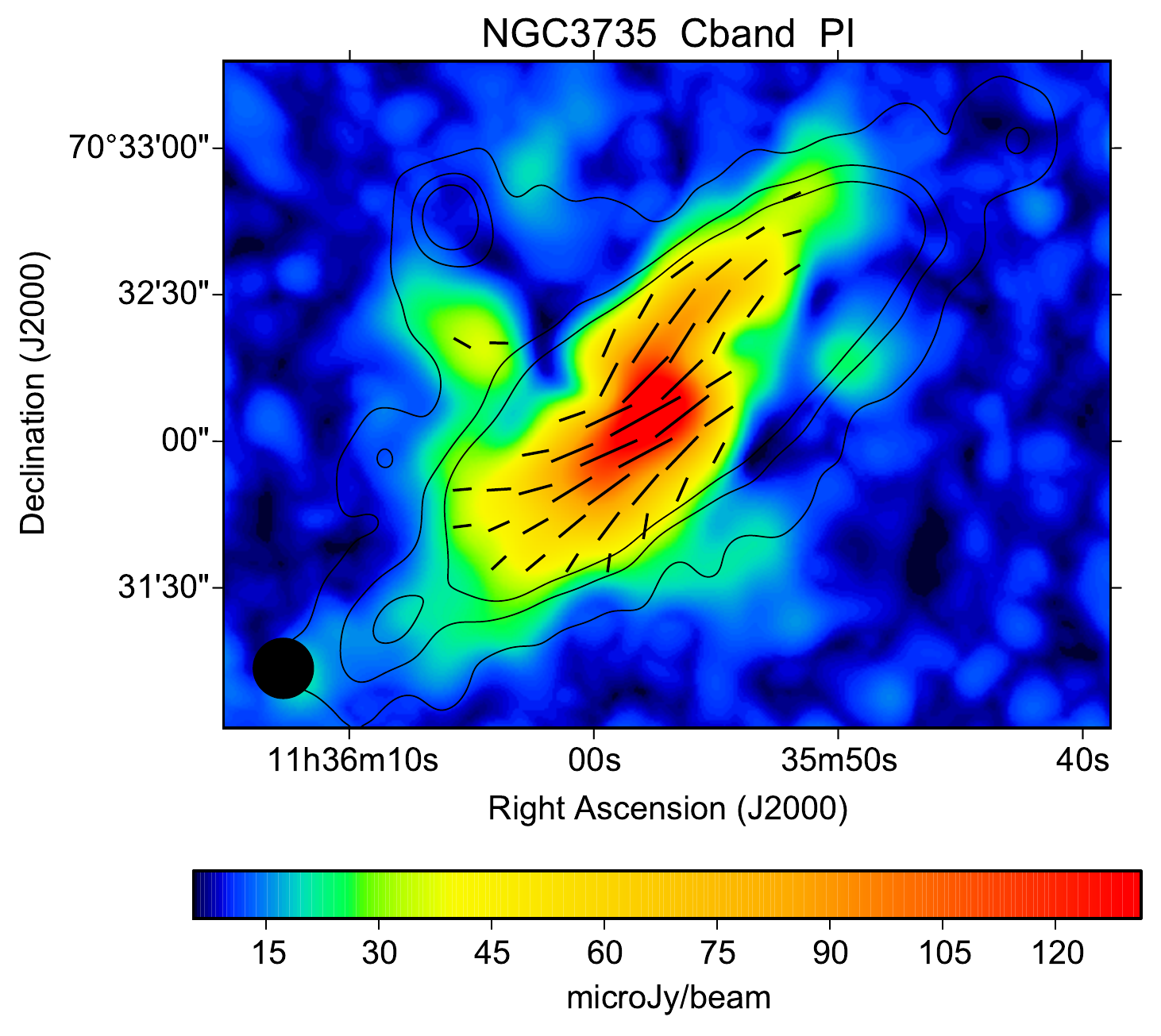}
\includegraphics[width=9.0 cm]{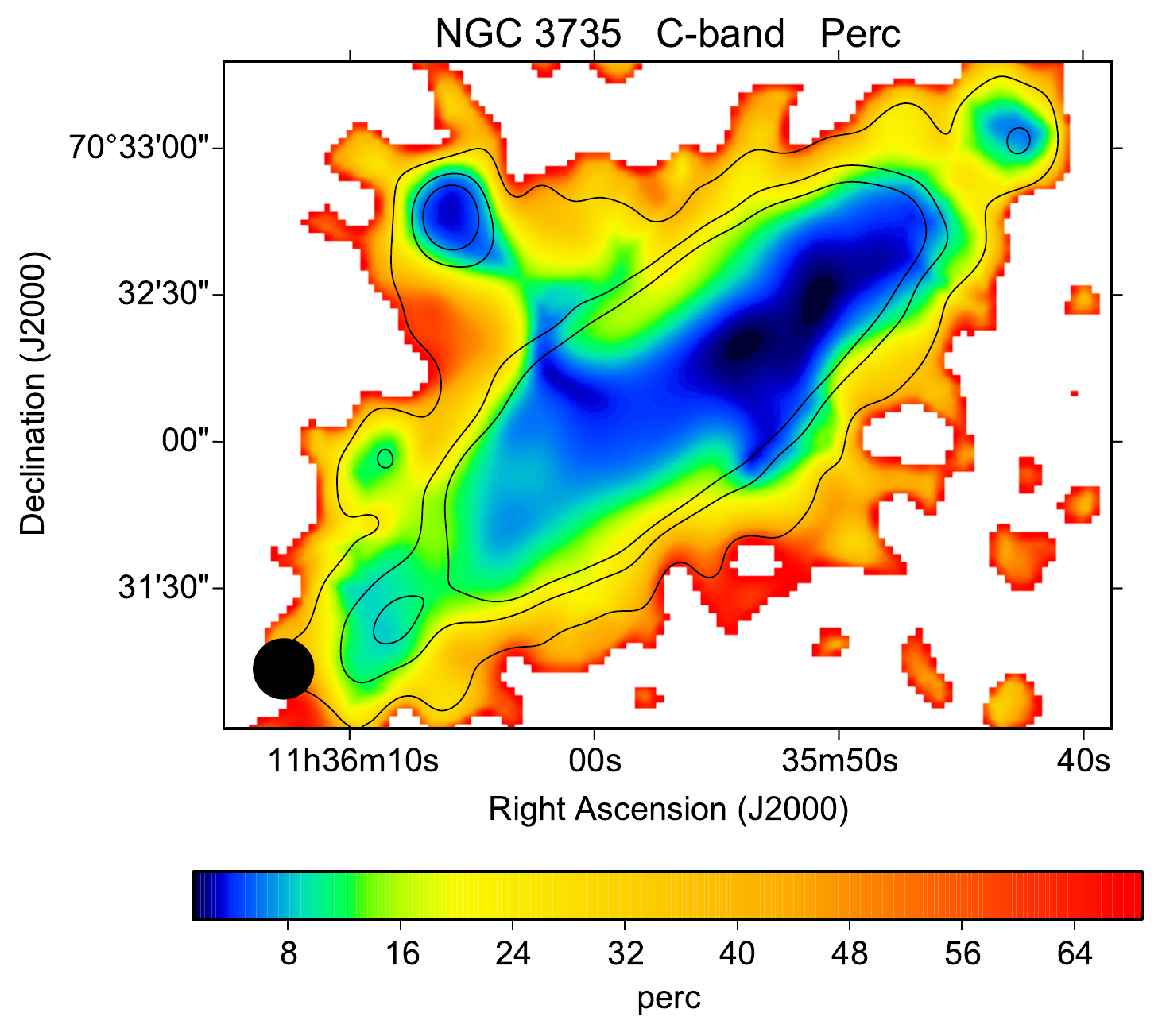}
\includegraphics[width=9.2 cm]{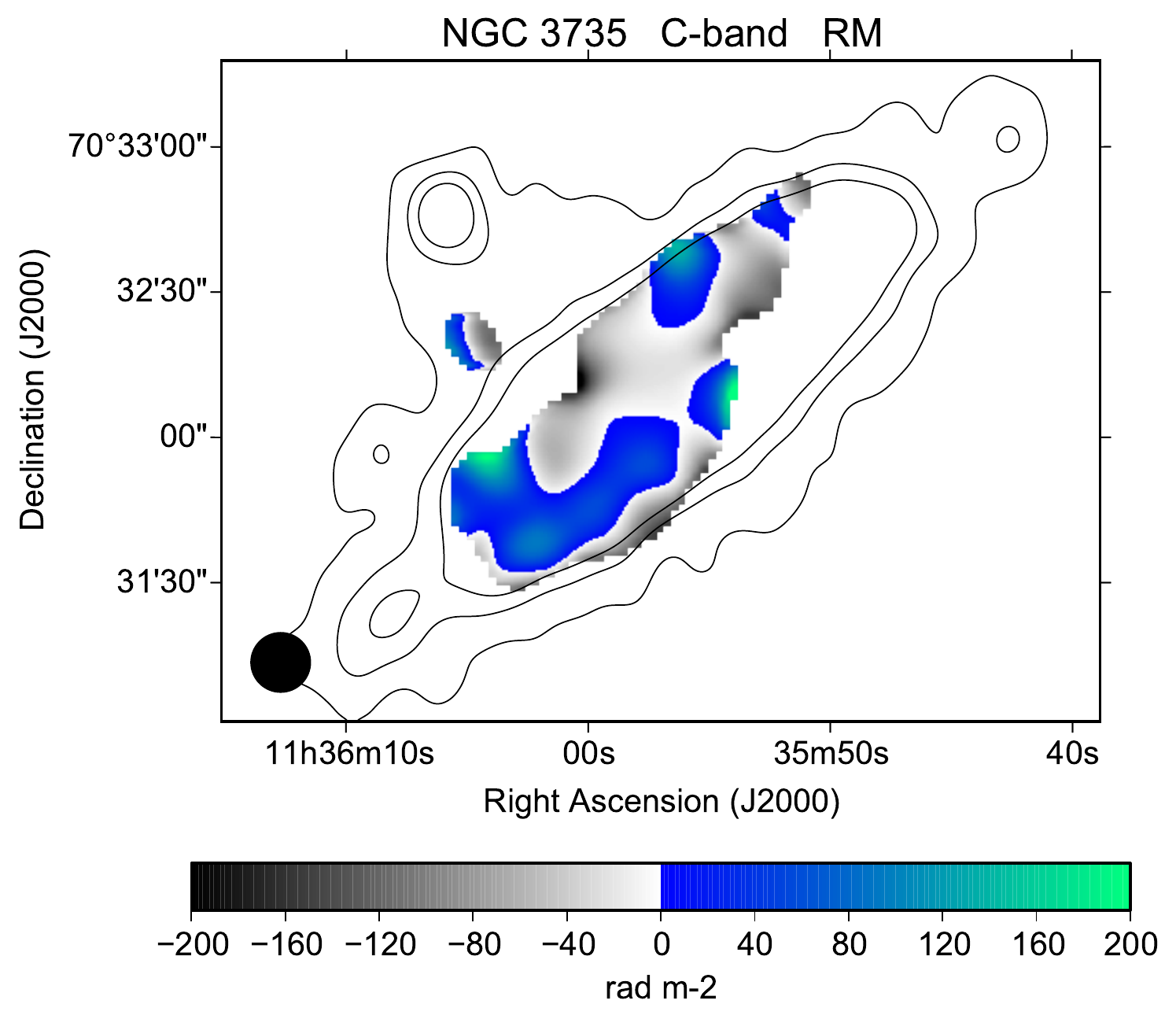}
\includegraphics[width=9.0 cm]{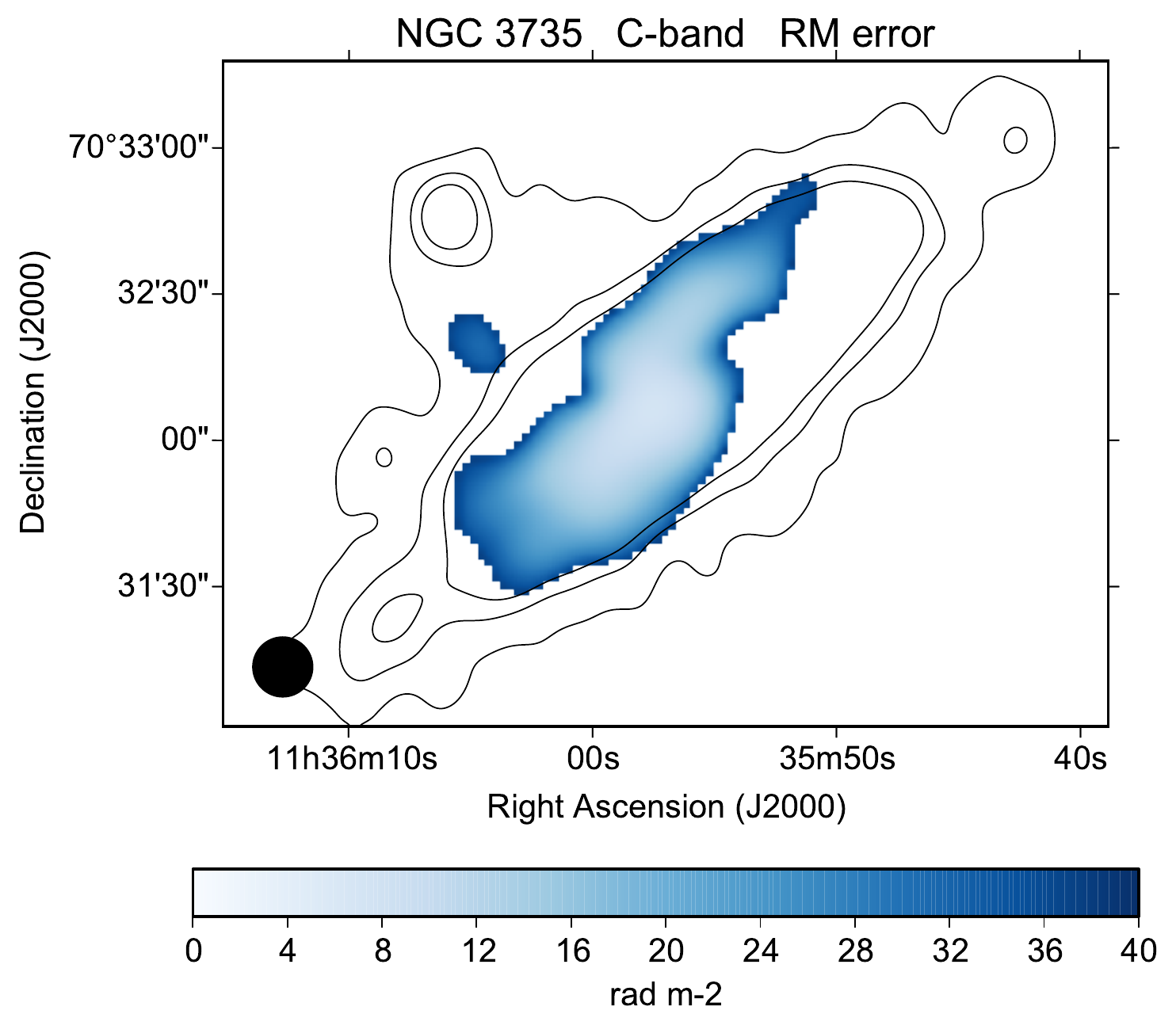}
\includegraphics[width=9.1 cm]{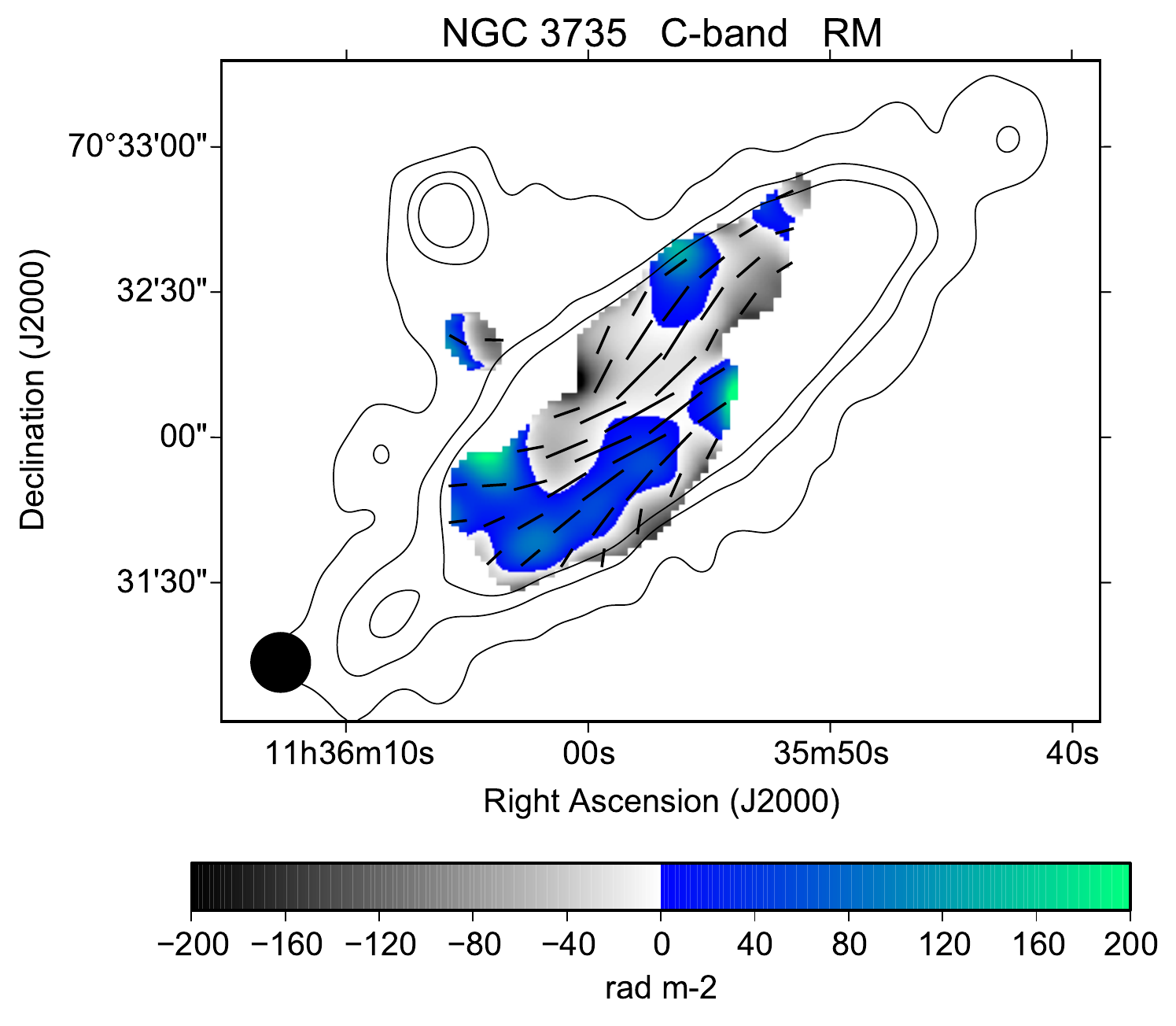}
\caption{Polarization results for NGC~3735 at C-band and $12 \arcsec$ HPBW, corresponding to $2440\,\rm{pc}$. The contour levels (TP) are 35, 105, and 175 $\mu$Jy/beam.
}
\label{n3735all}
\end{figure*}

\begin{figure*}[p]
\centering
\includegraphics[width=9.0 cm]{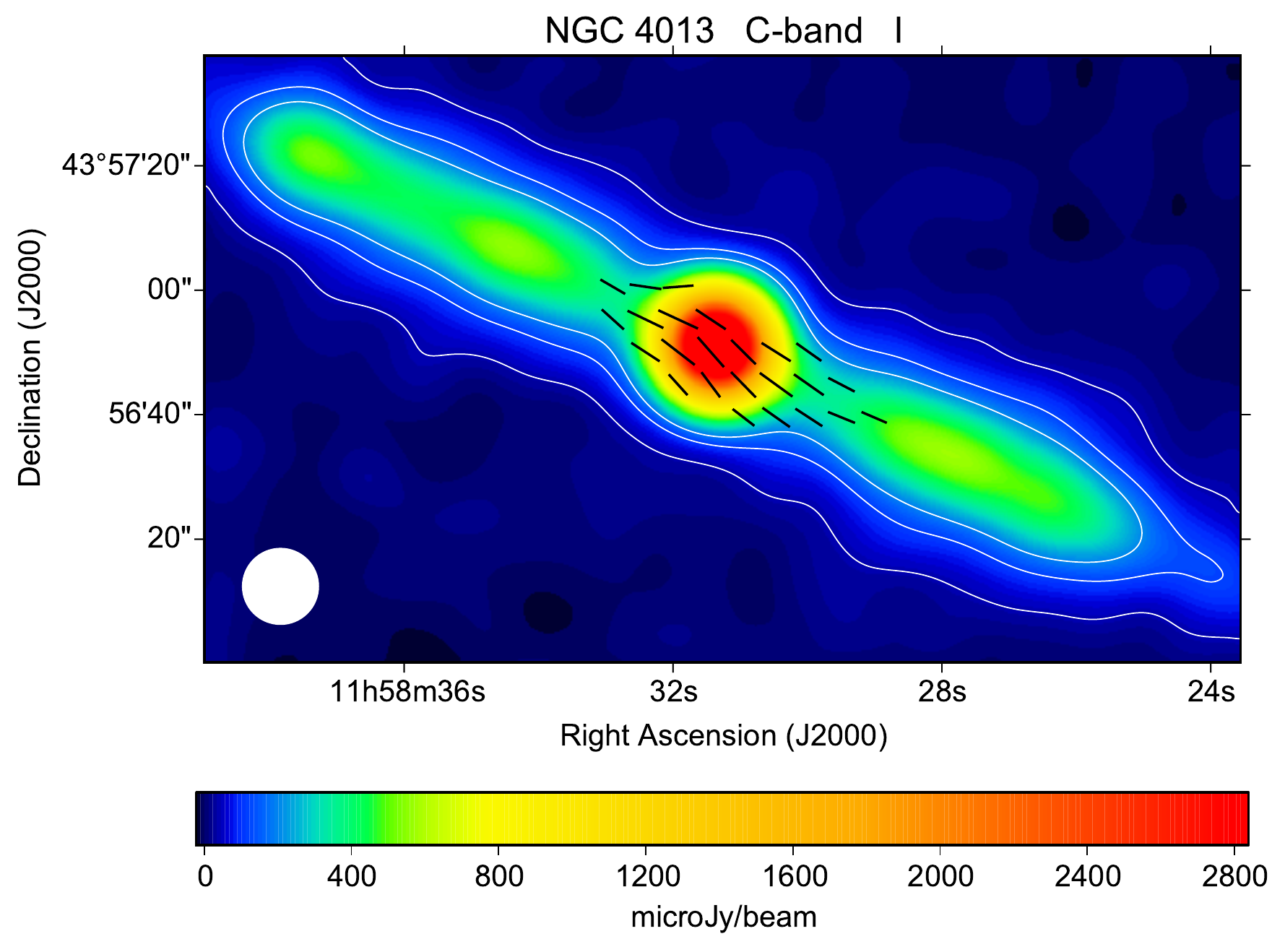}
\includegraphics[width=9.0 cm]{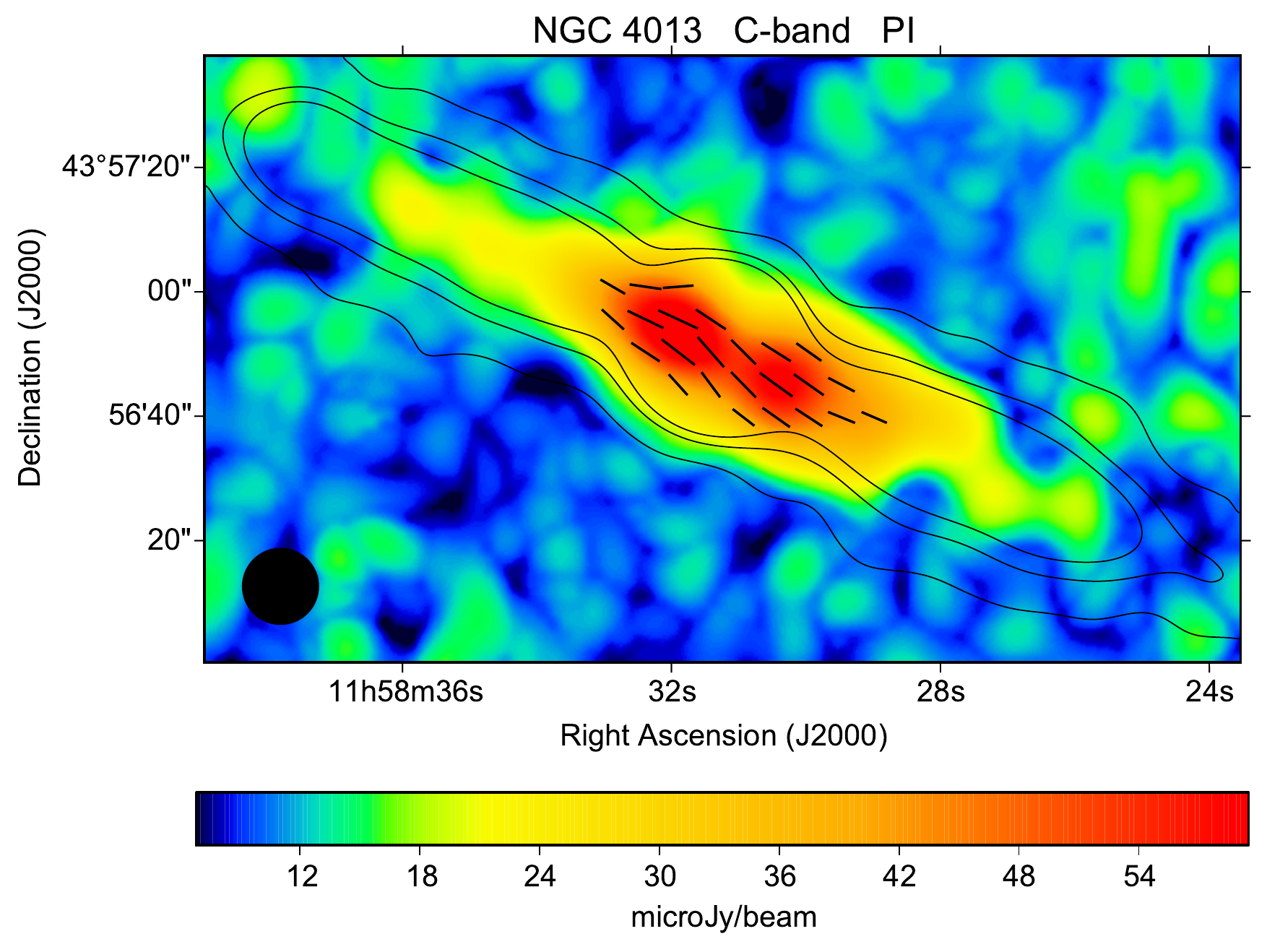}
\includegraphics[width=9.0 cm]{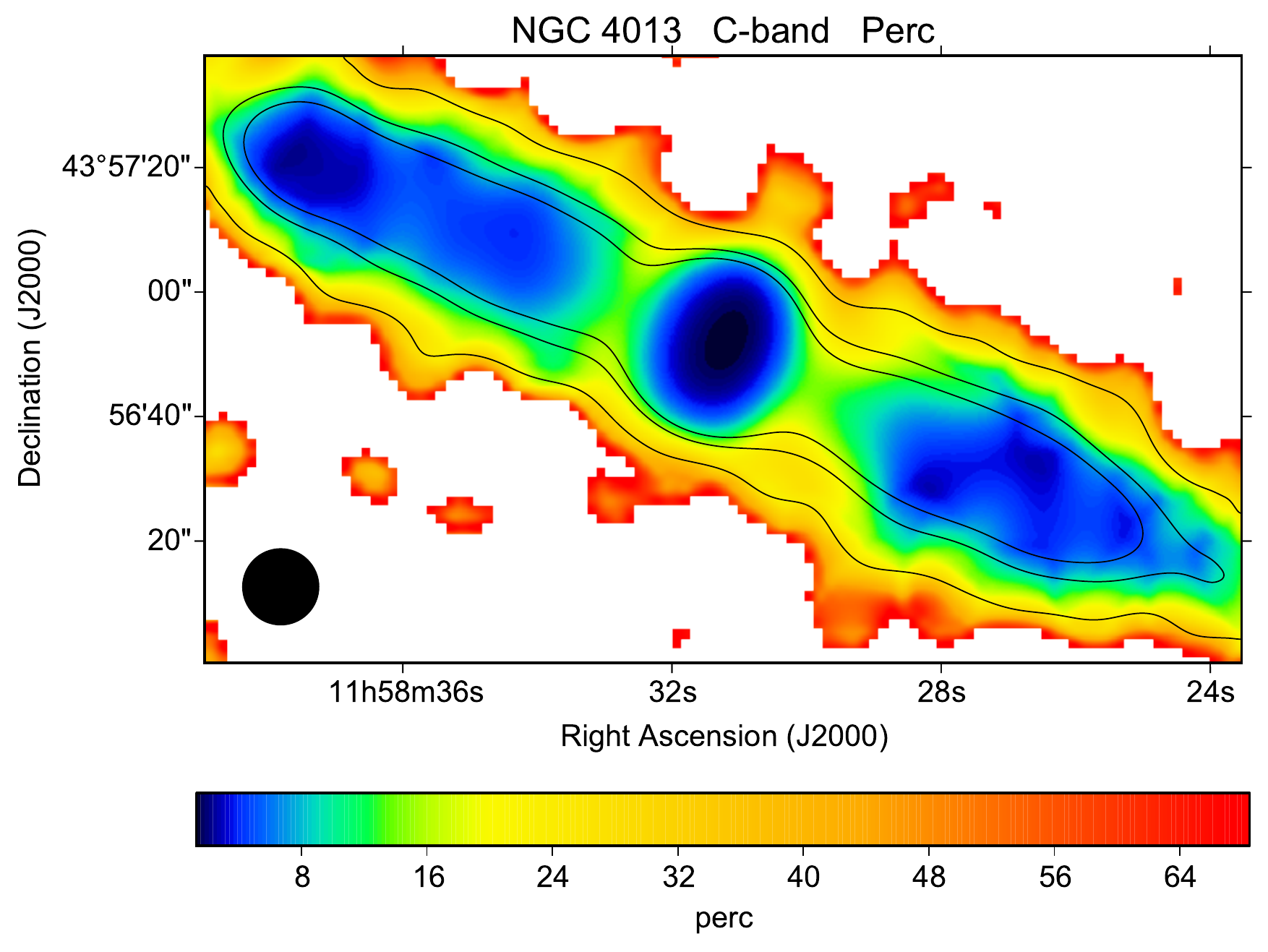}
\includegraphics[width=9.2 cm]{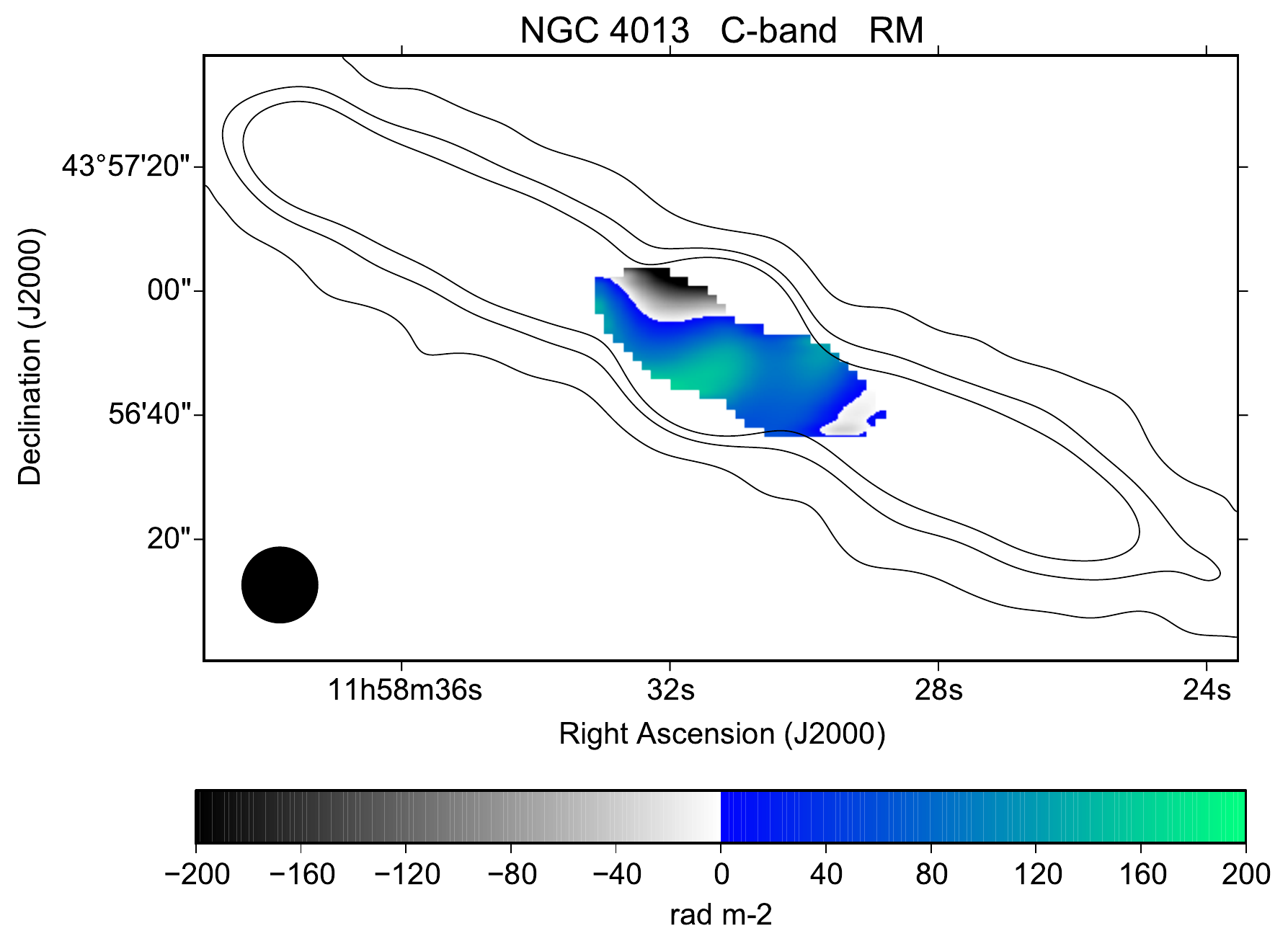}
\includegraphics[width=9.0 cm]{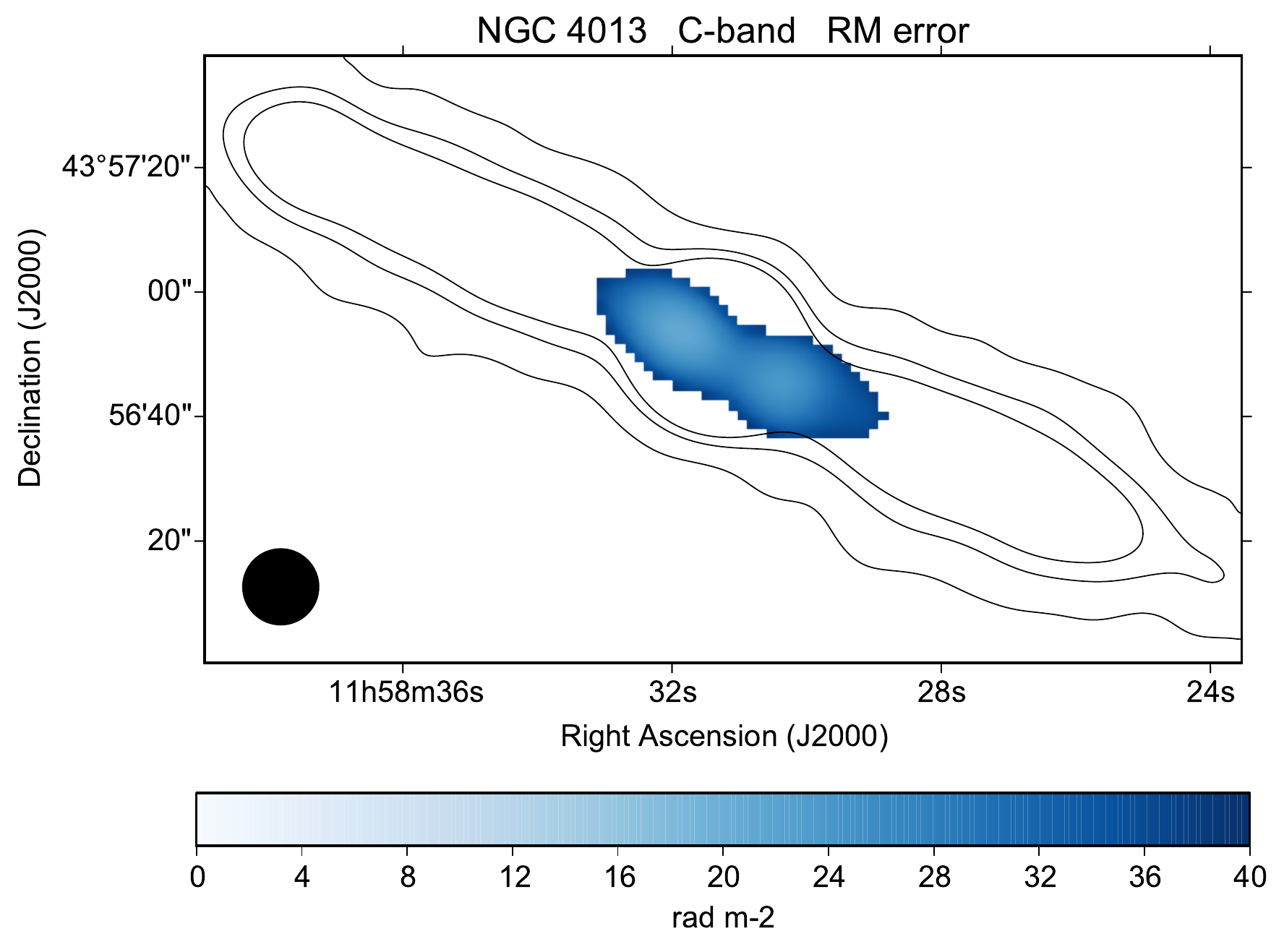}
\includegraphics[width=9.1 cm]{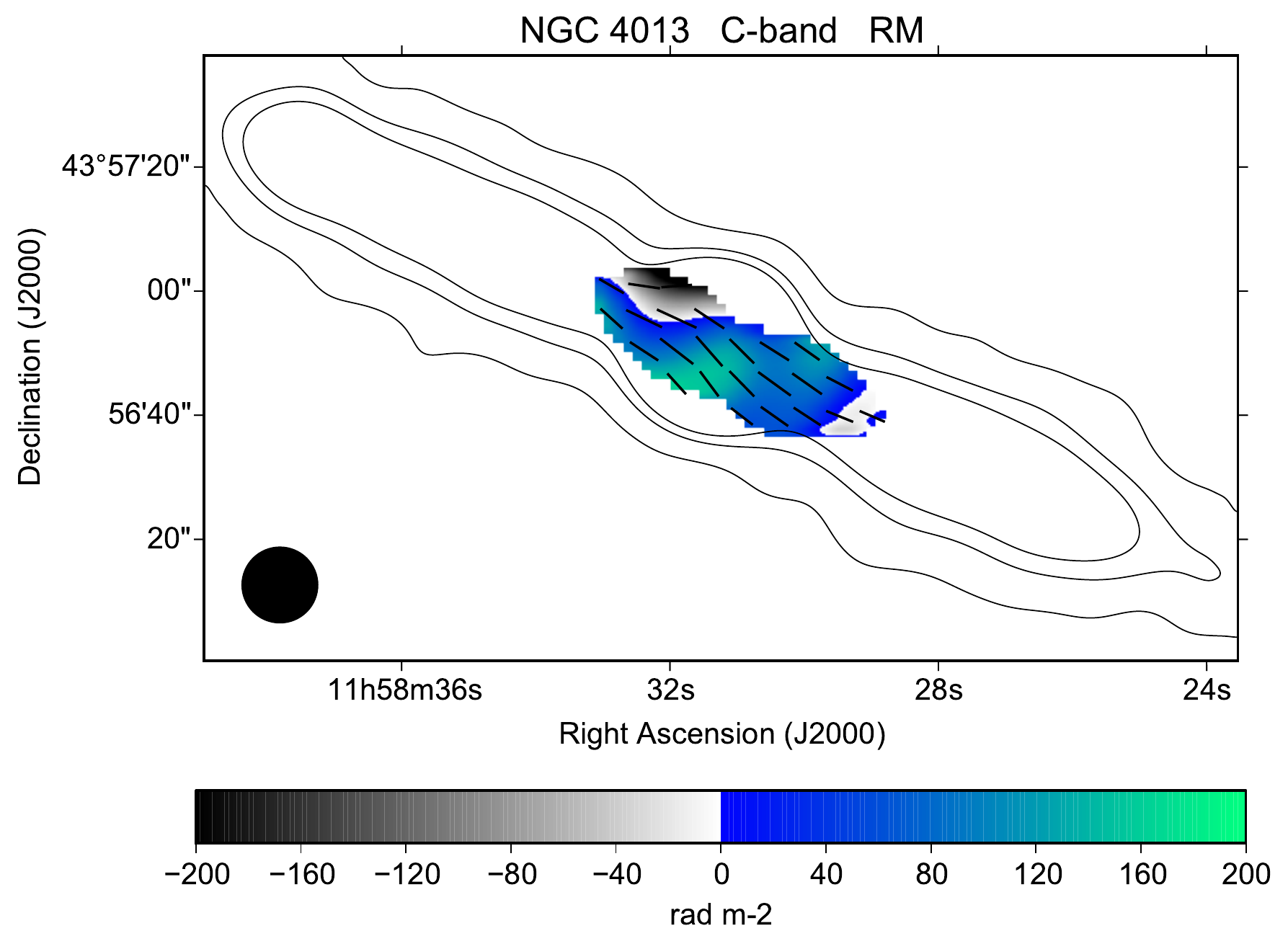}
\caption{Polarization results for NGC~4013 at C-band and $12 \arcsec$ HPBW, corresponding to $930\,\rm{pc}$. The contour levels (TP) are 40, 120, and 200 $\mu$Jy/beam.
}
\label{n4013all}
\end{figure*}

\begin{figure*}[p]
\centering
\includegraphics[width=9.0 cm]{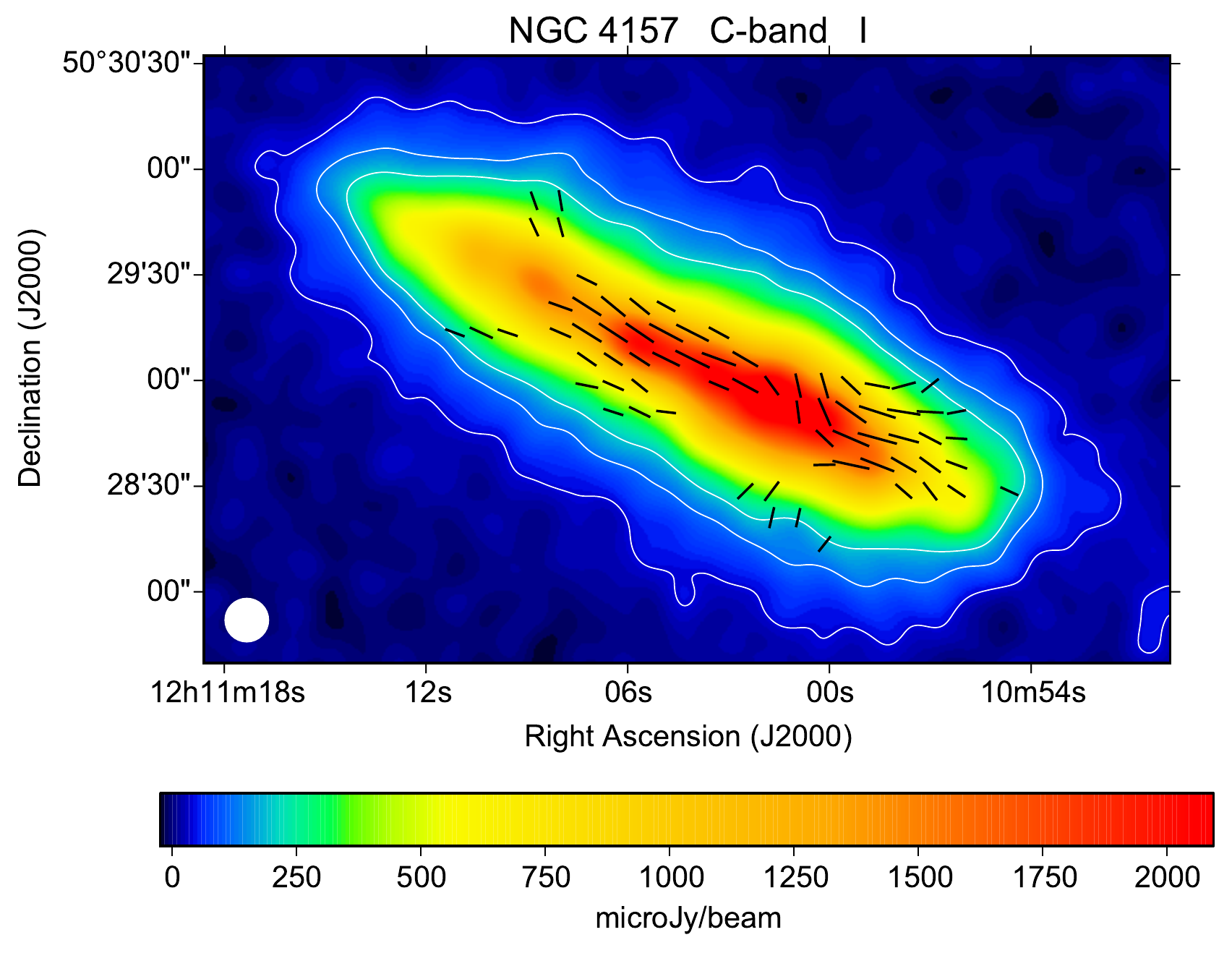}
\includegraphics[width=9.1 cm]{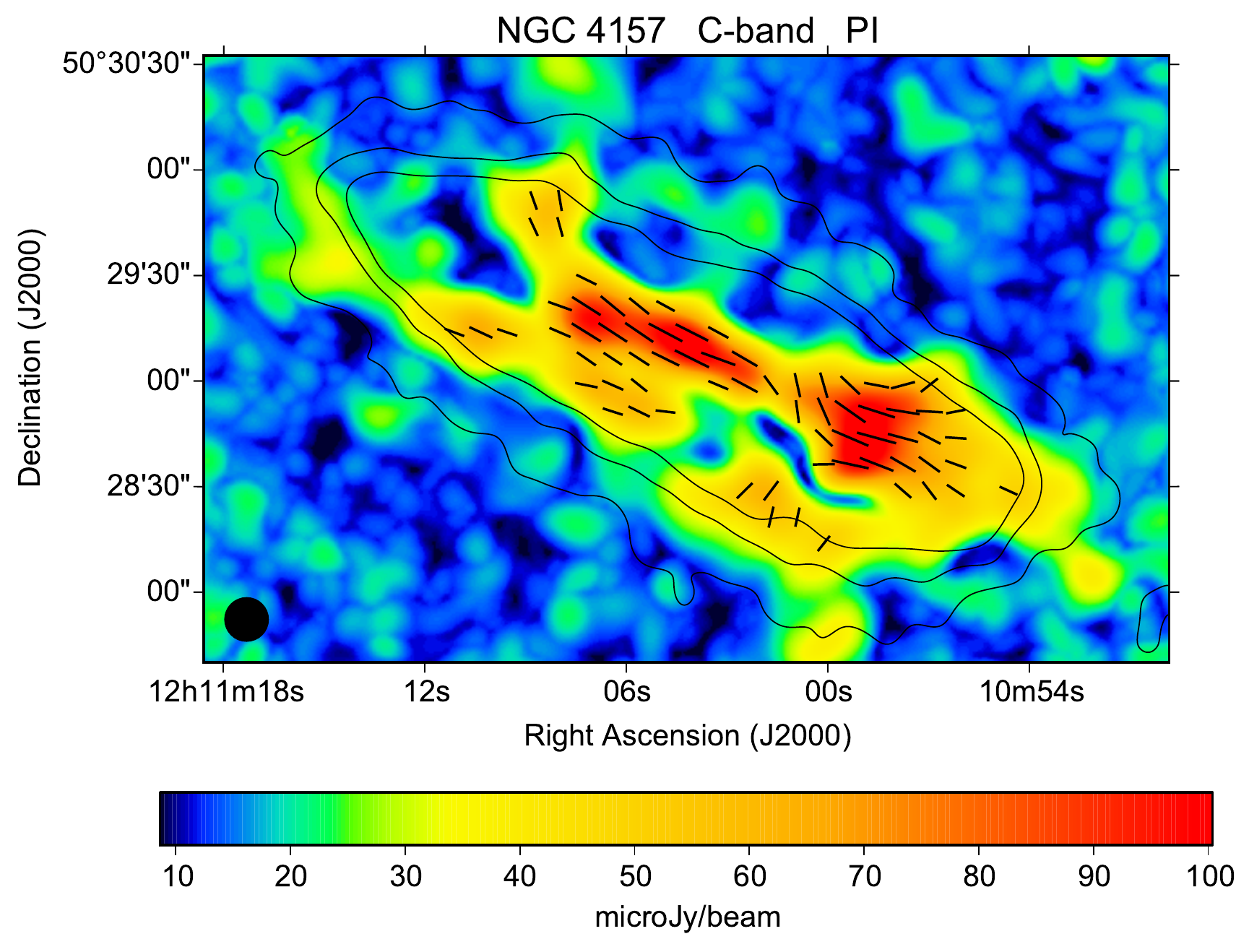}
\includegraphics[width=9.0 cm]{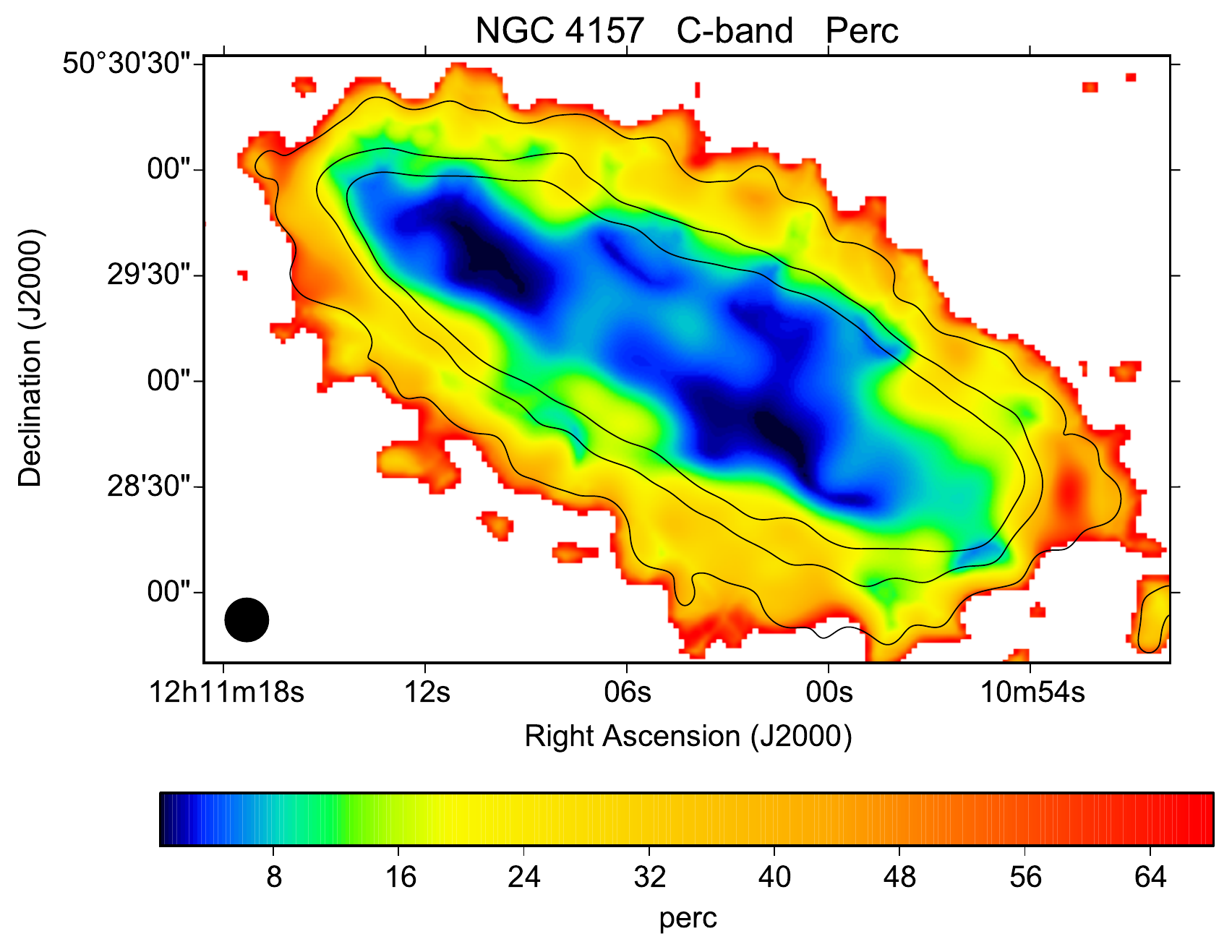}
\includegraphics[width=9.2 cm]{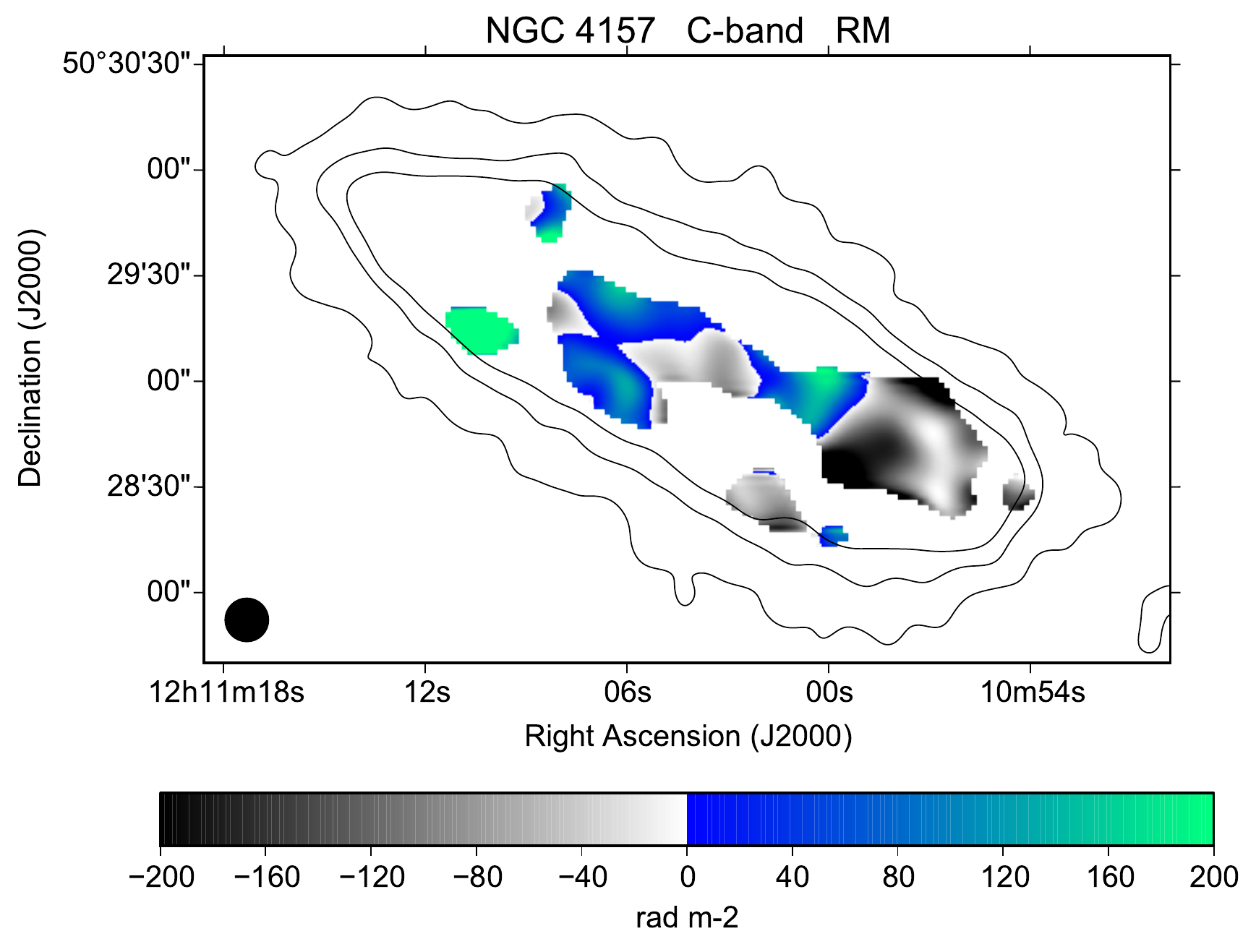}
\includegraphics[width=9.0 cm]{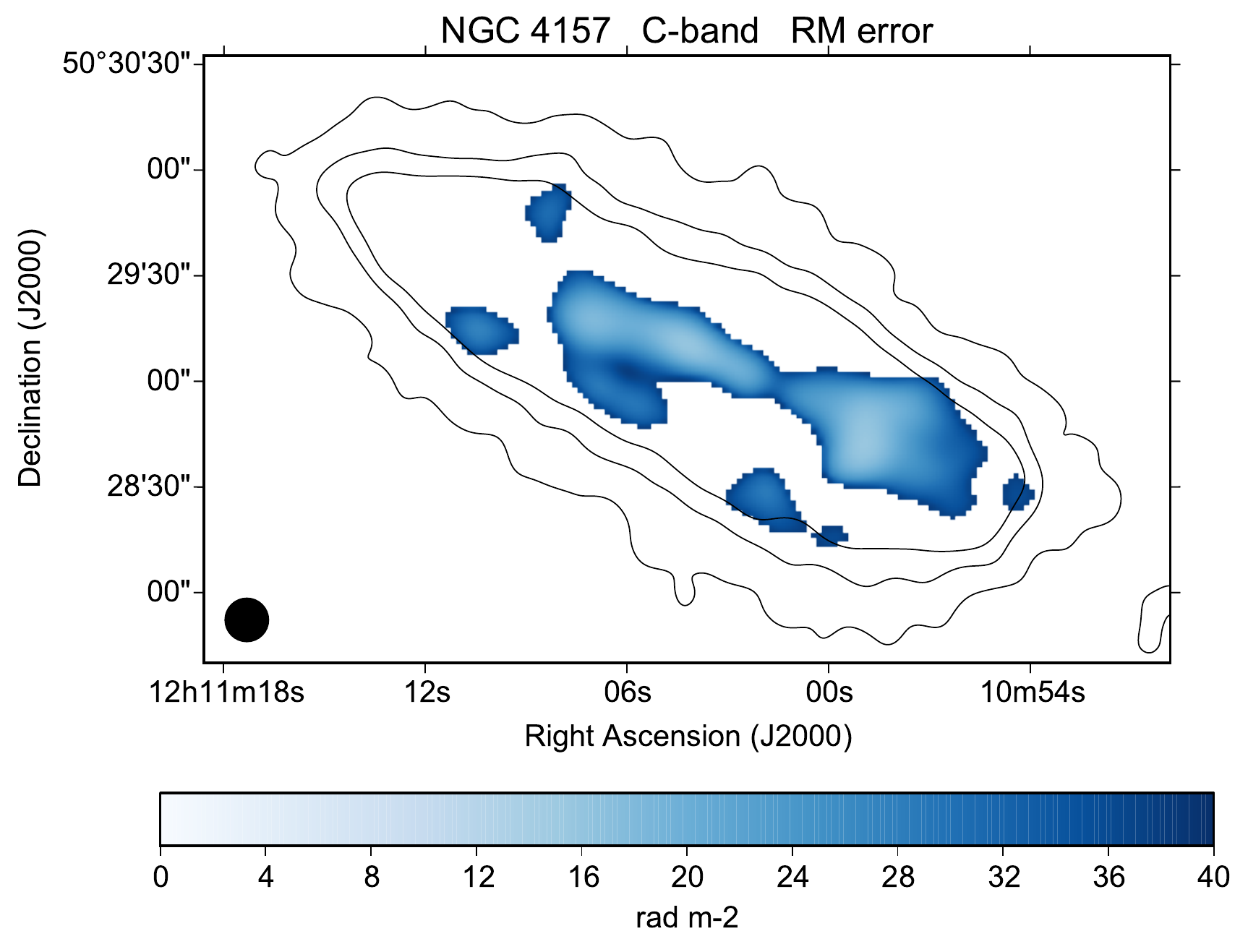}
\includegraphics[width=9.1 cm]{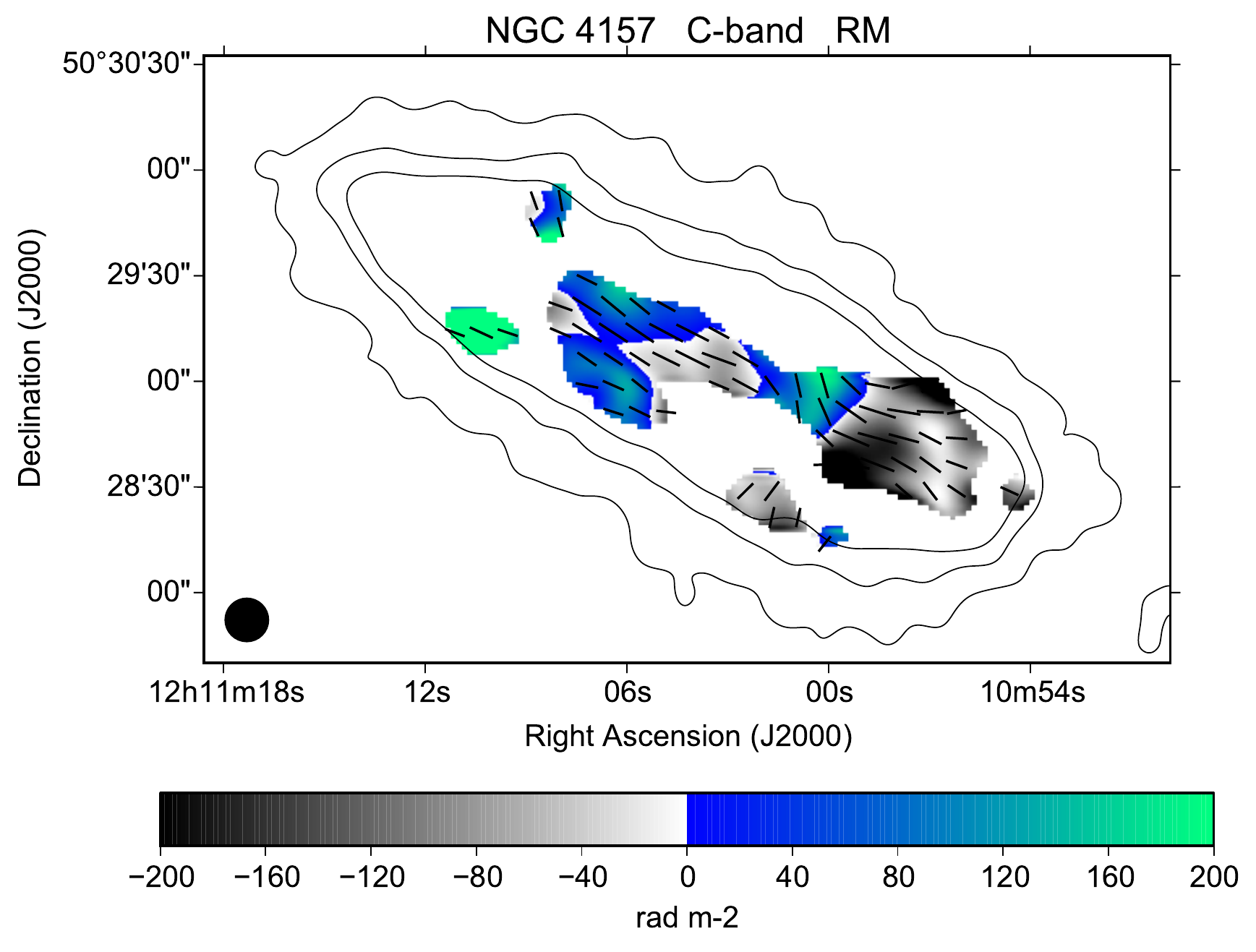}
\caption{Polarization results for NGC~4157 at C-band and $12 \arcsec$ HPBW, corresponding to $910\,\rm{pc}$. The contour levels (TP) are 40, 120, and 200 $\mu$Jy/beam.
}
\label{n4157all}
\end{figure*}

\begin{figure*}[p]
\centering
\includegraphics[width=7.8 cm]{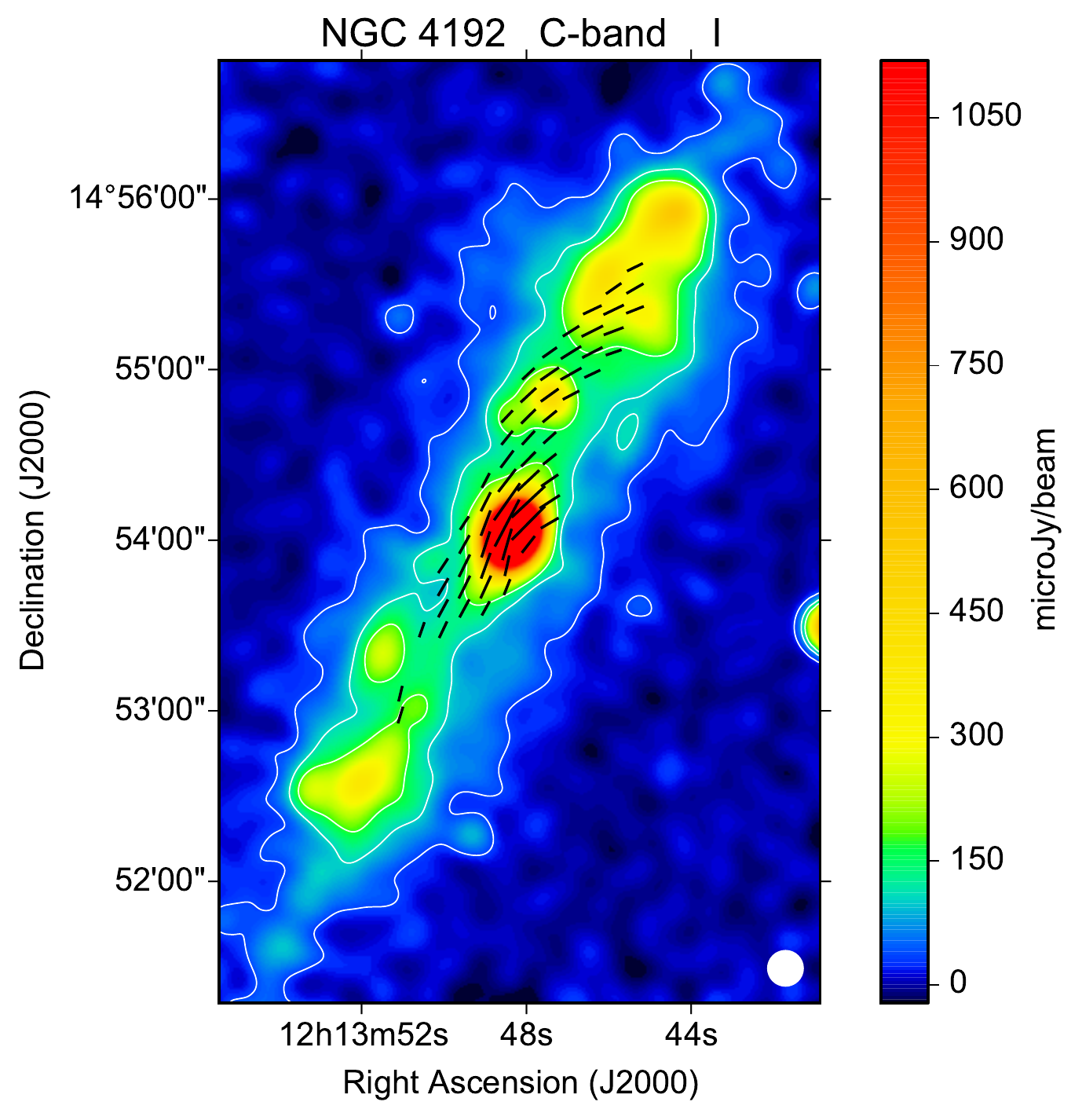}
\includegraphics[width=7.6 cm]{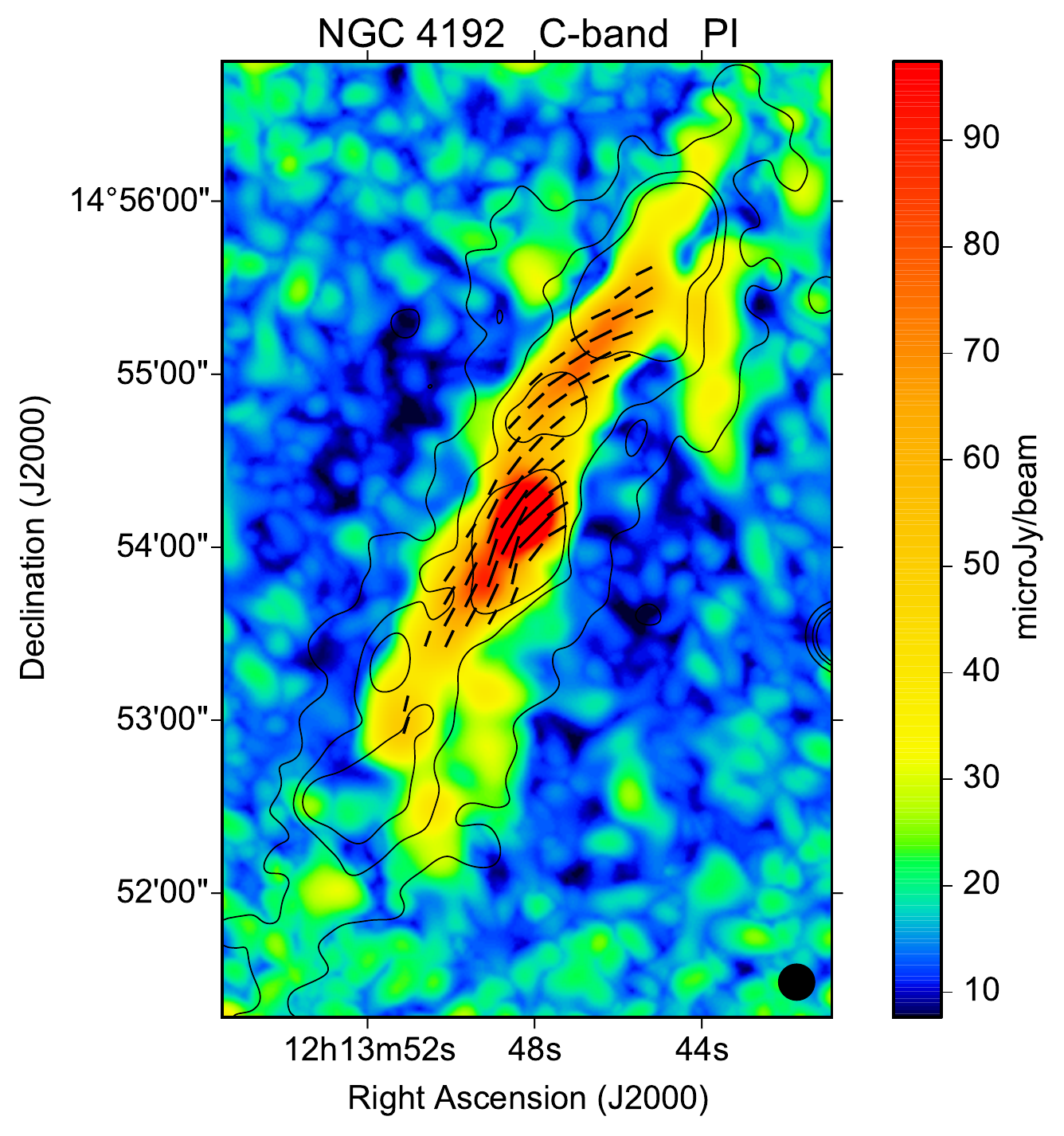}
\includegraphics[width=7.8 cm]{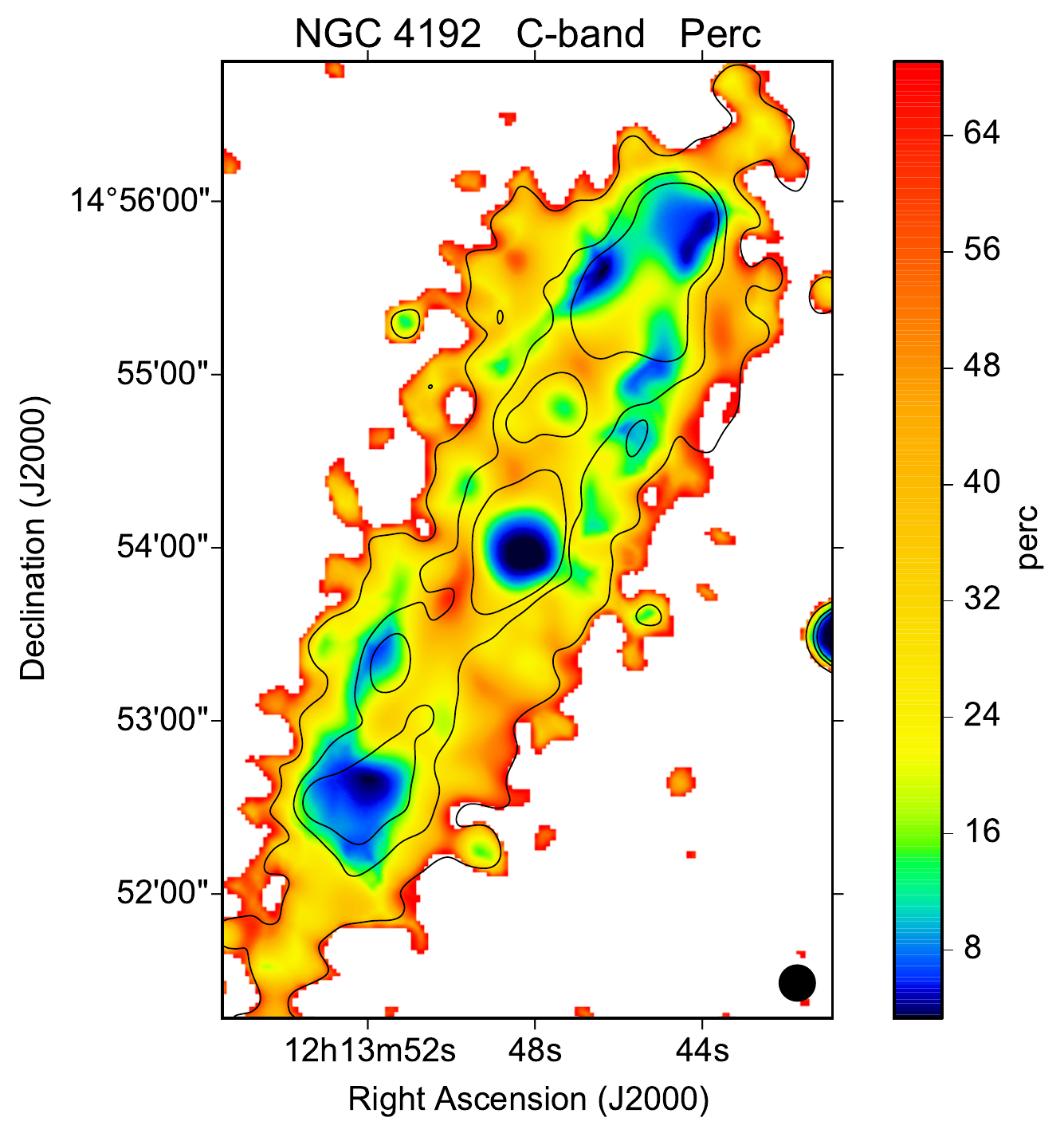}
\includegraphics[width=8.0 cm]{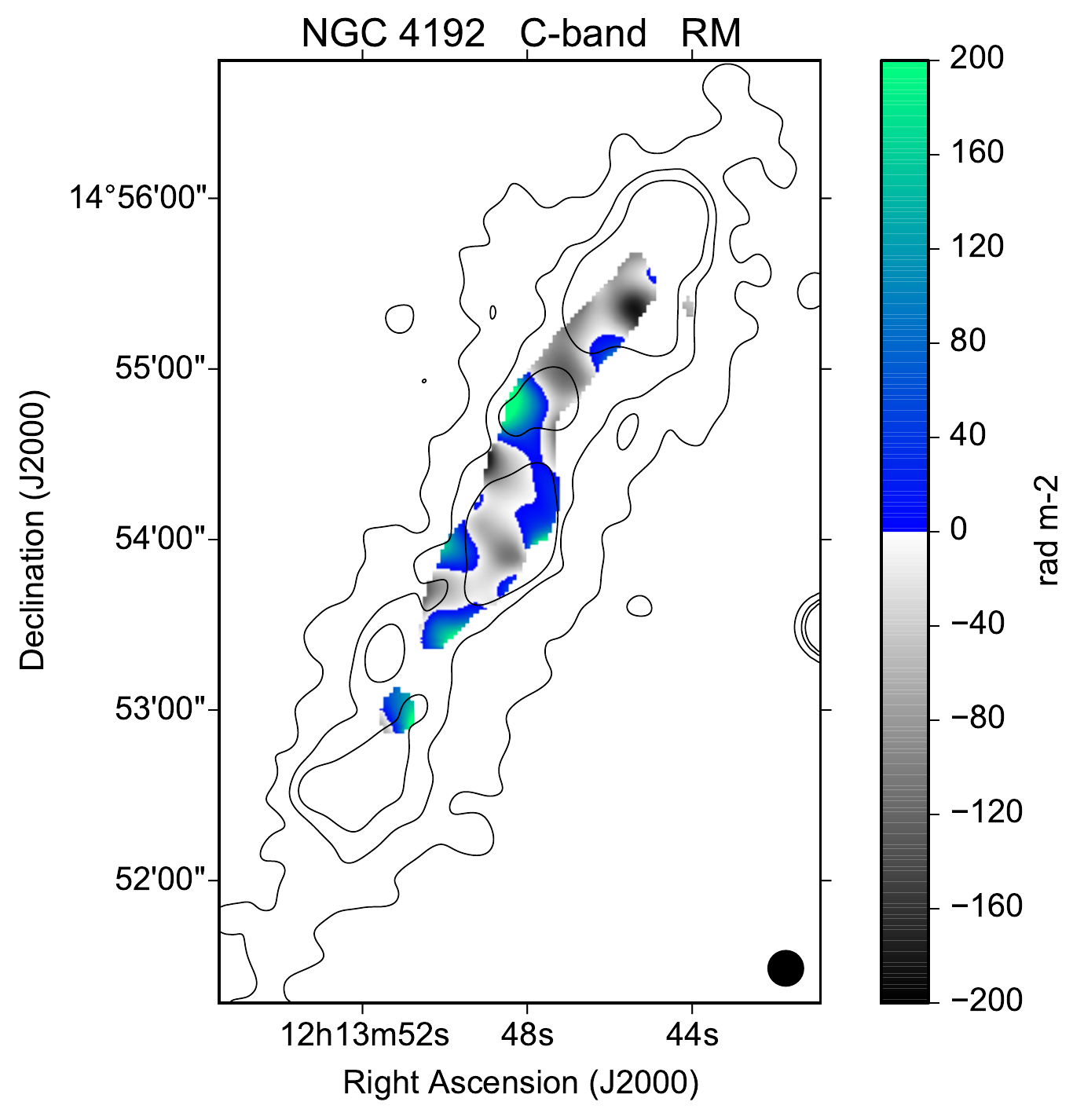}
\includegraphics[width=7.7 cm]{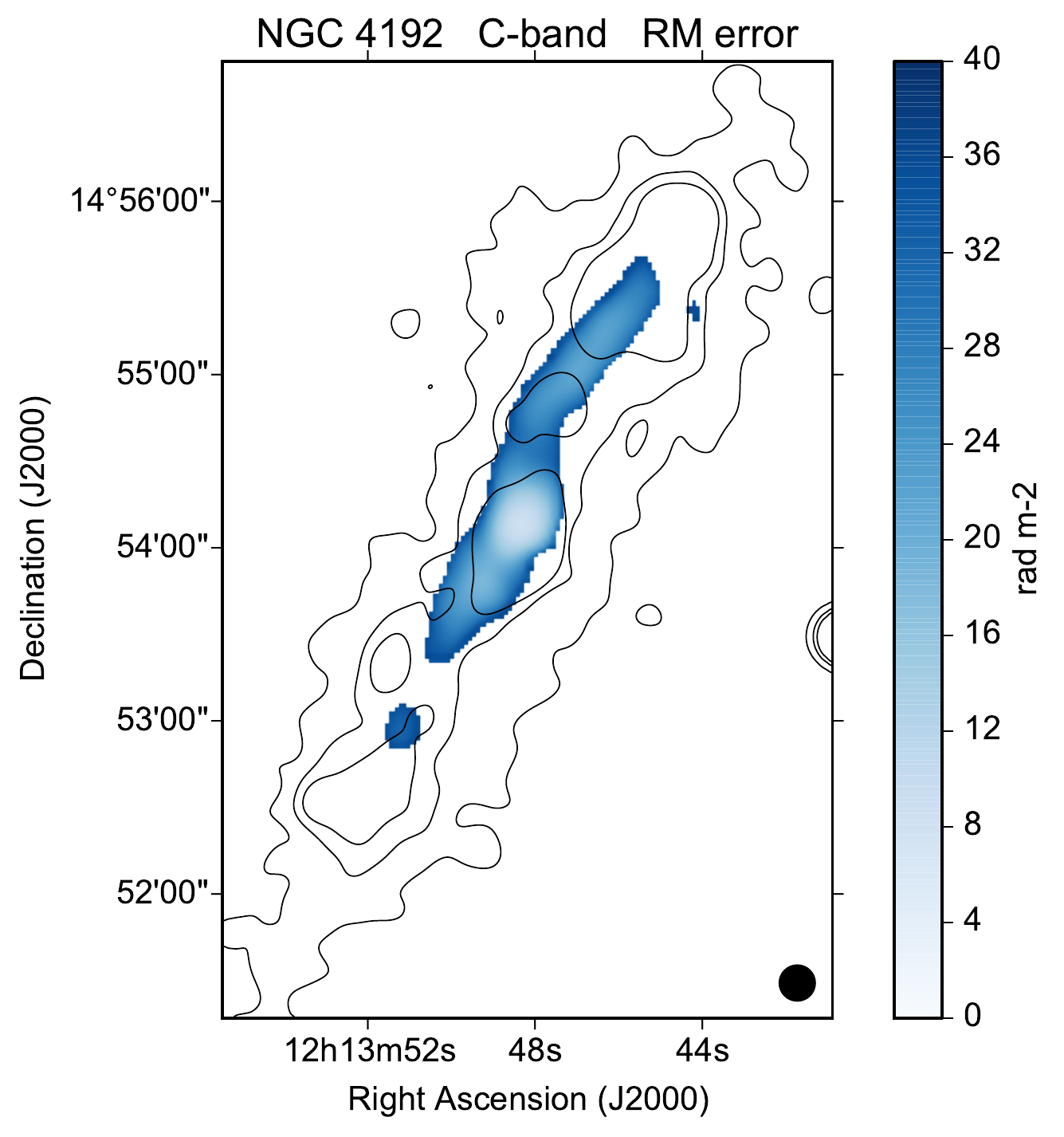}
\includegraphics[width=8.0 cm]{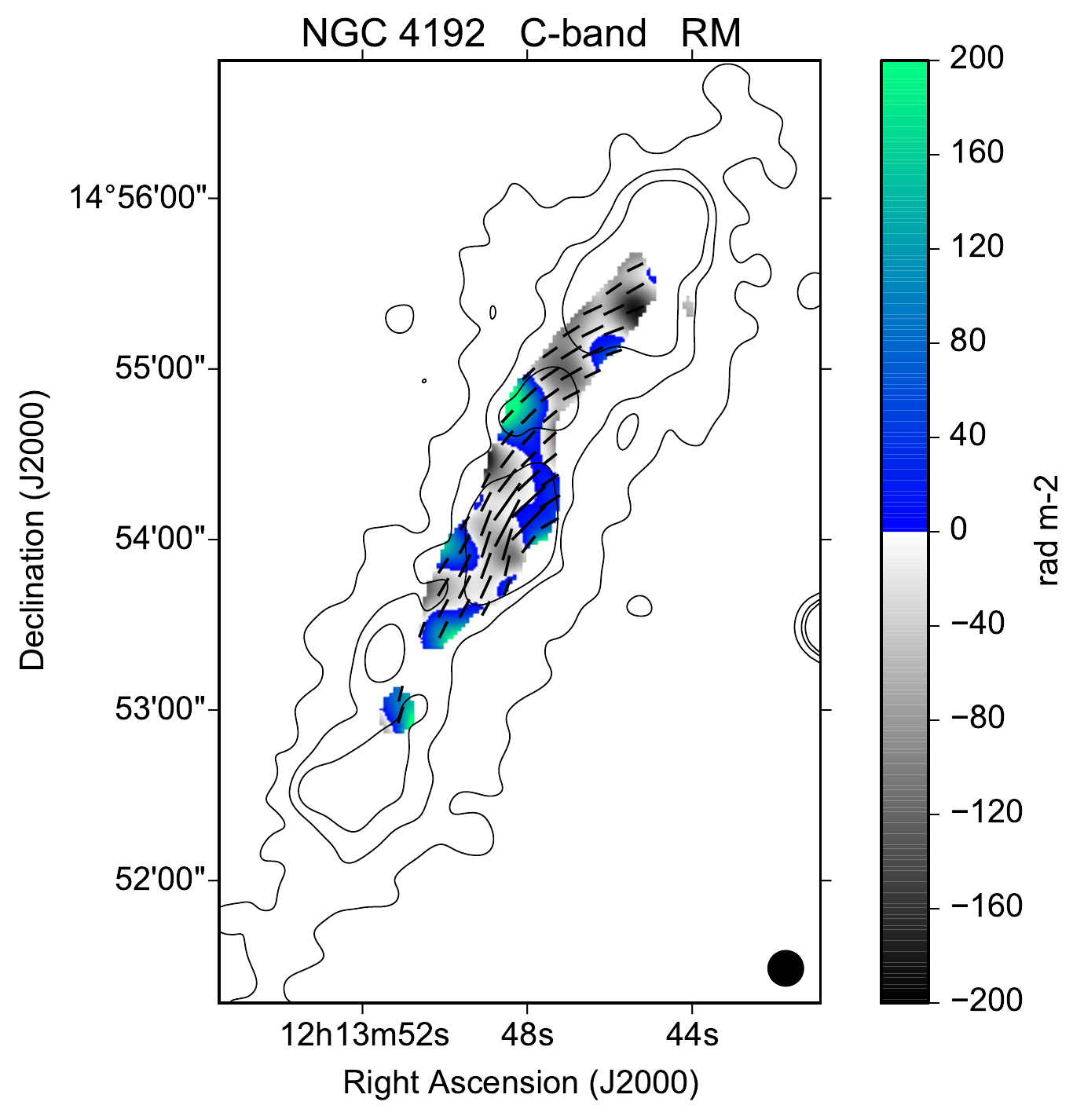}
\caption{Polarization results for NGC~4192 at C-band and $12 \arcsec$ HPBW, corresponding to $790\,\rm{pc}$. The contour levels (TP) are 35, 105, and 175 $\mu$Jy/beam.
}
\label{n4192all}
\end{figure*}

\begin{figure*}[p]
\centering
\includegraphics[width=9.0 cm]{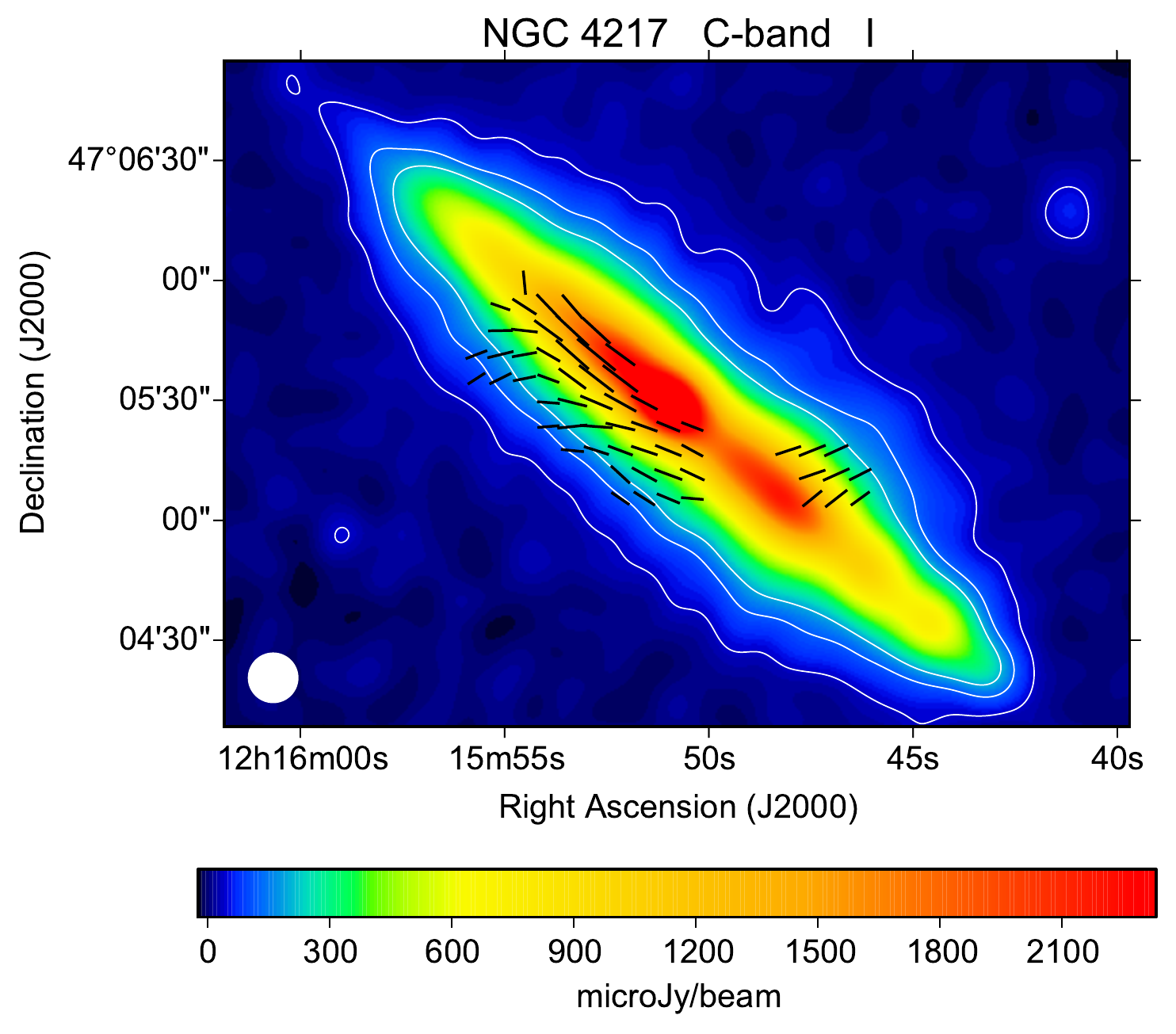}
\includegraphics[width=9.1 cm]{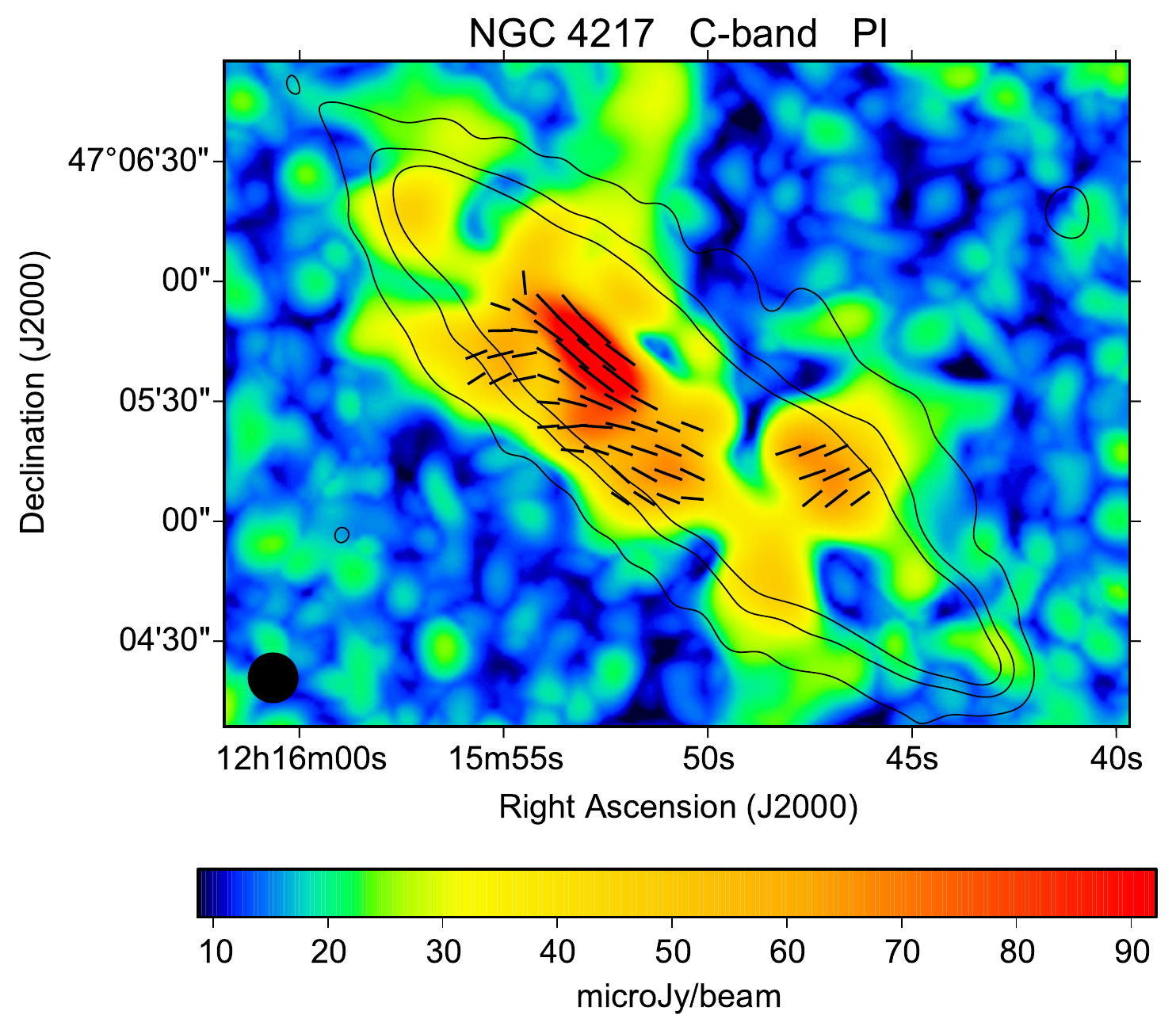}
\includegraphics[width=9.0 cm]{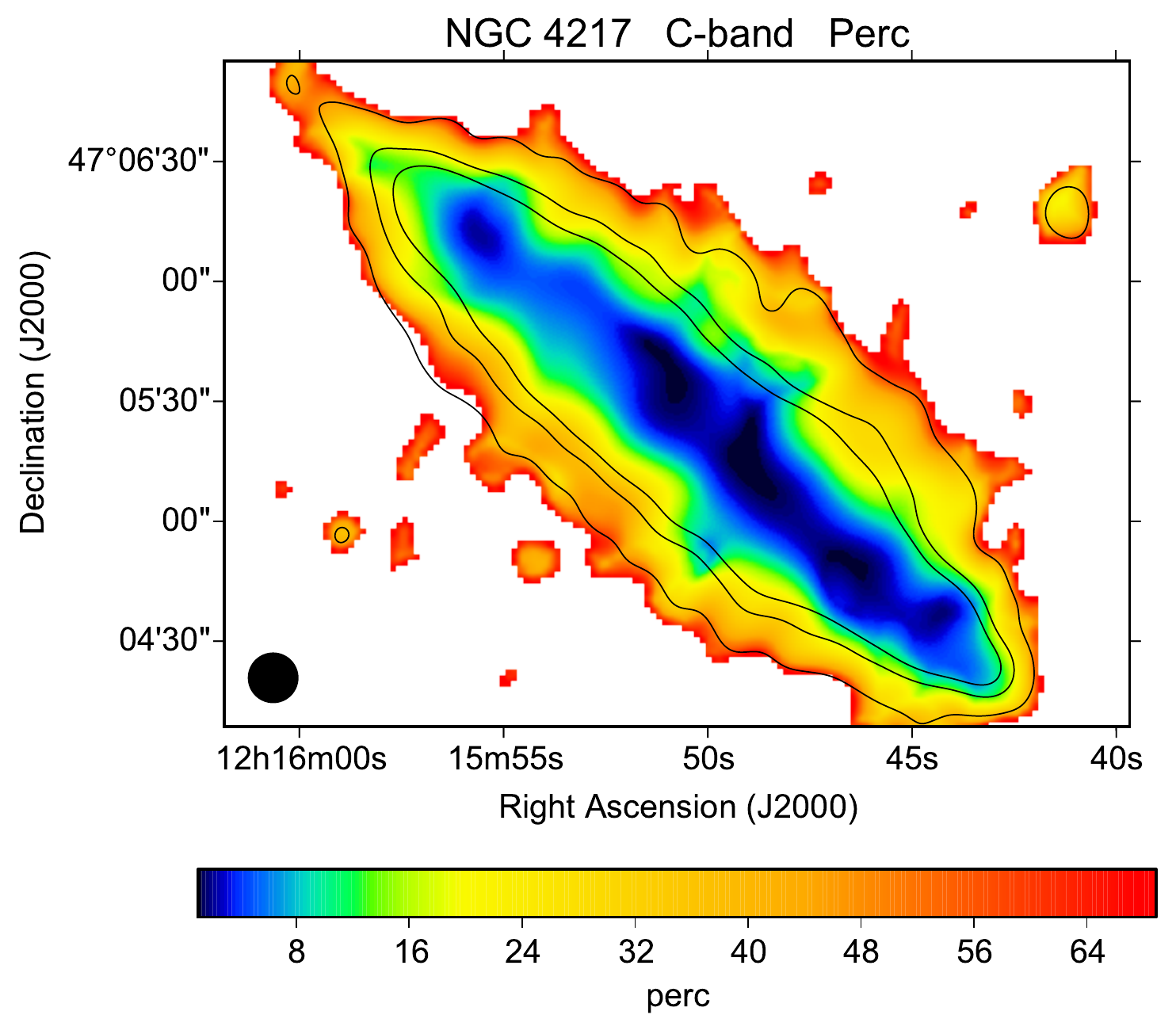}
\includegraphics[width=9.2 cm]{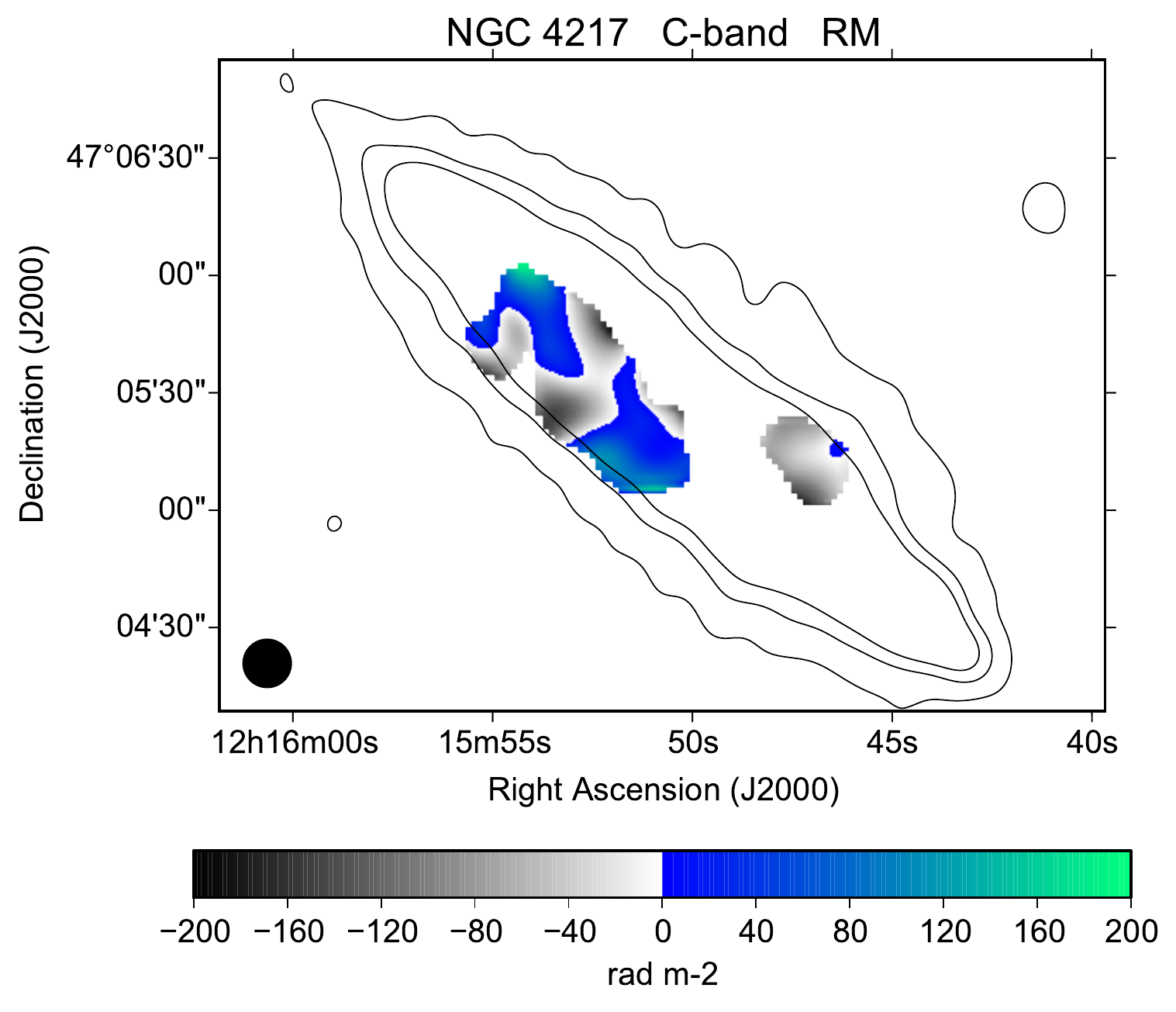}
\includegraphics[width=9.0 cm]{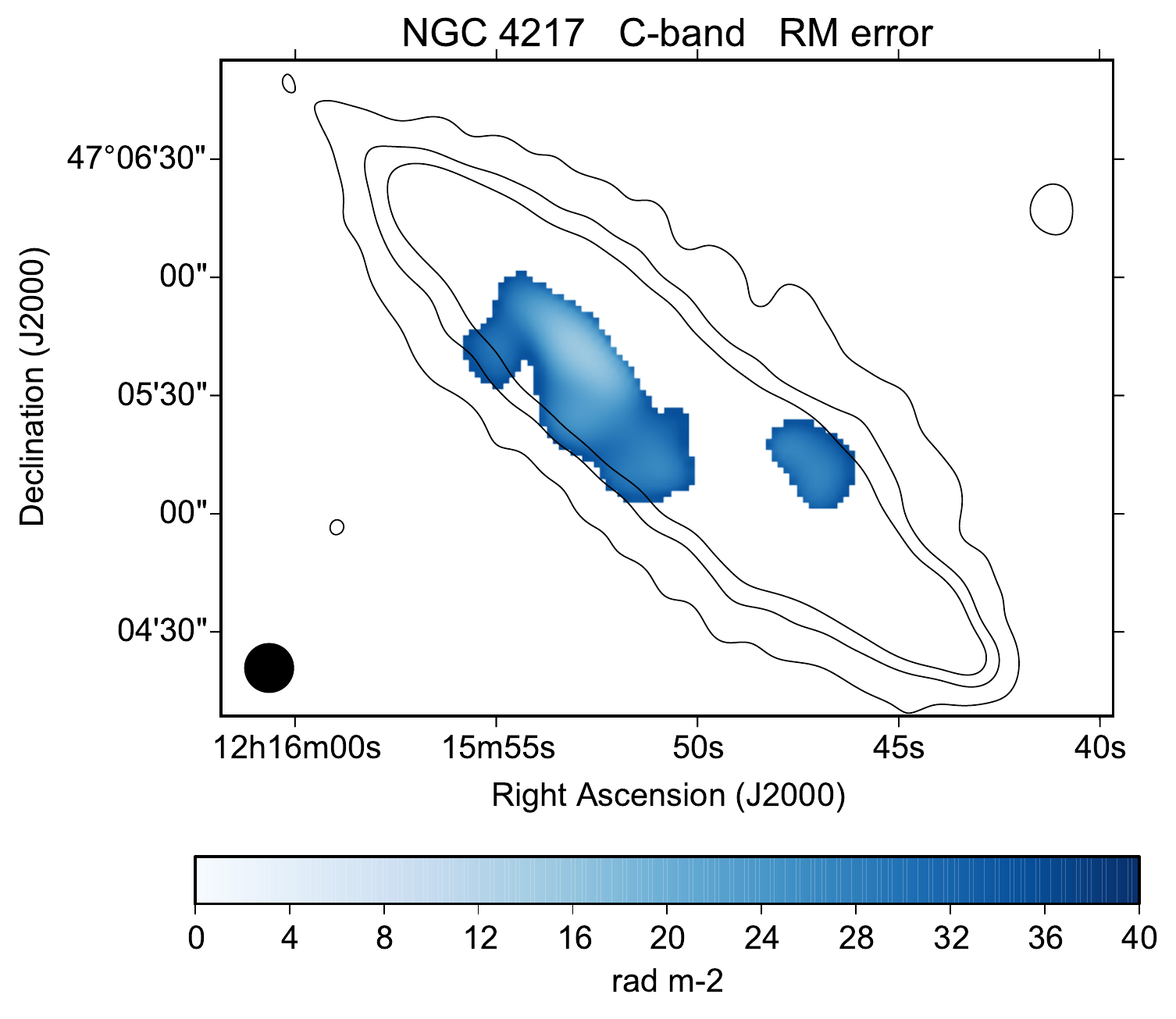}
\includegraphics[width=9.1 cm]{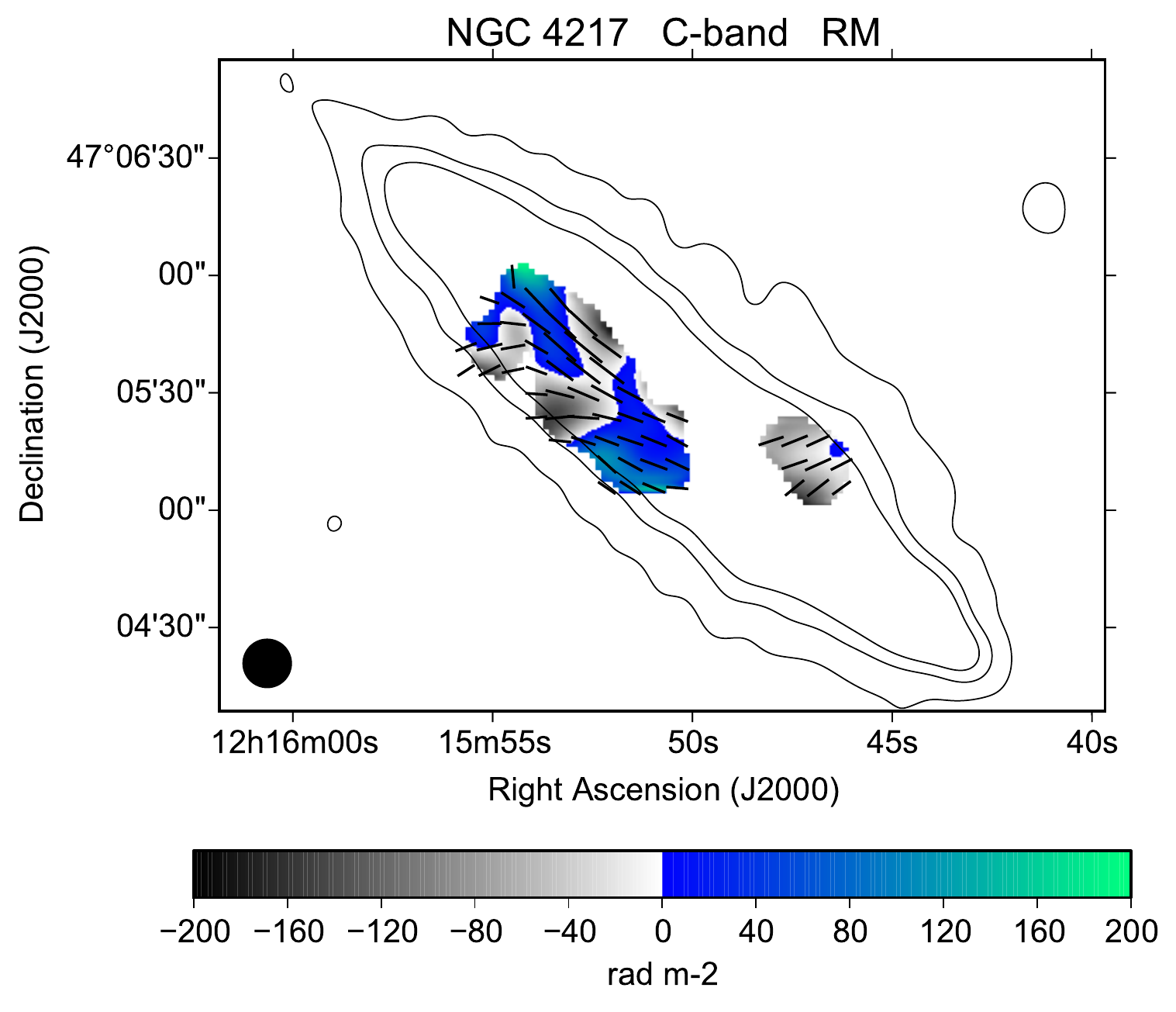}
\caption{Polarization results for NGC~4217 at C-band and $12 \arcsec$ HPBW, corresponding to $1200\,\rm{pc}$. The contour levels (TP) are 40, 120, and 200 $\mu$Jy/beam.
}
\label{n4217all}
\end{figure*}

\begin{figure*}[p]
\centering
\includegraphics[width=8.0 cm]{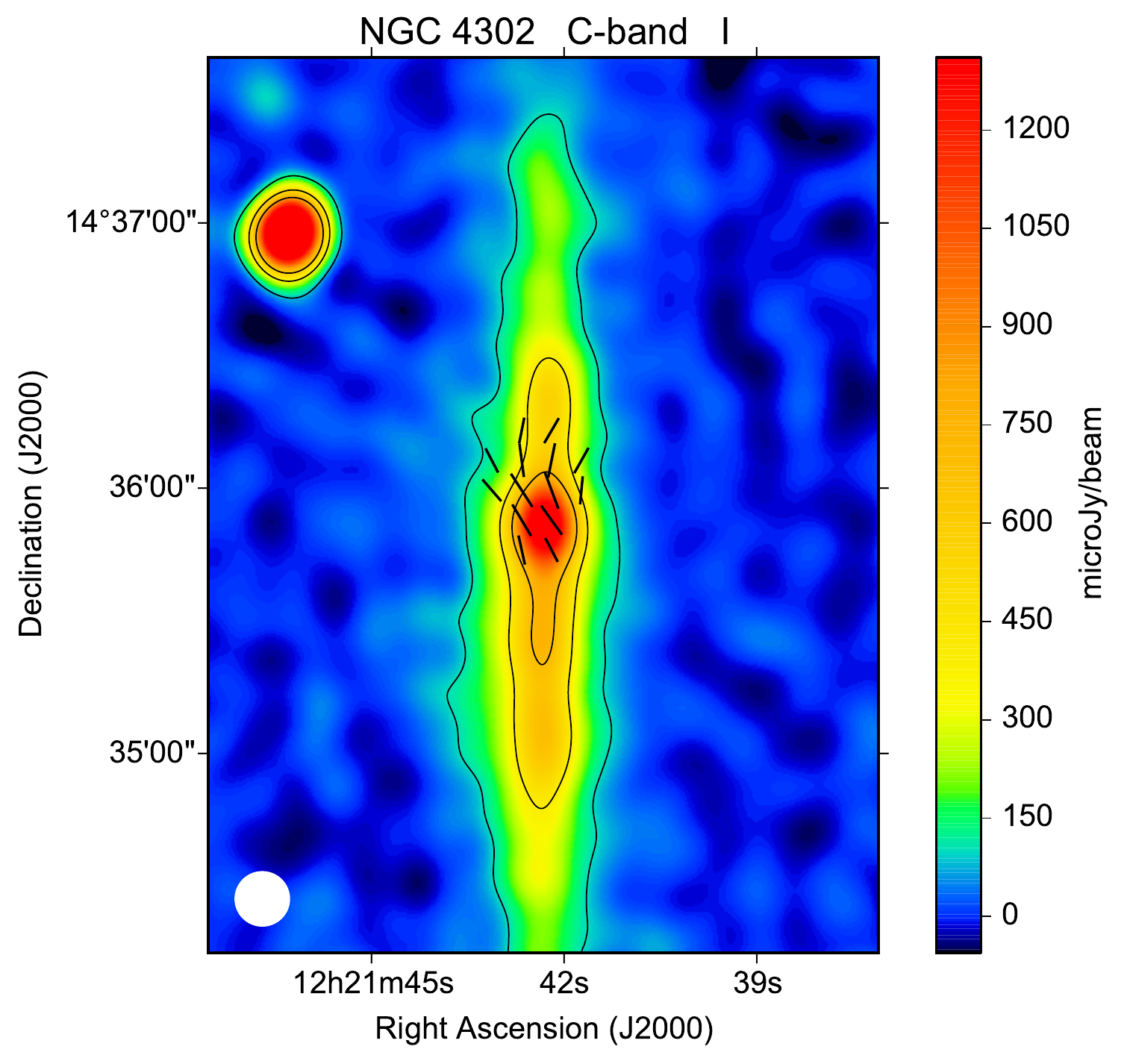}
\includegraphics[width=7.8 cm]{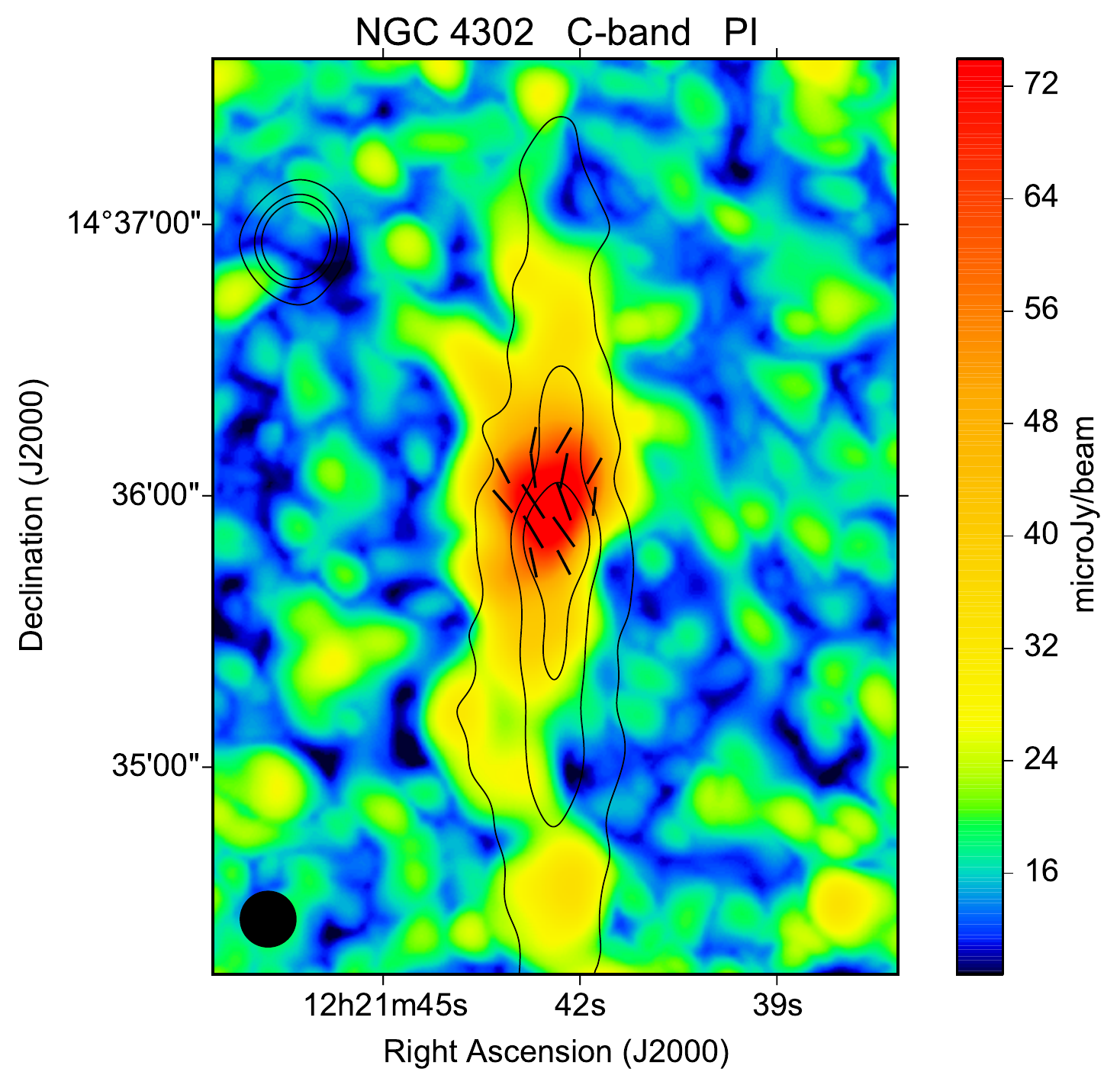}
\includegraphics[width=8.0 cm]{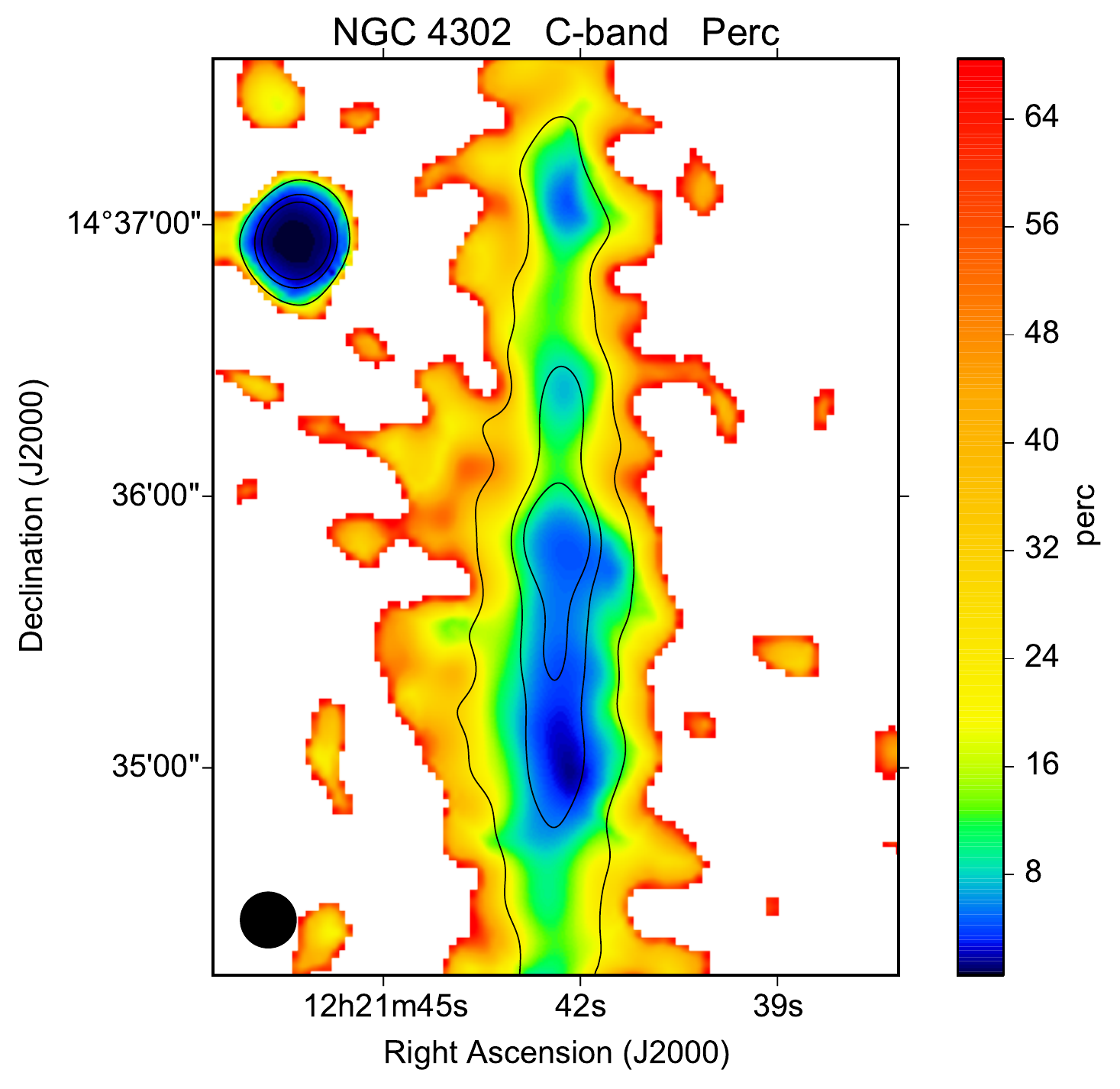}
\includegraphics[width=8.3 cm]{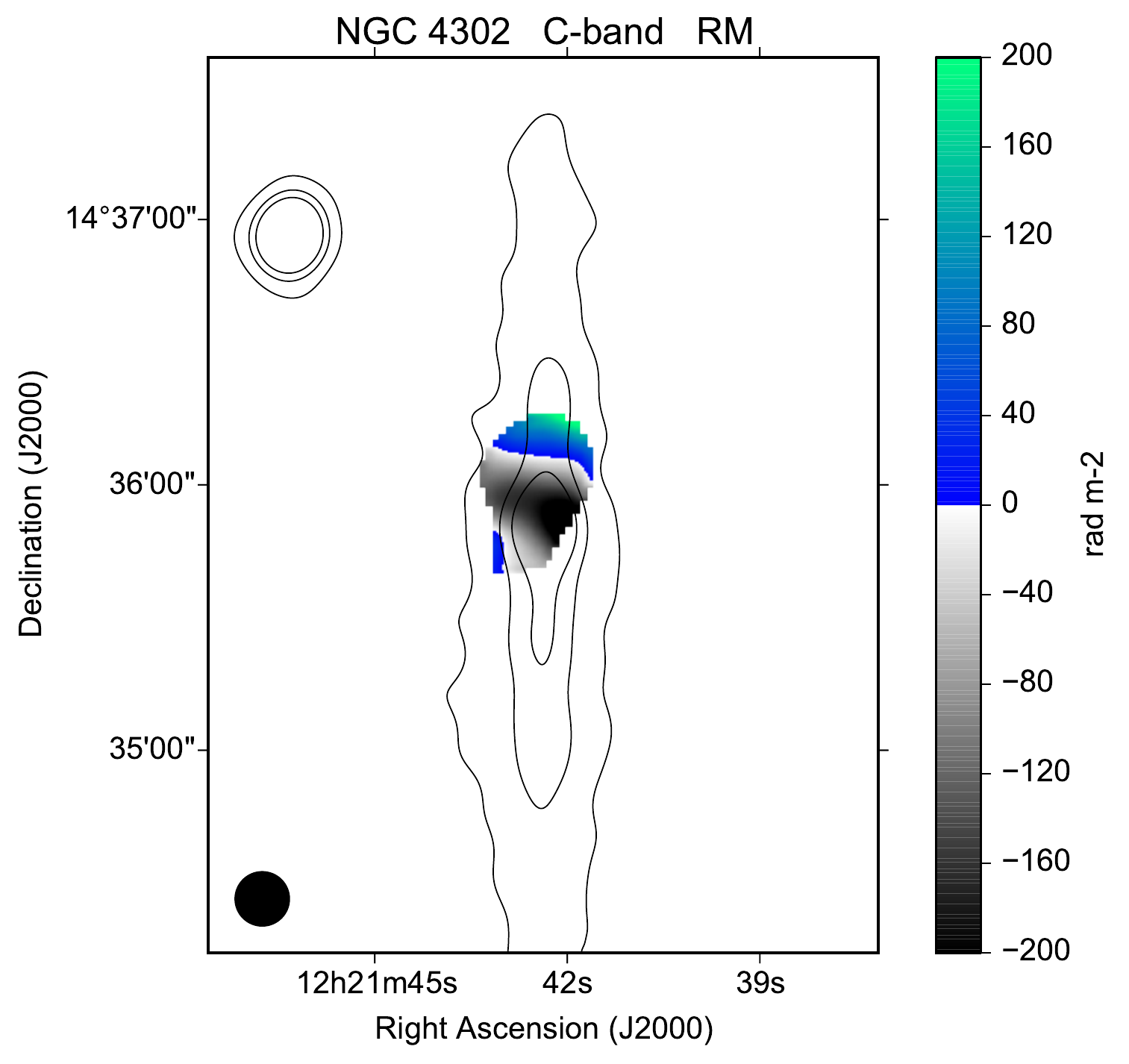}
\includegraphics[width=7.9 cm]{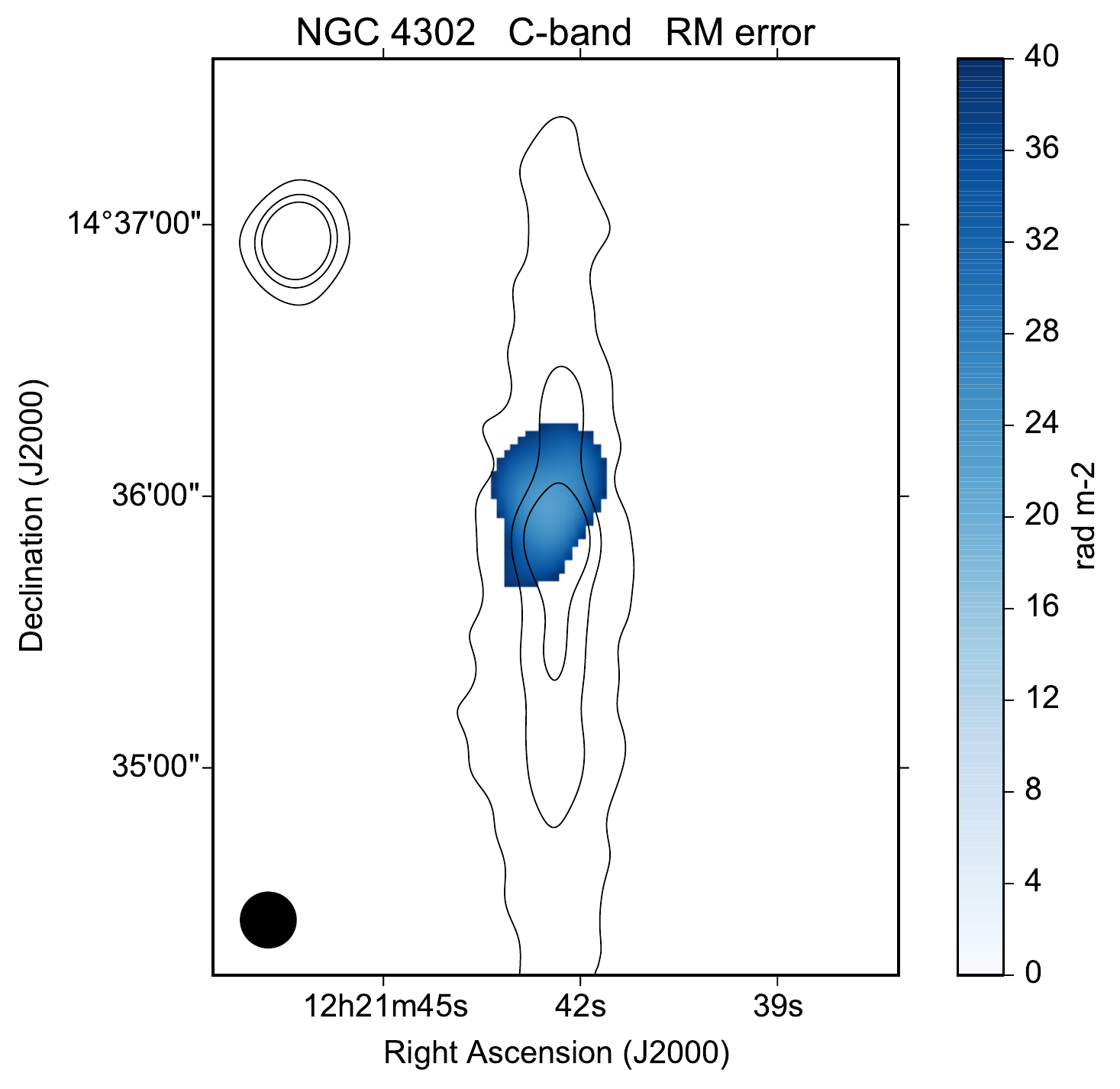}
\includegraphics[width=8.2 cm]{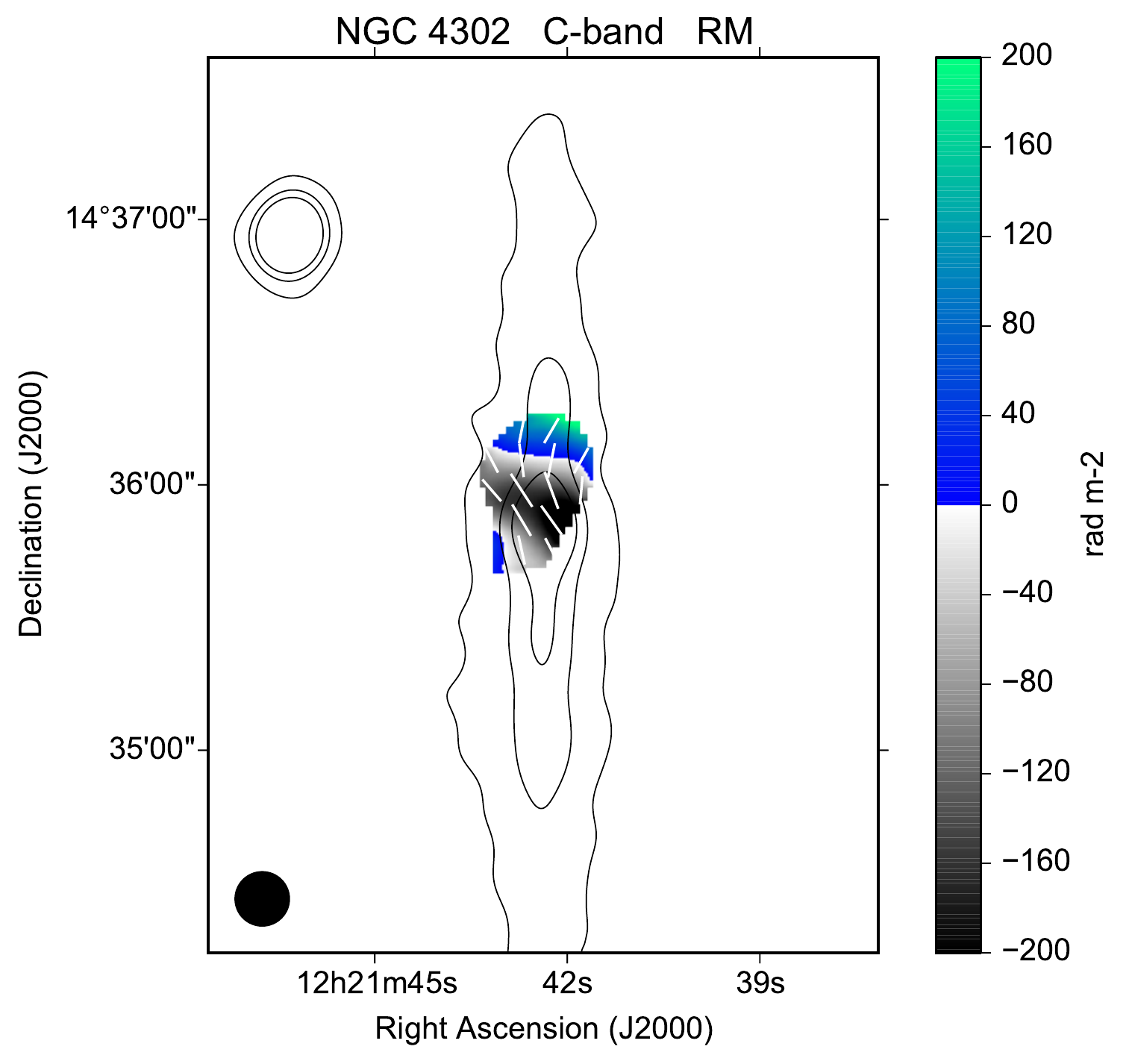}
\caption{Polarization results for NGC~4302 at C-band and $12 \arcsec$ HPBW, corresponding to $1130\,\rm{pc}$. The contour levels (TP) are 100, 300, and 500 $\mu$Jy/beam.
}
\label{n4302all}
\end{figure*}

\begin{figure*}[p]
\centering
\includegraphics[width=9.0 cm]{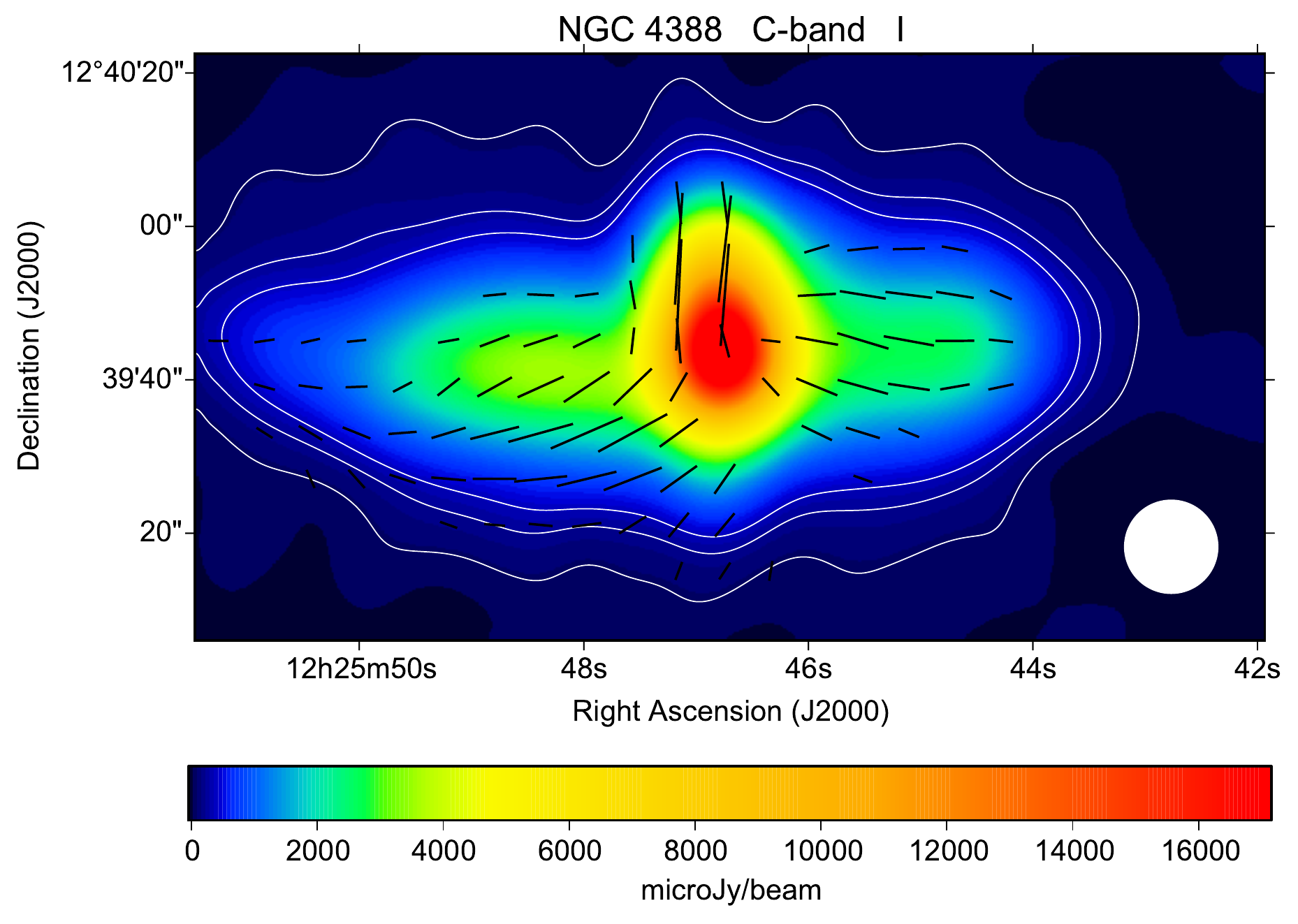}
\includegraphics[width=9.0 cm]{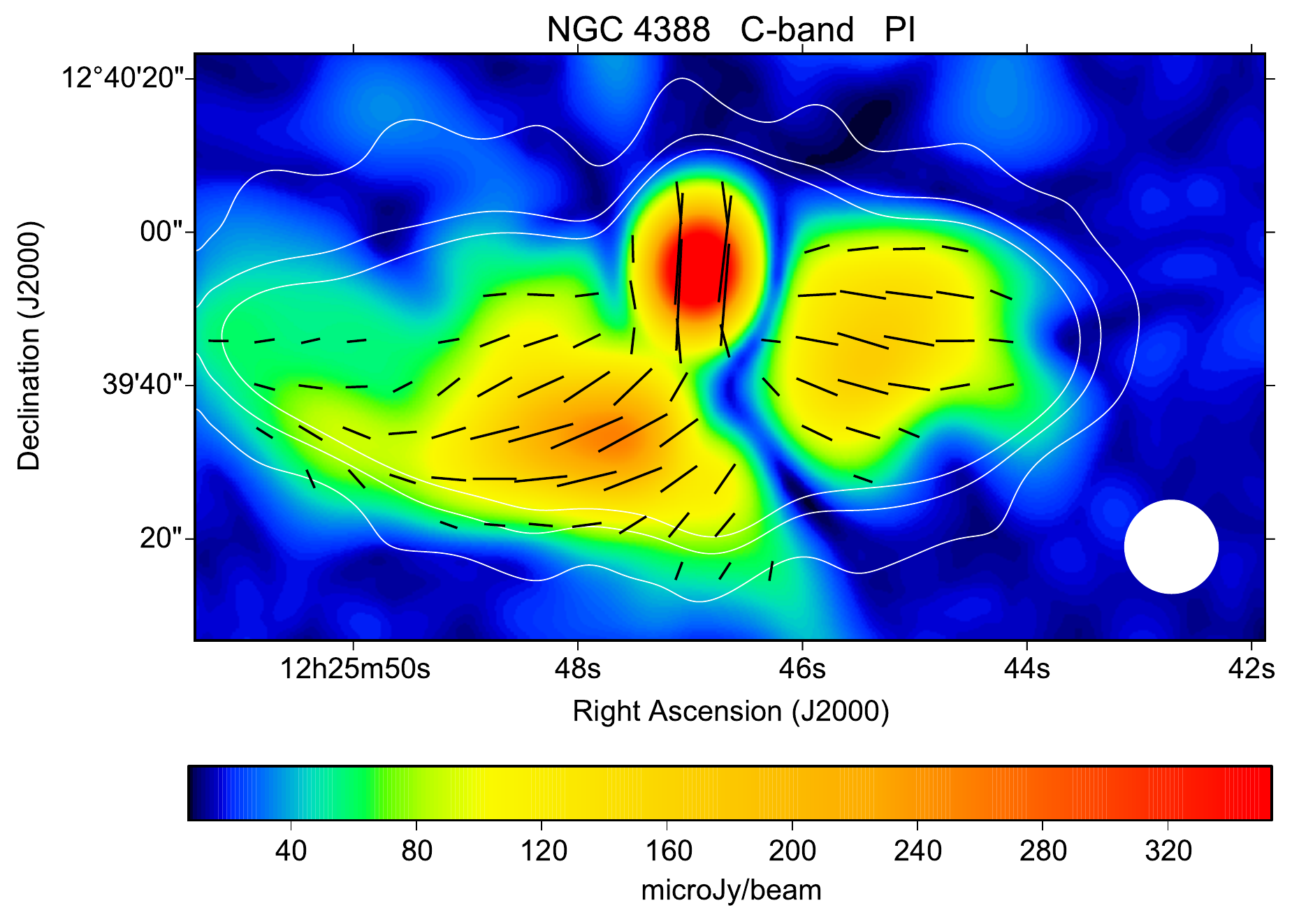}
\includegraphics[width=9.0 cm]{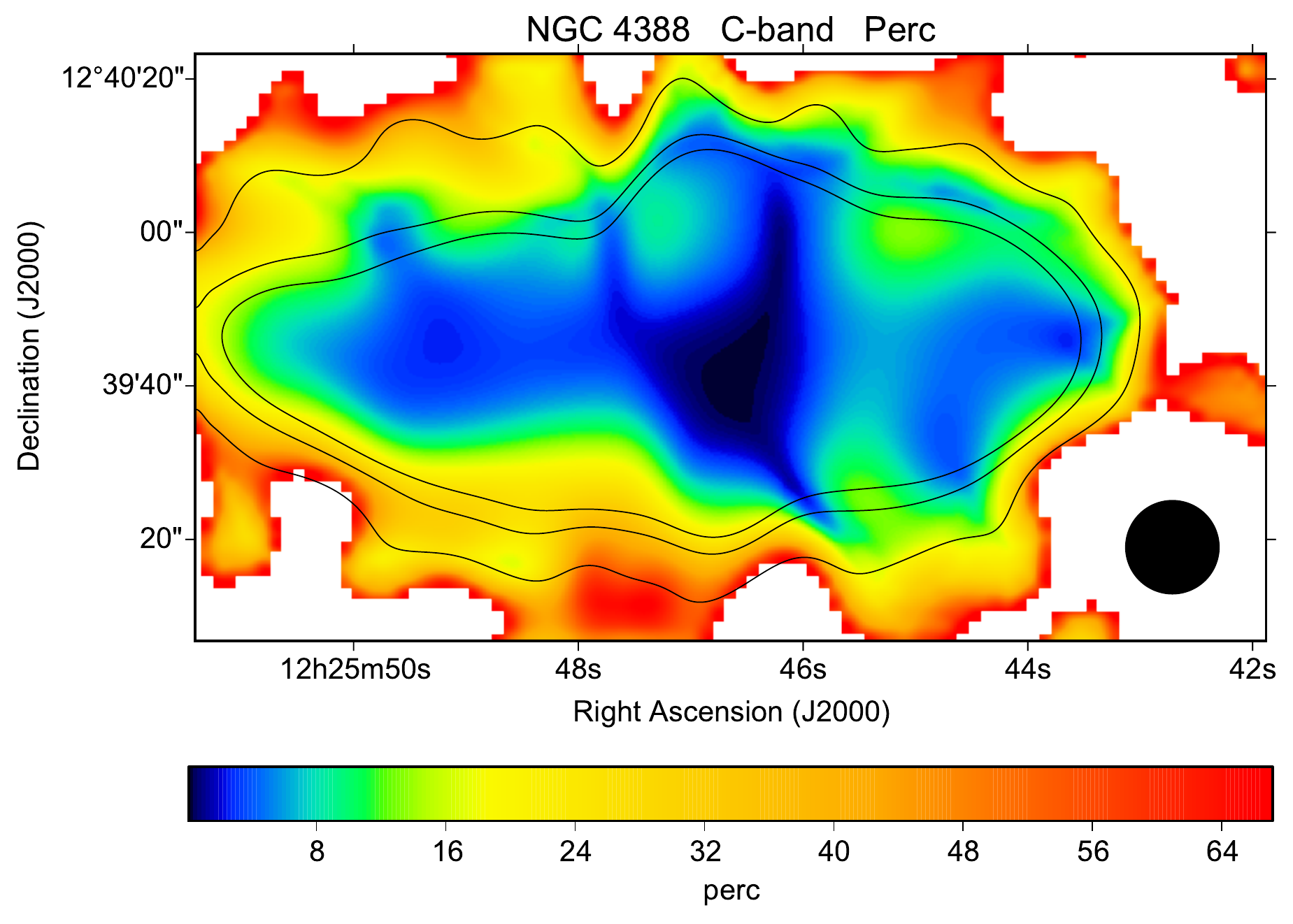}
\includegraphics[width=9.1 cm]{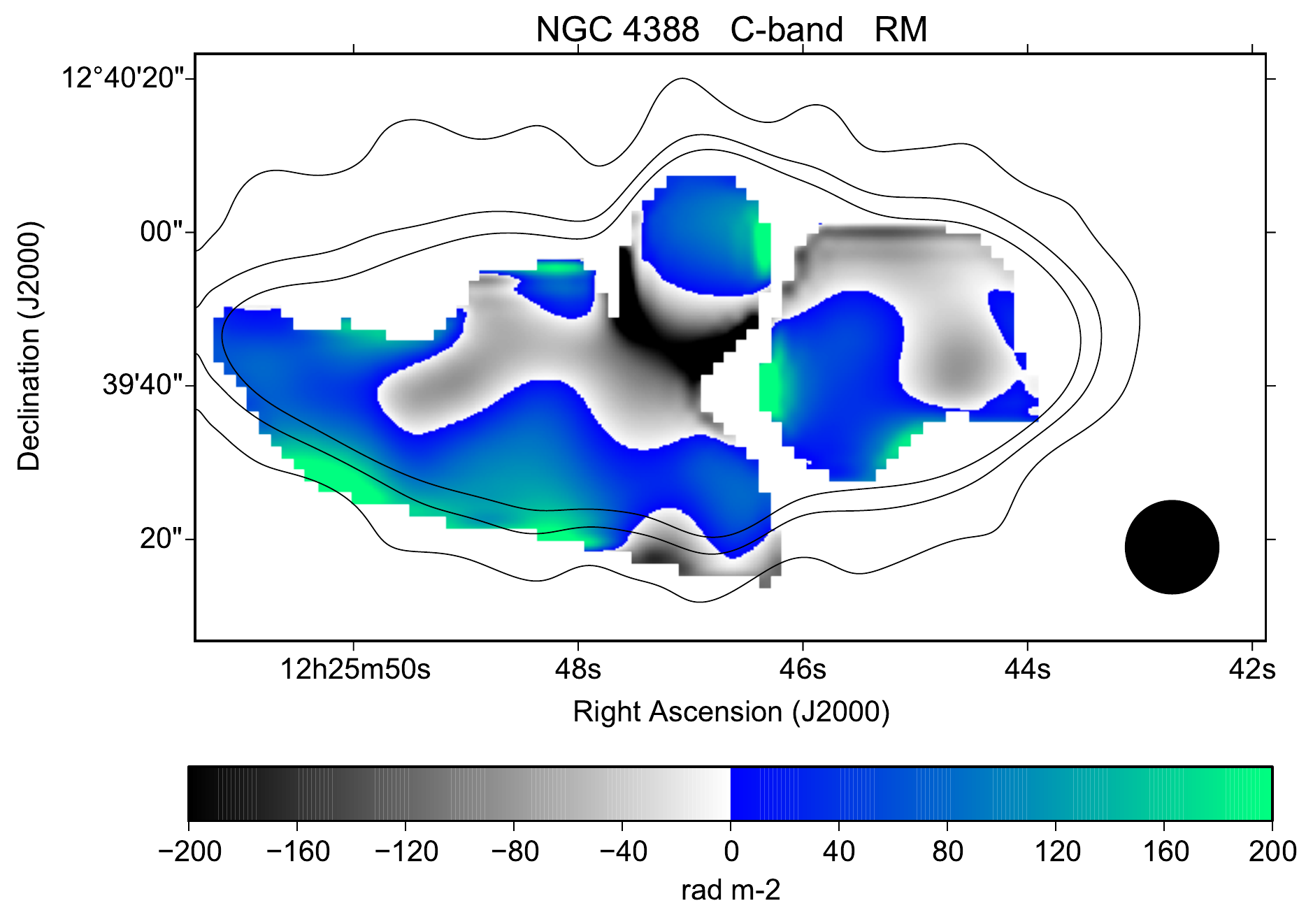}
\includegraphics[width=9.0 cm]{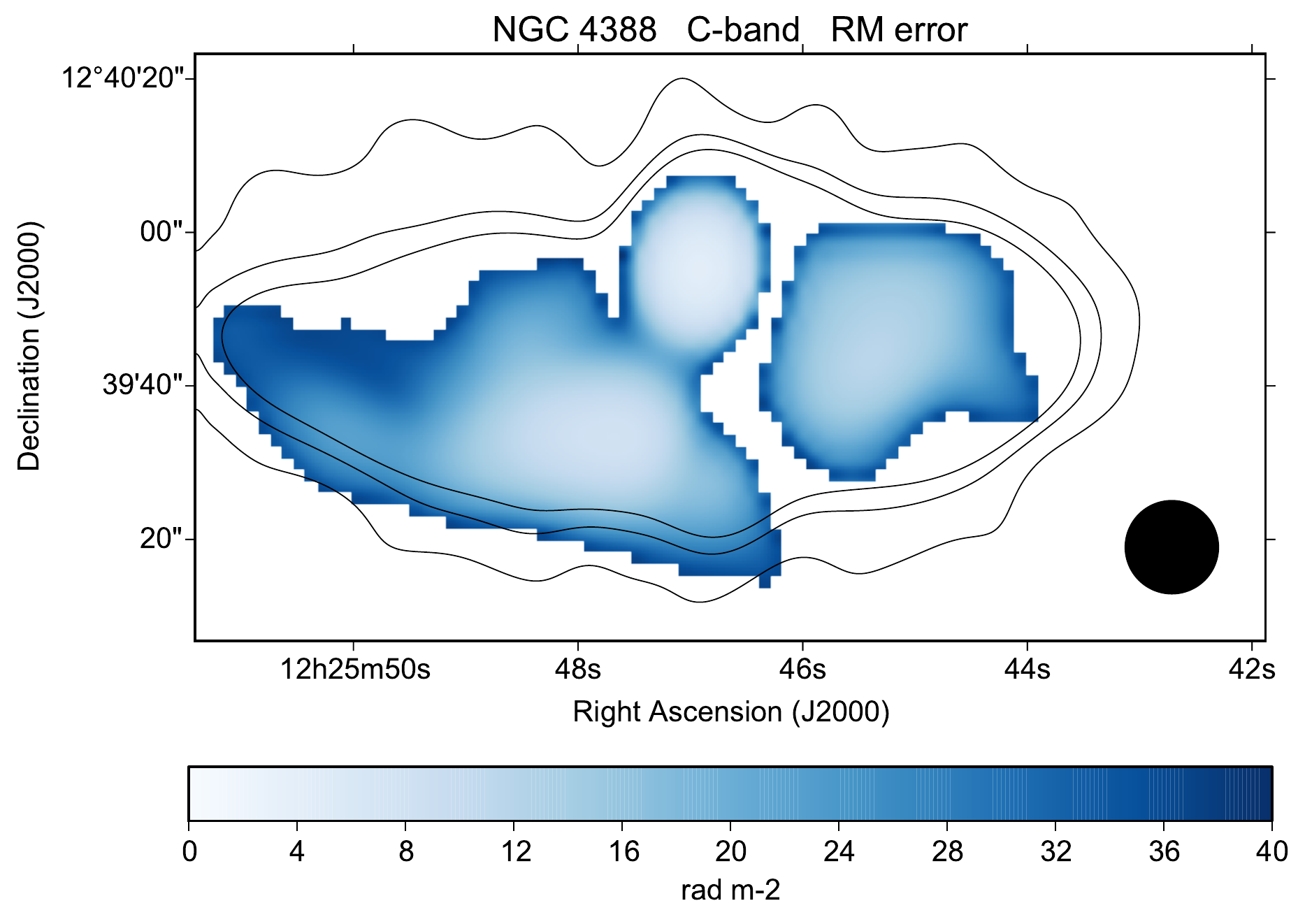}
\includegraphics[width=9.1 cm]{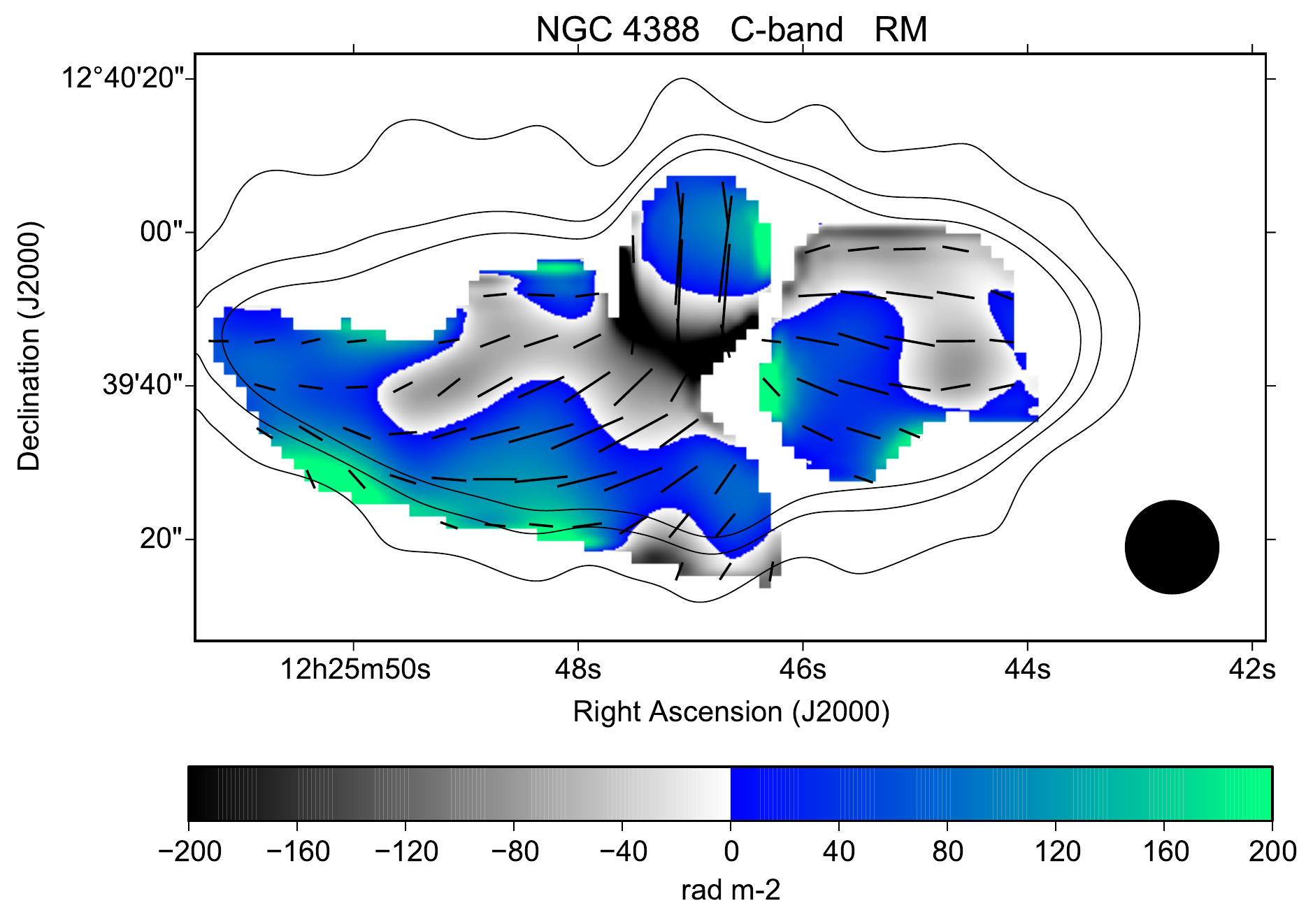}
\caption{Polarization results for NGC~4388 at C-band and $12 \arcsec$ HPBW, corresponding to $970\,\rm{pc}$. The contour levels (TP) are 75, 225, and 375 $\mu$Jy/beam.
}
\label{n4388all}
\end{figure*}

\begin{figure*}[p]
\centering
\includegraphics[width=9.0 cm]{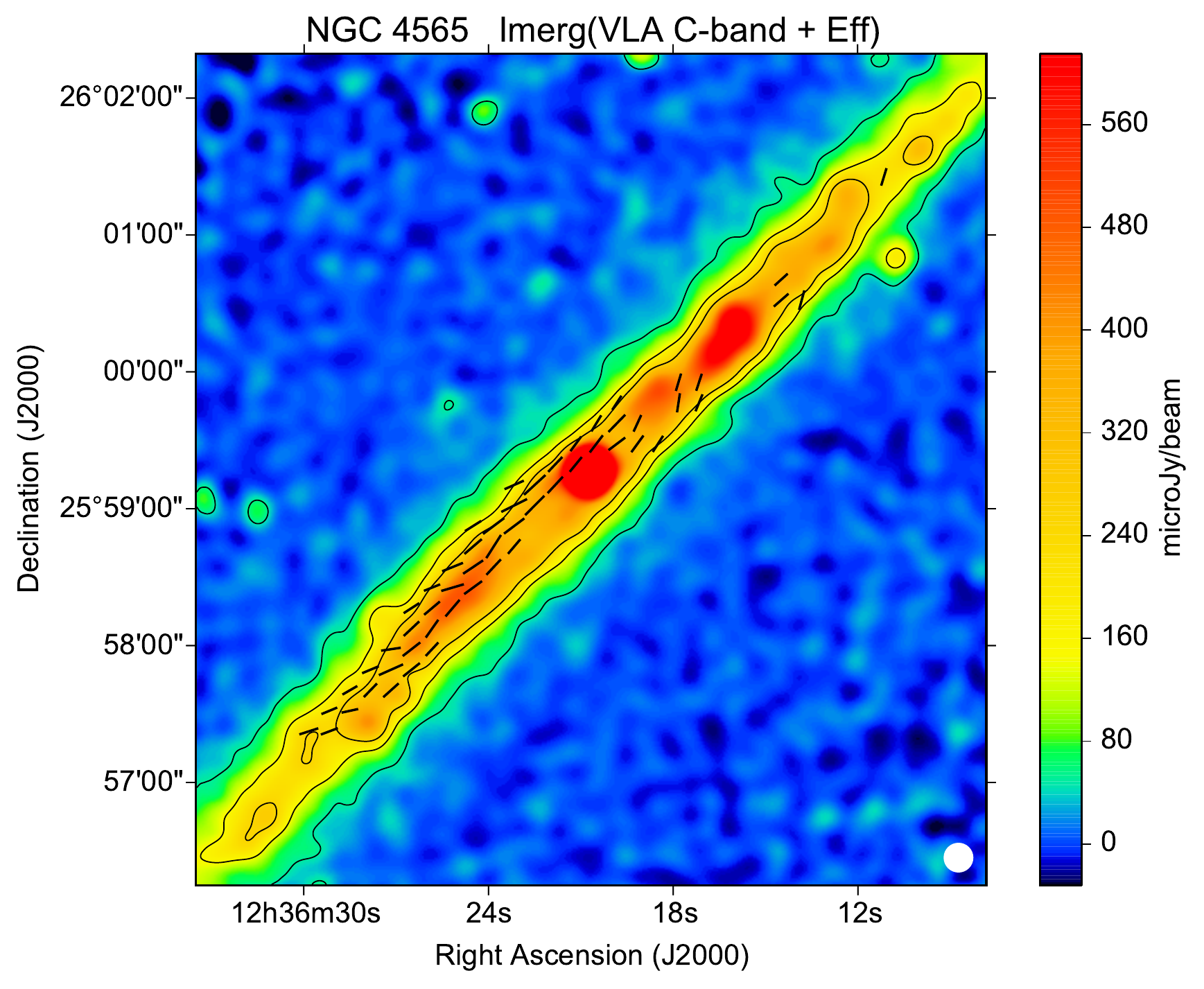}
\includegraphics[width=9.0 cm]{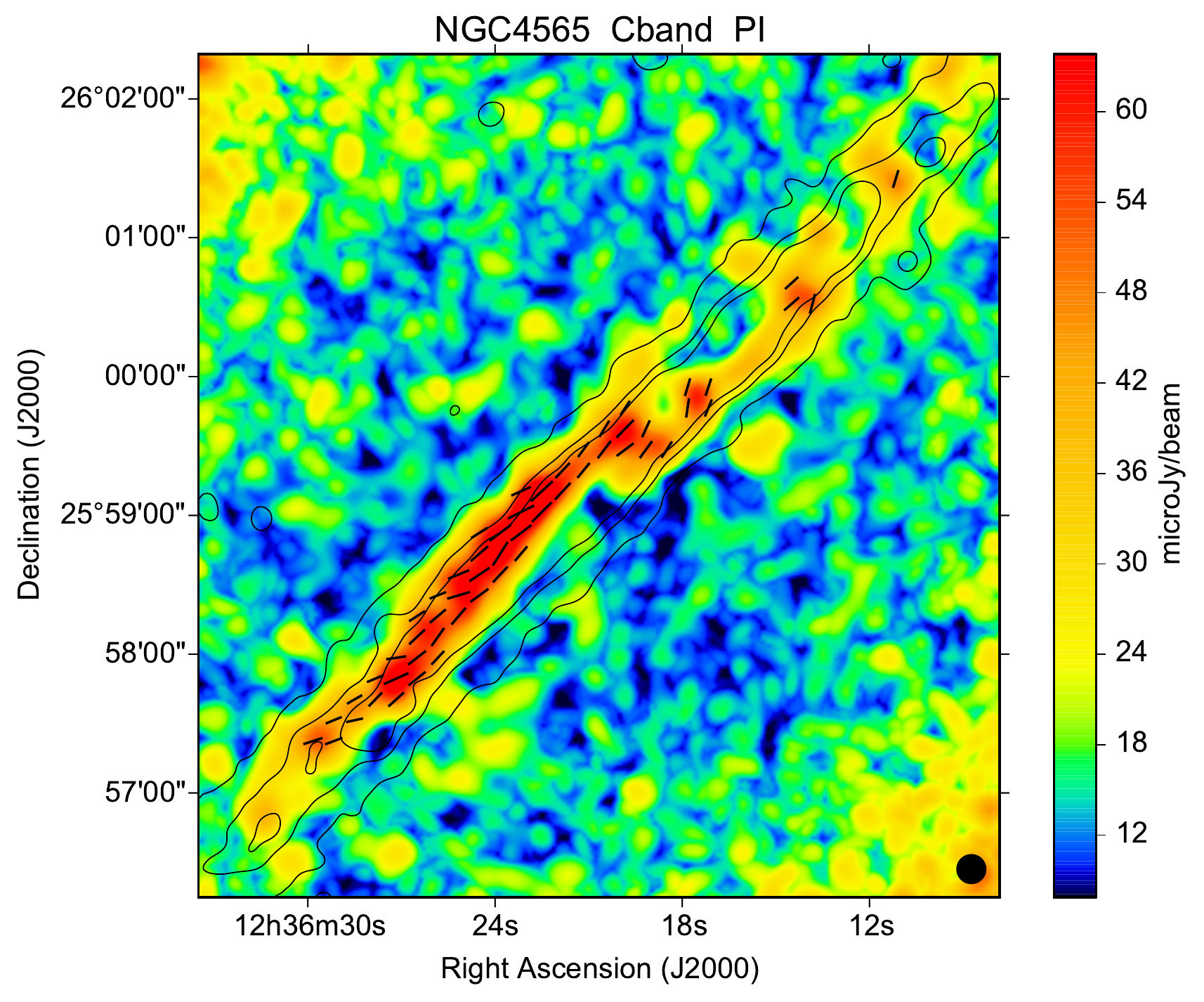}
\includegraphics[width=9.0 cm]{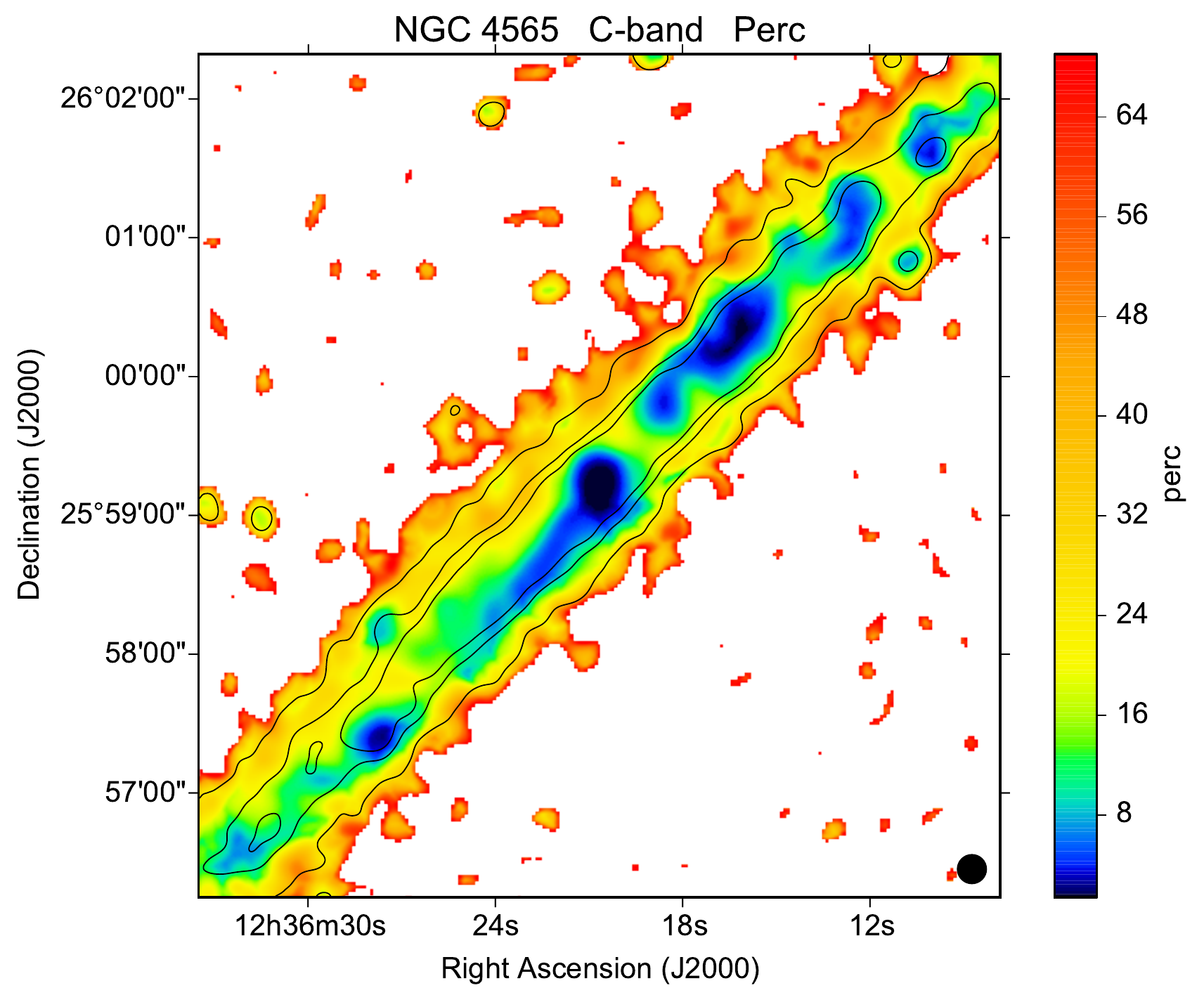}
\includegraphics[width=9.2 cm]{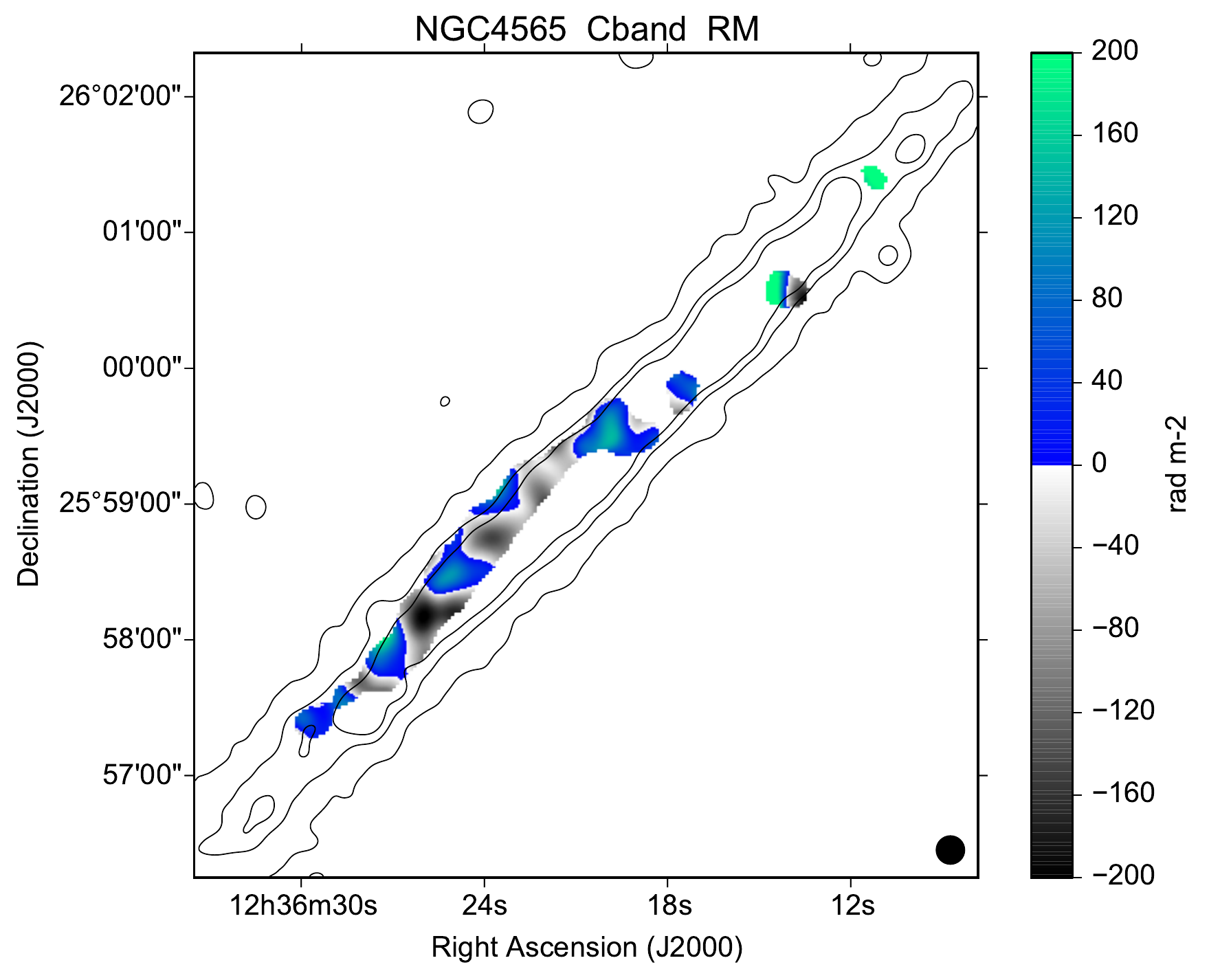}
\includegraphics[width=9.0 cm]{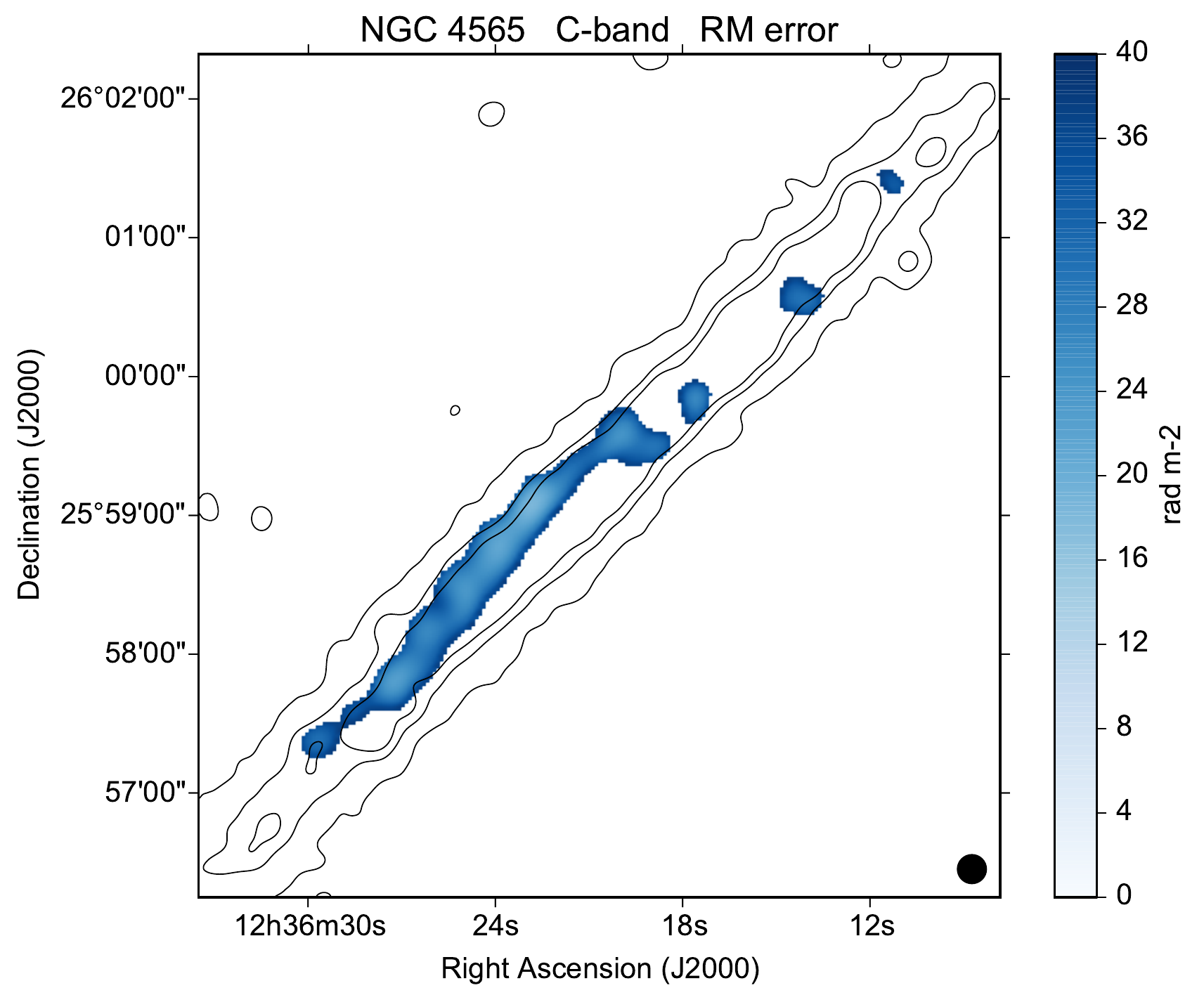}
\includegraphics[width=9.2 cm]{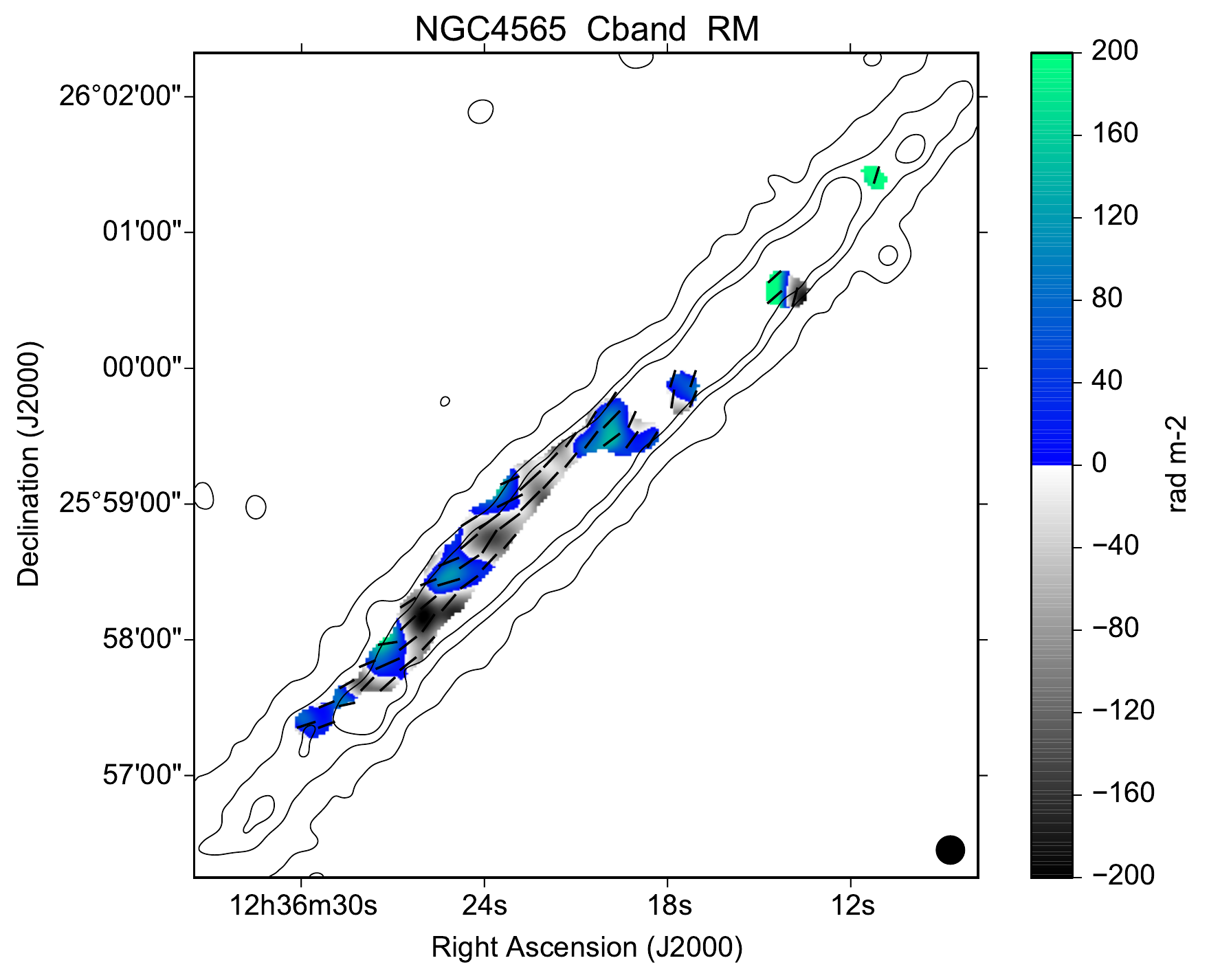}
\caption{Polarization results for NGC~4565 at C-band and $12 \arcsec$ HPBW, corresponding to $690\,\rm{pc}$. The contour levels (TP) are 50, 150, and 250 $\mu$Jy/beam.
}
\label{n4565all}
\end{figure*}

\begin{figure*}[p]
\centering
\includegraphics[width=9.0 cm]{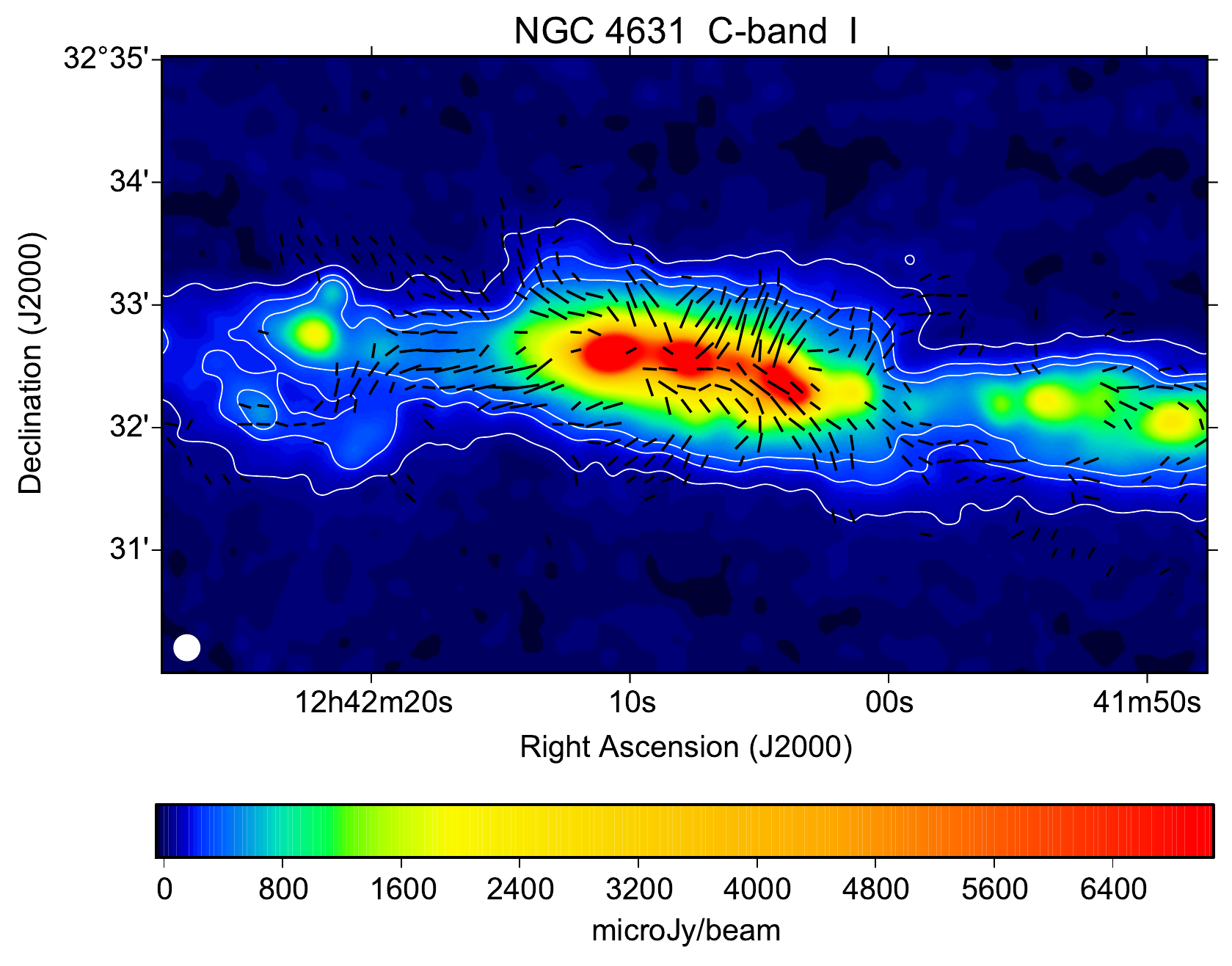}
\includegraphics[width=9.1 cm]{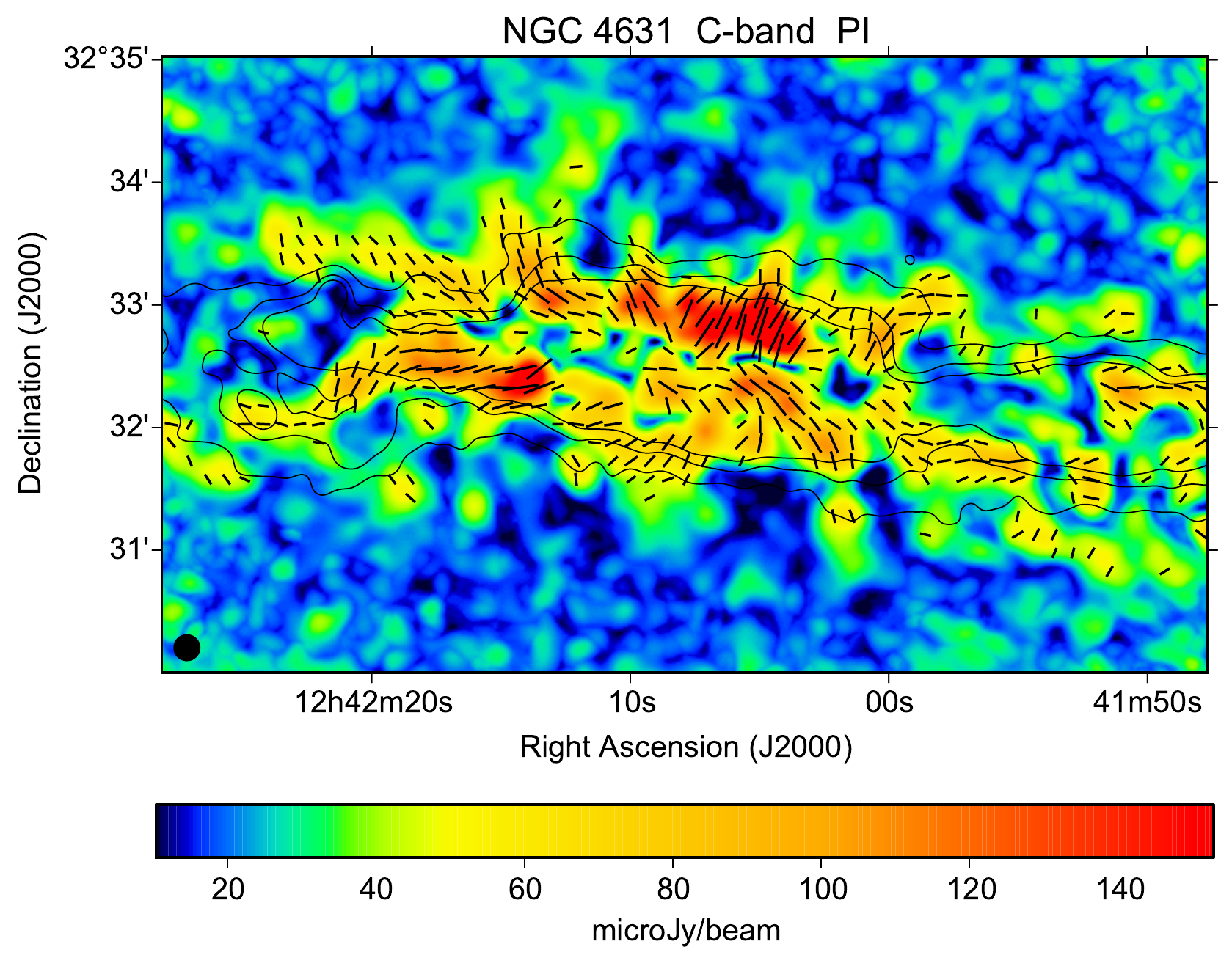}
\includegraphics[width=9.0 cm]{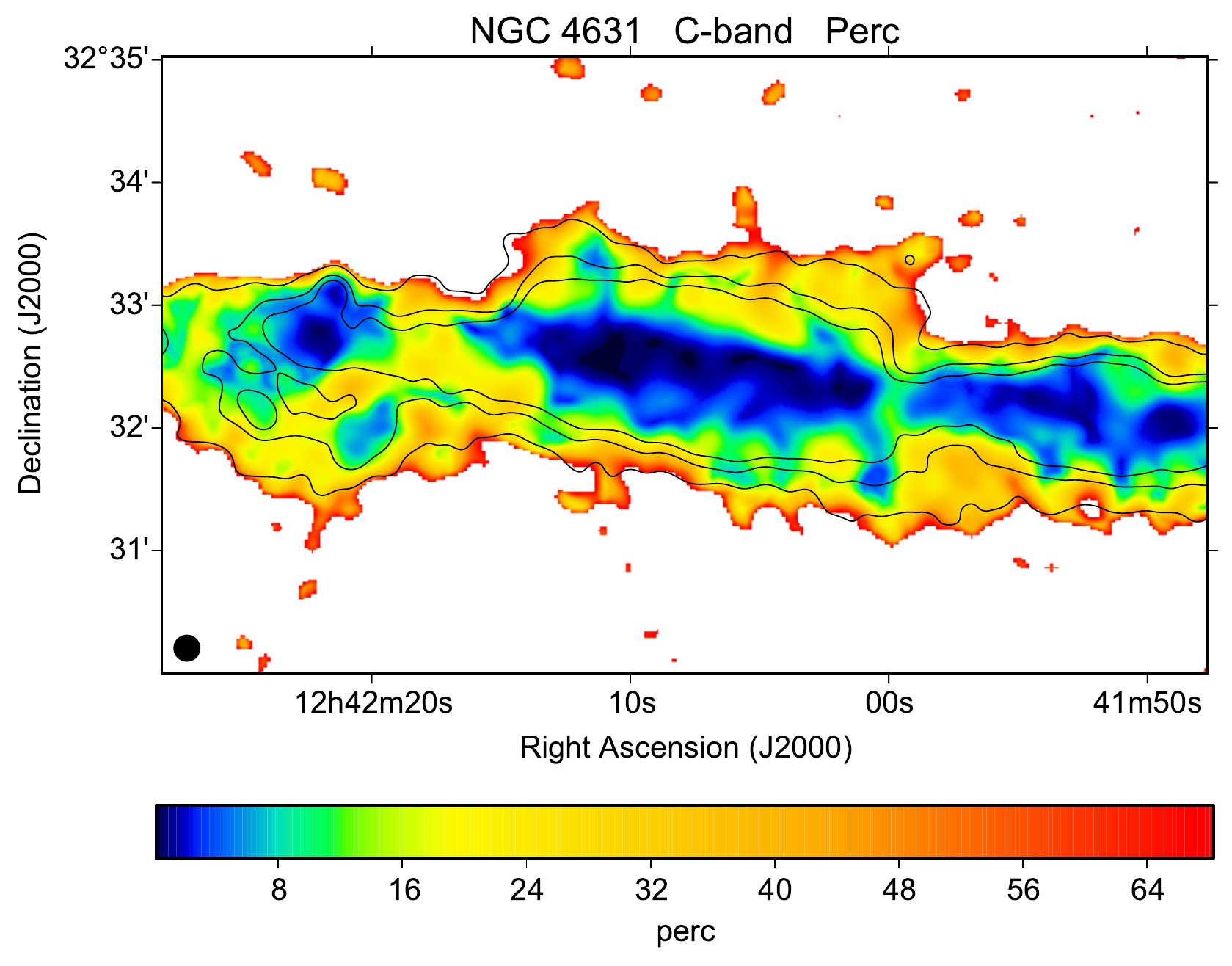}
\includegraphics[width=9.2 cm]{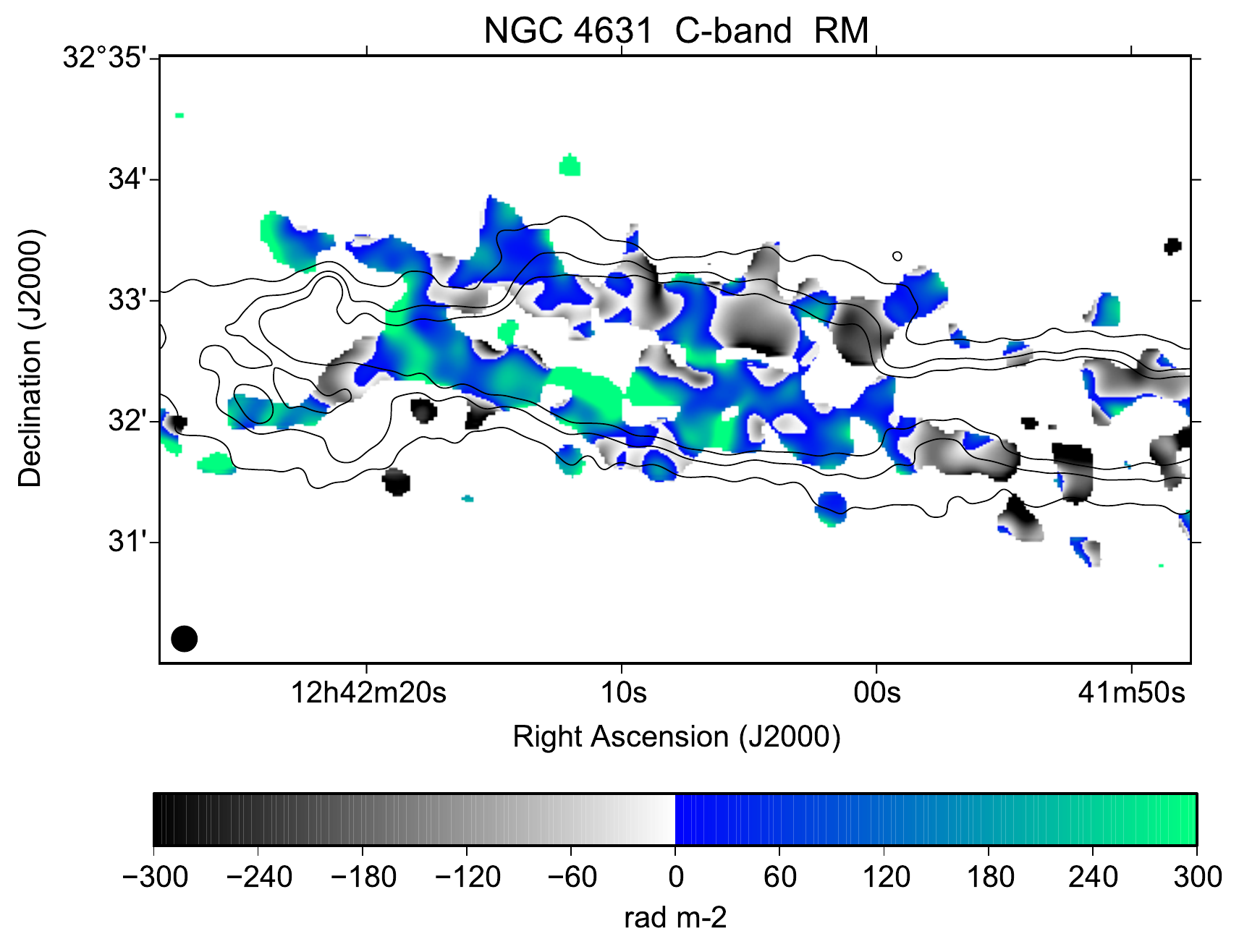}
\includegraphics[width=9.0 cm]{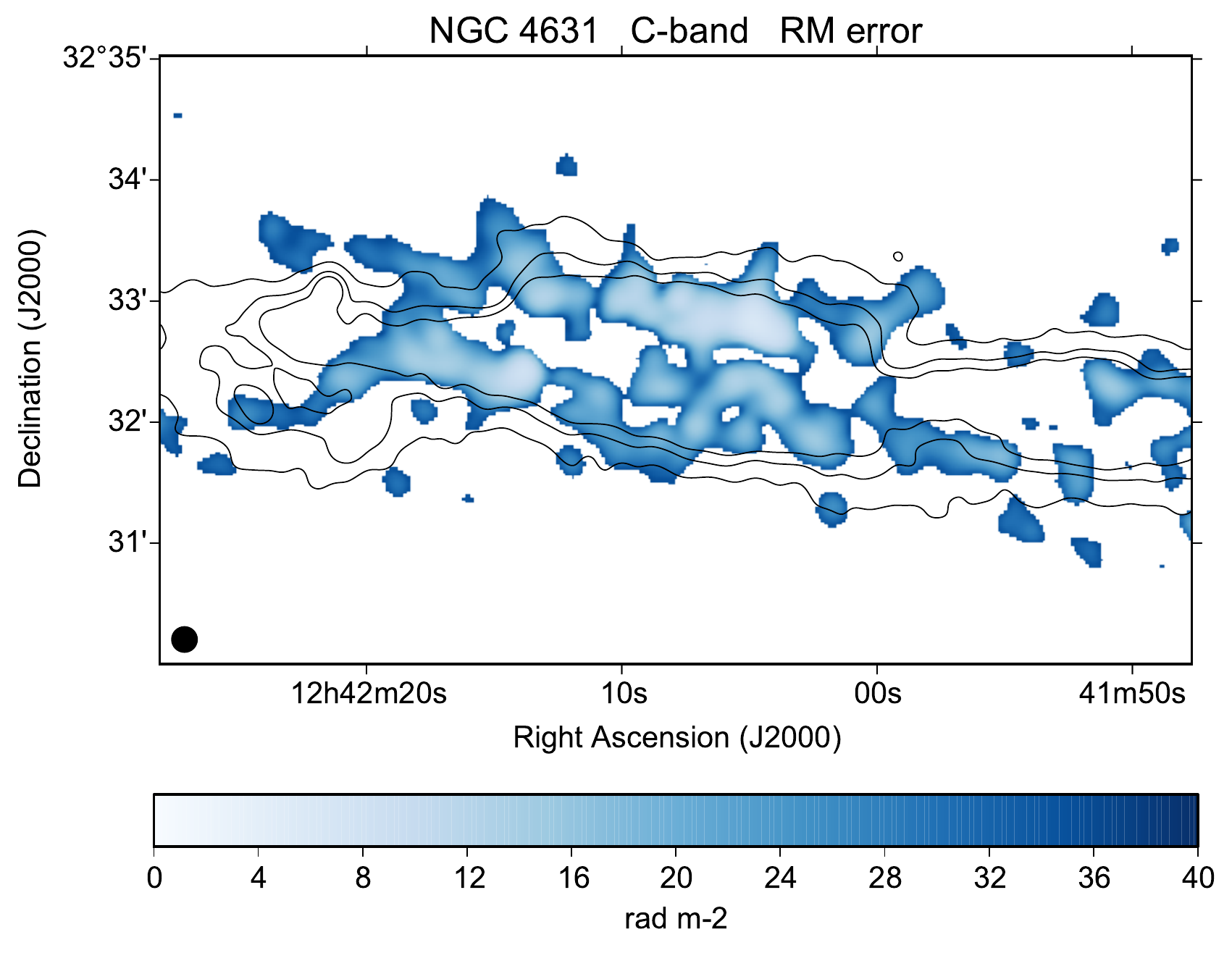}
\includegraphics[width=9.1 cm]{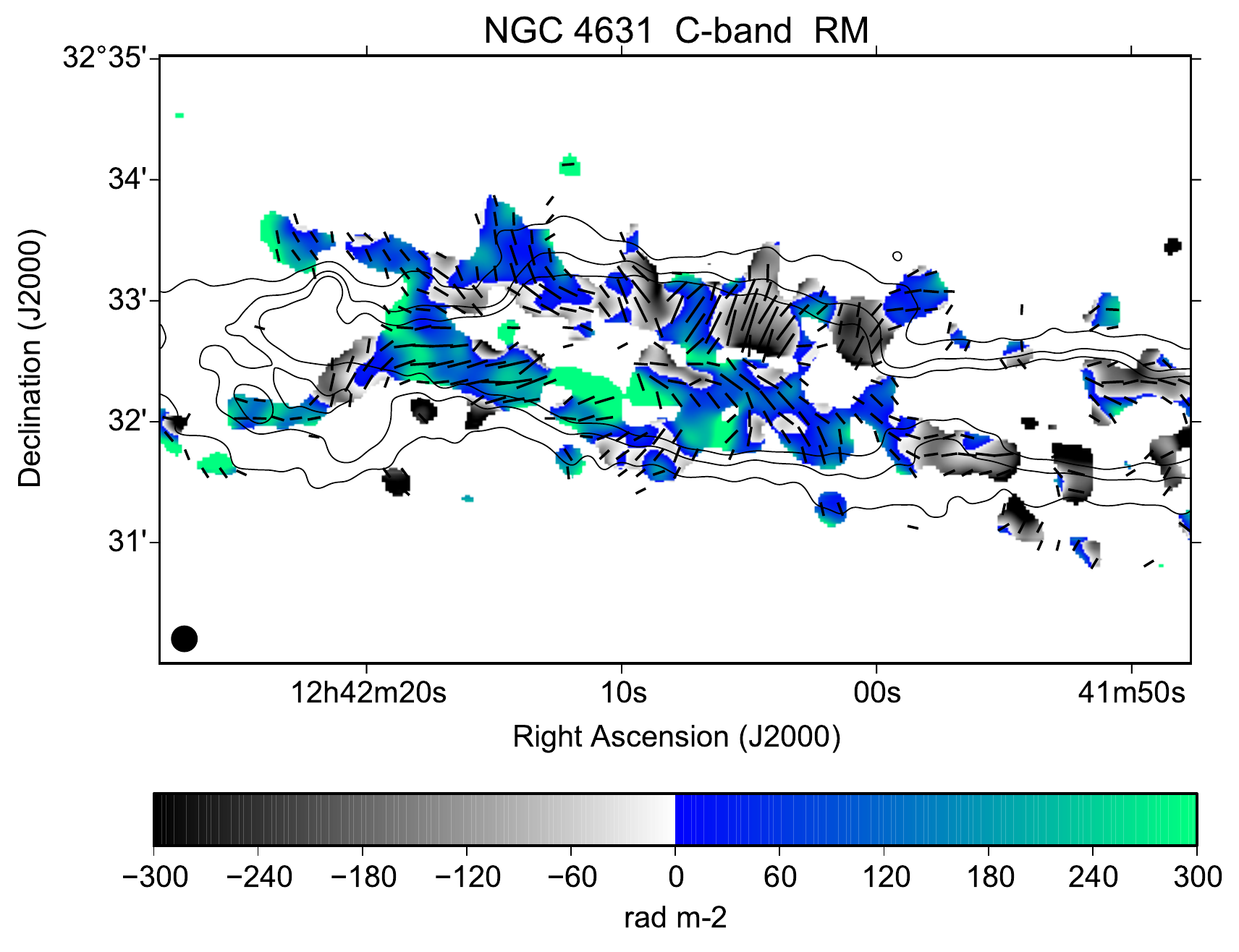}
\caption{Polarization results for NGC~4631 at C-band and $12 \arcsec$ HPBW, corresponding to $430\,\rm{pc}$. The contour levels (TP) are 75, 225, and 375 $\mu$Jy/beam.
}
\label{n4631_12all}
\end{figure*}

\begin{figure*}[p]
\centering
\includegraphics[width=9.0 cm]{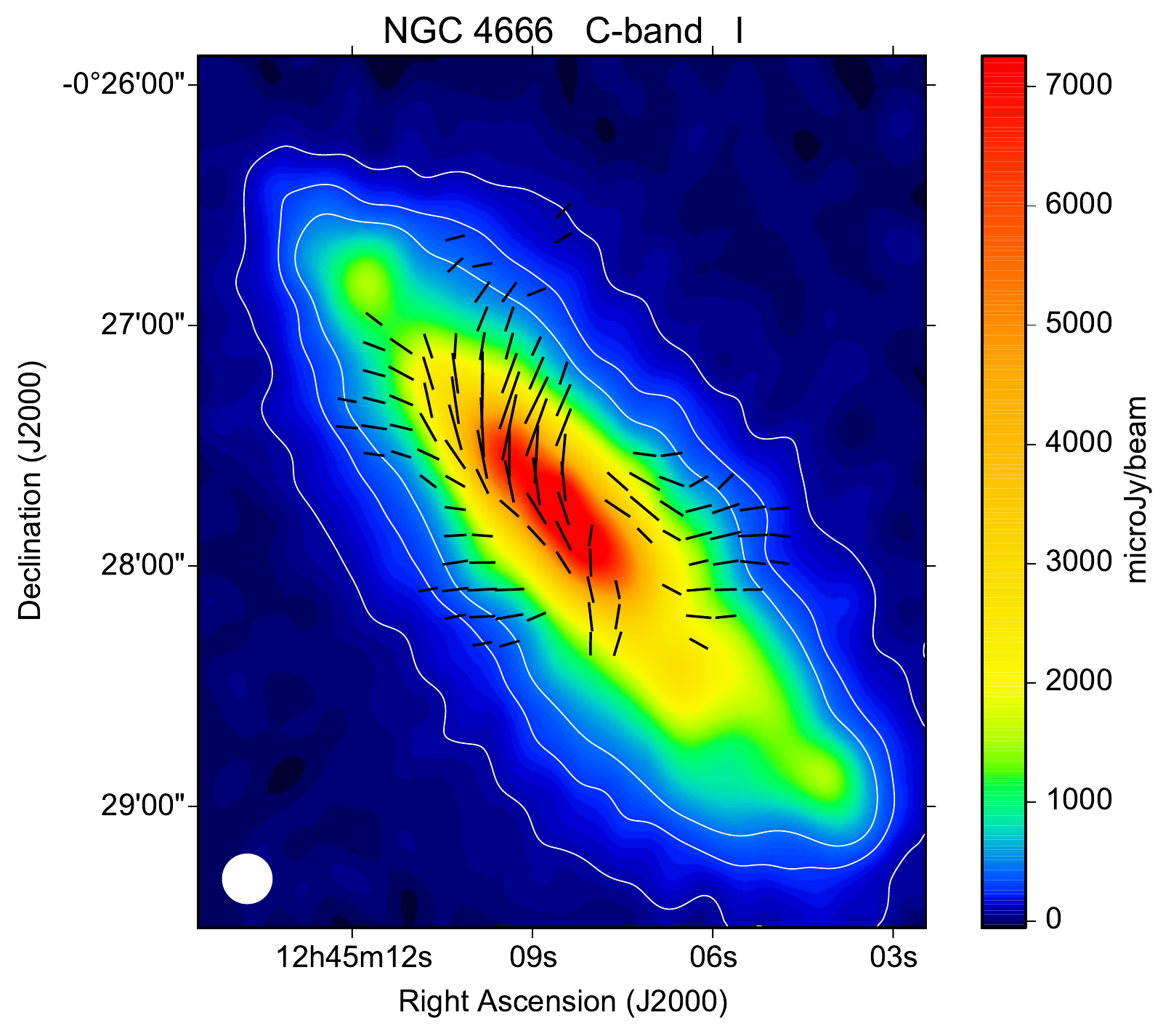}
\includegraphics[width=8.9 cm]{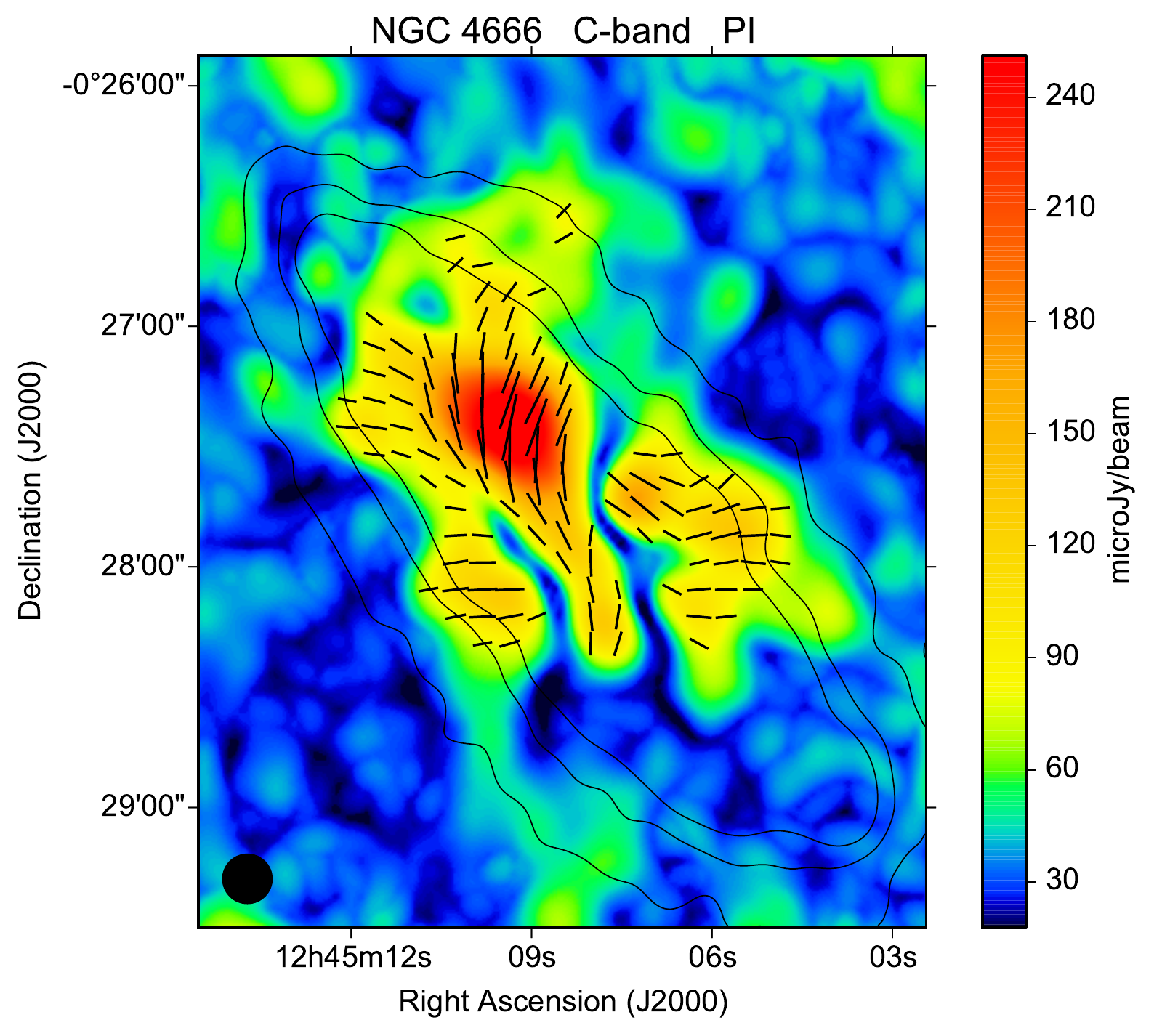}
\includegraphics[width=9.0 cm]{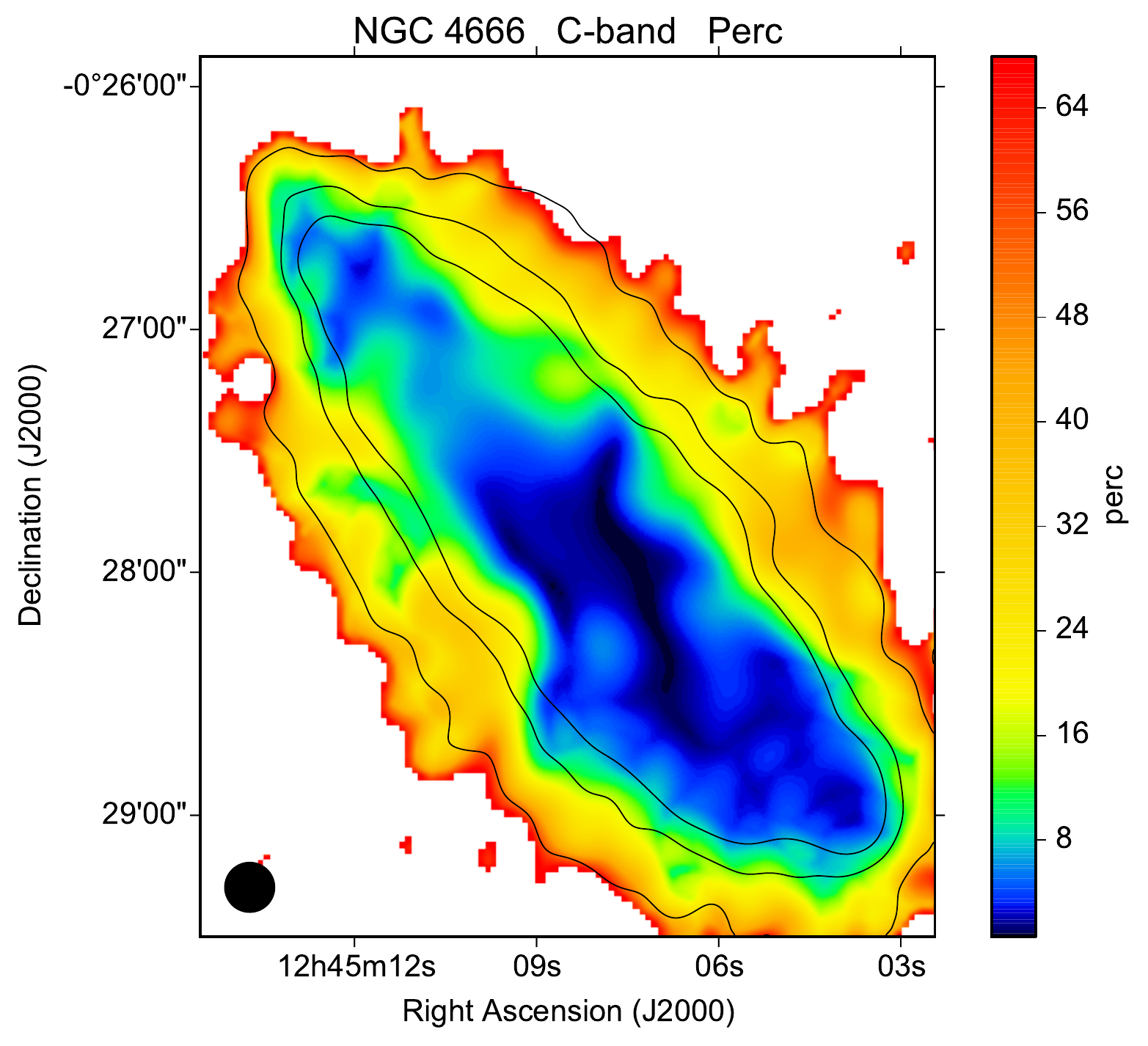}
\includegraphics[width=9.3 cm]{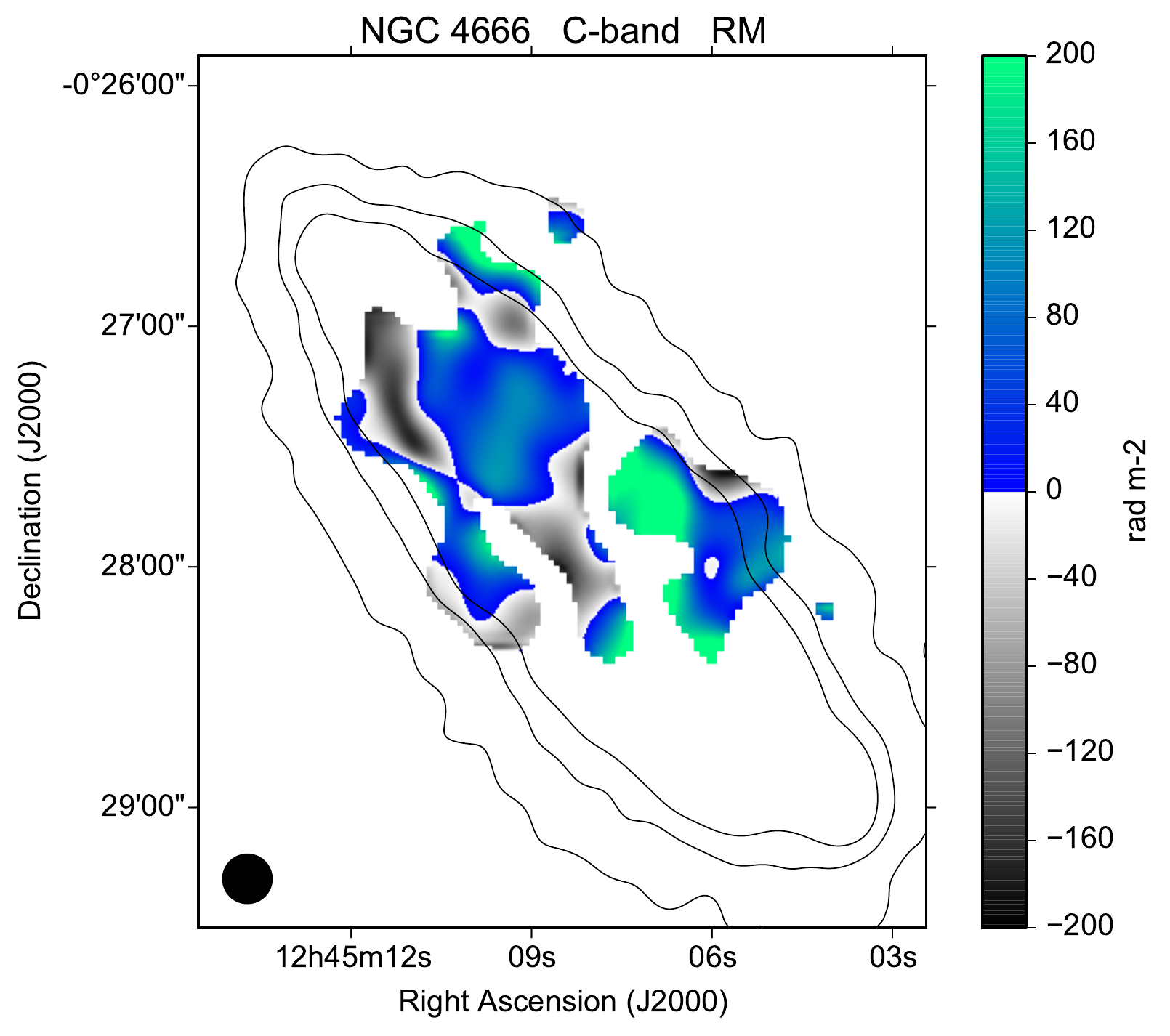}
\includegraphics[width=9.0 cm]{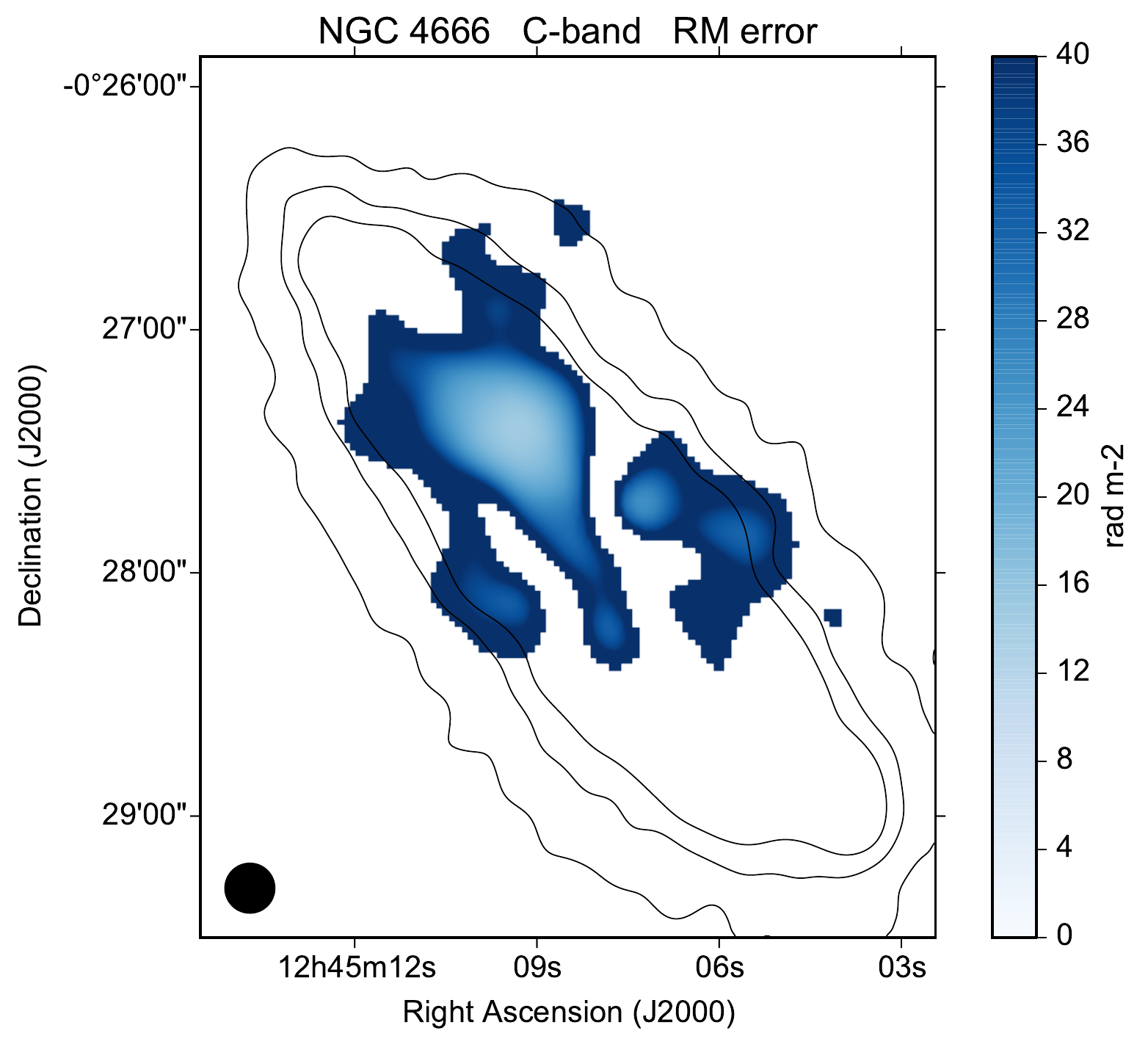}
\includegraphics[width=9.3 cm]{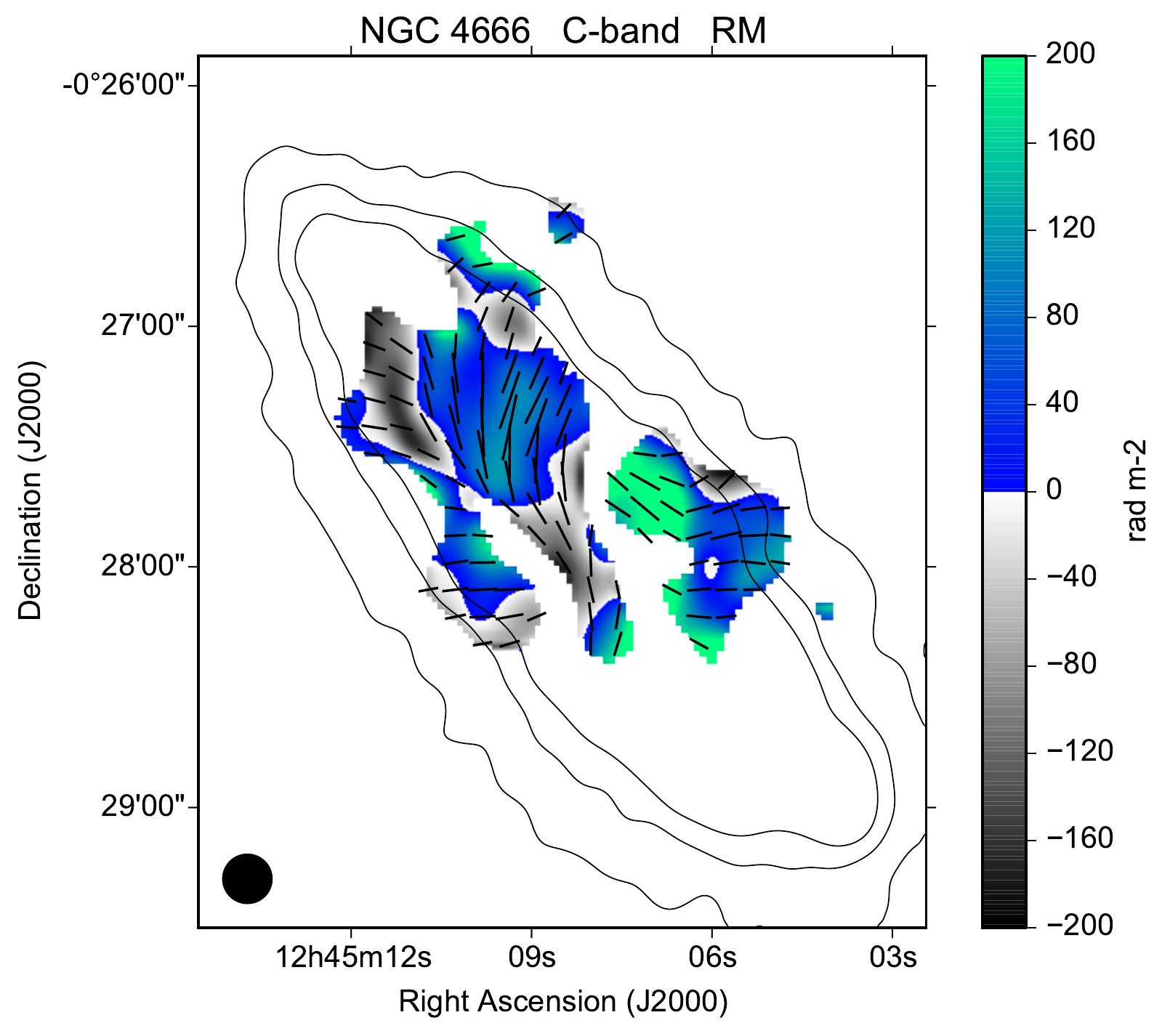}
\caption{Polarization results for NGC~4666 at C-band and $12 \arcsec$ HPBW, corresponding to $1600\,\rm{pc}$. The contour levels (TP) are 90, 270, and 450 $\mu$Jy/beam.
}
\label{n4666all4sigma}
\end{figure*}

\begin{figure*}[p]
\centering
\includegraphics[width=9.0 cm]{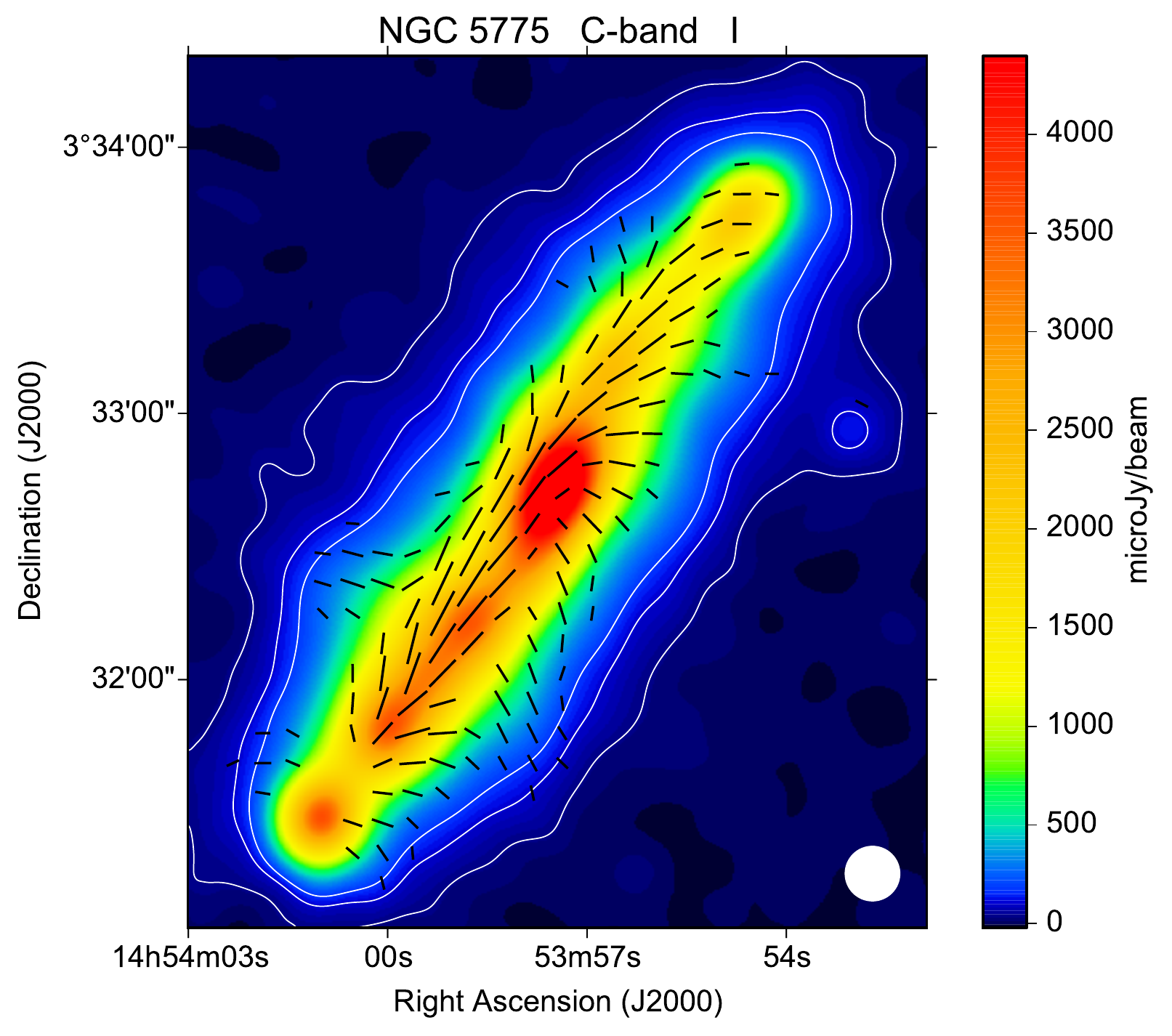}
\includegraphics[width=8.9 cm]{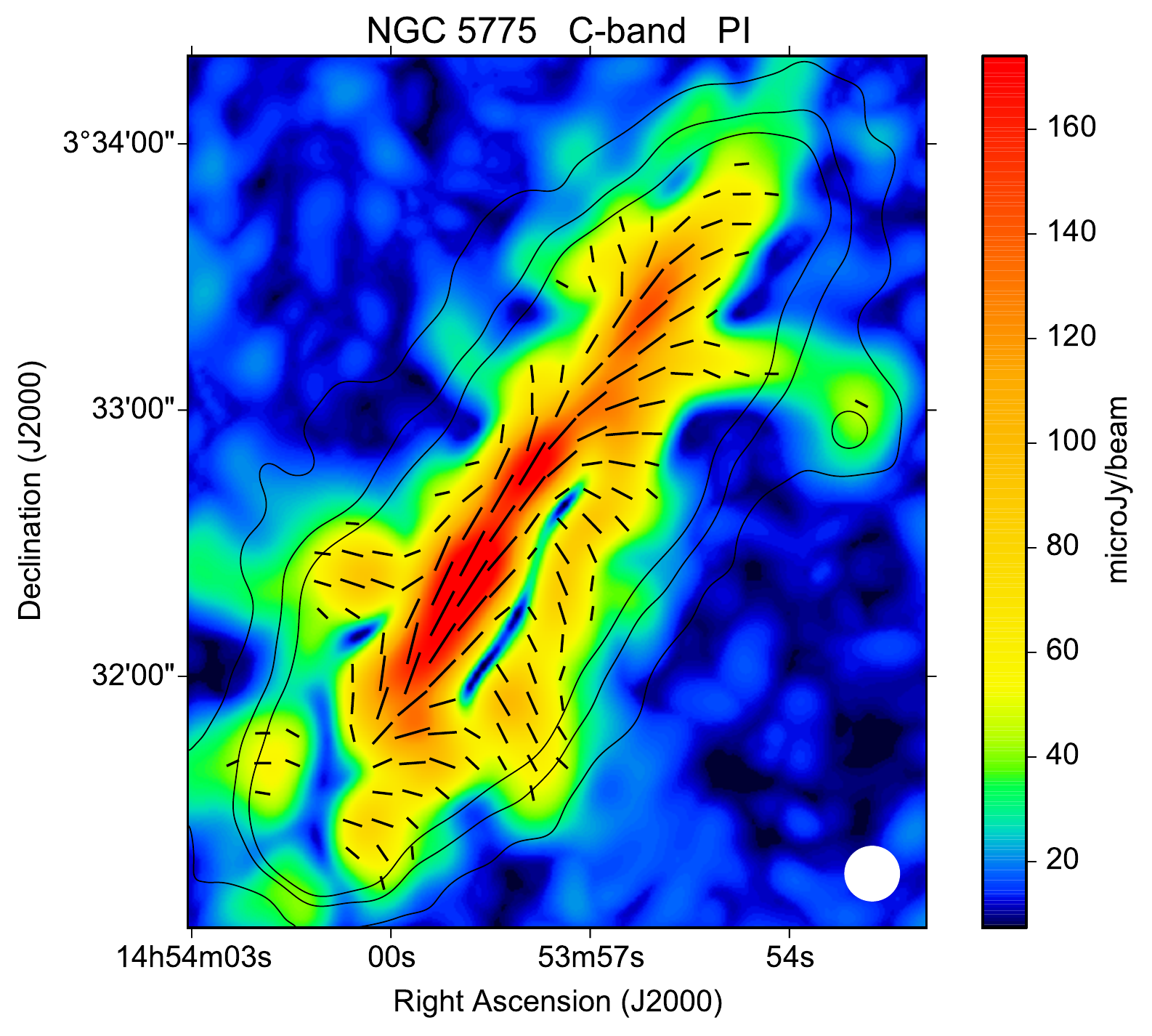}
\includegraphics[width=9.0 cm]{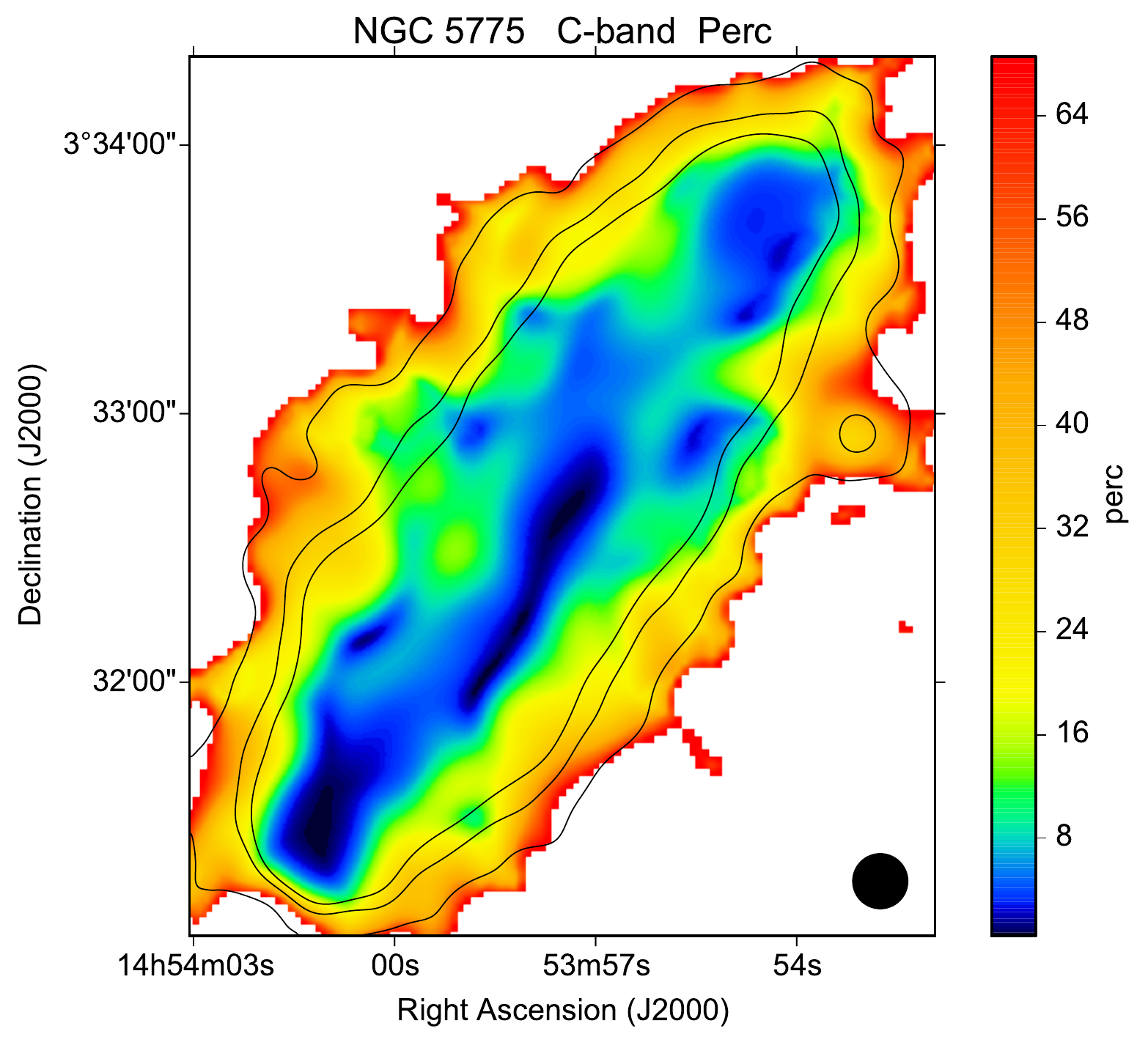}
\includegraphics[width=9.2 cm]{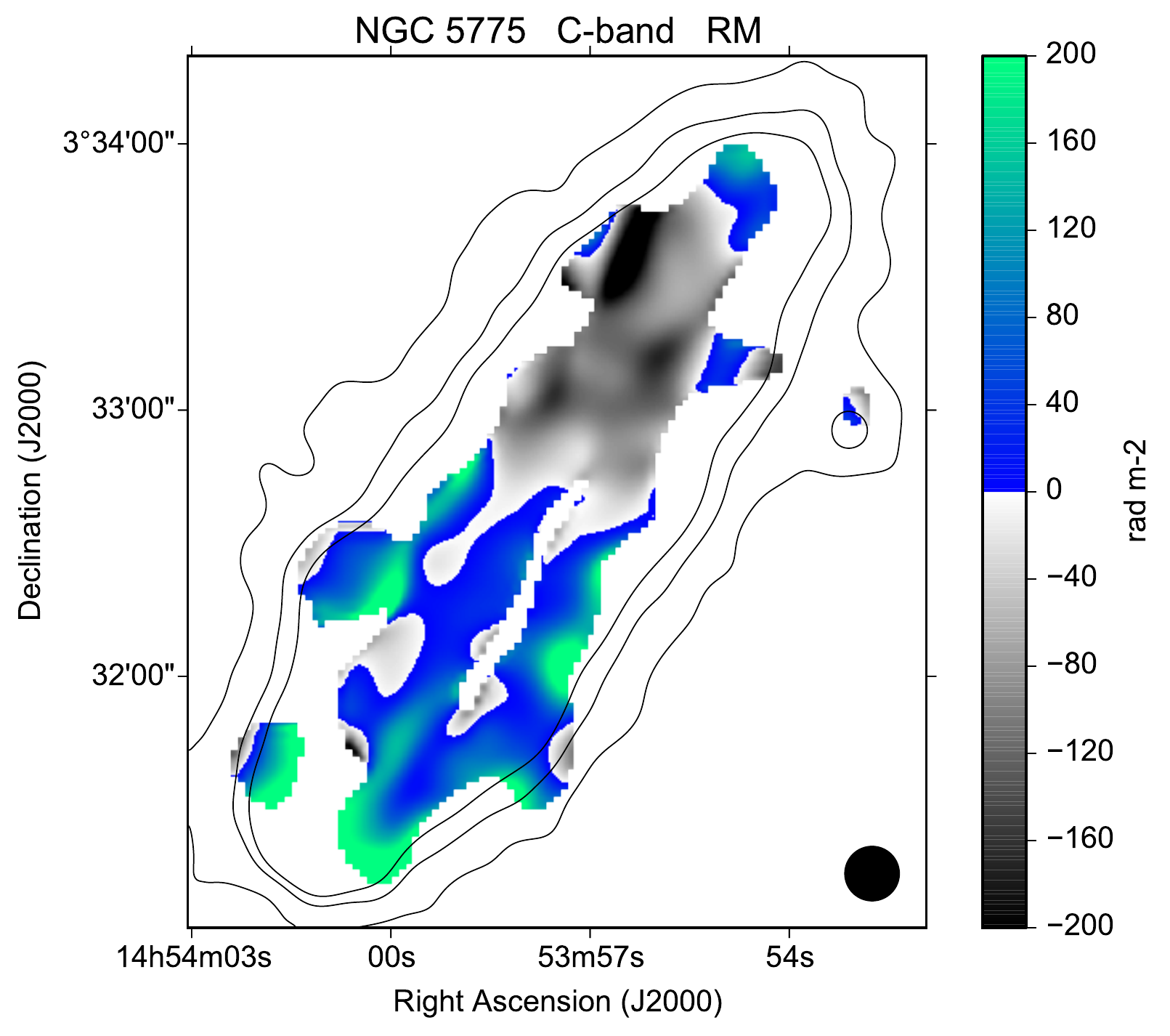}
\includegraphics[width=9.0 cm]{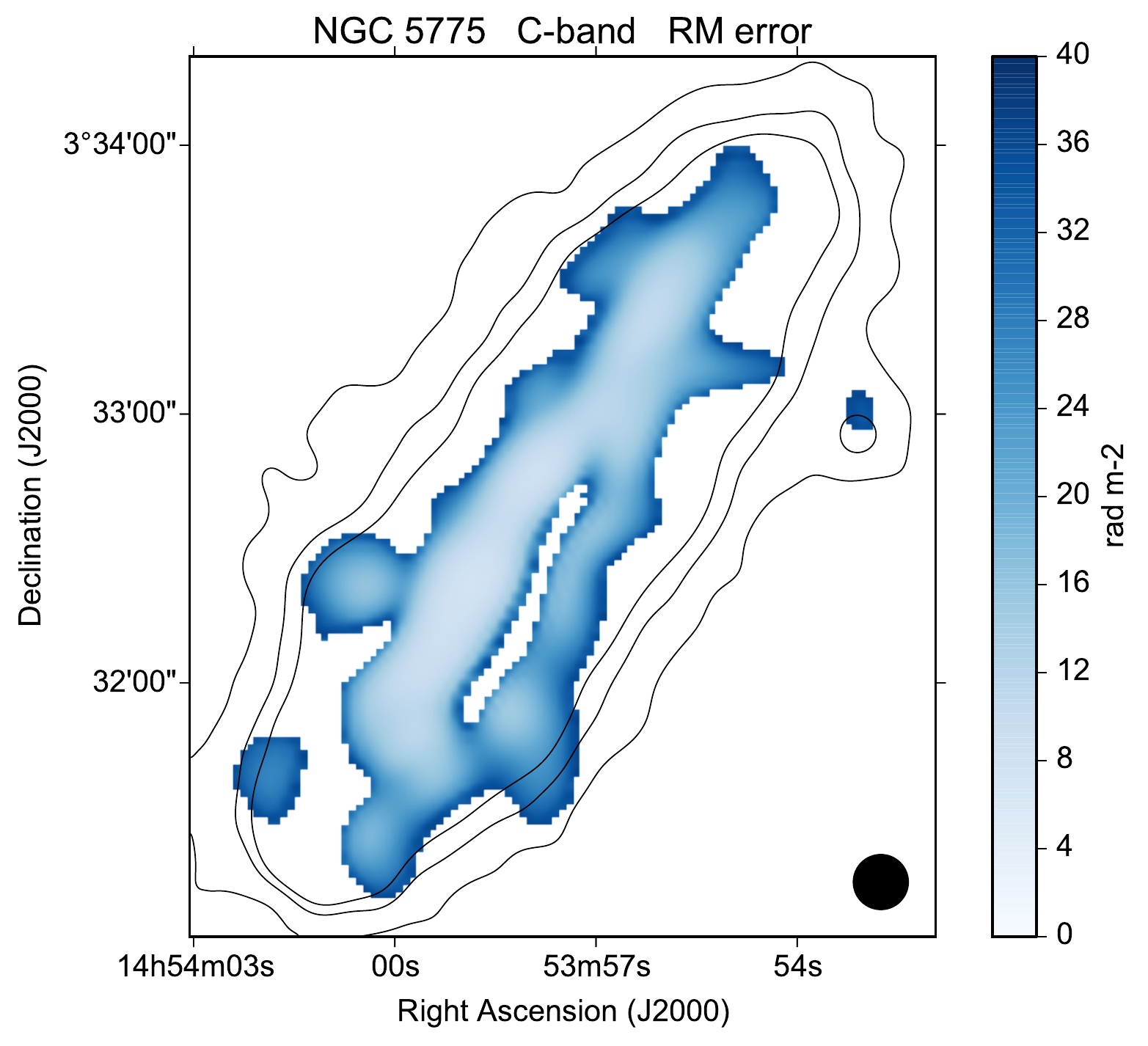}
\includegraphics[width=9.2 cm]{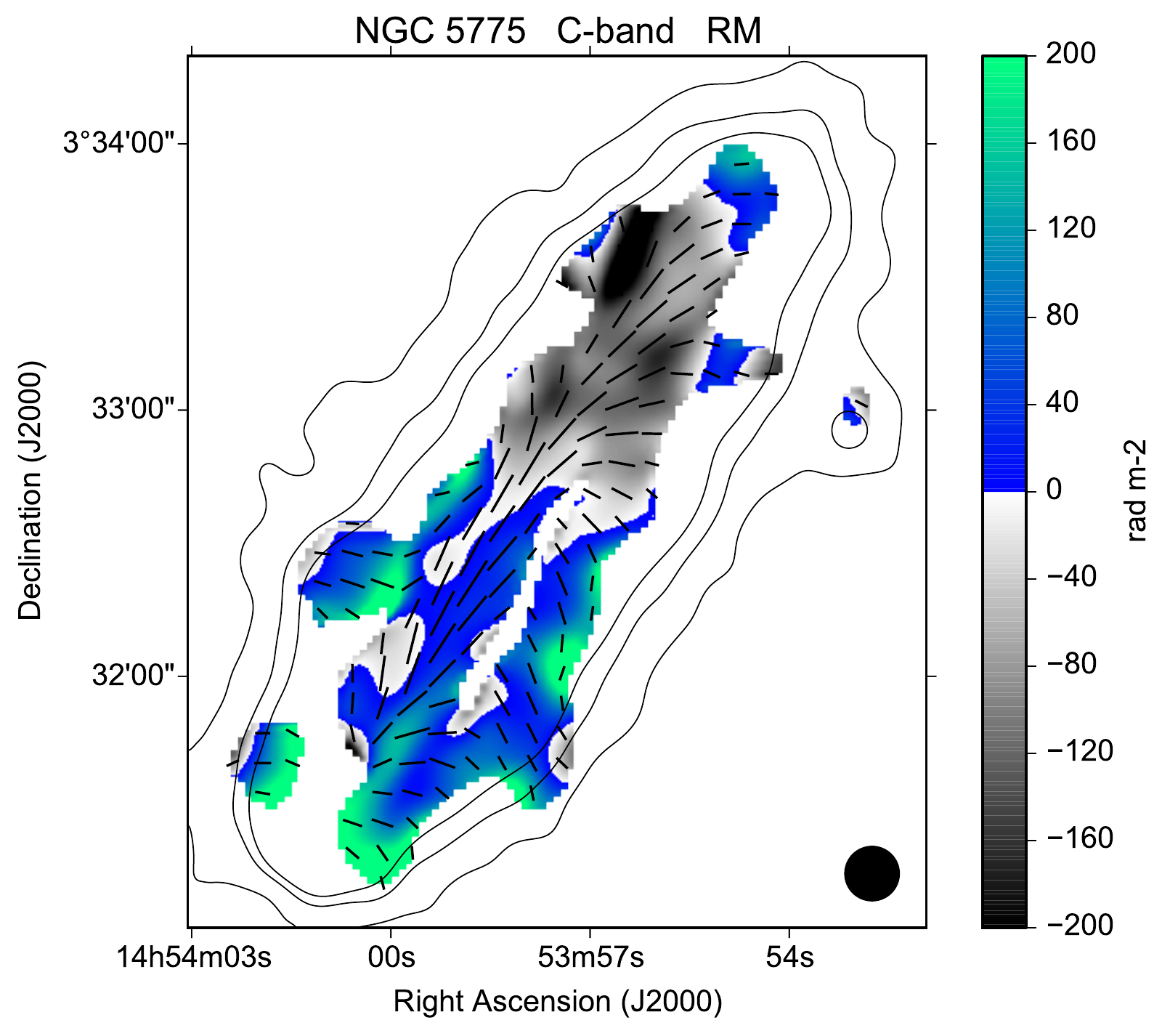}
\caption{Polarization results for NGC~5775 at C-band and $12 \arcsec$ HPBW, corresponding to $1680\,\rm{pc}$. The contour levels (TP) are 35, 105, and 175 $\mu$Jy/beam.
}
\label{n5775all}
\end{figure*}

\begin{figure*}[p]
\centering
\includegraphics[width=7.3 cm]{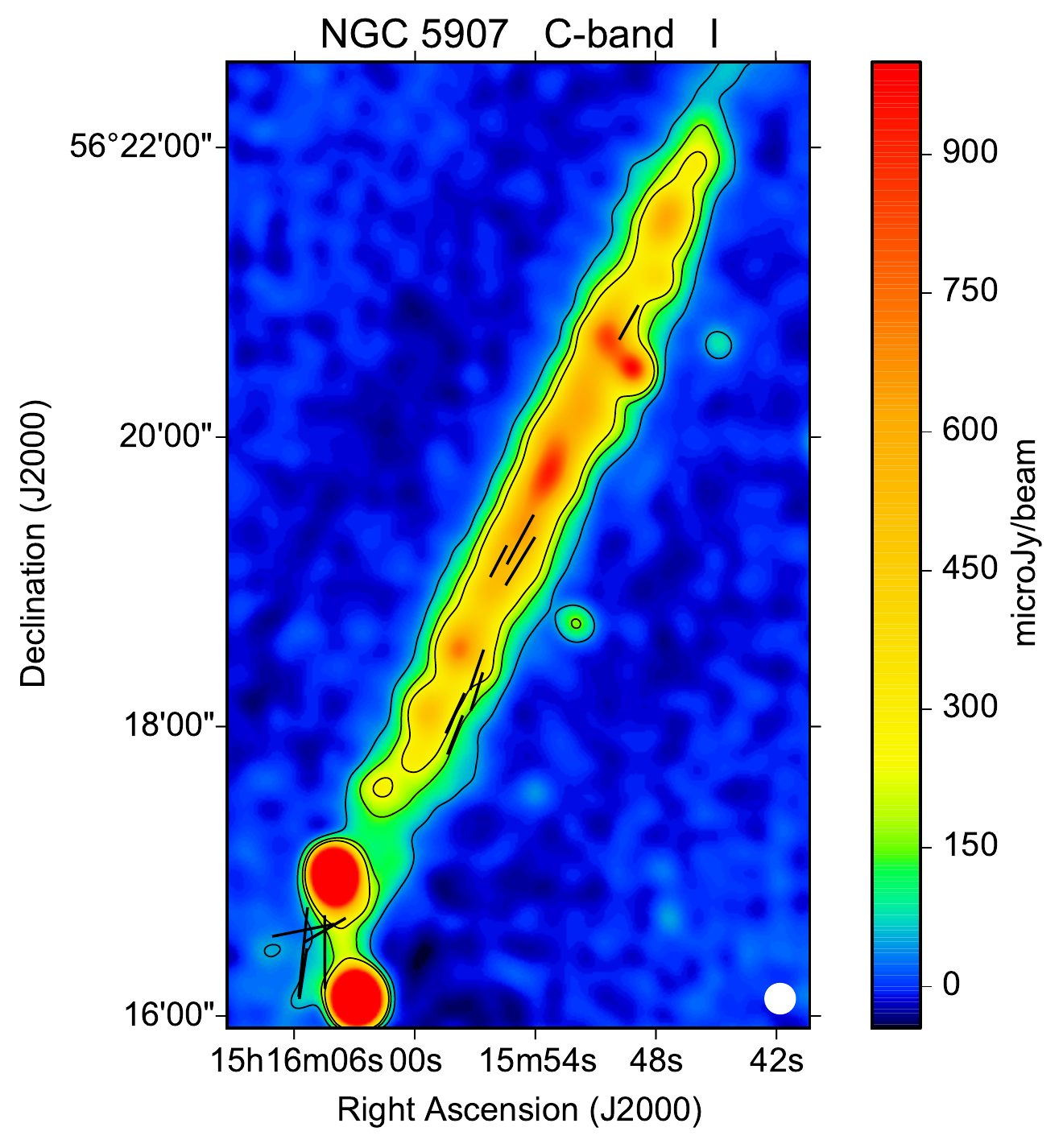}
\includegraphics[width=7.2 cm]{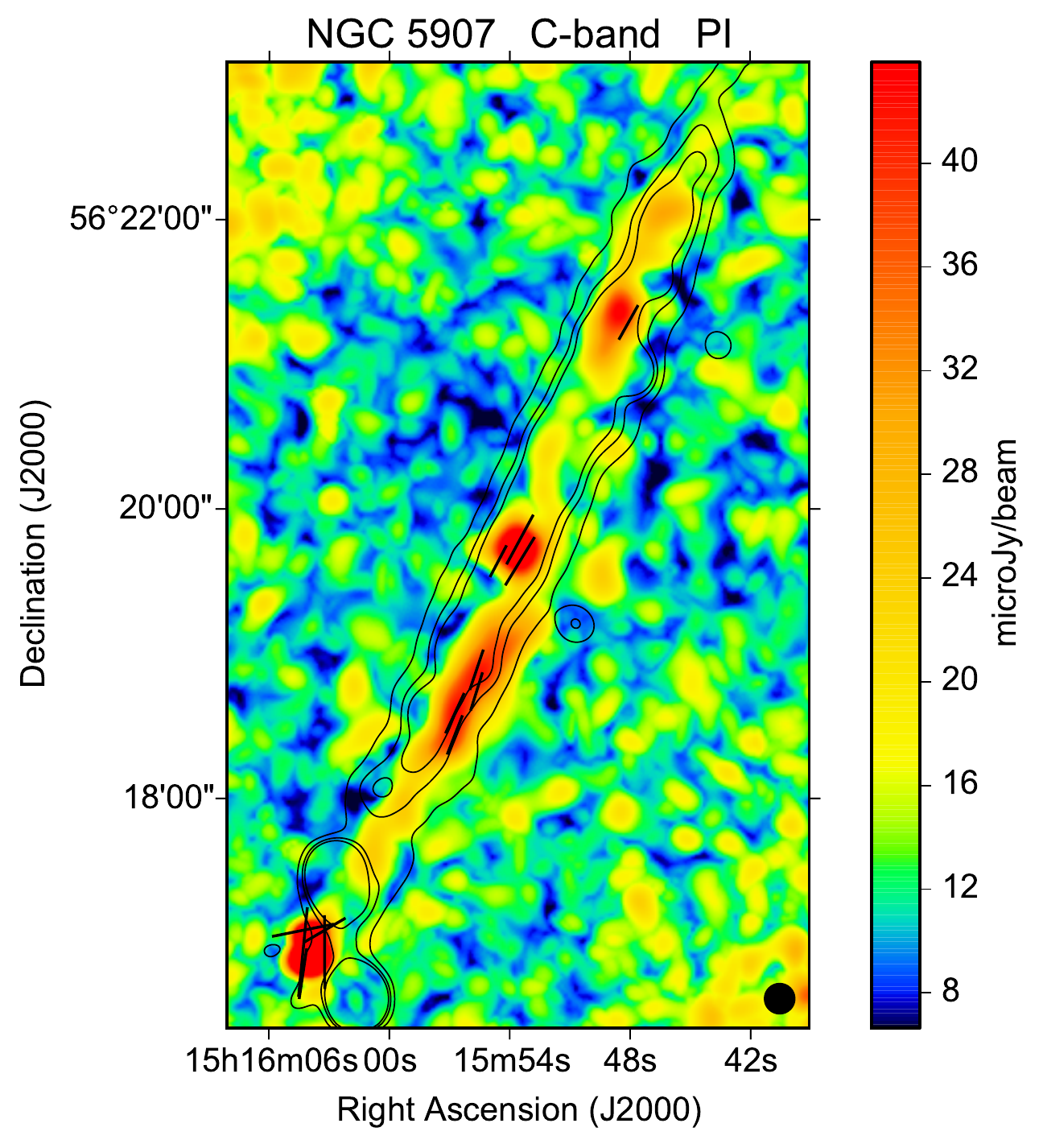}
\includegraphics[width=7.3 cm]{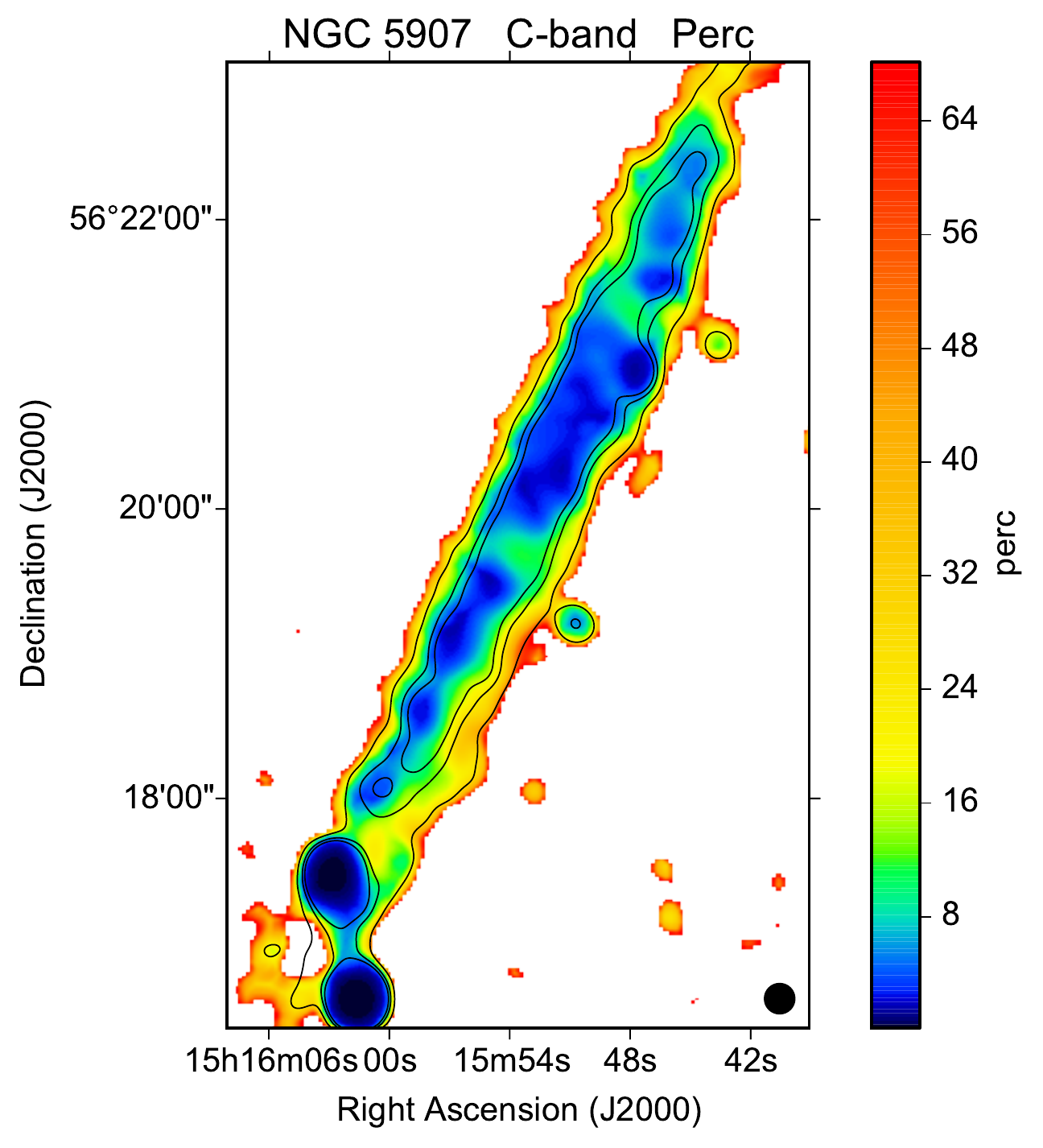}
\includegraphics[width=7.6 cm]{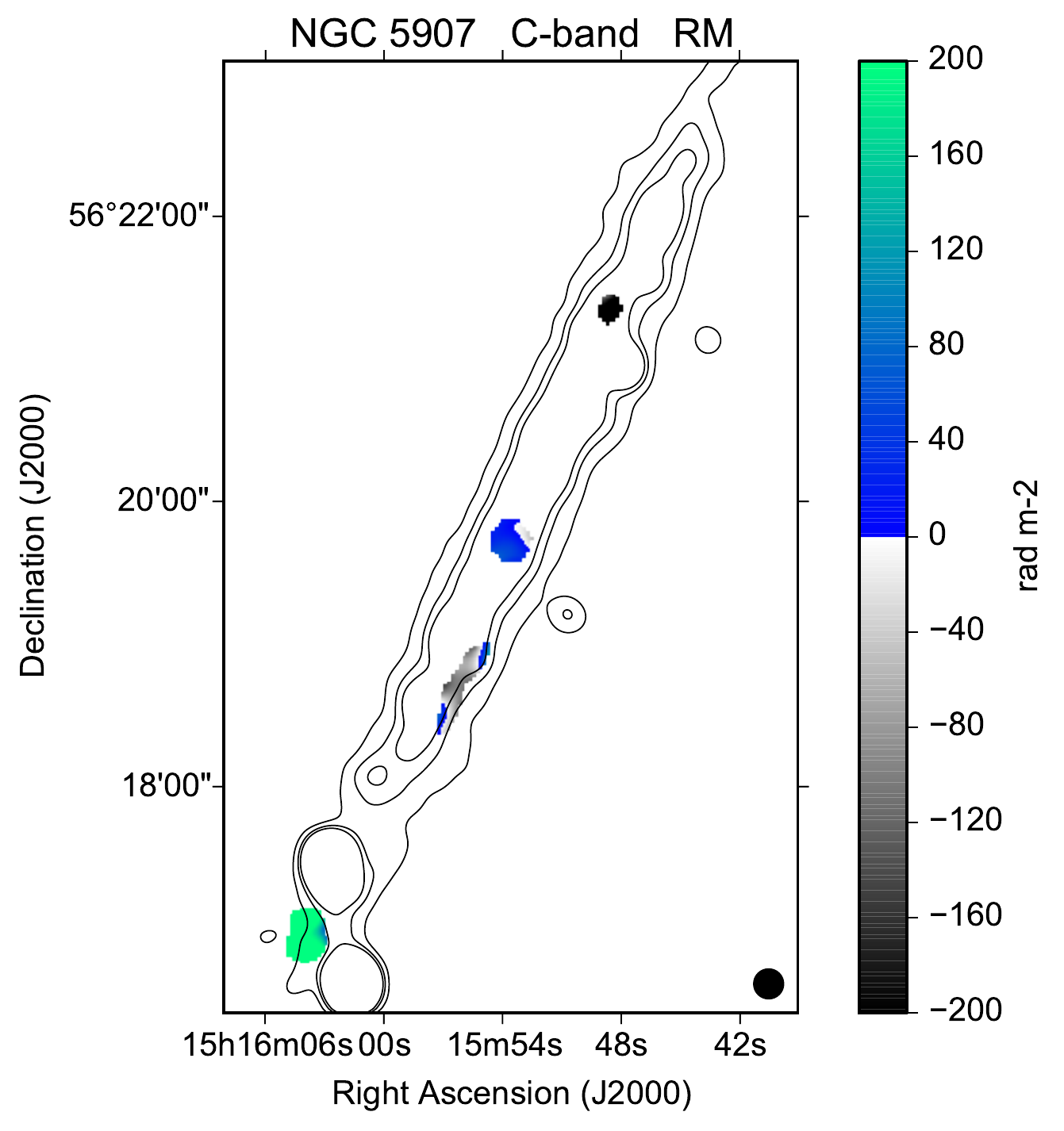}
\includegraphics[width=7.2 cm]{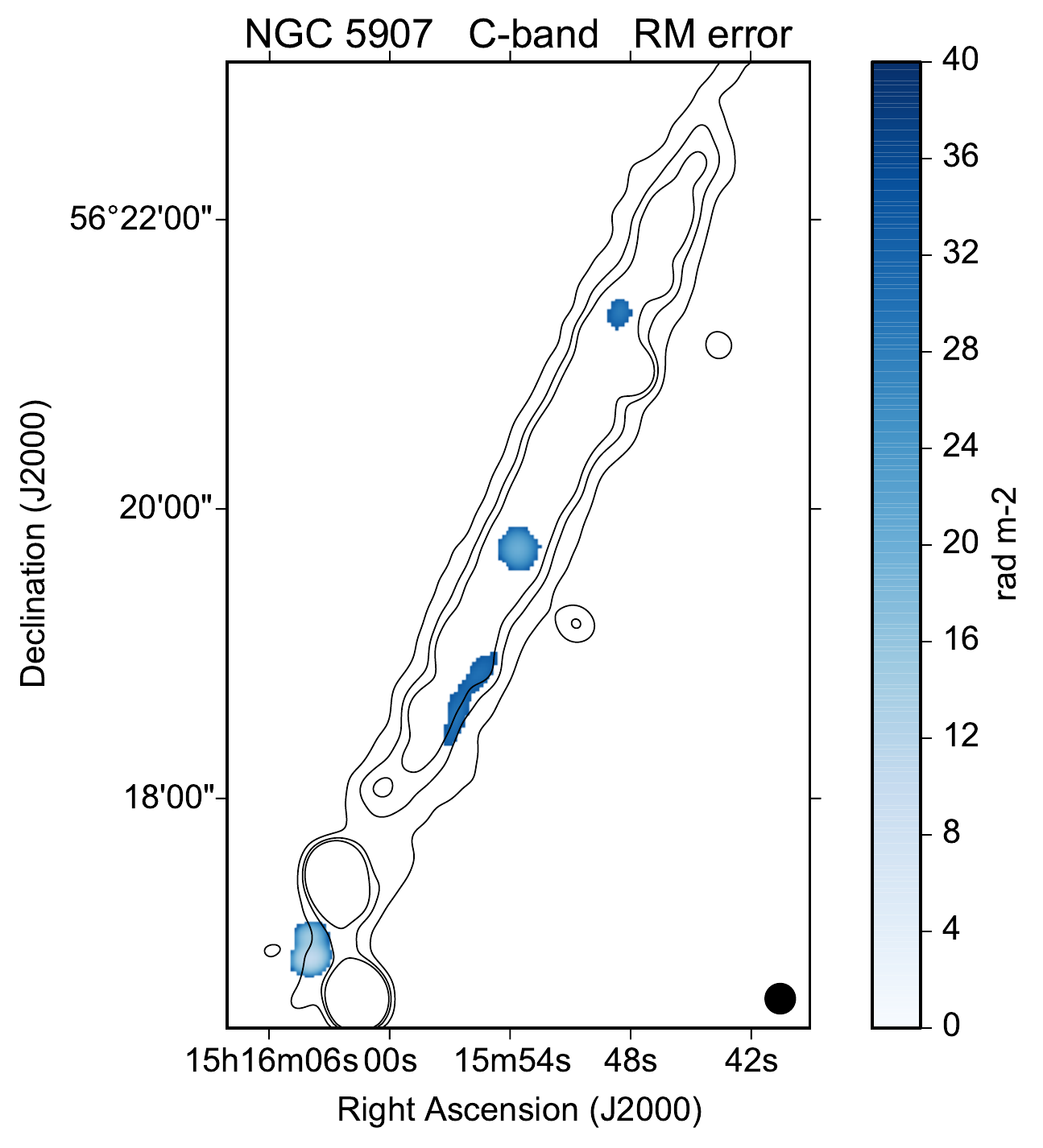}
\includegraphics[width=7.6 cm]{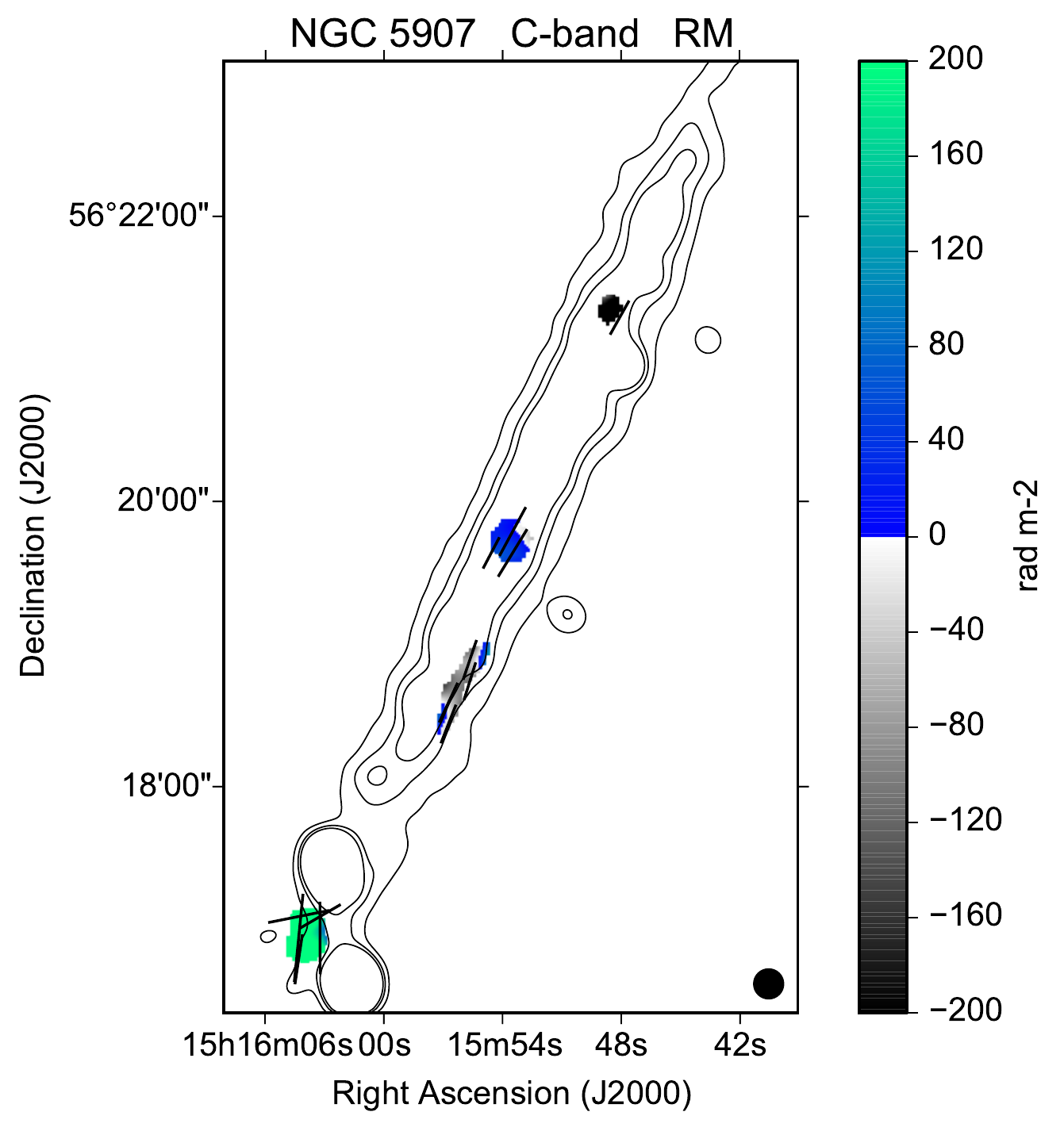}
\caption{Polarization results for NGC~5907 at C-band and $12 \arcsec$ HPBW, corresponding to $980\,\rm{pc}$. The contour levels (TP) are 50, 150, and 250 $\mu$Jy/beam.
The image of the TP map is cut at 1000~$\mu$Jy/beam in order to present the disk emission well. This just affects the two background sources in the south.
}
\label{n5907all}
\end{figure*}

\end{appendix}

\begin{appendix}
\section{HI observation of NGC~3735}
\label{N3735HI}

NGC~3735 was observed by the VLA under project code AC168 on 1987 March 8. The array was in D configuration, and the correlator was set up to deliver 31 channels over a 3~MHz bandwidth centred at 1407.75~MHz, in a single parallel-hand polarization (RR). The channel width corresponds to a velocity resolution of about $21~\mathrm{km\,s^{-1}}$. The total time on the target source was 57 minutes.

We used CASA version 5.4.1-31 to calibrate and image the data. Based on visual inspection of the visibilities, no flagging was required. The fluxscale was set based on the calibrator source 1634+628. Bandpass, delays, and gain phase and amplitude corrections were determined using the calibrator source 1203+645. Self-calibration was not implemented. An {\sc H\,i} image cube was produced using task {\tt tclean}, with Briggs weighting ({\tt robust=0}) and no deconvolution. Continuum subtraction was carried out in the image plane. An image of the velocity at the peak of the line spectrum in each pixel was produced using the {\sc miriad} task {\tt moment} with {\tt mom=-3}. On this basis, we determined the approaching side of the galaxy to correspond to the SE side.

\begin{figure}
   \centering
   \includegraphics[width=0.95\columnwidth]{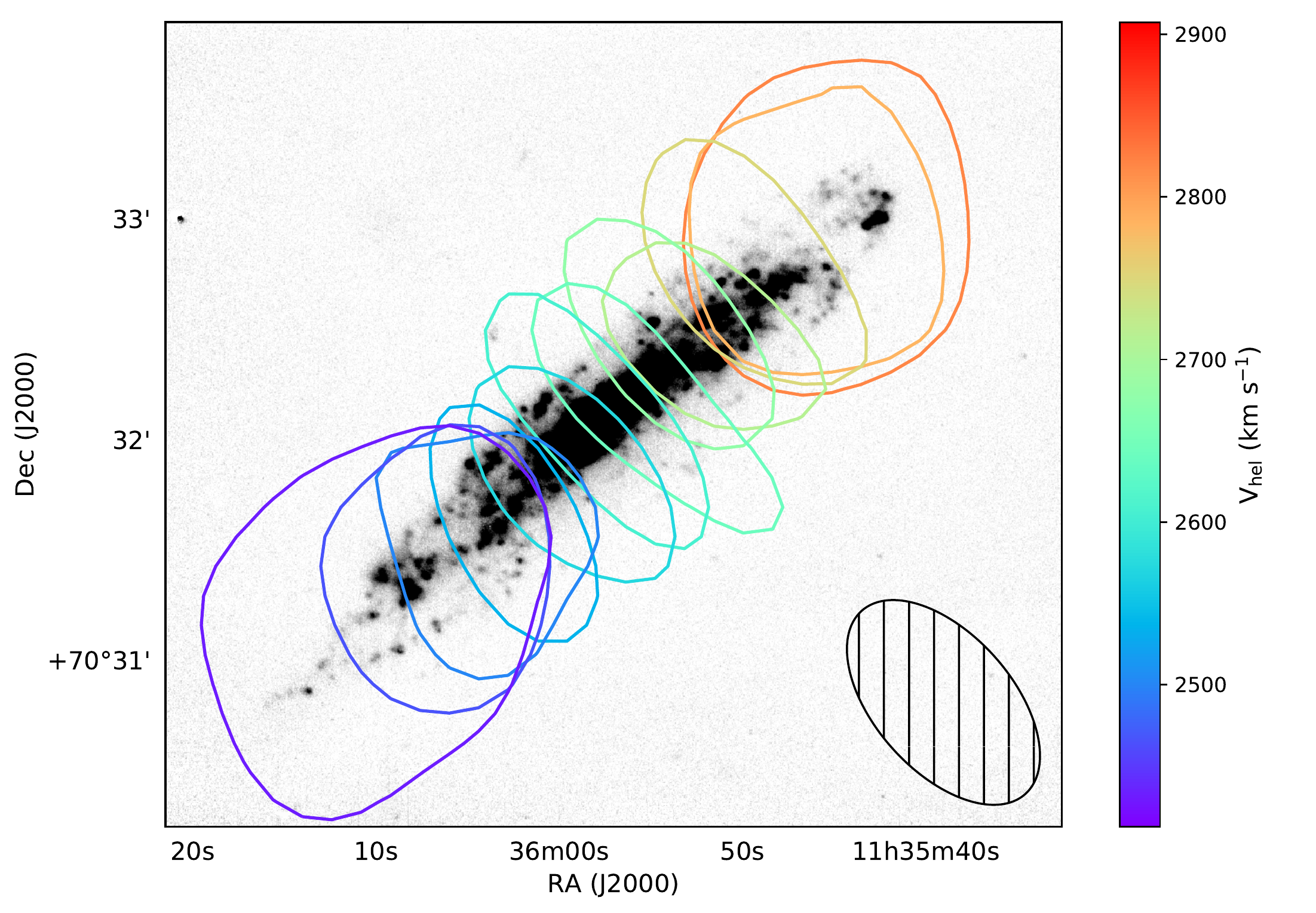}
      \caption{HI contour map of NGC3735 on top of its grey-scale H$\alpha$ map \citep{vargas+2019}. The contours are plotted at the 3 mJy/beam level, from every even channel in the HI cube, such that purple-blue colours correspond to the approaching side and orange-red colours correspond to the receding side. The velocity range is about 2380-3000 km/s. The synthesized beam is shown in the bottom right, and is 66.9x37.4 arcsec, with PA=41.8 deg.}
         \label{n3735HI}
\end{figure}

\end{appendix}

\end{document}